\documentclass[a4paper,11pt]{article}
\pdfoutput=1 

\usepackage{jheppub} 
\usepackage[T1]{fontenc} 
\usepackage{lmodern}
\usepackage{tikz}
\usepackage{subcaption}
\usepackage{amsmath,amssymb,xfrac,framed,verbatim,amsthm}
\usepackage{mathrsfs,accents,bbm,bbold} 
\usepackage{empheq}

\usepackage{microtype, ctable}
\usepackage{simplewick}
\usepackage{slashed,mathtools} 
\usepackage{sectsty} 
\usepackage{graphicx,enumerate,bm} 
\usepackage{comment}

\usepackage{young}
\usepackage[vcentermath]{youngtab}
\usepackage{ytableau}

\newcommand{\mc}[1]{\mathcal{#1}}
\newcommand{\mf}[1]{\mathfrak{#1}}
\newcommand{\msf}[1]{\mathsf{#1}}
\newcommand{\wt}[1]{\widetilde{#1}}

\newcommand{\mbb}[1]{\mathbb{#1}}
\newcommand{\tr}{\text{Tr}}
\newcommand{\ut}{\textbf{t}}
\newcommand{\uk}{\textbf{k}}

\newcommand{\tl}{\tilde}
\newcommand{\sgn}{\text{sgn}}
\renewcommand{\i}{\text{i}}
\newcommand{\te}{\text{e}}
\newcommand{\wh}[1]{\widehat{#1}}

\def\cC{\mathcal{C}}

\def\cS{\mathcal{S}}

\renewcommand{\a}{\alpha}

\newcommand{\e}{\epsilon}

\newcommand{\diag}{\text{diag}}
\newcommand{\ul}[1]{\underline{#1}}
\newcommand{\vgap}{\vphantom{\widehat{A}}}

\preprint{TIFR/TH/21-13} \title{\boldmath The Hilbert Space of large $N$ Chern-Simons matter theories}

\author[a,1]{Shiraz Minwalla,\note{minwalla@theory.tifr.res.in}}
\author[a,2]{Amiya Mishra,\note{amiya.mishra@theory.tifr.res.in}}
\author[a,c,3]{Naveen Prabhakar,\note{naveen.prabhakar@icts.res.in}}
\author[b,4]{Tarun Sharma\note{tks@physics.iitd.ac.in}}

\affiliation[a]{Department of Theoretical Physics, \\ Tata Institute
  of Fundamental Research, Homi Bhabha Rd, Mumbai 400005, India}
\affiliation[b]{Department of Physics, \\
  Indian Institute of Technology Delhi, Hauz Khas, New Delhi 110016,
  India} \affiliation[c]{International Centre for Theoretical
  Sciences,\\ Shivakote, Hesaraghatta Hobli, Bengaluru 560089, India}

\abstract{We demonstrate that the known expressions for the thermal
  partition function of large $N$ Chern-Simons matter theories admit a
  simple Hilbert space interpretation as the partition function of an
  associated ungauged large $N$ matter theory with one additional
  condition: the Fock space of this associated theory is projected
  down to the subspace of its \emph{quantum} singlets i.e.~singlets
  under the Gauss law for Chern-Simons gauge theory. Via the
  Chern-Simons / WZW correspondence, the space of quantum singlets are
  equivalent to the space of WZW conformal blocks. One step in our
  demonstration involves recasting the Verlinde formula for the
  dimension of the space of conformal blocks in $SU(N)_k$ and
  $U(N)_{k,k'}$ WZW theories into a simple and physically transparent
  form, which we also rederive by evaluating the partition function
  and superconformal index of pure Chern-Simons theory in the presence
  of Wilson lines. A particular consequence of the projection of the
  Fock space of Chern-Simons matter theories to quantum (or WZW)
  singlets is the `Bosonic Exclusion Principle': the number of bosons
  occupying any single particle state is bounded above by the
  Chern-Simons level. The quantum singlet condition (unlike its
  Yang-Mills Gauss Law counterpart) has a nontrivial impact on
  thermodynamics even in the infinite volume limit. In this limit the
  projected Fock space partition function reduces to a product of
  partition functions, one for each single particle state. These
  single particle state partition functions are $q$-deformations of
  their free boson and free fermion counterparts and interpolate
  between these two special cases. We also propose a formula for the
  large $N$ partition function that is valid for arbitrary finite
  volume of the spatial $S^2$ and not only at large volume.}

\begin{document}
\maketitle

{}\section{Introduction}

$SU(N)_k$ Chern-Simons theories (and their $U(N)$ cousins) coupled to
fundamental matter turn out to be effectively solvable in the 't Hooft
large $N$ limit
\begin{equation}\label{lartl}
  N \to \infty\ ,\quad k \to \infty\ ,\quad\text{with}\quad\lambda= \frac{N}{\kappa}\quad \text{fixed}\ ,\quad\text{with}\quad\kappa = k + \sgn(k) N \ . 
\end{equation} 
The study of these theories in the 't Hooft limit has led to several
qualitative insights including the discovery of Bose-Fermi duality in
these theories \cite{Sezgin:2002rt, Klebanov:2002ja, Giombi:2009wh,
  Benini:2011mf, Giombi:2011kc, Aharony:2011jz, Maldacena:2011jn,
  Maldacena:2012sf, Chang:2012kt, Jain:2012qi, Aharony:2012nh,
  Yokoyama:2012fa, GurAri:2012is, Aharony:2012ns, Jain:2013py,
  Takimi:2013zca, Jain:2013gza, Yokoyama:2013pxa, Bardeen:2014paa,
  Jain:2014nza, Bardeen:2014qua, Gurucharan:2014cva, Dandekar:2014era,
  Frishman:2014cma, Moshe:2014bja, Aharony:2015pla, Inbasekar:2015tsa,
  Bedhotiya:2015uga, Gur-Ari:2015pca, Minwalla:2015sca,
  Radicevic:2015yla, Geracie:2015drf, Aharony:2015mjs,
  Yokoyama:2016sbx, Gur-Ari:2016xff, Karch:2016sxi, Murugan:2016zal,
  Seiberg:2016gmd, Giombi:2016ejx, Hsin:2016blu, Radicevic:2016wqn,
  Karch:2016aux, Giombi:2016zwa, Wadia:2016zpd, Aharony:2016jvv,
  Giombi:2017rhm, Benini:2017dus, Sezgin:2017jgm, Nosaka:2017ohr,
  Komargodski:2017keh, Giombi:2017txg, Gaiotto:2017tne,
  Jensen:2017dso, Jensen:2017xbs, Gomis:2017ixy, Inbasekar:2017ieo,
  Inbasekar:2017sqp, Cordova:2017vab, Charan:2017jyc, Benini:2017aed,
  Aitken:2017nfd, Jensen:2017bjo, Chattopadhyay:2018wkp,
  Turiaci:2018nua, Choudhury:2018iwf, Karch:2018mer, Aharony:2018npf,
  Yacoby:2018yvy, Aitken:2018cvh, Aharony:2018pjn, Dey:2018ykx,
  Chattopadhyay:2019lpr, Dey:2019ihe, Halder:2019foo, Aharony:2019mbc,
  Li:2019twz, Jain:2019fja, Inbasekar:2019wdw, Inbasekar:2019azv,
  Jensen:2019mga, Kalloor:2019xjb, Ghosh:2019sqf, Inbasekar:2020hla,
  Jain:2020rmw, Minwalla:2020ysu, Jain:2020puw, Mishra:2020wos,
  Jain:2021wyn, Jain:2021vrv, Gandhi:2021gwn, toappear1}.

\subsection{The question posed in this paper}\label{quest}

One of the earliest explicit large $N$ computations in Chern-Simons
matter theories was that of the thermal partition function on an $S^2$
whose volume $\mc{V}_{2}$ is taken to scale like $N$ with all other
quantities held fixed (e.g.~masses, the temperature $T$, chemical
potential $\mu$). The large $N$ computations of the thermal partition
function (see \cite{Giombi:2011kc, Jain:2012qi, Yokoyama:2012fa,
  Aharony:2012ns, Jain:2013py, Yokoyama:2013pxa, Choudhury:2018iwf,
  Dey:2018ykx, Dey:2019ihe, Halder:2019foo, Minwalla:2020ysu} ) was
accomplished by evaluating the (chemical potential twisted) Euclidean
partition function $\mc{Z}$ of the relevant theories on
$S^2 \times S^1$. The purpose of this paper is to give a Hilbert space
interpretation of these known results, i.e.~to construct a Hilbert
space ${\cal H}$ and a simple effective Hamiltonian $H$ (and conserved
charge $Q$) so that the previously obtained path integral expression
for $\mc{Z}$ can be written as
\begin{equation}\label{traceoverhspace} 
\mc{Z} = {\rm Tr}_{{\cal H}}\ \te^{-\beta \left(H -\mu Q \right) } \ . 
\end{equation} 
Restated, the aim of this paper is to construct an effective Hilbert
space and Hamiltonian that reproduces the known thermodynamics of
large $N$ Chern-Simons matter theories.

\subsection{A simpler analogous question and its answer} \label{YMcase}

To explain the nature of the question posed in this paper (together
with elements of its answer) we briefly recall an analogous question
in a more familiar and simpler setting. Consider the $S^2\times S^1$
partition function of an $SU(N)$ Yang-Mills theory in the weak
coupling limit $g^2_{YM} \to 0$. In this (almost) free problem it is
easy to directly evaluate the $S^2 \times S^1$ path integral. This
problem was addressed by the authors of \cite{Sundborg:1999ue,
  Aharony:2003sx} who demonstrated that the final answer to this
partition function takes the form of a simple integral over a single
$N\times N$ unitary matrix $U$:
\begin{equation}\label{sumint} 
  \mc{I}_{\rm cl}= \int dU~ {\rm Tr}_{{\cal H}_{\rm Fock}} \left( \widehat{U} \te^{-\beta H} \right)  \ . 
\end{equation} 
$U$ in \eqref{sumint} is the zero mode (on $S^2$) of the holonomy of
the gauge field around the time circle. The integrand on the RHS of
\eqref{sumint} is the trace over the free Fock Space
$\mathcal{H}_{\rm Fock}$ which is obtained from zero coupling
Yang-Mills theory on $S^2$ {\it without} imposing the Gauss
law\footnote{More explicitly, the Hilbert space
  $\mathcal{H}_{\rm Fock}$ is a free Fock space over the one-particle
  Hilbert space, i.e. over the space of solutions of the linearized
  Yang-Mills equations on $S^2 \times S^1$.} of $\te^{-\beta {H}}$
twisted by $\widehat{U}$, the operator on $\mc{H}_{\rm Fock}$
corresponding to the holonomy $U$, and ${H}$ is the Hamiltonian on
this Fock space\footnote{The subscript `cl' in $\mc{I}_{\rm cl}$
	signifies that we are looking at a `classical' situation that arises
	in the limit $k \to \infty$ of the Chern-Simons matter theories that
	we study in this paper, where $k$ is the level of the Chern-Simons
	theory. The meaning of this notational point will become clearer in later sections (say, e.g. in subsection \ref{mainres}) when we discuss the corresponding claim in the finite $k$ Chern-Simons coupled matter theories.}.

The analog of the question posed in the previous subsection is the
following: can we identify a Hilbert Space ${\cal H}_{\rm cl}$ and a
Hamiltonian ${ H}_0$ such that
\begin{equation}\label{intou}
  {\rm Tr}_{{\cal H}_{\rm cl}} \left(
    \te^{-\beta { H}_0} \right) = \int dU ~ {\rm Tr}_{{\cal H}_{\rm Fock}}
  \left( \hat{U} \te^{-\beta H} \right) \ .
\end{equation}
In this simple context the answer to this question is both well known
(see \cite{Sundborg:1999ue, Aharony:2003sx}) and easy to guess. The
integral over $U$ in \eqref{sumint} simply projects the Fock space
$\mathcal{H}_{\rm Fock}$ onto its $SU(N)$ invariant sector
\cite{Sundborg:1999ue, Aharony:2003sx} (see Appendix \ref{classcount}
for a description of the formula for the number of singlets). It
follows that the Hilbert space ${\cal H}_{\rm cl}$ must be identified
with this $SU(N)$ invariant sector, and $H_0$ is the restriction of
the free Fock space Hamiltonian ${H}$ down to the $SU(N)$
singlet sector\footnote{Even though ${H}$ is originally
  defined as a Hamiltonian on $\mathcal{H}_{\rm Fock}$, as
  $H$ is gauge invariant, it commutes with the projector
  onto singlets and its restriction to the space ${\cal H}_{\rm cl}$
  is unambiguous and Hermitian.}.
 
We emphasize that while the Hilbert Space ${\cal H}_{\rm cl}$ is the
projection of a Fock space, it is not by itself a Fock space. The
projector onto singlets is an - indeed the only - `interaction' in the
system\footnote{This seemingly innocuous projection can have a
  dramatic impact on the value of the partition function e.g.~the
  large $N$ limit. In the large $N$ limit, the projection is
  responsible for the partition function \eqref{intou} to undergo
  phase transitions as a function of temperature, from a low
  temperature `glueball' phase to a high temperature gluonic phase.}.

\subsection{The Hilbert space of a Chern-Simons matter theory: an
  example} \label{fici}

In this paper we address a question similar to that posed in Section
\ref{YMcase}, but in the more complicated context of Chern-Simons
matter theories. In order to explain our answer in an intelligible
manner, in this subsection we outline the question in some technical
detail and work towards obtaining its answer in the context of an
example viz.~the regular boson theory in its unHiggsed phase.

\subsubsection{The theory}

The `regular boson' theory is a theory of $N$ complex bosons in three
spacetime dimensions whose (power counting renormalizable)
interactions preserve $U(N)$ invariance.  The $U(N)$ global symmetry
of this theory (or an $SU(N)$ subgroup thereof, depending on details)
is then gauged, and the self-interactions of the gauge field are
governed by a Chern-Simons action. For concreteness, we work with an
$SU(N)_k$ Chern-Simons gauge theory.

Working in the dimensional regularization scheme, the Euclidean action
for this theory is given by
\begin{equation}\label{eucact}
  \mc{S}[A,\phi] = \mc{S}_{\rm CS}[A] + 
  \int d^3x \left(  \overline{D_\mu \phi} D_\mu \phi 
    + V_{{\rm cl}}(\phi) \right)\ ,
\end{equation} 
where $D_\mu$ is the gauge covariant derivative, $\mc{S}_{\rm CS}[A]$
is the Chern-Simons action for the gauge field and the potential
$V_{{\rm cl}}(\phi)$ is given by
\begin{equation}\label{vcl}
V_{\rm cl}(\phi) = m_B^2 \bar{\phi}\phi + \frac{4\pi b_4}{\kappa_B}(\bar{\phi}\phi)^2+\frac{(2\pi)^2}{\kappa_B^2}(x_6^B+ 1) (\bar{\phi}\phi)^3\ .
\end{equation} 

\subsubsection{A theory of interacting bosons}
 
As Chern-Simons gauge fields have no propagating degrees of freedom in
the bulk, the regular boson theory on $S^2$ may be thought of as a
theory of matter fields with interactions induced by integrating out
the gauge fields\footnote{Indeed, this philosophy was quantitatively
  implemented in \cite{Giombi:2011kc} - and several subsequent papers
  - to initiate the program of solving Chern-Simons matter theories in
  the 't Hooft large $N$ limit.}. In particular, integrating out the
gauge fields results in a renormalization of the classical effective
potential of the theory.

In \cite{Dey:2018ykx}, the exact large $N$ quantum effective potential
for the operator ${\bar \phi} \phi$ was computed. This
renormalized potential in the unHiggsed phase can be read off\footnote{The exact quantum effective potential given in \cite{Dey:2018ykx} reads
  \begin{equation}
    V({\bar \phi} \phi)= m_B^2 \bar{\phi}\phi + \frac{4\pi b_4}{\kappa_B}(\bar{\phi}\phi)^2+\frac{(2\pi)^2}{\kappa_B^2}\left(x_6^B+ \tfrac{4}{3}(1-\lambda_B^2) \right) (\bar{\phi}\phi)^3\ ,\nonumber
\end{equation}
The above formula and \eqref{exactqepgb} differ in the coefficient of
$({\bar \phi} \phi)^3$ by a term $ -\tfrac{4\lambda_B^2}{3}$ where
$\lambda_B = N_B / \kappa_B$ is the 't Hooft coupling. This difference
is due to the contribution from the one-loop determinant of the
bosons. Specifically, this piece originates from the term proportional
to $c_B^3$ on the third line of \eqref{RBoffshellfe}, and is part of
the bosonic one-loop determinant; since this part will be accounted
for separately when we integrate over the bosons, we do not include
this contribution in \eqref{exactqepgb}.}  from the result of
\cite{Dey:2018ykx}:
\begin{equation}\label{exactqepgb}
V({\bar \phi} \phi)= m_B^2 \bar{\phi}\phi + \frac{4\pi b_4}{\kappa_B}(\bar{\phi}\phi)^2+\frac{(2\pi)^2}{\kappa_B^2}(x_6^B+ \tfrac{4}{3}) (\bar{\phi}\phi)^3\ .
\end{equation}
Consequently, the large $N_B$ regular boson theory may be viewed as a
theory of $N_B$ complex bosons, interacting via the potential
\eqref{exactqepgb}, plus other (as yet unspecified) nonlocal
interactions that have their origin in gauge boson exchange.

\subsubsection{$S^2$ thermal partition function ignoring the nonlocal interactions}

As a first approximation to the partition function of our theory, we
simply ignore all nonlocal interactions and compute the partition
function of the scalar theory interacting via the potential
\eqref{exactqepgb}, i.e. of the theory
\begin{align}\label{rblag}
&\mc{S}[\phi] =  \int d^3 x\Bigg( \partial_\mu \bar{\phi}\partial^\mu \phi + m_B^2 \bar{\phi}\phi + \frac{4\pi b_4}{\kappa_B}(\bar{\phi}\phi)^2+\frac{(2\pi)^2}{\kappa_B^2}(x_6^B+ \tfrac{4}{3}) (\bar{\phi}\phi)^3\Bigg)\ .
\end{align}
The computation is a simple exercise. We introduce two new Lagrange multiplier
fields $c_B^2$ and $\sigma_B$ and rewrite \eqref{rblag} as 
\begin{align} \label{lagare}
  &\mc{S}[c_B,\sigma_B, \phi] \nonumber\\
  &\quad=\int d^3 x \left( \partial_\mu \bar{\phi} \partial^\mu \phi
     +c_B^2 \bar{\phi} \phi + \frac{N_B}{2 \pi} \left((m_B^2 -c_B^2)
     \sigma_B + 2 \lambda_B b_4 \sigma_B^2
     + \lambda_B^2(x_6^B+ \tfrac{4}{3}) \sigma_B^3\right)  \right)\ , \nonumber\\
  &\quad=\int d^3 x \left( \partial_\mu \bar{\phi} \partial^\mu \phi
     +c_B^2 \bar{\phi} \phi +F_{\rm RB,int}(c_B, \sigma_B) \right)\ ,
\end{align} 
where $\lambda_B = N_B / \kappa_B$ and we have defined the quantity
$F_{\rm RB,int}$ by the last equality. Note that the term
$F_{\rm RB,int}(c_B, \sigma_B)$ in the last line of \eqref{lagare} is
a simple rewriting of the contact terms in \eqref{rblag} in terms of
the new variable $\sigma_B$.\footnote{In more detail, the equivalence
  of \eqref{lagare} and \eqref{rblag} may be seen as follows. Varying
  w.r.t.~$c_B^2$ in \eqref{lagare} gives
  $\sigma_B = 2 \pi {\bar \phi} \phi/N_B$. Substituting this solution
  back in \eqref{lagare} yields \eqref{rblag}. The $\sigma_B$ equation
  of motion determines $c_B$ in terms of $\sigma_B$ in a complicated
  manner, but as the Lagrangian is now independent of $c_B$ (see
  \eqref{rblag}) this does not matter.}

In the large $N$ limit, $c_B^2$ and $\sigma_B$ are frozen at their
saddle point values. Under the plausible assumption of translational
invariance of the thermal ensemble, these saddle point values are
constants. At fixed values of $c^2_B$ and $\sigma_B$, the scalar field
$\phi$ is a free boson of squared mass $c_B^2$. Integrating out the
field $\phi$ yields
\begin{equation}
\tr_{\mc{H}_{B,\rm Fock}}\left(\te^{-\beta (H - \mu Q)}\right)\ ,
\end{equation}
where $\mc{H}_{B,\rm Fock}$ is the Fock space of a free scalar field
of squared-mass $c_B^2$. It follows that the $S^2 \times S^1$
partition function of the theory governed by the action \eqref{lagare}
at large $N$ is given by
\begin{equation}\label{ZRBungauged}
\mc{Z}^{\rm ungauged\ RB}_{S^2 \times S^1} = \text{Ext}_{\{\sigma_B,c_B\}}\left[ \te^{-\mc{V}_2 \beta F_{\rm RB,int}(c_B, \sigma_B)}\ \tr_{\mc{H}_{B,\rm Fock}} \left(\te^{-\beta(H-\mu Q)}\right)\right]\ .
\end{equation}
where $\text{Ext}_{\{\sigma_B,c_B\}}$ denotes extremizing the result
over $c_B$ and $\sigma_B$.

\subsubsection{Accounting for the Gauss law} 

Equation \eqref{ZRBungauged} yields the partition function of the
regular boson theory after ignoring the effect of gauge boson mediated
nonlocal interactions. However gauge theories on compact manifolds
always include at least one nonlocal interaction viz.~the Gauss
law. Inspired by Section \ref{YMcase}, we might attempt to  account for the Gauss
law by replacing the Fock space partition function in
\eqref{RBoffshellfe} by its projection onto the space of `classical'
$SU(N)$ singlets, i.e. by
\begin{equation}\label{ZRBgauss}
  \mc{Z}^{\rm Gauss\ RB}_{S^2 \times S^1} = \text{Ext}_{\{\sigma_B,c_B\}}\left[ \te^{-\mc{V}_2 \beta F_{\rm RB,int}(c_B, \sigma_B)}\ \int dU ~\tr_{\mc{H}_{B,\rm Fock}} \left(\widehat{U}\te^{-\beta(H-\mu Q)}\right)\right]\ .
\end{equation}
where $U$ is a constant $SU(N)$ matrix and $\widehat{U}$ is the
operator corresponding to $U$ that acts on $\mc{H}_{B,\rm Fock}$, as
in Section \ref{YMcase}. However, \eqref{ZRBgauss} accounts for the
nonlocal gauge-mediated interactions of the Chern-Simons matter
theories only in a crude approximation, and so is not the exact answer
for the thermal partition function of the theory in the large $N_B$
limit.

\subsubsection{The actual partition function} 

Fortunately, as we have mentioned before, the exact large $N_B$ result
of the thermal partition function of this theory is known. While this
exact answer (presented in \eqref{RBoffshellfe}) differs from that
presented in \eqref{ZRBgauss}, it has many similarities to that naive
guess. Indeed in this paper we demonstrate that this exact answer,
\eqref{RBoffshellfe}, can be rewritten in the form
\begin{equation}\label{ZRBexact}
  \mc{Z}^{\rm RB}_{S^2 \times S^1} = \text{Ext}_{\{\sigma_B,c_B\}}\left[ \te^{-\mc{V}_2 \beta F_{\rm RB,int}(c_B, \sigma_B)}\ \wt{\mc{I}}_{B,k}(c_B)\right]\ .
\end{equation}
where $\wt{\mc{I}}_{B,k}(c_B)$ is the projection of the Fock space of
a free boson of squared mass $c_B^2$ down to its so-called
\emph{quantum} singlet sector that is precipitated by integrating out
the Chern-Simons gauge field (we describe the quantum singlet sector
briefly in Section \ref{mainres} below and describe it in detail in
the subsequent sections).

In other words, the crude guess \eqref{ZRBgauss} was not so far off
the mark. The exact answer takes the same form as \eqref{ZRBgauss}
with the single modification that the projector onto $SU(N)$ singlets
in the \eqref{ZRBgauss} is replaced by the projector onto $SU(N)_k$
\emph{quantum} singlets in the actual answer \eqref{ZRBexact}.

\subsection{The central result of this paper}\label{mainres}

The results outlined in the context of one example in the previous
subsection generalize to all studied Chern-Simons matter theories
coupled to vector-like matter in the large $N$ limit. For each of
these theories we demonstrate that the known expression for the
$S^2 \times S^1$ partition function can be re-expressed in the form
\begin{equation}\label{traceexp}
  {\rm Tr}_{{\cal H}_{k}} \left(\te^{-\beta \left( {H-\mu Q} \right) } \right)\ ,
\end{equation}
where 
\begin{itemize}
\item[(1)] The Hilbert space ${\cal H}_{k}$ in \eqref{traceexp} is
  obtained by projecting a Fock space $\mathcal{H}_{\rm Fock}$ onto
  the space of singlets under a modified Gauss law that arises from
  the Chern-Simons action which we refer to as the \emph{quantum}
  singlet condition. This projection is implemented in the quantity
  $\wt{\mc{I}}_{B,k}(c_B)$ in \eqref{ZRBexact} in the example of
  Section \ref{fici}.

  The Chern-Simons Gauss law can be applied effectively to the Fock
  space by appealing to the well-known correspondence between
  $SU(N)_k$ Chern-Simons theory and the $SU(N)_k$ WZW model. In the
  WZW model, the number of singlets is given by the number of times
  the identity representation appears in the fusion of the
  representations that correspond to the states in the Fock
  space. This is also the dimension of the space of conformal blocks
  involving the representations that correspond to Fock space
  states. We sometimes use the alternate term `WZW singlets' for
  quantum singlets to emphasize this connection.

  Using the correspondence between the $SU(N)_k$ WZW model and the
  quantum group $SU(N,q)$ with $q = \te^{2\pi\i/\kappa}$, the WZW
  singlets are also equivalent to the invariants in the tensor product
  of representations of the quantum group that correspond to the
  states in the Fock space. This also lends substance to the term
  `quantum' singlet in our description above.

\item[(2)] $H$ in \eqref{traceexp} is the free Hamiltonian on the Fock
  space $\mathcal{H}_{\rm Fock}$, corrected by mean field or
  forward-scattering-type four and six particle interactions. The
  precise form of these energy renormalizations depends on the details
  of the contact interactions in the theories under study in a simple
  way and explicitly known manner (see Section \ref{energy} for
  details). In the example discussed in Section \ref{fici}, these
  interactions are accounted for by dressing $\wt{\mc{I}}_{B,k}(c_B)$
  with $\te^{-\mc{V}_2 \beta F_{\rm RB,int}(c_B, \sigma_B)}$ and
  extremizing over $c_B$ and $\sigma_B$.
\end{itemize}
As may be clear from the example of Section \ref{fici}, the energy
renormalizations described in (2) above are familiar - they are
identical in structure to those that occur in much more familiar
ungauged large $N$ theories like the large $N$ Wilson-Fisher theory
(see eq.~\eqref{ZRBungauged}). All qualitatively new features in
Chern-Simons matter theories have their origin in the interaction of
matter with gauge fields. Integrating these out induces both local as
well as non-local interactions between the matter fields. The local
interactions effectively renormalize the matter Lagrangian and are
easily accounted for\footnote{In the example of Section \ref{fici}
  this effect resulted in replacement of the classical potential
  \eqref{vcl} with the renormalized potential
  \eqref{exactqepgb})}. The effect of the nonlocal interactions is
more interesting.  At intermediate stages of all computations these
interaction are complicated, at least in the lightcone gauge of
\cite{Giombi:2011kc}. Remarkably, however, the entire effect of these
messy looking non-local interactions on the final physical observable
- namely the large $N$ $S^2 \times S^1$ partition function - is
strikingly elegant and simple. These interactions simply enforce the
\emph{quantum} singlet condition on an effective Fock space of an
otherwise completely local theory - see (1) above.

The fact that the Chern-Simons path integral enforces the quantum
singlet condition is not un-intuitive; in the world line
representation of quantum field theory, any given configuration of
matter trajectories is a particular configuration of Wilson lines in
Chern-Simons theories. However, away from large $N$, would expect the
quantum singlet condition to be only part of the story. Particles
should also interact with each other via (gauge mediated) monodromies
as one particle goes around another. Apparently the effects of these
monodromies is subleading in the partition function at large $N$.
\footnote{We suspect that monodromy effects do renormalize the
  energies of some states at order $N^0$, but the number of states for
  which this happens is a small fraction of the total number in the
  large $N$ limit. To see roughly how this might happen,  recall that the two particle state consisting of a fundamental and an antifundamental has an anyonic phase of order unity in the singlet channel, but this channel makes up only a small fraction of the total number of states ($1$ out of $N^2$). On the other hand the in the adjoint channel, which makes up most of the states of the two particle system ($N^2-1$ out of $N^2$ states), the anyonic phase is parametrically small at large $N$. We thank D. Gaiotto, S. Giombi and D. Tong for
  related discussions.}

\subsection{Three interesting consequences}

Perhaps the most striking consequence of the projection of the Fock
space down to the space of quantum singlets is the Bosonic Exclusion
Principle; no single particle state can be occupied by more than $k$
particles ($k$ the level of the $SU(N)$ Chern-Simons theory). The
bosonic exclusion principle follows from the projection to WZW
singlets because the symmetrization of $k$ or more fundamentals yields
a non-integrable representation, and it was demonstrated long ago by
Gepner and Witten that conformal blocks involving a mix of integrable
and non-integrable representations all vanish
\cite{Gepner:1986wi}. This striking principle (first conjectured in
\cite{Minwalla:2020ysu}) is the image of the Fermi exclusion principle
under Bose-Fermi duality, and may turn out to be one aspect of a
richer structure (see Section \ref{disc} for some discussion).  The
bosonic exclusion principle follows from an interesting interplay
between Bose statistics and the quantum singlet condition. As of yet,
we have an incomplete understanding of its derivation directly from
the Chern-Simons path integral (see Sections \ref{nonint} and
\ref{disc} for some discussion).

In Section \ref{hccb} we show that the Verlinde fusion algebra of WZW
theories has a striking universality in a limit in which the number of
fused operators becomes parametrically large. One consequence is that
the Fock space partition function, projected down to quantum singlets,
takes a very simple, effectively free form presented in Section
\ref{winf}. In fact, in this limit, the full partition function
reduces to a product of single state partition functions. The formulae
for these single state partition functions are $q$-deformations of
their free bosonic or free fermionic counterparts, see \eqref{fbinlvo}. They map to each other under duality; moreover there is a sense in which each of them,
in turn, also interpolates between the formulae for Bose-Einstein and
Fermi-Dirac statistics (see subsection \ref{interp} ) as the 't Hooft coupling varies between zero
and one.

Standard treatments of Fermi liquids introduce an entropy functional
as a functional of the occupation numbers of the single particle
spectrum. Occupation numbers (and other thermodynamic quantities) may
then be computed by extremizing this functional separately w.r.t.~each
occupation number at fixed total energy. The dependence of this
entropy functional on occupation numbers is dictated by Fermi
statistics, and takes the simple universal form \eqref{newent}. This
method is easily generalized to the study of bosons (see e.g.~equation
\eqref{entfrbos}). In Section \ref{entropy} of this paper we follow
\cite{Geracie:2015drf} to formulate an analogous entropy functional as
a function of occupation numbers for the new emergent (effectively
free) statistics described in the previous paragraph. We also
demonstrate that the extremization of this entropy functional
reproduces the partition functions of Chern-Simons theories coupled to
fundamental matter. Unfortunately, while our functionals are well
defined, they are not completely explicit since they are defined in
terms of the solutions to an algebraic equation for which we have not
yet found a closed form solution.  We hope this defect will be
remedied in future work.

Apart from these physically interesting results, as part of the
analysis of this paper, in Section \ref{counting} we have also recast
the Verlinde formula for the dimension of the space of conformal
blocks into a simple and completely explicit form in the case of
$SU(N)_k$ and $U(N)_{k,k'}$ theories (see Section \ref{verlinde} for a
listing of results). We have also re-derived these these formulae (1)
by evaluating a path integral following \cite{Blau:1993tv}, and also
(2) by evaluating the same path integral using the methods of
supersymmetric localization. The three methods, each of which has its
own advantages, all yield exactly the same answer. Finally, we have
also explained the interplay between the results obtained in this
section and level-rank duality.

\subsection{Outline of the paper} 

The outline of the rest of this long paper is as follows. In Section
\ref{dsr}, we present a brief review of the large $N$ thermal
partition function of Chern-Simons matter theories and spell out the
logic of the main computation we do in this paper. In Section
\ref{hccb} we perform the main computation of this paper which is to
massage the large $N$ expression for the thermal partition function
into a trace over the Hilbert space of quantum singlets. We also
discuss the validity of this Hilbert space interpretation beyond large
$N$ and large volume in this section. In Section \ref{winf} we
specialize the results of Section \ref{hccb} to the infinite volume
limit and exhibit the simplification of this limit described in the
previous subsection.  In Section \ref{energy} we review the fact that
the Hilbert space of ungauged large $N$ matter theories is a Fock
Space with precisely defined mean field or forward scattering type
energy renormalizations, and explain how this fact (plus the reduction
to quantum singlets) may be used to reproduce the partition function
of Chern-Simons matter theories.  In Section \ref{entropy}, we write
down an (unfortunately, as of yet, not completely explicit) expression for an `entropy
functional' which captures the thermodynamics of Chern-Simons matter
theories. Section \ref{counting} is the longest and most technical
section in this paper. In this section we present and analyse explicit
formulae for the number of quantum singlets described in the previous
subsection. Finally, in Section \ref{disc}, we conclude the paper with
a discussion of unresolved puzzles and future directions. In the
appendices to this paper we supply technical details related to the
analysis in the main text.

\section{Setting up the problem} \label{dsr}

\subsection{Notation and terminology for Chern-Simons
  levels} \label{ntcs} In this paper we study Chern-Simons theories
coupled to matter in the fundamental representation of the gauge group
which we will always take to be either $SU(N)_k$ or $U(N)_{k, k'}$. As
in Appendix A of \cite{Minwalla:2020ysu} the levels $k$ and $k'$ have
the following meaning: they are the levels of the pure Chern-Simons
theory obtained by massing up all matter fields with the convention
that fermion masses have the same sign as $k$ and $k'$ and then
integrating out the matter fields\footnote{While the notation we use
  to label $SU(N)_k$ theories is standard, our notation $U(N)_{k, k'}$
  are labelled by two levels; $k$ is the level for the $SU(N)$ part of
  the gauge group (in a Yang-Mills regularization scheme), while $Nk'$
  is the level for the overall $U(1)$ part of the gauge group (working
  in a normalization in which every fundamental field carries overall
  $U(1)$ charge unity).See Appendix A of \cite{Minwalla:2020ysu} for a
  detailed explanation of this notation. The levels $k$ and $k'$ are
  taken to have the same sign.}. We use the notation
\begin{equation}\label{kpdef} 
  \kappa = {\rm sgn}(k)(|k|+N )\ ,\quad\text{and}\quad k' = \sgn(k)(|\kappa| + s  N)\ .
\end{equation}
As reviewed in Appendix A of \cite{Minwalla:2020ysu}, consistency
demands that $k$ and $s$ are integers. In this paper we will be
especially interested in two values of $s$: the case $s=0$
($k'=\kappa$) which we call the Type I $U(N)$ theory and the case
$s=-1$ ($k'=k$) which we call the Type II $U(N)$ theory.

\textbf{Note:} While the levels $k$, $k'$ and $\kappa$ could possibly
have either sign in the definitions above, we restrict them to be
positive in Sections \ref{verlinde}, \ref{lwcoo} and \ref{tln} to
avoid clutter of notation. The reader interested in the case of
negative $k$ and $k'$ can obtain the results for her theory by making
the replacement $k \rightarrow |k|$ and $k' \rightarrow |k'|$ in all
the formulae in the sections mentioned above.

\subsection{The thermal partition function of Chern-Simons matter
  theories}
\label{rtpf} 

In this paper, we are interested in the large $N$ 't Hooft limit which
is described as
\begin{equation}
  N \to \infty\ , \quad k \to \infty\ ,\quad \mc{V}_2 \to \infty\quad\text{with}\quad \lambda = \frac{N}{\kappa}\ ,\quad V_2 = \frac{\mc{V}_2}{N}\ ,\ \beta\ ,\ \mu\quad\text{fixed.}
\end{equation}
where $\mc{V}_2$ is the volume of the two dimensional space and
$\beta = 1/ T$ is the inverse temperature and $\mu$ is the chemical
potential. It was demonstrated in \cite{Aharony:2012ns, Jain:2013py}
that the final expression for the $S^2 \times S^1$ partition function
of large $N$ Chern-Simons matter theories is given by an expression of
the form
\begin{equation}\label{pfum} 
  \mathcal{Z}_{S^2\times S^1} = \int [dU]_{\rm CS}\ 
  \te^{-\mc{V}_{2} T^2 v[\rho]} \ ,
\end{equation}
(note the similarity with the Yang-Mills case \eqref{sumint}). The
unitary matrix $U$ is the constant mode of the gauge holonomy around
the thermal circle.  $\rho(\alpha)$ is the (effectively continuous)
`eigenvalues density function' of $U$ defined in terms of the
eigenvalues $\te^{\i \alpha_{j}}$, $j = 1,\ldots,N$, of $U$ by
\begin{equation}\label{rhdef} 
\rho(\alpha) = \lim_{N\to \infty}  \frac{1}{N} \sum_{j=1}^{N}  \delta(\alpha-\alpha_{j})\ .  
\end{equation}
The symbol $\int [dU]_{\rm CS}$ denotes an integral with the usual
Haar measure over unitary matrices subject to the constraint that we
integrate only over those matrices whose eigenvalue density functions
obey the constraint
\begin{equation}\label{inequality}
\rho(\alpha)  \leq \frac{1}{2\pi {|\lambda|}} \ . 
\end{equation} 

The quantity $v[\rho]$ in \eqref{pfum} encodes the dynamical details
of the theory in question (e.g.~the matter content and the details of
their interactions). The evaluation of $v[\rho]$ is a nontrivial
computational task which, quite remarkably, is solvable
\cite{Giombi:2011kc, Jain:2012qi, Yokoyama:2012fa, Aharony:2012ns,
  Jain:2013py, Yokoyama:2013pxa, Choudhury:2018iwf, Dey:2018ykx,
  Dey:2019ihe, Halder:2019foo,
  Minwalla:2020ysu}. In particular, the result for $v[\rho]$ for all four classes of
theories of interest to this paper was presented in \cite{Dey:2018ykx,
  Dey:2019ihe} in the following form. The authors of
\cite{Dey:2018ykx, Dey:2019ihe} presented an `off-shell' free energy
$F[\varphi_{\rm aux}; \rho]$. This free energy is a function of a few
auxiliary variables $\{\varphi_{\rm aux}\}$ in addition to the
holonomy distribution $\rho$. It was then demonstrated that the
extremization of this free energy w.r.t.~the variables
$\{\varphi_{\rm aux}\}$, at fixed $\rho$, yields
$v[\rho]$\footnote{Apart from simplifying all formulae, the use of
  this `off-shell' free energy allows for the incorporation of the
  different phases of the Chern-Simons matter theory under
  consideration in one unified expression. Upon extremization,
  different solutions of the extremization equations correspond to the
  different phases of the theory.}. In the large $N$ limit, in other
words, the thermal partition function $\mc{Z}_{S^2 \times S^1}$
equals\footnote{In going from the second to the third expression in
  \eqref{offshellZ} we have used the fact that an integration over
  $\varphi_{\rm aux}$ reduces, in the large $N$ limit, to an
  extremization of the integrand over $\varphi_{\rm aux}$.}
\begin{equation}\label{offshellZ}
  \mc{Z}_{S^2 \times S^1} =\int [d U]_{\rm CS} \left(  {\rm Ext}_{ \varphi_{\rm aux} } \te^{-\mc{V}_2\beta F[\varphi_{\rm aux}; \rho]} \right) \ =  \int [d U]_{\rm CS} \int d \varphi_{\rm aux}\ \te^{-\mc{V}_2\beta F[\varphi_{\rm aux}; \rho]}\ .
\end{equation}
In this paper, we will find it useful to interchange the order of
integration in the last expression in \eqref{offshellZ}; i.e.~to first
evaluate the integral over $U$ at fixed values of auxiliary variables
and then extremize the resultant expression over auxiliary
variables\footnote{Alternatively, the exchange of orders of
  integration can be justified as follows. Like the integral over
  $\varphi_{\rm aux}$, the integral over $U$ may also be evaluated in
  the saddle point approximation in the large $N$ limit. To evaluate
  \eqref{offshellZ}, we are therefore instructed to extremize the
  integrand over both $\varphi_{\rm aux}$ and $\rho(\alpha)$. As the
  process of extremizing w.r.t.~different variables commutes, the
  extremization over $U$ and over the auxiliary variables can be
  performed in any order.}. This strategy is useful because the
off-shell free energies for all theories of interest to this paper
have a simple, universal dependence on the matrix $U$ (equivalently,
the holonomy distribution $\rho$) in the various Chern-Simons matter
theories of interest. To explain this fact - and also to prepare the
ground for the subsequent analysis of this paper - we present the
previously obtained explicit (all-orders in $\lambda$) results for the
off-shell free energies in the rest of this subsection\footnote{The
  results for the off-shell free energy that we quote from
  \cite{Minwalla:2020ysu} are those for the bosonic `upper cap' and
  dual fermionic `lower gap' phases, listed in equations 3.34 and 3.35
  of \cite{Minwalla:2020ysu}. These formulae apply only for
  sufficiently large values of $\mc{V}_2 / N$. Nonetheless we suspect
  that the final Hamiltonian formula, \eqref{extmatrb}, presented in
  this paper applies to the partition function of our theory in every
  phase and even at finite values of the volume. See Section
  \ref{discoo} and \ref{disc} for some discussion of this conjecture,
  whose clear justification (or negation) we leave to further work.}.

\textbf{A few remarks on notation:} Throughout this paper we give the
parameters $N$, $k$, $\kappa$ and $\lambda$ an extra subscript $F$ or
$B$ depending on the whether the Chern-Simons theory is coupled to
fermionic or bosonic matter respectively. Also, in presenting our
results for the off-shell free energies, following the literature
\cite{Dey:2018ykx, Dey:2019ihe, Minwalla:2020ysu}, we quote the
results in terms of dimensionless parameters. As in the cited papers,
our notation is the following. Every dimensionful quantity has an
associated, hatted, non-dimensional analog obtained by scaling with
appropriate powers of temperature. So, for instance, ${\hat \mu}$, the
hatted version of the (mass dimension one) chemical potential $\mu$,
is defined by $\hat\mu = \mu / T$. See \cite{Dey:2018ykx, Dey:2019ihe,
  Minwalla:2020ysu} for more details of the notation.

\subsubsection{Fermionic theories} \label{ftpf}

In the case of the regular fermion (RF) theory (defined, e.g.~in
Eq. $2.1$ of \cite{Minwalla:2020ysu}), the off-shell free energy
\cite{Dey:2018ykx, Dey:2019ihe} depends on the auxiliary variables
${c}_F$ and $\tl\cC$ and is given by\footnote{The extremization of the
  expression \eqref{RFofe} over $\tl{\mc{C}}$ and ${c}_F$ determines
  both the actual value of the thermal mass $c_F$ (in terms of the UV
  parameters of the theory and the temperature and chemical potential)
  as well as the shift in energy away from free values.}:
\begin{align}\label{RFofe}
  &F_{\rm RF}[\tilde{\mathcal{C}},{c}_F;\rho_F]\nonumber\\
  &= \frac{N_F T^3}{2\pi} \bigg[ - \frac{8}{3}\lambda_{F}^2 \tl{\mc{C}}^3 - \tl{\mc{C}} \left( \hat{c}_{F}^2 - \big(2\lambda_F \tl{\mc{C}} +\hat{m}_F \big)^2 \right)  - 2\lambda_{F} \hat{m}_F \tl{\mc{C}}^2
    \nonumber  \\
  &\qquad + \frac{1}{3}\hat{c}_{F}^3 - \int_{\hat{c}_F}^{\infty}  d\hat{\e}\ \hat{\e}  \int_{-\pi}^{\pi}  d\alpha\, \rho_F(\alpha)\left(\log\big(1+\te^{-\hat{\e}-\hat{\mu}-\i \alpha }\big)+\log\big(1+\te^{-\hat{\e}+ \hat{\mu}+\i \alpha }\big)  \right)  \bigg]  \ .
\end{align}
Here, $\mu$ is the chemical potential, $m_F$ is the mass parameter
that appears in the Lagrangian, ${c}_F$ is a variable whose extremal
value is the thermal mass and the physical interpretation of
$\tl{\mc{C}}$ is as yet unclear.

In the case of the more general critical fermion (CF) theory (see
e.g.~Eq.~3.12 of \cite{Minwalla:2020ysu}), the off-shell free energy
is a function of the three auxiliary variables $c_F$, $\zeta_F$ and
$\tl\cC$:
\begin{align}\label{CFoff}
  &F_{\rm CF}[{\zeta}_F,\tilde{\cC},c_F;\rho_F]\nonumber\\
  & = \frac{N_F T^3}{2\pi} \bigg[ -\frac{8}{3}\lambda_{F}^2 \tl{\mc{C}}^3 - \tl{\mc{C}} \bigg( \hat{c}_{F}^2 - \Big(2\lambda_F \tl{\mc{C}} - \frac{4\pi \hat{\zeta}_F}{\kappa_F} \Big)^2 \bigg) + 2\lambda_{F} \tl{\mc{C}}^2 \Big(\frac{4\pi \hat{\zeta}_F}{\kappa_F} \Big) \nonumber  \\ 
  &\qquad\qquad\ \   +   \frac{\hat{y}_2^2}{2\lambda_F} \frac{4\pi \hat{\zeta}_F}{\kappa_F} - \frac{\hat{y}_4}{2\lambda_F} \Big(\frac{4\pi \hat{\zeta}_F}{\kappa_F} \Big)^2 +\frac{x_6^F}{8\lambda_F} \Big(\frac{4\pi \hat{\zeta}_F}{\kappa_F} \Big)^3 \nonumber   \\
  &\qquad\qquad\ \ +  \frac{1}{3}\hat{c}_{F}^3 - \int_{\hat{c}_F}^{\infty} d\hat{\e}\ \hat{\e} \int_{-\pi}^{\pi}d\alpha\ \rho_F(\alpha) \left(\log\big(1+\te^{-\hat{\e}-\hat{\mu}-\i\alpha }\big)+\log\big(1+\te^{-\hat{\e}+ \hat{\mu}+\i\alpha }\big)  \right)  \bigg]  \ .
\end{align}
Note that the last lines of the off-shell free energies of the RF and
CF theories \eqref{RFofe} and \eqref{CFoff} are identical; the
expression on these lines is simply the free energy at temperature $T$
and chemical potential $\mu$ of a system of $N_F$ fermions of mass
$ { c}_F$ whose boundary conditions around the thermal circle are
twisted by the background holonomy $U$. We find it convenient to give
a new notation to this free energy\footnote{The quantity
  $F_{\rm Fock}$ is denoted as $F_{\rm det}$ in
  \cite{Minwalla:2020ysu} signifying that it arises as a one-loop
  determinant from the quadratic action of the fermionic theory. We
  have included the term $\frac{1}{3}\hat{c}_F^3$ in
  $F_{F, {\rm Fock}}$ - rather than pushing it to the remaining part
  of the free energy since the free fermionic determinant evaluated in
  the dimensional regularization scheme includes this piece; see e.g.
  \cite{Minwalla:2020ysu}. Similar remarks apply to the free boson
  determinant presented below.}:
\begin{align}\label{ferdet}
  &F_{F,\rm Fock}[{c}_F;\rho_F]  =\nonumber\\
  &\quad\frac{N_F T^3}{2\pi}\bigg[ \frac{1}{3}\hat{c}_{F}^3 - \int_{\hat{c}_F}^{\infty} d\hat{\e}\ \hat{\e} \int_{-\pi}^{\pi}d\alpha\ \rho_F(\alpha) \left(\log\big(1+\te^{-\hat{\e}-\hat{\mu}-\i\alpha }\big)+\log\big(1+\te^{-\hat{\e}+ \hat{\mu}+\i\alpha }\big)  \right)  \bigg]\ .
\end{align}
The rest of the RHS of \eqref{RFofe} and \eqref{CFoff} (the first line
of the RHS of \eqref{RFofe} and the first two lines on the RHS of
\eqref{CFoff}) differ from each other. As we will explain in more
detail later, these differences reflect the differences between the
contact interactions between the fermions in these two theories.

\subsubsection{Bosonic theories} \label{btpf}
In the case of the critical boson (CB) theory the off-shell free
energy is a function of the auxiliary variables ${ c}_B$ and
$\tilde{\mc{S}}$ (see Eq. 3.35 of \cite{Minwalla:2020ysu} for details):
\begin{align}\label{boseoff}
&F_{\rm CB}[\tilde{\mc{S}},c_B;\rho_B]\nonumber\\
&=\frac{N_B T^3}{2\pi} \bigg[\frac{1}{2} \hat{c}_B^2\hat{m}_B^{\text{cri}} - \frac{4}{3}\lambda_B^2\left(\tilde{\mc{S}}-\tfrac{1}{2}\hat{m}_{B}^{\text{cri}} \right)^3  + 2|\lambda_B| \hat{c}_B\left(\tilde{\mc{S}}-\tfrac{1}{2}\hat{m}_{B}^{\text{cri}}\right)^2   \nonumber \\
&\qquad\qquad\ \
-\frac{1}{3}\hat{c}_B^3 + \int_{\hat{c}_B}^{\infty} d\hat{\e}\ \hat{\e} \int_{-\pi}^{\pi} d\alpha \ \rho_B(\alpha) \ \left(\log(1-\te^{-\hat{\e}+\hat{\mu}+\i \a})+\log(1-\te^{-\hat{\e}-\hat{\mu}-\i \alpha})\right)\nonumber\\
&\qquad\qquad\ \  - \Theta(|{ \mu}|-{ c}_B ) \frac{(|\hat{\mu}| - {\hat c}_B)^2(|\hat{\mu}| + 2 {\hat c}_B)}{6 |\lambda_B|} \bigg]\ .
\end{align}
Just like in the fermionic theories, the variable $c_B$ has as its
extremal value the thermal mass of the boson while the variable
$\tl\cS$ does not yet have a clear physical interpretation.

In the case of the more general regular boson (RB) theory (see
Eq. $2.5$ of \cite{Minwalla:2020ysu}), the off-shell free energy is a
function of the auxiliary variables $c_B$, $\sigma_B$ and
$\tl\cS$\footnote{The variable ${ \sigma_B}$ in the regular boson free
  energy \eqref{RBoffshellfe} has a simple physical
  interpretation. It is related to the expectation value of the
  lightest gauge-invariant operator, ${\bar \phi} \phi$, of the
  regular theory as
  ${\sigma}_{B} = \frac{2\pi }{N_B} \langle \bar{\phi}\phi \rangle$.}:
\begin{align}\label{RBoffshellfe}
  &F_{\rm RB}[{\sigma}_B,\tilde\cS,c_B;\rho_B]\nonumber\\
  & =\frac{N_B T^3}{2\pi} \bigg[-\hat{c}_B^2\hat{\sigma}_B  -\frac{4}{3}\lambda_B^2(\tilde{\cS}+\hat{\sigma}_{B})^3 + 2|\lambda_B| \hat{c}_B(\tilde{\cS}+\hat{\sigma}_B)^2 \nonumber \\
  &\qquad\qquad\ \ + \hat{m}_B^2\hat{\sigma}_B+2\lambda_B\hat{b}_4\hat{\sigma}_B^2+(x_6^B + \tfrac{4}{3}) \lambda_B^2 \hat{\sigma}_B^3\nonumber \\
  &\qquad\qquad \ \ 
    -\frac{1}{3}\hat{c}_B^3 + \int_{\hat{c}_B}^{\infty} \hat{\e} \ d\hat{\e} \int_{-\pi}^{\pi} d\alpha \ \rho_B(\alpha) \ \left(\log(1-\te^{-\hat{\e}+\hat{\mu}+\i \a})+\log(1-\te^{-\hat{\e}-\hat{\mu}-\i \alpha})\right) \nonumber\\
  &\qquad\qquad \ \
    - \Theta(|{ \mu}|-{ c}_B ) \ \frac{(|\hat{\mu}| - {\hat c}_B)^2(|\hat{\mu}| + 2 {\hat c}_B)}{6 |\lambda_B|} \ \bigg]\ .
\end{align}
\footnote{The equation of motion that results from varying w.r.t. ${\tilde \cS}$ in 
	\eqref{RBoffshellfe} is quite simple. One solution to this equation of motion is given  by setting ${\tilde \cS}=-{\tilde c}_B$. This solution puts us in the unHiggsed phase \cite{Dey:2018ykx}. After making this choice, the terms in the first and second line of \eqref{RBoffshellfe} reduce to the function $F_{\rm int}(c_B^2, \sigma)$ of 
	\eqref{ZRBexact}.} 
Once again the expression on the last two lines of the off-shell free
energies for the critical and regular boson theories \eqref{boseoff}
and \eqref{RBoffshellfe} are identical and equal to the partition
function of $N_B$ free bosons of mass $c_B$ twisted by the holonomy
$U$, but corrected with the strange looking term proportional to
$\Theta(|{ \mu}|-{ c}_B )$. Again, we find it convenient to give a new
symbol for the above free energy of the effectively free system of
bosons:
\begin{align}\label{bosdet}
  &F_{B,\rm Fock}[c_B;\rho_B]  =\nonumber\\
  &\quad-\frac{N_B T^3}{2\pi}\bigg[ \frac{1}{3}\hat{c}_{B}^3 - \int_{\hat{c}_B}^{\infty} d\hat{\e}\ \hat{\e} \int_{-\pi}^{\pi}d\alpha\ \rho_B(\alpha) \left(\log\big(1 - \te^{-\hat{\e}-\hat{\mu}-\i\alpha }\big)+\log\big(1 - \te^{-\hat{\e}+ \hat{\mu}+\i\alpha }\big)  \right) \nonumber\\
  &\quad+ \Theta(|{ \mu}|-{ c}_B ) \ \frac{(|\hat{\mu}| - {\hat c}_B)^2(|\hat{\mu}| + 2 {\hat c}_B)}{6 |\lambda_B|}\bigg]\ .
\end{align}
It was already suggested in \cite{Minwalla:2020ysu} - and we will
explain more completely in this paper - that the term proportional to
$\Theta(|\mu| - c_B)$ actually implements the `Bosonic Exclusion
Principle' which forbids any particular free particle state from being
occupied more than $|k_B|$ times.

The remaining terms on the RHS of \eqref{boseoff} and
\eqref{RBoffshellfe} are different, capturing the difference between
contact interactions in the critical boson and regular boson theories.

\subsection{Interchanging the order of integration}\label{inter}

With the expressions for the free energies of the various bosonic and
fermionic theories at hand, we return to the evaluation of the
partition function $\mc{Z}_{S^2 \times S^1}$ in \eqref{offshellZ}. After interchanging the order of integration,  \eqref{offshellZ} takes the form 
\begin{equation}\label{offshellZagain}
  \mc{Z}_{S^2 \times S^1} = \int d \varphi_{\rm aux}\  \int [d U]_{\rm CS}  
  \te^{-\mc{V}_2\beta F[\varphi_{\rm aux}; \rho]}\ .
\end{equation}
 We also learned from the explicit
expressions for the off-shell free energies in the previous
subsubsections that the off-shell free energy $F$ can be written as
\begin{equation}\label{fsplit} 
  F[\ldots,c;\rho] = F_{\rm int}[\ldots,c] + F_{\rm Fock}[c;\rho]\ ,
\end{equation}
where $\ldots$ indicates all the auxiliary variables other than the
thermal mass $c$, $F_{\rm Fock}$ is the free energy of free bosons /
fermions of mass $c$ at temperature $T$ and twisted by the holonomy
$U$ with eigenvalue distribution $\rho$ (plus an at-first strange
looking term proportional to $\Theta(\mu-c_B)$ in the case of bosons,
see \eqref{bosdet}), and $F_{\rm int}$ is the part of the off-shell
free energy which depends on the detailed contact interactions of the
specific theory under consideration\footnote{Note also that the
  interaction part $F_{\rm int}$ is independent of the chemical
  potential, a fact that was useful in the analysis of
  \cite{Minwalla:2020ysu}.}.  Inserting \eqref{fsplit} into
\eqref{offshellZagain} we find
\begin{equation}\label{offshellZagainn}
\mc{Z}_{S^2 \times S^1} = \int d \varphi_{\rm aux}\  \te^{-\mc{V}_2\beta F_{\rm int}[\varphi_{\rm aux}]}\  \int [d U]_{\rm CS}  \te^{-\mc{V}_2 \beta F_{\rm Fock}[c;\rho]}
\ .
\end{equation}
Evaluating the integral over $\varphi_{\rm aux}$ by saddle points in
the large $N$ limit, it follows that
\begin{align}\label{extmatrb}
  &\mc{Z}_{S^2 \times S^1} = \text{Ext}_{\{\varphi_{\rm aux}\}}\left[ \te^{-\mc{V}_2\beta F_{\rm int}[\varphi_{\rm aux}]}\ \wt{\mc{I}}_k(c)\right]\ ,\quad \text{with}\quad \wt{\mc{I}}_k(c) = \int [dU]_{\rm CS}\ \te^{-\mc{V}_2 \beta F_{\rm Fock}[c;\rho]}\ ,
\end{align}
where the operation $\text{Ext}_{\{\varphi_{\rm aux}\}}$ denotes
extremization w.r.t.~the auxiliary variables $\varphi_{\rm aux}$
including the thermal mass $c$ and the expression $F_{\rm Fock}$ is
given by \eqref{ferdet} for fermions and \eqref{bosdet} for
bosons. The subscript $k$ on $\mc{I}_k$ stands for the level of the
Chern-Simons gauge theory (this is accurate only for the $SU(N)_k$
theory whose level is $k$. The $U(N)$ theory has two levels $k$ and
$k'$ but we suppress such detail in the subscript).

The two different pieces that appear in the partition function
\eqref{extmatrb} are
\begin{enumerate}
\item the prefactor that involves $F_{\rm int}$  and
\item the matrix integral $\wt{\mc{I}}_k$.
\end{enumerate}
As we have seen in Section \ref{fici}, $F_{\rm int}$ encodes the
effective local contact interactions of the matter fields in the
theory. We return to a discussion of these interactions in Sections
\ref{energy} and \ref{entropy}. In the next two sections -
i.e.~Sections \ref{hccb} and \ref{winf} - we turn to a detailed study
of $\wt{\mc{I}}_k$.

\section{A Hilbert space interpretation of the matrix integral
  $\wt{\mc{I}}$}\label{hccb}

In this section we will discuss the Hilbert space interpretation of
the matrix integral $\wt{\mc{I}}$ in the partition function
\eqref{extmatrb}. We designate this quantity as $\wt{\mc{I}}_F$ and
$\wt{\mc{I}}_B$ for the fermionic and bosonic theories
respectively. In the first few subsections, we discuss path integrals
of pure Chern-Simons theories with Wilson line insertions on
$\Sigma \times S^1$ where $\Sigma$ is a two dimensional Riemann
surface of genus $g$. In particular, we discuss the correspondence
between observables in pure Chern-Simons theories and the WZW
models. The formulae that we present here will be useful in our
analysis of the integrals $\wt{\mc{I}}_B$ and $\wt{\mc{I}}_F$.

As mentioned in Section \ref{ntcs} we study $SU(N)_k$ and
$U(N)_{k, k'}$ Chern-Simons theories in this paper. Most of the
results presented in this section are valid for all finite values of
$N$ and $k$, $k'$. With an eye on later use, however, we also
specialize our results to the large $N$ 't Hooft limit in Section
\ref{tln}.

\textbf{Note:} Only in Sections \ref{verlinde}, \ref{lwcoo} and
\ref{tln}, we take $k$, $k'$ and $\kappa$ to be positive integers to
avoid cluttering of notation. In the rest of this section, the levels
can be positive or negative and will have subscripts $F$ or $B$
depending on whether the corresponding Chern-Simons theories are
coupled to fermions or bosons.

\subsection{The Chern-Simons/WZW correspondence and the Verlinde
  formula}\label{verlinde}

Consider the correlation function on $\Sigma \times S^1$ of $n$ Wilson
lines with $SU(N)$ or $U(N)$ representations $R_1$, $R_2$, \ldots,
$R_n$ placed at $n$ points on $\Sigma$ and winding once around the
thermal circle $S^1$. This correlation function evaluates to give the
dimension of the Hilbert space $\mc{H}_{\Sigma}(R_1,\ldots,R_n)$ of
the Chern-Simons theory on $\Sigma \times S^1$ \cite{Witten:1988hf}
which is what we designate as the space of \emph{quantum} singlets in
the tensor product of representations $R_1,\ldots,R_n$. According to
the correspondence between Chern-Simons theories and (chiral-)WZW
models first elucidated in \cite{Witten:1988hf}, the Hilbert space
$\mc{H}_{\Sigma}(R_1,\ldots, R_n)$ is the space of conformal blocks of
the WZW model on $\Sigma$ involving the representations
$R_1$,\ldots,$R_n$. The dimension of this space of conformal blocks is
given by the Verlinde formula \cite{Verlinde:1988sn} which we present
below.

First we establish some notation. Let the $SU(N)$ (or $U(N)$) highest
weights of the representations $R_1$,\ldots,$R_n$ be denoted as
$\mu_1$,\ldots,$\mu_n$; we also assume that these correspond to
integrable representations of the WZW model. Then, the Verlinde
formula states that the dimension of the space of conformal blocks is
given by
\begin{equation}\label{verfor}
  \mc{N}_{g,n}(R_1,\ldots,R_n) =  \sum_\lambda (\mc{S}_{\lambda 0})^{2-2g}\ \prod_{I=1}^n \frac{\mc{S}_{\lambda\mu_I}}{\mc{S}_{\lambda 0}}\ ,
\end{equation}
where $\mc{S}_{\lambda\mu}$ is the Verlinde $\mc{S}$-matrix (see
Section \ref{counting} for details), $\lambda$ runs over the set of
highest weights corresponding to integrable representations of the WZW
model and $0$ corresponds to the trivial representation. Sometimes, we
suppress the dependence on the representations $R_1,\ldots,R_n$ in the
notation $\mc{N}_{g,n}(R_1,\ldots,R_n)$.

In Section \ref{counting}, we have massaged the Verlinde formula using
well-known results on the Verlinde $\mc{S}$-matrix to obtain an
explicit formula for the dimension of the space of conformal
blocks. In particular, we encounter a well-known map
\cite{Elitzur:1989nr,Zuber:1995ig} between integrable representations
of $SU(N)_k$ and distinguished $SU(N)$ conjugacy classes in our
computations. We extend these considerations to $U(N)$ Chern-Simons
theories as well in Section \ref{counting} and present the formulae
for both the $SU(N)$ and $U(N)$ groups.

\subsubsection{$SU(N)_k$}
Consider $N$ distinct phases $\ul{w} = \{w_1,\ldots, w_N\}$ which
satisfy
\begin{equation}\label{suwconst}
  \prod_{i = 1}^N w_i = 1\ ,\quad  w_i^\kappa = w_j^\kappa\ ,\quad\text{for all}\quad i,j=1,\ldots,N\ .
\end{equation}
Let the set of solutions of the above equations up to permutation of
the $w_i$ be denoted by $\mc{P}_k$. It is easy to check that that
number of solutions is given by
\begin{equation}
  \binom{N + k - 1}{N -1}\ .
\end{equation}
The Verlinde formula is then given by
\begin{equation}\label{verlindesu}
  \mc{N}_{g,n}(R_1,\ldots, R_n) = \frac{1}{
    \left( N \kappa^{N-1} \right)^{1-g}} \sum_{\ul{w} \in \mc{P}_k}\ \prod_{1 \leq i < j \leq N} |w_i - w_j|^{2-2g}\ \prod_{I=1}^n \chi_{\vgap R_I}({\ul{w}})\ ,
\end{equation}
where the sum is over the set of solutions $\mc{P}_k$ to the equations
in \eqref{suwconst}, $\chi_{\vgap R_I}(\ul{w})$ is the $SU(N)$
character of the representation $R_I$ evaluated on the diagonal matrix
with diagonal entries $\ul{w} = \{w_1,\ldots, w_N\}$.

We emphasize again that $\chi_{\vgap R_I}({\ul{w}})$ is an ordinary
Lie algebra character (as opposed to a WZW character). The same is
true of all the characters that appear anywhere in this section, and
indeed anywhere in this paper outside Section \ref{qdnote}.

If a collection of distinct eigenvalues $\ul{w} = \{w_1,\ldots, w_N\}$
obey the condition \eqref{suwconst} then the rephased collection
\begin{equation}\label{rephase} 
  \ul{w}^{v} = \te^{ \frac{2 \pi \i v}{N}} \{ w_1,\ldots, { w_N}\}\ ,\quad v=0, 1, \ldots, N-1\ .
\end{equation} 
also obeys the same equation. It follows that the space of solutions
to \eqref{suwconst} may be decomposed into orbits $S_a$ of
$\mbb{Z}_N$, the centre of $SU(N)$, i.e.
\begin{equation}\label{pkdecompsu} {\cal P}_k = \bigcup_{a} \mc{S}_a\ .
\end{equation} 
The summation in \eqref{verlindesu} can be re-organized into a sum over
orbits and the sum over the elements of each orbit, that is
\begin{equation}\label{sumreorg} 
  \sum_{\ul{w} \in \mc{P}_k} = \sum_{a}  \sum_{\ul{w} \in \mc{S}_a}
\end{equation} 
Now if $R$ is a representation corresponding to a Young tableau with
$t$ boxes, then we have
\begin{equation}\label{chartransf}
  \chi_{\vgap R}({\ul{w}^v})= \te^{ \frac{2 \pi \i v t}{N}}\chi_{\vgap R}({\ul{w}})\ .
\end{equation} 	
It follows that the summation $\sum_{\ul{w} \in \mc{S}_a}$ in
\eqref{sumreorg} vanishes unless the summand in \eqref{verlindesu},
$\prod_{I=1}^n \chi_{\vgap R_I}({\ul{w}})$, is a $\mathbb{Z}_N$
singlet. Restated, the Verlinde formula \eqref{verlindesu} conserves
$\mbb{Z}_N$ centre charges.

\subsubsection{$U(N)_{k,k'}$}
Recall that $k' = \kappa + s N$ with $s = 0$ for Type I theories and
$s = -1$ for Type II theories. Consider $N$ distinct phases
$\ul{w} = \{w_1,\ldots, w_N\}$ which satisfy the
equations\footnote{Note that \eqref{uwconst} implies that
  $w_i^\kappa = w_j^\kappa$, for all $i,j=1,\ldots,N$.}
\begin{equation}\label{uwconst}
  (w_i)^\kappa \left(\prod_{j=1}^N w_j\right)^s = (-1)^{N-1} \ ,\quad\text{for every}\quad i = 1,\ldots,N\ . 
\end{equation} 
Let the set of solutions of the above equations (up to permutations of
the $w_i$) be denoted $\mc{P}_{k,k'}$. The number of such solutions
can be counted explicitly (see Section \ref{unintrep} or Appendix
\ref{csu}) and is given by
\begin{equation}\label{Unumintrep}
  \binom{N + k - 1}{N-1} \times \frac{k'}{N}\ .
\end{equation}
This matches precisely with the number of integrable representations
of $U(N)_{k,k'}$ WZW model. The Verlinde formula for the number of
conformal blocks is then given by
\begin{equation}\label{verlindeu}
  \mc{N}_{g,n}(R_1,\ldots, R_n) = \frac{1}{\left(k' \kappa^{N-1} \right)^{1-g}} \sum_{\ul{w} \in \mc{P}_{k,k'}}\ \prod_{1 \leq i < j \leq N} |w_i - w_j|^{2-2g}\ \prod_{I=1}^n \chi_{\vgap R_I}({\ul{w}})\ ,
\end{equation}
where the sum is over the set of solutions $\mc{P}_{k,k'}$ to the
equations \eqref{uwconst}, $\chi_{\vgap R_I}$ is now a $U(N)$
character corresponding to the $U(N)$ representations $R_I$ and
$\ul{w} = \{w_1,\ldots, w_N\}$ are the diagonal entries of a diagonal
$U(N)$ matrix on which the character $\chi_{\vgap R_I}$ is evaluated.

As in the previous subsubsection, if a collection of distinct
eigenvalues $\ul{w} = \{w_1,\ldots, w_N\}$ obey the condition
\eqref{uwconst}, then the rephased collection
\begin{equation}\label{rephaseu}
  \ul{w}^{v} = \te^{ \frac{2 \pi \i v}{k'}} \{ w_1,\ldots, { w_N}\}\ ,\quad v=0,1, \ldots, k'-1 \ ,
\end{equation} 
also obeys the same equation. It follows that the space of solutions
to \eqref{uwconst} may be decomposed into orbits of $\mbb{Z}_{k'}$:
\begin{equation}\label{pkdecompu}
  {\cal P}_{k, k'} = \bigcup_{a} \mc{S}_a\ ,
\end{equation} 
where $\mc{S}_a$ are the orbits of $\mbb{Z}_{k'}$. Once again the
summation over solutions to \eqref{uwconst} can be reorganized in a
manner analogous to \eqref{sumreorg}. In this case if $R$ is an
integrable representation corresponding to a Young tableau with $t$
boxes then
\begin{equation}\label{chartransfu}
  \chi_{\vgap R}({\ul{w}^v})= \te^{ \frac{2 \pi \i v t}{k'}}\chi_{\vgap R}({\ul{w}}).
\end{equation} 	
Again, it follows that the summation $\sum_{\ul{w} \in \mc{S}_a}$ in
\eqref{sumreorg} vanishes unless the summand in \eqref{verlindeu},
$\prod_{I=1}^n \chi_{\vgap R_I}({\ul{w}})$, is a $\mathbb{Z}_{k'}$
singlet. We conclude that the Verlinde formula \eqref{verlindeu}
conserves $\mbb{Z}_{k'}$ charges.  

We explicitly present the Verlinde formulae for the Type I and Type II
$U(N)$ theories of principal interest in this paper:
\begin{equation}\label{verlindeuI}
\text{Type I}:\quad  \mc{N}_{g,n}(R_1,\ldots, R_n) = \frac{1}{\kappa^{N(1-g)}} \sum_{\ul{w} \in \mc{P}_{k,\kappa}}\ \prod_{1 \leq i < j \leq N} |w_i - w_j|^{2-2g}\ \prod_{I=1}^n \chi_{\vgap R_I}({\ul{w}})\ ,
\end{equation}
with the $\ul{w} \in \mc{P}_{k,\kappa}$ satisfying
$(w_i)^\kappa = (-1)^{N-1}$, and
\begin{equation}\label{verlindeuII}
  \text{Type II}:\quad  \mc{N}_{g,n}(R_1,\ldots, R_n) = \frac{1}{(k\kappa^{N-1})^{1-g}} \sum_{\ul{w} \in \mc{P}_{k,k}}\ \prod_{1 \leq i < j \leq N} |w_i - w_j|^{2-2g}\ \prod_{I=1}^n \chi_{\vgap R_I}({\ul{w}})\ ,
\end{equation}
with the $\ul{w} \in \mc{P}_{k,k}$ satisfying
$(w_i)^\kappa (\prod_j w_j)^{-1} = (-1)^{N-1}$.

\subsubsection{Non-integrable representations}\label{nonint}

The Verlinde formula \eqref{verfor} defined in terms of
$\mc{S}$-matrices is clearly only defined whenever the representations
$R_1,\ldots,R_n$ are all integrable representations of the WZW model
since the $\mc{S}$-matrix is, by definition, the matrix which
implements the $\tau \to -1/\tau$ transformation for torus characters
of the WZW model. However, the equivalent formulae \eqref{verlindesu}
and \eqref{verlindeu} (that will be derived in Section \ref{counting}
using well-known expressions for $\mc{S}$-matrices in terms of Lie
algebra characters and also by using path integral methods to evaluate
pure Chern-Simons partition functions and indices in the presence of
Wilson lines (see Sections \ref{suNpath}, \ref{UNpath} and Appendix
\ref{oneloop}) are well-defined for arbitrary representations $R_i$,
including non-integrable ones.

Putting aside the question of the physical interpretation of the
formulae \eqref{verlindesu} and \eqref{verlindeu} applied to
non-integrable representations, for the moment, let us investigate the
results obtained from this procedure. As is well-known
\cite{Elitzur:1989nr,Moore:1989vd}(also cf.~the Kac-Walton formula
e.g.~in \cite[Chapter 16.2.1]{di1996conformal}) and as we explain in
Appendices \ref{sucount}, \ref{to}, \ref{tt}, characters of
non-integrable representations - evaluated on any of the special
holonomies with eigenvalues \eqref{suwconst} or \eqref{uwconst} - can
be re-expressed as the characters of other integrable representations,
sometimes with a negative sign. Consequently, if we allow ourselves to
insert non-integrable representations into the Verlinde formulae
\eqref{verlindesu}, \eqref{verlindeu}, we find an integer that is not
always zero and is sometimes negative.

What is the physical interpretation of \eqref{suwconst} or
\eqref{uwconst} with non-integrable insertions? In order to answer
this, we note that - at least naively - the direct evaluation of the
path integral on $\Sigma_g \times S^1$ for the Chern-Simons correlator
with Wilson lines in representations $R_1,\ldots,R_n$ (performed for
semisimple gauge groups including $SU(N)$ in \cite{Blau:1993tv} and
which we extend to $U(N)$ in Section \ref{counting}) appears to yield
the formulas \eqref{verlindesu}, \eqref{verlindeu}. Naively, the
derivation of \cite{Blau:1993tv} seems completely insensitive to
whether the representations $R_1,\ldots,R_n$ are integrable or
non-integrable. Consequently, the most straightforward interpretation
of \eqref{verlindesu} and \eqref{verlindeu} with some insertions in
non-integrable representations is simply the following; these formulae
compute the expectation value of the corresponding Wilson lines of
Chern-Simons theory on $\Sigma_g\times S^1$. Similar conclusions
follow from the semi-classical study of Chern-Simons theories in the
presence of Wilson lines.  From this point of view the quantum
identities of Appendices \ref{sucount}, \ref{to}, \ref{tt} are simply
quantum consequences of the following semi-classical fact: Wilson
lines in representations that have identical quantum characters (upto
a sometimes puzzling sign and small - presumably quantum - shifts in
angular momentum quantum numbers) produce the same semi-classical
holonomy for the Chern-Simons coupled gauge field, and so, appear to
yield the same result in a Chern-Simons path integral.
 
While the conclusion described in the paragraph above may seem
satisfactory at first sight, there is something about this resolution
that is troubling. Recall that Gepner and Witten \cite{Gepner:1986wi}
established that WZW correlators that contain at least one integrable
representation and at least one non-integrable representation evaluate
to zero and thus, the dimension of the space of such conformal blocks
is zero. Consequently, if the conclusion of the previous paragraph is
correct, it would imply that the correspondence between WZW conformal
blocks and Chern-Simons Wilson line correlators is a partial one; the
two structures agree perfectly when all insertions lie in `integrable'
representations, but disagree when one or more of the representations
are taken to be non-integrable. Relatedly, it would appear to imply
that level-rank duality, which is an exact symmetry of WZW conformal
blocks, is only a partial symmetry of Chern-Simons path integrals in
the presence of Wilson lines (it works when all insertions are
integrable, but can fail when one or more of them is
non-integrable). This conclusion - while conceivably correct - has an
ugly feel to it, at least to the authors of this paper.

A second (at the moment completely wishful) possibility is that the
current evaluations of path integrals with insertions of Wilson lines
in Chern-Simons theory are missing a subtlety (something like a Gribov
ambiguity), and that when this new effect is taken into account, will
turn out to evaluate to zero, in agreement with the WZW result of
Gepner and Witten. The aesthetic appeal of this possibility was noted
by Witten in his original paper on Jones polynomials (see e.g.~the
remarks at the end of Pg. 372 in
\cite{Witten:1988hf}). 

The resolution of this question appears to be of relevance to the
current paper. As we have already explained in the introduction, and
as we explain in detail in Section \ref{bep}, we find that previously
obtained results for the thermal partition function of matter
Chern-Simons theories admit a simple interpretation in terms of a Fock
space partition function restricted to WZW singlets. In more detail
\begin{itemize} 
\item We note that each single particle state - with any occupation
  number - always transforms in a single irrep of $SU(N)$. In
  particular a single particle state occupied by $n$ bosons transforms
  in the representation $S_n$ with $n$ boxes in the first row of the
  Young tableau. Consequently the number of states with $n_1$, $n_2$
  particles in the first, second $\ldots$ state generate factors of
  $\chi_{\vgap S_{n_1}}(U)$, $\chi_{\vgap S_{n_2}}(U), \ldots $, in
  the holonomy integral and the summation over holonomy eigenvalues
  determines the weight of such states to equal the dimension of
  conformal blocks with primary operators in representations
  $S_{n_1}$, $S_{n_2},\ldots$.
\item Given that non-integrable representations decouple from their
  integrable counterparts, states with any $n_i>k_B$ contribute with
  weight zero (i.e. do not contribute) to the partition function sum,
  and so can be set to zero (see around \eqref{ZBfocktrunc} ).
\end{itemize}   
While this logic feels compelling, there is a potential fly in the
ointment. As we have discussed in detail in this section, while the
formulae in \eqref{verlindesu} and \eqref{verlindeu} correctly compute
the number of conformal blocks when all representations are
integrable, these formulae do not correctly compute the number of
conformal blocks (i.e. 0 ) when some of the insertions are
non-integrable. However the fact that we get the answer zero for
non-integrable representations is precisely the Bosonic exclusion
principle. As we have explained in this section, if we replace the
`number of conformal blocks' with the formulae \eqref{verlindesu}
\eqref{verlindeu} with insertions $\chi_{\vgap S_{n_1}}(U)$,
$\chi_{\vgap S_{n_2}}(U),\ldots$ which in general gives a non-zero
integer, then the Bosonic exclusion principle no longer holds.

There is compelling independent evidence for the Bosonic exclusion
principle. First, it is necessary to ensure Bose-Fermi duality of
Chern-Simons matter theories. Second, explicit results large $N$
results for the Bosonic partition function imply this principle;
indeed it was in this context that this principle was first
encountered \cite{Minwalla:2020ysu}.  \footnote{The argument provided
  for this phenomenon in the large $N$ theory (see
  \cite{Minwalla:2020ysu}) was somewhat indirect - involving an
  analytic continuation in chemical potential. For this reason this
  computation does not provide us with direct insight into how the
  path integral enforces the Bosonic exclusion principle. It would be
  interesting to revisit this computation and its physical
  interpretation.}

How, then, does the Bosonic exclusion principle emerge from the
Chern-Simons matter path integral? We see two possible resolutions to
this puzzle. First, as we have mentioned above, it is (just) possible
that the path integral evaluation of Chern-Simons Wilson lines have
missed a subtlety, the accounting for which will set the expectation
value of non-integrable Wilson lines to zero. It would be very nice of
this turns out to be the case, but we see no evidence for this
possibility at the moment.

Another possibility, that seems more likely to us is the following.
It may turn out that modelling the contribution of $n$ bosons in a
single state by a single effective insertion of
$\chi_{\vgap S_{n}}(U)$ is too crude. The fact that this contribution
arises from a collection of $n$ particles (rather than a single
effective particle in a higher representation) is relevant, and the
careful analysis of the relevant Schrodinger equation (and, in
particular, the consequence of imposing Bose symmetry on the wave
function) will lead to the Bosonic exclusion principle. Once the
validity of the Bosonic Exclusion principle has been understood in the
Hamiltonian language of this paragraph, its explanation from a path
integral point of view will then, also, hopefully follow.

It is clearly very important to clear this matter up and we hope to
return to it in future work. In the rest of this paper, however, we
simply proceed by interpreting the expectation values of products of
$\chi_{\vgap S_{n}}(U)$ as the number of WZW conformal blocks with the
corresponding insertions. This interpretation then enforces the
bosonic exclusion principle by hand. We leave the question of finding
a convincing justification of this prescription to future work.
\footnote{We thank D. Gaiotto, O. Aharony, G. Moore for discussions
  related to this subsection. We especially thank E. Witten for
  extensive correspondence on every aspect of this question.}

\subsection{A large number of Wilson lines} \label{lwcoo} 

The Verlinde formulae \eqref{verlindesu} and \eqref{verlindeu} apply
for arbitrary values of $n$, the number of Wilson line insertions. It
is well known\footnote{We thank O. Parrikar for alerting us to this
  fact, and for explaining the argument outlined in this paragraph to
  us.} (see e.g.\cite{Tong:2016kpv}) that this formula simplifies in
the limit that $n$ is taken to infinity with $N$ and $k$ held fixed
(more generally this is the case when $n$ is parametrically larger
than $N$ and $k$). The argument for this goes roughly as follows. By
successively fusing representations, the number of singlets in the
fusion of representations $\mu_1$, $\mu_2, \ldots, \mu_n$ may easily
be seen to be given by the $(\mu_1, \mu_n)^{\rm th}$ matrix element of
the product of fusion matrices
$N_{\mu_2} N_{\mu_3} \ldots N_{\mu_{n-1}}$ (see \eqref{numsing} in
Appendix \ref{proofsub}). As we review in Appendix \ref{proof}, the
fusion matrices commute with each other, and so are simultaneously
diagonalizable. In Appendix \ref{proof} we explain that the largest
eigenvalue of the matrix $N_{\mu_p}$ is ${\cal D}(R_{\mu_p})$ - the
quantum dimension of the representation $R_p$.\footnote{See \cite[Chapter 16]{di1996conformal} and also section
  \ref{qdnote} for a discussion of quantum dimensions.} Moreover, the
eigenvector corresponding to this largest eigenvalue is the same for
all $\mu_p$. It follows that, in the large $n$ limit, the product of
fusion matrices receives its dominant contribution from the product of
these largest eigenvalues (all other contributions are exponentially
suppressed in the large $n$ limit) and the formula for the number of
singlets reduces to a single term proportional to the quantum
dimensions of all representations that participate in the fusion
process.

In this subsection we explain how the simplification described above
manifests itself in the explicit formulae \eqref{verlindesu} and
\eqref{verlindeu}, and in the process also make the argument of the
previous paragraph more precise (in particular we derive a precise
value for the numerical proportionality constant of this leading
term).

Consider the following eigenvalue configuration that satisfies
\eqref{suwconst} as well as \eqref{uwconst} for every value of $N$,
$k$ and $k'$.
\begin{equation}\label{confoo}
\ul{w}_{0} = \left\{q^{-(N-1)/2},\ q^{-(N-3)/2}, \ldots,\ q^{(N-3)/2},\ q^{(N-1)/2}\right\}\ ,\quad q = \te^{2\pi\i/\kappa}\ ,
\end{equation}
corresponding to the holonomy matrix
\begin{equation}\label{configon1}
  U^{(0)} = \diag\left\{q^{-(N-1)/2},\ q^{-(N-3)/2}, \ldots,\ q^{(N-3)/2},\ q^{(N-1)/2}\right\}.
\end{equation}
The analogue of the dominance of largest eigenvalues in the product of
fusion matrices is that the summation over discrete holonomies of the
product of characters
\begin{equation}\label{propso}
\prod_{I=1}^n \chi_{\vgap R_I}({\ul{w}})\ ,
\end{equation} 
in \eqref{verlindesu}, \eqref{verlindeu} is maximized in absolute
value - and so is dominated in the large $n$ limit - by the $SU(N)$
(or $U(N)$) holonomy eigenvalue configuration $\ul{w}_0$ and its
$\mbb{Z}_N$ images (resp.~$\mbb{Z}_{k'}$ images for $U(N)_{k,k'}$).

\noindent This result is a consequence of the following theorem proved
in Appendix \ref{proof}.

\noindent \textbf{Theorem}: {\it The eigenvalue configuration
  \eqref{configon1} and its $\mbb{Z}_N$ images (resp.~$\mbb{Z}_{k'}$
  images for $U(N)_{k,k'}$) maximize the absolute value of the
  character of all integrable representations $R$ of $SU(N)_k$ (resp.~
  $U(N)_{k,k'}$) evaluated on the distinguished eigenvalue
  configurations \eqref{suwconst} (resp.~\eqref{uwconst})~i.e.}
\begin{equation}\label{QCmaxconjec}
  \chi_{\vgap R}(\ul{w}_{0}) \geq \left|\chi_{\vgap R}(\ul{w})\right|\ ,
\end{equation} 
{\it where the equality holds when $\ul{w}$ is a $\mbb{Z}_N$ image (resp.~$\mbb{Z}_{k'}$ image for $U(N)_{k,k'}$) 
  of $\ul{w}_{0}$.} 

The theorem stated above follows immediately from the following well known statements, explained in detail in Appendix \ref{proof}.
\begin{enumerate}
\item The \emph{quantum} characters i.e.~characters
  $\chi_{\vgap R}(\ul{w})$ evaluated on the various distinguished
  $SU(N)$ eigenvalue configurations that satisfy \eqref{suwconst}
  (resp.~\eqref{uwconst} for $U(N)$) are eigenvalues of fusion
  matrices of the $SU(N)_k$ WZW model (resp.~$U(N)_{k,k'}$ WZW
  model),\footnote{See for instance \cite[Exercise
    3.7]{Moore:1989vd},\cite{Zuber:1995ig}.}\footnote{From the
    viewpoint of the formula \eqref{numsing} for the number of
    singlets, the summations in \eqref{verlindesu} and
    \eqref{verlindeu} reflects the summation over the $r$ eigenspaces
    of the simultaneously diagonal $r \times r$ fusion matrices, where
    $r$ is the number of integrable representations.}
\item The character $\chi_{\vgap R}(\ul{w}_0)$ is the \emph{quantum dimension}
  of the representation $R$, i.e.
  \begin{equation}\label{qdchar}
  \mc{D}(R) = \chi_{\vgap R}(\ul{w}_{0})\ .
  \end{equation}
   The quantum dimension is real, positive and it turns out to
  be largest eigenvalue of the fusion matrix corresponding to the
  representation $R$ i.e. $\chi_{\vgap R}(\ul{w}_0) \geq |\chi_{\vgap R}(\ul{w})|$ for
  every $SU(N)$ eigenvalue configuration $\ul{w}$ that satisfies
  \eqref{suwconst} (resp.~\eqref{uwconst} for $U(N)$).
\end{enumerate}

The fact that \eqref{configon1} dominates the summations in \eqref{verlindesu} and \eqref{verlindeu} is an almost obvious consequence of the theorem stated above (see the next subsection for  more detail). Before turning to this point, however, we pause to provide some intuition for  \eqref{QCmaxconjec} for the benefit of those readers who may be more familiar with classical group theory than WZW fusion algebras.

 In classical group theory it is well known (and trivial 
to see) that the absolute values of characters are maximized on group
elements proportional to identity\footnote{In the case of $SU(N)$
	there are $N$ such configurations: the group elements corresponding
	to the $\mbb{Z}_N$ centre of the group. In the case of $U(N)$ it
	corresponds to the central $U(1)$.}. Given this fact, it is thus natural to guess that, among those the holonomy eigenvalue configurations that appear in the summations in \eqref{verlindesu} and 
\eqref{verlindeu}, the ones that maximize the characters of all integrable representations are those that lie nearest to a configuration proportional to the identity matrix.

The holonomy eigenvalue configuration which lies nearest to the
identity matrix is the one whose eigenvalues are all as close as they
can possibly be to unity. It is easy to convince oneself that the
discrete eigenvalue configuration that meets this description both in
the case of the $SU(N)_k$ theory and in the case of the $U(N)_{k, k'}$
theory is given by \eqref{configon1}\footnote{ That \eqref{configon1}
  is a rather special eigenvalue configuration can be seen as follows.
  As we explain in Section \ref{counting} there exists a one-to-one
  map between eigenvalue configurations that appear in
  \eqref{verlindesu} (resp.~\eqref{verlindeu}) and integrable
  representations of $SU(N)_k$ (resp.~$U(N)_{k,k'})$. In Section
  \ref{counting} we demonstrate that this map takes the configuration
  \eqref{configon1} to the trivial (or identity)
  representation.\label{verfoot}}. The $\mbb{Z}_N$ images
(resp.~$\mbb{Z}_{k'}$ images in the $U(N)_{k,k'}$ theory) of the above
configuration also give the same value of the product of characters
that appear in \eqref{verlindesu} and \eqref{verlindeu} since the
representations $R_I$ in \eqref{propso} are chosen such that the
product of characters is invariant under the action of $\mbb{Z}_N$
(resp.~$\mbb{Z}_{k'}$); see the discussion around equations
\eqref{pkdecompsu} and \eqref{pkdecompu}.

We also pause to note that while the proof of \eqref{QCmaxconjec} presented in Appendix \ref{proof} uses fairly sophisticated structures, the final result is an easy to state inequality for elementary functions.
For instance, in the case of the $SU(2)_k$ theory, consider the
quantity $\chi_{j}(n)$
\begin{equation}\label{sutmaxcl}
\chi_{j}(n) =\frac{\displaystyle\sin \frac{(2j+1)n\pi}{k+2}}{\displaystyle\sin \frac{n\pi}{k+2}}\ ,
\end{equation} 
at any fixed value of the $SU(2)$ spin $j$ and level $k$. Upon varying
$n$ over the range $$ n \in \{1,2, \ldots, k+1\}\ ,$$ the theorem
\eqref{QCmaxconjec} asserts that $|\chi_{j}(n)|$ attains
its maximum at $n=1$ and $n=k+1$. 

In Appendix \ref{sutch}, we have
plotted $|\chi_{j}(n)|$ as a function of $n$ at fixed $j$ and $k$ for
several different values of $j$ and $k$; for every set of values we have looked at,  it
is indeed true that $|\chi_j(n)|$ is maximum at $n=1,k+1$.

\subsubsection{Consequences of our result for large $n$}

The character for any representation, evaluated on any allowed
eigenvalue configuration is a number of order the (classical)
dimension of the representation. Recall that the set of integrable
representations is finite. It follows that ${f}(\ul{w})$ defined by
the equation
\begin{equation}\label{propsof}
{f}(\ul{w})^n = \prod_{I=1}^n \chi_{\vgap R_I}({\ul{w}})\ ,
\end{equation}
has a modulus of order somewhere between the classical dimension of
the smallest nontrivial integrable representation\footnote{In the
	current context this representation these are the fundamental and
	antifundamental representations with classical dimension $N$.}, and
the classical dimension of the largest integrable representation. In
particular ${f}(\ul{w})$ remains bounded, both from below and from
above, in the limit $n\to\infty$ for any nontrivial choice of representations.

The RHS of both \eqref{verlindesu} and \eqref{verlindeu} involve a
summation over a finite number of distinct eigenvalue
configurations. As the contribution of the configuration ${\ul{w}}$
is proportional to ${f}({\ul{w}})^n$, it follows that when $n$ is
parametrically larger than $N$ and $k$, the configuration(s) with the
largest value of $|f({\ul{w}})|$ will dominate the summations in
\eqref{verlindesu} and \eqref{verlindeu}. In the limit $n\to\infty$ we
can thus accurately estimate the summations in \eqref{verlindesu} and
\eqref{verlindeu} by retaining only the contribution of the eigenvalue
configuration(s) with the maximum value of $f({\ul{w}})$. The error
we make by dropping the contribution of all other eigenvalue
configurations is exponentially small in $n$.

From \eqref{propso} we see that the eigenvalue configuration(s) that
maximize the absolute values of characters will also maximize
$|f(\ul{w})|$.

The theorem of the previous subsubsection implies that the configuration \eqref{configon1} and
its $\mbb{Z}_N$ images (resp.~$\mbb{Z}_{k'}$ images for $U(N)_{k,k'}$)
dominate the summations in \eqref{verlindesu} (resp.~\eqref{verlindeu}
for $U(N)_{k,k'}$) when $n$ is parametrically larger than $N$ and
$k$. Since the contribution of these $\mbb{Z}_N$ (resp.~$Z_{k'}$)
images of \eqref{configon1} is equal to the contribution of
\eqref{configon1} itself, the sum over images simply gives an
additional factor of $N$ (resp.~$k'$).

It follows then that, in the limit $n\to\infty$, the
$SU(N)_k$ Verlinde formula reduces to
\begin{equation}\label{verlindesularge}
  \mc{N}_{g,n}(R_1,\ldots, R_n) = \frac{N}{
    \left( N \kappa^{N-1} \right)^{1-g}} \prod_{1 \leq i < j \leq N} \left|w_{0i} - w_{0j}\right|^{2-2g}\ \prod_{I=1}^n \mc{D}(R_I)\ ,
\end{equation}
while the $U(N)_{k,k'}$ Verlinde formula \eqref{verlindeu} reduces to
\begin{equation}\label{verlindeularge} 
\mc{N}_{g,n}(R_1,\ldots, R_n) = \frac{k'}{(k'\kappa^{N-1})^{1-g} } \prod_{1 \leq i < j \leq N} \left|w_{0i} - w_{0j}\right|^{2-2g}\ \prod_{I=1}^n \mc{D}(R_I)\ ,
\end{equation}
where, once again, we have used \eqref{qdchar} (i.e the fact that  $\chi_{\vgap R_I}(\ul{w}_0)$  equals the quantum dimension $\mc{D}(R)_I$).

\subsubsection{Decomposition of a `sea' of representations}

As above, let us
consider a genus $g$ Riemann surface with $n$ fixed insertions where
$n$ is parametrically larger than $k$ and $N$. Let us then add one
more Wilson line in the representation $R$ to the same Riemann
surface. The formulae \eqref{verlindesularge} and
\eqref{verlindeularge} imply that in both the $SU(N)_k$ and the
$U(N)_{k, k'}$ theories, we have
\begin{equation}\label{ratofsing} 
\mc{N}_{g,n+1}(R, R_1,\ldots, R_n)
={\cal D}(R) ~\mc{N}_{g,n}(R_1,\ldots, R_n)\ .
\end{equation} 
The formula \eqref{ratofsing} can be interpreted as follows. Let us
call the collection of representations $R_1, \ldots, R_n$ `the
sea'. The quantity $\mc{N}_{g,n}(R_1,\ldots, R_n)$ is the number of
singlets produced by the repeated fusion of the integrable
representations $R_1, \ldots, R_n$. The reader may find herself
unsatisfied with this partial information; she may wish to know how
many of each integrable representation (and not just the trivial
representation) are produced in the fusion of the representations
present in the sea. The answer to this question is easily given: using
the fact that the fusion of the representation $R$ with $R'$ produces
the trivial representation if and only of $R'={\bar R}$, it follows
that the number $n(R)$ of the representation $R$ produced by the
fusion of the representations in the sea is given by
$\mc{N}_{g,n+1}({\bar R}, R_1,\ldots, R_n)$. Using \eqref{ratofsing}
it then follows that
\begin{equation}\label{ratodim}
  \frac{n(R_1)}{n(R_2)}=\frac{{\cal D}({\bar R }_1)}{{\cal D}({\bar R}_2)}= \frac{{\cal D}({ R }_1)}{{\cal D}({R}_2)}\ ,
\end{equation} 
where the last equality is due to the fact that the quantum dimension
of $\bar{R}$ is the same as that of $R$ for any representation $R$
(this can be see from the fact that
$\chi_{\vgap \bar{R}} = \chi^*_{\vgap R}$ and that
$\chi_{\vgap R}(\ul{w}_0)$ is real for any $R$).

Let us summarize. The formula for the number of times a representation
$R$ produced by the fusion of the representations in the sea is
relatively complicated and depends of the details of the precise
representations that make up the sea even when the number of
representations $n$ is large (see \eqref{verlindesularge},
\eqref{verlindeularge} and \eqref{ratofsing}). However, the formula
for the ratios of these quantities is completely universal and very
simple in this limit. This fact will have important implications for
the thermodynamics of Chern-Simons matter theories in the large volume
limit.

\subsubsection{Large volume limit of Chern-Simons matter partition
  functions} \label{lvmc}

The fact that the matrix \eqref{configon1} dominates the Verlinde
formula in the limit of a large number of insertions has already been
indirectly encountered in previous studies of the partition function of
large $N$ matter Chern-Simons theories, as we now briefly review.

  As discussed in detail in
\cite{Aharony:2003sx,Jain:2013py} (and as we have already reviewed in subsection \ref{rtpf}) the large $N$ 't Hooft limit the $S^2 \times S^1$ thermal partition function of large $N$ fundamental matter Chern-Simons theories is determined
by extremization of the effective potential $v[\rho]$ w.r.t.~the holonomy
eigenvalue density function $\rho(\alpha)$ generated by integrating out the matter fields. This matter potential which scales (in the 't Hooft limit) like the dimension of the representation of the matter
fields, as well as the volume of spatial manifold and temperature,
generically turns out to be a confining potential which tries to
squeeze the eigenvalues to unity i.e.~towards the density
$\rho(\alpha) = \delta(\alpha)$ (see subsection \ref{ftpf} for explicit expressions in several examples).  A second  universal contribution
(independent of the matter fields, volume and temperature) to the
potential comes from the Vandermonde factor which, being proportional
to modulus of difference of eigenvalues, tries to spread the
eigenvalues uniformly over the circle i.e.~towards the density
$\rho(\alpha)=1/(2\pi)$. This contribution scales likes $N^2$ in the
't Hooft limit.

The competition between the two factors leads to interesting phase
transitions which have been extensively studied in the literature (see
\cite{Aharony:2003sx,Jain:2013py} and references therein). In particular, in the large volume limit, the matter contribution to the effective potential dominates over the repulsive force from the Vandermonde factor, and tries to squeeze the eigenvalues 
as much as possible. In limit $\lambda\to 0$
the eigenvalue density function is squeezed all the way down to a $\delta$ function 
$\delta(\alpha)$; this reflects the fact, reviewed above, that the classical Weyl integral formula, \eqref{wif}, receives its dominant contribution from the identity (more generally from central elements) in the limit of a large number of insertions.

In Chern-Simons theories at finite $\lambda$, however, the summation
over the flux sector discretizes the eigenvalues
(\cite{Blau:1993tv,Jain:2013py}, see subsection \ref{UNpath} and
Appendix \ref{oneloop} for more on this) which, in conjunction with
exclusion of configurations in which two or more eigenvalues coincide,
enforces an upper bound on the eigenvalue density
$\rho(\alpha) \leq 1/(2\pi|\lambda|)$ \eqref{inequality}. As a
consequence, it was noted in \cite{Aharony:2003sx,Jain:2013py}, that
in the large volume limit the holonomy eigenvalue distribution is
squeezed down not to $\delta(\alpha)$ but, instead, to the universal
`table top' distribution
\begin{equation}\label{rhoalphe} 
  \rho(\alpha) = \left\{\renewcommand{\arraystretch}{1.5}\begin{array}{cc} \frac{1}{2\pi|\lambda|} & |\alpha|< \pi |\lambda|\ , \\ 0 & |\alpha|> \pi |\lambda|\ .\end{array}\right.
\end{equation} 
Of course \eqref{rhoalphe} is simply the large $N$ holonomy eigenvalue
distribution function associated with the more precise holonomy
configuration \eqref{confoo} (see subsection \ref{tln}).

Partition functions in the large volume limit are dominated by states
with a very large number of matter particles, i.e. states for which
the Verlinde formula reduces to \eqref{verlindesularge} or
\eqref{verlindeularge}. Consequently, the previously observed
universality of the holonomy eigenvalue distributions in the partition
functions of large volume matter Chern-Simons theories may be viewed
as a special case of the more general formulae \eqref{verlindesularge}
or \eqref{verlindeularge}.

\subsection{The 't Hooft large $N$ limit} \label{tln} 

In this subsection we describe the large $N$ limit of the formulae
\eqref{verlindesu} and \eqref{verlindeu} for genus zero i.e.~for
Chern-Simons theory on $S^2 \times S^1$. Let us focus on the Type I
$U(N)$ theory for simplicity. The Verlinde formula becomes
\begin{equation}\label{verlindeuIg0}
  \mc{N}_{0,n}(R_1,\ldots, R_n) = \frac{1}{\kappa^{N}} \sum_{\ul{w} \in \mc{P}_{k,\kappa}}\ \prod_{1 \leq i < j \leq N} |w_i - w_j|^{2}\ \prod_{I=1}^n \chi_{\vgap R_I}({\ul{w}})\ .
\end{equation}
Let us write $w_i = \te^{\i \theta_i}$ with
$0 \leq \theta_i < 2\pi$. Since the solution set
$\mc{P}_{k,\kappa}$ consists of solutions $\ul{w}$ to the equations
\eqref{uwconst} up to permutations, it is useful to order the $w_i$
such that
\begin{equation}
0 \leq  \theta_1 < \theta_2 < \cdots < \theta_N < 2\pi\ .
\end{equation}
The solutions then correspond to discrete values of the $\theta_i$
such that the gap between different $\theta_i$ is at least
$2\pi / \kappa$ (this follows from the second condition in
\eqref{uwconst}).

Recall that, in the 't Hooft large $N$ limit, $N$ and $k$ are both
taken to infinity with the ratio $N / \kappa$ held fixed. Consider
\eqref{verlindeuI} in this limit. It is easy to see that the sum over
the solutions $\mc{P}_{k,\kappa}$ should be replaced the following
integral:
\begin{equation}\label{reprulle}
\frac{1}{\kappa^N} \sum_{\ul{w} \in \mc{P}_{k,\kappa}} \longrightarrow 
\int_0^{2 \pi - \frac{2 \pi}{\kappa} }\frac{d \theta_1 }{2\pi}
\int_{\theta_1+ \frac{2 \pi}{\kappa}}^{2 \pi - \frac{2 \pi}{\kappa} } \frac{d\theta_2}{2\pi} \ldots 
	\int_{\theta_{N-1}+ \frac{2 \pi}{\kappa}}^{2 \pi - \frac{2 \pi}{\kappa} } \frac{d\theta_N}{2\pi}\ .
\end{equation} 
It follows that \eqref{verlindeuI} reduces to the following in the 't
Hooft large $N$ limit:
\begin{align}\label{verlindeuIln} 
  &\mc{N}_{0,n}(R_1,\ldots,R_n) =\nonumber\\
  &\quad \int_0^{2 \pi - \frac{2 \pi}{\kappa} }\frac{d \theta_1 }{2\pi}
\int_{\theta_1+ \frac{2 \pi}{\kappa}}^{2 \pi - \frac{2 \pi}{\kappa} } \frac{d\theta_2}{2\pi} \ldots 
	\int_{\theta_{N-1}+ \frac{2 \pi}{\kappa}}^{2 \pi - \frac{2 \pi}{\kappa} } \frac{d\theta_N}{2\pi} \prod_{1 \leq i < j \leq N} 4 \sin^2\left(\frac{\theta_i - \theta_j}{2}\right)\ \prod_{I=1}^n \chi_{\vgap R_I}({\ul{w}})  \ . 
\end{align}
The lower and upper limits for the integral over $\theta_i$ in
\eqref{verlindeuIln} are
\begin{equation}\label{lolimthetai}
\Big(\theta_{i-1}+ \frac{2\pi}{\kappa}, 2 \pi - \frac{2 \pi}{\kappa} \Big)\ .
\end{equation} 
Even though \eqref{lolimthetai} tends to
\begin{equation}\label{lolimthetainaive}
  (\theta_{i-1}, 2 \pi)\ ,
\end{equation}
as $\kappa \to \infty$, the limits \eqref{lolimthetainaive} and
\eqref{lolimthetai} are not actually equivalent in the 't Hooft limit
though they are equivalent in the limit $k \to \infty$ at fixed
$N$ \footnote{If we could use the limits
  \eqref{lolimthetainaive}, \eqref{verlindeuIln} would reduce to the
  simpler expression
\begin{equation}\label{verlindeuIlnnaive} 
  \frac{1}{N!} \prod_{I=1}^N \int_0^{2\pi}
  \frac{d \theta_i}{2 \pi} 
  \prod_{1 \leq i < j \leq N}  4 \sin^2\Big(\frac{\theta_i-\theta_j}{2}\Big) \prod_{I = 1}^n \chi_{\vgap {R_I}}(\ul{\theta}) \ ,\nonumber
\end{equation}
which has a simple interpretation: comparing to \eqref{nsingets} in
the Appendix, we see that \eqref{verlindeuIlnnaive} is the same as
$\mc{I}_{\rm cl}$ in \eqref{sumint}, the partition function of the
zero coupling Yang-Mills theory with Wilson line insertions. This
conclusion is indeed correct for the limit $k \to \infty$ at fixed
$N$, but is not correct for the 't Hooft large $N$ limit as we explain
below. }. In order to see this, it is useful to change integration
variables from $\theta_i$ to the eigenvalue distribution
$\rho(\theta)$ defined in \eqref{rhdef} in the 't Hooft limit:
\begin{equation}
  \rho(\theta) = \lim_{N \to\infty} \frac{1}{N} \sum_{i = 1}^N \delta(\theta - \theta_i)\ .
\end{equation}
The fact that the minimum gap between two $\theta_i$ cannot be smaller
than $\frac{2 \pi}{\kappa}$ tells us that the eigenvalue density
function $\rho(\theta)$ must obey the inequality
\eqref{inequality}
\begin{equation}\label{inequality1}
  \rho(\theta) \leq \frac{1}{2\pi|\lambda|}\ ,
\end{equation}
where $\lambda$ is the 't Hooft coupling $\lambda = N /
\kappa$. Therefore, when reformulated as a functional integral over
the $\rho(\theta)$, the measure is identical to what it would have
been in the classical formula \eqref{nsingets} (see
\eqref{verlindeuIlnnaive}) along with the additional constraint that
the integral over eigenvalue distributions is taken only over those
$\rho(\theta)$ that obey \eqref{inequality1} at every value of
$\theta$.\footnote{See \cite{Jain:2013py} for further explanations and
  also the explicit exact solution of the large $N$ Gross-Witten-Wadia
  unitary matrix integral for eigenvalue distributions that obey the
  inequality \eqref{inequality1}.} It follows that, in the large $N$
't Hooft limit, \eqref{verlindeuI} becomes
\begin{align}\label{verlindeuIfinal}
  &\mc{N}_{0,n}(R_1,\ldots, R_n) \longrightarrow \wt{\mc{N}}_{0,n}(R_1,\ldots,R_n) = \int [dU]_{\rm CS} \prod_{I=1}^n \chi_{\vgap R_I}(\ul{w})  \ . 
\end{align}
where the measure $[dU]_{\rm CS}$ in \eqref{verlindeuIfinal} is the
usual measure over eigenvalue distributions $\rho(\alpha)$ of the
eigenvalues $\ul{w} = \{w_1,\ldots,w_N\}$, with a further restriction
on the range of this integral to $\rho(\alpha)$ that obey
\eqref{inequality1}.

Since the $SU(N)$ and $U(N)$ matrix integrals are equivalent in the
large $N$ limit, the Verlinde formulae for the $SU(N)$ theory and Type
II $U(N)$ theory also reduce to \eqref{verlindeuIfinal} in the 't
Hooft large $N$ limit.

\textbf{Note:} The quantity $\mc{N}_{0,n}$ defined for finite $N$ and
$k$ is a positive integer since it counts the dimension of the space
of conformal blocks. However, the large $N$, large $k$ limit
$\wt{\mc{N}}_{0,n}$ might not necessarily be an integer due to the
various continuum approximations that were made in the matrix integral
over $U$.

\subsection{The fermionic integral $\wt{\mc{I}}_F$}\label{ferint}

With all necessary technical preliminaries out of the way we now
tackle the question we set out to address in this section, namely to
find a clear Hilbert space interpretation of the quantities
$\wt{\mc{I}}$ in \eqref{extmatrb}, Section \ref{inter}. In this
subsection we focus on the case of Chern-Simons gauged fermions.

Recall from equation \eqref{extmatrb} that
\begin{equation}\label{unimm}
  \wt{\mc{I}}_{F,k} = \int [dU]_{\rm CS}\ \wt{\mc{I}}_{F,\rm Fock}(U)\ ,\quad\text{with}\quad \wt{\mc{I}}_{F,\rm Fock} = \te^{-\mc{V}_2 \beta F_{F,\rm Fock}[c_F;\rho_F]} \ , 
\end{equation} 
where $F_{F,\rm Fock}$ is given by \eqref{ferdet}\footnote{We have
  used $\hat{c}_F= \beta c_F$ and have made a change of integration
  variables $\epsilon = T {\hat \epsilon}$ in \eqref{ferdet} to get
  the result \eqref{zns} above.}
\begin{align}\label{zns}
  &F_{F,\rm Fock}[{c}_F;\rho_F]  =  \frac{N_F c_F^3}{6\pi}\nonumber\\
  &\quad - \frac{N_F T}{2\pi}\int_{{c}_F}^{\infty} d{\e}\, {\e} \int_{-\pi}^{\pi}d\alpha\ \rho_F(\alpha) \left(\log\big(1+\te^{-\beta({\e}+{\mu})-\i\alpha }\big)+\log\big(1+\te^{-\beta({\e} - {\mu}) +\i\alpha }\big)  \right) \ .
\end{align}

As we have explained above, we wish to check whether
$\wt{\mc{I}}_{F,k}$ can be identified with the thermal trace over a
Hilbert space, and, if so, determine the nature of this Hilbert space
and its Hamiltonian.

One way to check whether an expression can be identified with a
thermal trace over a genuine Hilbert space is to expand that quantity
in powers of $\te^{-\beta E}$, and check whether the coefficient of
every term in this expansion is a positive integer.  Recall, however,
that the expression \eqref{unimm} is valid only when $N_F$ and the
volume $\mc{V}_2$ of the $S^2$ are both parametrically large. The
integral nature of coefficients is usually obscured in such
thermodynamical limits (i.e.~large $N$ as well as large volume). For
this reason, in this subsection we first identify a more precise
finite $N$ and finite volume generalization of $\wt{\mc{I}}_{F,k}$
which we call ${\mc{I}}_{F,k}$. By construction ${\mc{I}}_{F,k}$
becomes $\wt{\mc{I}}_{F,k}$ in the large volume and large $N_F$
limit. We then demonstrate that ${\mc{I}}_{F,k}$ is a trace over a
Hilbert space, and characterize this space precisely.

To begin our analysis we note the expression \eqref{unimm} has two
components: $\wt{\mc{I}}_{F,\rm Fock}(U)$ and the integral over $U$
with measure $[dU]_{\rm CS}$. We study and simplify each of these
components in turn.

\subsubsection{The twisted partition function at finite volume} \label{tpfv}

From a path integral point of view, the quantity
$\wt{\mc{I}}_{F,\rm Fock}(U)$ is simply the  large $N$ and large volume
limit of the $S^2 \times S^1$ partition function of $N_F$ free
fermions $\psi$ of mass $c_F$, evaluated using dimensional
regularization, and subject to the following twisted boundary
conditions around the thermal circle:
\begin{equation}\label{tbac}
  \psi(\beta)=-U \psi(0)\ ,\quad {\bar \psi}(\beta)=- {\bar \psi} (0) U^\dag\ .
\end{equation}
One natural finite $N$ and finite $\mc{V}_2$ generalization of this
quantity is the same path integral, computed in the same
regularization scheme, but at finite $N$ and $\mc{V}_2$. This path
integral has a standard Hamiltonian interpretation; in order to be
completely accurate with the vacuum energy, (and also to gain some additional insight into the formula, see below) it is useful to re-obtain
this Hamiltonian interpretation starting from \eqref{unimm}.

In the dimensional regularization scheme, we have a non-trivial zero
point energy in $F_{F,\rm Fock}$ \eqref{zns} which is the term
${N_F c_F^3}/{6\pi}$. When the theory is on an $S^2$ of finite volume,
there is a further contribution to $F_{F,\rm Fock}$ which is the
Casimir energy density of $N_F$ Dirac fermions $N_F E_{F,\rm Cas}$
where $E_{F,\rm Cas}$ is the Casimir energy density\footnote{It
  follows from dimensional analysis that
  $E_{F,\rm Cas} = \frac{1}{\mc{V}_2}\frac{g(R c_F)}{R}$ where $R$ is
  the radius of the $S^2$ and $g(x)$ is a function to be
  determined. We leave the evaluation of the function $g(x)$ as an
  exercise for the interested reader.} of one free Dirac fermion of
mass $c_F$ on $S^2$. Thus, the finite volume version of
$F_{F,\rm Fock}$ has total vacuum energy density
\begin{equation}
N_F E_{F,\rm vac} = \frac{N_F c_F^3}{6\pi} + N_F E_{F,\rm Cas}\ ,
\end{equation}
and is given by
\begin{align}\label{znsfv}
  &{\cal V}_2 \beta F_{F,\rm Fock}[{c}_F;\rho_F]  = {\cal V}_2 \beta  N_F E_{F,\rm vac}\nonumber\\
  &\quad - {N_F} \sum_a \int_{-\pi}^{\pi}d\alpha\ \rho_F(\alpha) \left(\log\big(1+\te^{-\beta(E_a+\mu)-\i\alpha }\big)+\log\big(1+\te^{-\beta(E_a - \mu) + \i\alpha }\big)  \right)\ .
\end{align}
where the summation over the index $a$ runs over the single particle
states of the fermion on an $S^2$ with the $a^{\rm th}$ state carrying
an energy $E_a$.

The Hamiltonian perspective helps us understand that the twisted
fermionic determinant on a finite volume $S^2$ is only one of an
infinite class of physically meaningful finite volume regularizations
of $\wt{\mc{I}}_{F,k}$. This finite volume determinant is given by
\eqref{znsfv} with the sum over the index $a$ running over the
spectrum of single particle fermionic states on $S^2$ and
$E_{F,\rm Cas}$ is the fermionic Casimir energy. However we get an
equally good finite volume regulator of $\wt{\mc{I}}_{F,k}$ if we
choose $E_a$ to be any spectrum of energies whose density of states
has the correct large volume limit\footnote{The density of states that
  the $E_a$ are constrained to have, in the large volume limit, may be
  deduced as follows.  In flat space the energy $\epsilon$ is given by
  $\epsilon = \sqrt{c_F^2 + \vec{k}{}^2}$ where $\vec{k}$ is the
  spatial momentum of the single particle state. The large volume
  density thus takes the form
\begin{equation}
{\cal V}_2\int \frac{d^2 k}{(2\pi)^2}=
\ {\cal V}_2\int_0^\infty \frac{k dk}{2\pi} \int_0^{2\pi} \frac{d\theta}{2\pi} =
{\cal V}_2\int_{c_F}^\infty \frac{\epsilon d\epsilon}{2\pi}\ , \nonumber 
\end{equation} 
in agreement with, for instance, the integral over energies in
\eqref{zns}.\label{dosfootnote}} and the Casimir energy is taken to be
any quantity that scales in the appropriate manner with the volume of
the sphere. In this paper we use a finite volume regulator of the form
\eqref{znsfv}, remaining agnostic about the precise choice of single
particle spectrum (or Casimir energy) apart from the general
properties that they are constrained to obey as outlined above. One
plausible choice of spectrum of one-particle states is the finite
volume spectrum of the particular matter theory whose free energy
contains the determinant \eqref{znsfv} e.g.~the critical fermion or
the regular fermion theories.

\subsubsection{The twisted partition function at finite $N$ }

To proceed to obtain the finite $N$ version of $F_{F,\rm Fock}$, we
recall the definition of the eigenvalue distribution
\begin{equation}\label{infiniterho}
  \rho_F(\alpha) = \lim_{N_F\to\infty} \frac{1}{N_F}\sum_{i=1}^{N_F} \delta(\alpha - \alpha_i)\ ,
\end{equation}
where $\te^{\i\alpha_i}$ are the eigenvalues of the holonomy matrix
$U$. In the finite $N$ situation, the integral over $\alpha$ is then
replaced by a trace over the fundamental indices $I$ as is clear from
\eqref{infiniterho} and the following manipulation:
\begin{equation}\label{finiterho}
  \int_{-\pi}^\pi d\alpha \rho_F(\alpha) f(\te^{\i\alpha}) = \frac{1}{N_F} \sum_{i=1}^{N_F} f(\te^{\i\alpha_i})= \frac{1}{N_F} \tr\, f(U)\ .
\end{equation}
It then follows that $F_{F,\rm Fock}$ can be simplified to
\begin{equation}\label{oneloopdet}
 {\cal V}_2 \beta F_{F,\rm Fock} = N_F {\cal V}_2 \beta E_{F,\rm vac} - {N_F}  \sum_a \left( \tr\log(1 + y_a U) + \tr\log (1 + \tl{y}_a U^\dag)\right)\ ,
\end{equation}
where the $\tr$ is over the fundamental gauge indices,
$y_a = \te^{-\beta(E_a - \mu)}$ and
$\tl{y}_a = \te^{-\beta(E_a + \mu)}$ are the Boltzmann factors for
particles and antiparticles at energy $E_a$. 

It follows from the discussion above that the natural finite $N$ and
finite $\mc{V}_2$ generalization of the partition function
$\wt{\mc{I}}_{F,\rm Fock}$ \eqref{unimm} of $N_F$ free fermions with mass
$c_F$ is
\begin{equation}\label{actraceaa}
  \mc{I}_{F,\rm Fock}(U) \equiv \exp\left(-\beta N_F \mc{V}_2 E_{F,\rm vac}\right) \tr_{\mathcal{H}_{F,\rm Fock}}  \left( \widehat{U} \te^{-\beta(H-\mu Q)} \right)\ ,
\end{equation}
where the trace is taken over the Fock space $\mc{H}_{F,\rm Fock}$ of
$N_F$ fermions of mass $c_F$ on $S^2$, the operator $\widehat{U}$
represents the action of $U$ on the Fock space, and the operator $H$
is the free multiparticle Hamiltonian on this Fock space with the
convention that the vacuum energy on this Fock space is zero. The
operator $Q$ is the global $U(1)$ charge described in detail in
\cite{Minwalla:2020ysu}.

For brevity in subsequent calculations, we define the quantity
$\mc{Z}_{F,\rm Fock}(U)$ to be just the Fock space trace in
\eqref{actraceaa}:
\begin{align}\label{ZFfock}
  \mc{Z}_{F,\rm Fock}(U) &= \tr_{\mathcal{H}_{F,\rm Fock}} \left( \widehat{U} \te^{-\beta(H-\mu Q)} \right) \ ,\nonumber\\
                         &= \prod_{a=1}^\infty \prod_{i=1}^{N_F} \left(1 + \te^{-\beta(E_a - \mu)} w_i\right) \left(1 + \te^{-\beta(E_a + \mu)} \bar{w}_i\right)\ ,
\end{align}
where $w_i$ are the eigenvalues of the holonomy $U$, the index $a$ in
\eqref{ZFfock} runs over the space of positive definite energy
solutions of a free Dirac fermion of mass $c_F$ on $S^2$ and $E_a$ is
the energy of the solution. Expanding the product over $i$ in
\eqref{ZFfock}, we get an expression for $\mc{Z}_{F,\rm Fock}$ in
terms of products of sums of characters:
\begin{align} \label{zinpfi}
  &\mc{Z}_{F,\rm Fock}(U)= \prod_{a=1}^\infty  \left(\sum_{n_a=0}^{N_F} \te^{-n_a \beta \left( E_a-\mu \right) }
    \ \chi_{\vgap A_{n_a}}(\ul{w})\right)\left(\sum_{\tl{n}_a=0}^{N_F}
    \te^{-\tl{n}_a \beta \left( E_a+\mu \right) } \
    \bar\chi_{\vgap A_{\bar{n}_a}}(\ul{w})\right)\ ,
\end{align}
where $\ul{w} = \{w_1,\ldots,w_N\}$ are the eigenvalues of $U$, the
index $n_a$ (resp.~$\bar{n}_a$) counts the occupation number of the
state of energy $E_a$ by fundamental (resp.~antifundamental) fermions
with distinct colours, and $A_{n_a}$ (resp.~$A_{\bar{n}_a}$) is the
totally antisymmetric $SU(N)$ representation with $n_a$ boxes
(resp.~$\bar{n}_a$ boxes).

\subsubsection{The integral over $U$}

Now that we have suitably defined the finite $N_F$ and finite
$\mc{V}_2$ quantity $\mc{I}_{F,\rm Fock}$ \eqref{actraceaa}, we move
on to the finite $N$ version of the integral over the holonomy matrix
$U$. To see what the finite $N$ version should be, we first replace
the large $N_F$ and $\mc{V}_2$ quantity $\wt{\mc{I}}_{F,\rm Fock}$ in
\eqref{unimm} by $\mc{I}_{F,\rm Fock}$. This yields
\begin{align}\label{iformF}
&\te^{-\beta \mc{V}_2 N_F E_{F,\rm vac}} \int [dU]_{\rm CS}\
\prod_{a=1}^\infty \left( \sum_{n_a=0}^{N_F} \te^{-n_a \beta \left( E_a-\mu \right) } \ \chi_{\vgap A_{n_a}}(\ul{w}) \right) 
\left( \sum_{{\bar n}_a=0}^{N_F} \te^{-{\bar n}_a \beta \left( E_a+\mu \right) } \  \bar\chi_{\vgap A_{\bar{n}_a}}(\ul{w}) \right)   \ ,\nonumber \\
 &\equiv \te^{-\beta \mc{V}_2 N_F  E_{F,\rm vac}}\sum_{\{n_a\}, \{\bar{n}_a\}} \wt{\mc{N}}{}^{k}\left(\{A_{n_a}\}, \{A_{{\bar n}_a}\}\right)\ \te^{ -\beta \sum_a  \left(  
 n_a (E_a -\mu) + {\bar n}_a (E_a +\mu) \right) }\ ,
\end{align}
with
\begin{equation}\label{NsingF}
  \wt{\mc{ N}}^k(\{A_{n_a}\}, \{A_{{\bar n}_a}\}) \equiv \int [dU]_{\rm CS}\ \prod_{a=1}^\infty  \chi_{\vgap A_{n_a}}(\ul{w}) \bar\chi_{\vgap A_{{\bar n}_a}}(\ul{w}) \  .
\end{equation}
The summation in the second line of \eqref{iformF} is over the
infinite strings of integers $\{n_a\}$, $\{\bar{n}_a\}$,
$a = 1,2,\ldots$ where for each $a$, $n_a$ and $\bar{n}_a$ take the
values $0,\ldots,N_F$. Recall that the notation $A_{n}$ stands for the
totally antisymmetric $SU(N)$ representation with $n$ boxes and that
$\ul{w} = \{w_1,\ldots, w_N\}$ are the eigenvalues of $U$.

Observe that \eqref{NsingF} is simply a particular case of the
Verlinde formula in the large $N$ limit
\eqref{verlindeuIfinal}\footnote{In this particular case the
  representations that appear on the RHS of the Verlinde formula are
  set of representations $A_{n_a}$, $A_{\bar{n}_a}$ appearing on the
  RHS of \eqref{NsingF}.}. As we have already remarked in the
discussion around \eqref{verlindeuIfinal}, the
$\wt{\mc{ N}}^{k}(\{A_{n_a}\}, \{A_{{\bar n}_a}\})$ are generally not
integers and that this is an artefact of the continuum nature of the
integral over $U$ in the large $N$ limit. It follows that
\eqref{iformF} does not define a genuine partition function.

However, it is now clear what the finite $N$ resolution of the
integral over $U$ should be. The finite $N$ Verlinde formulae
presented in \eqref{verlindesu} and \eqref{verlindeu} instruct us to
replace $\wt{\mc{ N}}^{k}(\{A_{n_a}\}, \{A_{{\bar n}_a}\})$ with its
finite $N$ version $\mc{N}^k(\{n_a\},\{\bar{n}_a\};A)$ defined by
\begin{equation}\label{finiteNsingF}
  \mc{ N}^{k}(\{A_{n_a}\}, \{A_{{\bar n}_a}\}) \equiv \frac{1}{|k_F'|{|\kappa_F|^{N_F-1}}} \sum_{\ul{w} \in \mc{P}} \prod_{1 \leq i < j \leq N_F} |w_i - w_j|^{2} \prod_{a=1}^\infty  \chi_{\vgap A_{n_a}}({\ul{w}}) \bar\chi_{\vgap A_{{\bar n}_a}}({\ul{w}}) \ ,
\end{equation}
In \eqref{finiteNsingF}, the level $k'_F$ is given by
$k'_F = N_F, \kappa_F, k_F$ for $SU(N_F)$, Type I $U(N_F)$ and Type II
$U(N_F)$ theories respectively, and the solutions sets
$\mc{P} = \mc{P}_{|k_F|}$ or $\mc{P}_{|k_F|,|k'_F|}$ for $SU(N_F)$ and
$U(N_F)$ theories are defined under \eqref{suwconst} and
\eqref{uwconst} respectively (see Section \ref{verlinde} for more
details).

As we have explained in Section \ref{hccb}, the quantities
$\mc{N}^k(\{A_{n_a}\},\{A_{\bar{n}_a}\})$ are indeed integers. It
follows that the finite $N_F$ version of $\wt{\mc{I}}_{F,k}$ given by
\begin{equation}\label{finiteiformF}
  \mc{I}_{F,k}  \equiv  \te^{-\beta \mc{V}_2 N_F E_{F,\rm vac}}\!\!\! \sum_{\{n_a\},\{\bar{n}_a\}}    \mc{ N}^{k}(\{A_{n_a}\}, \{A_{\bar{n}_a}\})\ \te^{ -\beta \sum_a  \left(n_a (E_a -\mu) + {\bar n}_a (E_a +\mu) \right) }\ ,
\end{equation}
is a genuine partition function. Like earlier, we define the quantity
$\mc{Z}_F$ to denote the content of \eqref{finiteiformF} without the
vacuum energy factor:
\begin{equation}\label{ZfiniteiformF}
  \mc{Z}_{F,k} \equiv \sum_{\{n_a\},\{\bar{n}_a\}}    \mc{ N}^{k}(\{A_{n_a}\}, \{A_{\bar{n}_a}\};A)\ \te^{ -\beta \sum_a  \left(n_a (E_a -\mu) + {\bar n}_a (E_a +\mu) \right) }\ .
\end{equation}
The integers $\mc{N}^k(\{A_{n_a}\},\{A_{\bar{n}_a}\})$ count the
number of singlets in the tensor product of the representations
$\{A_{n_a}\}$, $\{A_{\bar{n}_a}\}$ subject to the Chern-Simons Gauss
law constraint (which is the same as the dimension of the space of
conformal blocks with representations $\{A_{n_a}\}$,
$\{A_{\bar{n}_a}\}$) which we call the \emph{quantum} singlet
constraint. The quantity $\mc{Z}_{F,k}$ in \eqref{ZfiniteiformF} is
then the trace of $\te^{-\beta(H - \mu Q)}$ over a Hilbert space
$\mc{H}_{F,k}$ which consists of the states in the Fock space
$\mc{H}_{F,\rm Fock}$ subject to the \emph{quantum} singlet constraint
i.e.~the space of conformal blocks with representations $\{A_{n_a}\}$,
$\{A_{\bar{n}_a}\}$. Thus, we can write
\begin{equation}\label{IFtrace}
  \mc{Z}_{F,k} = \tr_{\mc{H}_{F,k}}\left( \te^{-\beta(H - \mu Q)}\right)\ ,\quad \mc{I}_{F,k} = \te^{-\beta \mc{V}_2 N_F E_{F,\rm vac}} \mc{Z}_{F,k}\ .
\end{equation}
The matrix integral $\wt{\mc{I}}_{F,k}$ is then the large $N$ and
large $\mc{V}_2$ limit of the genuine partition $\mc{I}_{F,k}$ defined
for the Hilbert space $\mc{H}_{F,k}$ which is the projection of the
finite $N$ and finite $\mc{V}_2$ Fock space onto the sector of quantum
singlets.

\subsection{The bosonic integral $\wt{\mc{I}}_B$ and the bosonic
  exclusion principle} \label{bep} Recall from Section \ref{dsr},
eq.~\eqref{extmatrb} that the quantity $\wt{\mc{I}}_B$ is given by
\begin{equation}\label{uintebb} 
  \wt{\mc{I}}_{B,k} = \int [dU]_{\rm CS}\  \wt{\mc{I}}_{B,\rm Fock}(U) \ , \quad\text{with}\quad \wt{\mc{I}}_{B,\rm Fock}(U) = \te^{-\mc{V}_2\beta F_{B,\rm Fock}[c_B;\rho_B]}\ ,
\end{equation}
and from \eqref{bosdet},
\begin{align}\label{znsb}
  &F_{B,\rm Fock}[c_B;\rho_B]  =\nonumber\\
  &\quad-\frac{N_B T^3}{2\pi}\bigg[ \frac{1}{3}\hat{c}_{B}^3 - \int_{\hat{c}_B}^{\infty} d\hat{\e}\ \hat{\e} \int_{-\pi}^{\pi}d\alpha\ \rho_B(\alpha) \left(\log\big(1 -\te^{-\hat{\e}-\hat{\mu}-\i\alpha }\big)+\log\big(1 -\te^{-\hat{\e}+ \hat{\mu}+\i\alpha }\big)  \right) \nonumber\\
  &\quad\qquad\qquad\ \ + \Theta(|{ \mu}|-{ c}_B ) \ \frac{(|\hat{\mu}| - {\hat c}_B)^2(|\hat{\mu}| + 2 {\hat c}_B)}{6 |\lambda_B|}\bigg]\ .
\end{align}
The new element in the bosonic formula - as compared to its fermionic
counterpart - is the term proportional to $\Theta(|{ \mu}|-{ c}_B )
$. We will explain in this subsection that the formulae
\eqref{uintebb} and \eqref{znsb} including the term proportional to
$\Theta(|{ \mu}|-{ c}_B )$ can be obtained from the large $N_B$, large
volume limit of a finite $N_B$ and finite volume partition function
which is the natural bosonic analog of its finite $N_F$ and finite
volume fermionic counterpart discussed in the previous subsection.

As in the case of fermions we first construct the finite $N_B$ and
$\mc{V}_2$ analogue of $\wt{\mc{I}}_{B,\rm Fock}(U)$ and then turn to
the finite $N_B$ analog of the integral over $U$. As for the fermions
above in \eqref{actraceaa}, it is natural to guess that the finite
$N_B$, finite volume partition function $\mc{I}_{B,\rm Fock}(U)$ is
the path integral on $S^2 \times S^1$ of $N_B$ complex bosons of
squared mass $c_B^2$ in the dimensional regularization scheme, subject
to the twisted boundary conditions
\begin{equation}\label{tbacb}
  \phi(\beta)=U \phi(0)\ ,\quad {\bar \phi}(\beta)= {\bar \phi} (0) U^\dag\ .
\end{equation}
In Hamiltonian language this path integral is written as
\begin{equation}\label{actraceaaf}
  \mc{I}_{B,\rm Fock}(U)= \exp\left(-\beta \mc{V}_2 N_B E_{B,\rm vac}\right)  \mc{Z}_{B,\rm Fock}(U)\ ,\quad \mc{Z}_{B,\rm Fock}(U) = {\rm Tr}_{\mathcal{H}_{B,\rm Fock}} \left( \widehat{U} \te^{-\beta(H-\mu Q)} \right)\ ,
\end{equation}
where 
\begin{equation}\label{vacenb}
  N_B E_{B,\rm vac} = -\frac{N_B c_B^3}{6 \pi} + N_B E^B_{\rm Cas}\ ,
\end{equation}
with $E_{B,\rm Cas}$ being the Casimir energy density for a single
complex scalar on $S^2$. \footnote{As in the fermionic case the
  spectrum of energies of this finite volume Fock space does not have
  to precisely match those of the free boson theory, as long as their
  density of states matches the flat space density of states (see
  Footnote \ref{dosfootnote}) in the infinite volume limit.}

 As in the fermionic case, we can write
\begin{align}\label{ZBfock}
  \mc{Z}_{B,\rm Fock}(U)=&
                           \prod_{a=1}^\infty \prod_{i=1}^{N_B} \left(\frac{1}{1 - \te^{-\beta(E_a - \mu)} w_i}\right) \left(\frac{1}{1 - \te^{-\beta(E_a + \mu)} \bar{w}_i}\right)\ ,\nonumber\\
                         &=\prod_{a=1}^\infty \left( \sum_{n_a=0}^{\infty} \te^{-n_a \beta
                           \left( E_a-\mu \right) } \chi_{\vgap S_{n_a}}(\ul{w}) \right)
                           \left( \sum_{{\bar n}_a=0}^{\infty} \te^{-{\bar n}_a \beta \left( E_a+\mu \right) } \bar\chi_{\vgap S_{{\bar n}_a}}(\ul{w}) \right) \ ,
\end{align}
where $\ul{w} = \{w_1,\ldots,w_N\}$ are the eigenvalues of the
holonomy $U$, the index $a$ counts the solutions of a free boson of
mass $c_B$ on $S^2$, the index $n_a$ (resp.~$\bar{n}_a$) counts the
occupation number of the particle (resp.~antiparticle) state with
energy $E_a$ and $S_{n}$ is the totally symmetric $SU(N)$
representation with $n$ boxes.

We now turn to the finite $N_B$ analog of the integral over $U$ in
\eqref{uintebb} with measure $[dU]_{\rm CS}$. As in the case of the
fermionic theory of the previous subsection, the integral over $U$ has
to be replaced by a discrete sum over eigenvalue configurations $U$
that arises in the finite $N$ Verlinde formula \eqref{verlindesu} and
\eqref{verlindeu}. The summand of the discrete sum in the fermionic
case was the infinite product of characters corresponding to the
representations $\{A_{n_a}\}$, $\{A_{\bar{n}_a}\}$. In the current
context the representation content that appears in this infinite
product needs to be modified for the following reason.

The Verlinde formulae \eqref{verlindesu} and \eqref{verlindeu}
correctly count the number of quantum singlets (i.e.~dimension of WZW
conformal blocks) {\it only} when all primary operator insertions are
integrable representations.  In fact, as Gepner and Witten
\cite{Gepner:1986wi} demonstrated long ago, primary operators of WZW
theory corresponding to `non-integrable representations' decouple from
the correlators. That is, conformal blocks involving one or more
insertions of `non-integrable' representations all vanish and hence
there are no singlets in the fusion of representations which contain
at least one non-integrable representation.  Since the Verlinde
formula does not seem to automatically reproduce this result, we need
to perform the truncation to integrable representations in
\eqref{ZBfock} by hand\footnote{As discussed in Section \ref{nonint},
  the procedure known and used at present to evaluate the Chern-Simons
  path integral with Wilson lines in representations $R_1,\ldots,R_n$
  (see \cite{Blau:1993tv} and Section \ref{counting}) is insensitive
  to whether the representations are integrable or non-integrable and
  are thus in direct contradiction to Gepner and Witten's result. We
  wish to address this issue in future work.}.

The non-integrable representations that appear in \eqref{ZBfock} are
the totally symmetric representations $S_{n_a}$ of $SU(N)$ with
$n_a > |k_B|$ i.e.~the Young tableau corresponding to $S_{n_a}$
contains more than $|k_B|$ boxes in its first (and only)
row\footnote{The same statement holds for $\text{U}(N)_{k,k'}$
  theories since the primary operators are described by the product of
  an $SU(N)$ representation $\lambda$ and a $U(1)$ representations
  with charge $q$ which satisfies $q = |\lambda|\ \text{mod}\ N$ where
  $|\lambda|$ is the number of boxes in the Young tableau of the
  representation $\lambda$. The Gepner-Witten selection rule continues
  to apply for the $SU(N)$ part of the primary and hence, one
  continues to truncate the $SU(N)$ part of the primary. There are
  further relations between different primaries of the $U(N)$ theory
  due to a $\mbb{Z}_N$ `gauge' symmetry. See Appendix \ref{cpu} for
  more details.}. It follows that we must first truncate each of the
summations in the second line of \eqref{ZBfock} to $n_a \leq |k_B|$:
\begin{equation}\label{ZBfocktrunc} 
  \mc{Z}_{B,\rm Fock}(U) = \prod_{a=1}^\infty \left(
    \sum_{n_a=0}^{|k_B|} \te^{-n_a \beta \left( E_a-\mu \right) }
    \chi_{\vgap S_{n_a}}(\ul{w}) \right)
  \left( \sum_{{\bar n}_a=0}^{|k_B|} \te^{-{\bar n}_a \beta \left( E_a+\mu \right) } \bar\chi_{\vgap S_{\bar{n}_a}}(\ul{w}) \right)\ .
\end{equation}
This replacement has striking physical consequences: it ensures that
no particular single particle state can be occupied more that $|k_B|$
times i.e.~it enforces the Bosonic Exclusion principle discussed in
\cite{Minwalla:2020ysu}.

The remaining analysis closely parallels that for the fermionic
theory. The quantity
\begin{align}\label{iformB}
  \sum_{\{n_a\}, \{\bar{n}_a\}} \wt{\mc{N}}{}^{k}\left(\{S_{n_a}\}, \{S_{\bar{n}_a}\}\right)\ \te^{ -\beta \sum_a  \left(n_a (E_a -\mu) + {\bar n}_a (E_a +\mu) \right) }\ ,
\end{align}
with
\begin{equation}\label{NsingB}
  \wt{\mc{ N}}^{k}(\{S_{n_a}\}, \{S_{\bar{n}_a}\}) \equiv \int [dU]_{\rm CS}\ \prod_{a=1}^\infty  \chi_{\vgap S_{n_a}}(\ul{w}) \bar\chi_{\vgap S_{\bar{n}_a}}(\ul{w}) \  ,
\end{equation}
is not a genuine partition function since the numbers $\wt{\mc{N}}^k$ are
generally not integers due to the continuum nature of the integral
over $U$. Like in the fermionic theory, the natural finite $N$ version
is given by the discrete sum over eigenvalue configurations of the
Verlinde formulae \eqref{verlindesu} and \eqref{verlindeu}:
\begin{equation}\label{finiteNsingB}
  \mc{ N}^{k}(\{S_{n_a}\}, \{S_{\bar{n}_a}\}) \equiv \frac{1}{|k_B'|{|\kappa_B|^{N_B-1}}} \sum_{\ul{w} \in \mc{P}} \prod_{1 \leq i < j \leq N_B} |w_i - w_j|^{2} \prod_{a=1}^\infty  \chi_{\vgap S_{n_a}}({\ul{w}}) \bar\chi_{\vgap S_{{\bar n}_a}}({\ul{w}}) \ ,
\end{equation}
where $k'_B = N_B, \kappa_B, k_B$ for $SU(N_B)$, Type I $U(N_B)$ and
Type II $U(N_B)$ theories respectively, and the set
$\mc{P} = \mc{P}_{|k_B|}$ or $\mc{P}_{|k_B|,|k'_B|}$ for $SU(N_B)$ and
$U(N_B)$ theories respectively. 

It follows that the finite $N_B$, finite volume version of
$\wt{\cal I}_B$ in \eqref{uintebb} is given by
\begin{equation}\label{ibfirst}
  \mc{I}_{B,k} =  \te^{-\beta N_B \mc{V}_2 E_{B,\rm vac} }\mc{Z}_{B,k} \ ,
\end{equation} 
where $E_{B,\rm vac}$ is the vacuum energy density of a single free scalar on
an $S^2$ of finite volume $\mc{V}_2$ \eqref{vacenb} and
\begin{equation}\label{finiteiformB}
  \mc{Z}_{B,k} = \sum_{\{n_a\},\{\bar{n}_a\}}    \mc{ N}^{k}(\{S_{n_a}\}, \{S_{\bar{n}_a}\})\ \te^{ -\beta \sum_a  \left( n_a (E_a -\mu) + {\bar n}_a (E_a +\mu) \right) }\ .
\end{equation}
The quantity $\mc{N}^k(\{S_{n_a}\},\{S_{\bar{n}_a}\})$
\eqref{finiteNsingB} is just the number of quantum singlets in the
fusion of the totally symmetric $SU(N_B)$ representations
$\{S_{n_a}\}$, $\{S_{\bar{n}_a}\}$ and $\mc{Z}_{B,k}$ is the trace of
$\te^{-\beta(H - \mu Q)}$ over the Hilbert space $\mc{H}_{B,k}$ which
is obtained by imposing the quantum singlet constraint on the Fock
space $\mc{H}_{B,\rm Fock}$ (equivalently, it is the space of
conformal blocks with representations $\{S_{n_a}\}$,
$\{S_{\bar{n}_a}\}$):
\begin{equation}\label{IBtrace}
 \mc{Z}_{B,k} = \tr_{\mc{H}_{B,k}}\left(\te^{-\beta(H - \mu Q)}\right)\ ,\quad \mc{I}_{B,k} = \te^{-\beta \mc{V}_2 N_B E_{B,\rm vac}} \mc{Z}_{B,k}\ .
\end{equation}
In the next subsection we will demonstrate that $\mc{I}_{B,k}$
reduces, in the large $N_B$ and large $\mc{V}_2$ limit, to
$\wt{\mc{I}}_{B,k}$ defined in \eqref{uintebb}, \eqref{znsb} including
the terms proportional to $\Theta(c_B - |\mu|)$. It thus follows that
$\mc{I}_{B,k}$ is the finite $N_B$ and ${\cal V}_2$ version of
$\wt{\mc{I}}_{B,k}$.

\subsubsection{Reproducing the $\Theta(|\mu|-c_B)$ term in the large $N$ limit}
We now show that the cutoff in the summation over $n_a$ in
\eqref{ZBfocktrunc} indeed reproduces the $\Theta$ term in
\eqref{znsb} in the large $N$ limit. Let us define the quantity $Q(y)$
\begin{equation}
  Q(y) = \sum_{n = 0}^{|k_B|} y^n \chi_{\vgap S_{n}}(\ul{w})\ ,
\end{equation}
where $ y = \te^{-\beta(E - \mu)}$ is the usual Boltzmann factor for a
particle state with energy $E$, and $U$ is a diagonal $SU(N)$
(resp.~$U(N)$) matrix with entries $w_i$, $i=1,\ldots, N$ satisfying
the conditions \eqref{suwconst} (resp.~\eqref{uwconst}). It is
straightforward to check that $Q(y)$ is also given by
\begin{equation}\label{losims}
  Q(y) = \prod_{i = 1}^{N_B} \frac{1}{1 - y w_i }\ \bigg|_{|k_B|}\ .
\end{equation}
We will now manipulate \eqref{losims} to obtain a version of this
formula in which the truncation is automatic.

Let us assume that we are working in the Type I theory (this assumption
is convenient, but does not change the final answer in the large $N$
limit). In that case the eigenvalues $w_i$ of the $U(N)$ matrix $U$
are $N_B$ distinct $|\kappa_B|^{\rm th}$ roots of $(-1)^{N_B-1}$.  We
use the symbol $w'_{i'}$, $i' = 1,\ldots,|k_B|$ to denote the remaining
$|\kappa_B|^{\rm th}$ roots of $(-1)^{N_B-1}$. Multiplying the
numerator and denominator of \eqref{losims} by
\begin{equation}\label{expinla}
\prod_{j'=1}^{|k_B|} \left( 1-y w'_{j'} \right) \ , 
\end{equation} 
we see that \eqref{losims} can be rewritten as
\begin{equation}\label{losnex}
  Q(y)= \frac{ \prod_{j'=1}^{|k_B|} \big( 1- y w'_{j'} \big)}{\prod_{a=1}^{|\kappa_B|}\big( 1- y q_a \big) }\ \Bigg|_{|k_B|} \ , 
\end{equation}
where the product in the denominator runs over all
$|\kappa_B|^{\rm th}$ roots of $(-1)^{N_B-1}$ denoted by $q_a$,
$a =1,\ldots,|\kappa_B|$.  Using the identity
\begin{equation} \label{identoo}
\prod_{a=1}^{|\kappa_B|}\left( 1-y q_a \right)= 1 - y^{|\kappa_B|} (-1)^{N_B-1} \ , 
\end{equation} 
we find that \eqref{losnex} simplifies to 
\begin{equation}\label{losnexen}
Q(y)=  \frac{1}{1- y^{|\kappa_B|} (-1)^{N_B-1}}\prod_{j'=1}^{|k_B|} \big( 1- y w'_{j'} \big)\ \Bigg|_{|k_B|}  \ . 
\end{equation}
Now the product over $j'$ in \eqref{losnexen} is a polynomial of
degree $|k_B|$, while the prefactor has a power series expansion in
$y^{|\kappa_B|}$. As $|\kappa_B|$ is greater than $|k_B|$, it follows
that it does not contribute to the truncation of the RHS to order
$|k_B|$. In other words, we have the following explicit expression for
the degree $|k_B|$ polynomial $Q(y)$ \eqref{losims}:
\begin{equation}\label{losfinbose}
   \prod_{i = 1}^{N_B} \frac{1}{1 - y w_i }\ \bigg|_{|k_B|} = \prod_{j'=1}^{|k_B|} ( 1-y w'_{j'}) \ .
\end{equation}
Multiplying and dividing the right hand side by
$\prod_{i = 1}^{N_B} (1 - y w_i)$ and using \eqref{identoo} we get
\begin{equation}
  Q(y) = \left(1 - y^{|\kappa_B|}(-1)^{N_B-1}\right) \prod_{i=1}^{N_B}\frac{1}{1-y w_i}\ . 
\end{equation}
We will now use the exact expression \eqref{losfinbose} to obtain a
convenient large $N$ approximation for $\log Q(y)$. Let us first note
that
\begin{equation}\label{losimf}
 \prod_{i=1}^{N_B}\frac{1}{1-y w_i} = \exp \Big(-\tr\log\big(1-y U\big) \Big) \ ,
\end{equation}
where $\tr$ is a trace over the fundamental gauge indices on
$U$. Inserting \eqref{losimf} into \eqref{losfinbose} we conclude that
\begin{equation} \label{logvor}
Q(y) = \exp \Big(-{\rm \tr}\log\big(1-y U\big)
+ \log \left(1- y^{|\kappa_B|} (-1)^{N_B-1}\right) \Big) \ . 
\end{equation} 
The derivation of \eqref{logvor} is correct only for $y<1$ because
\eqref{losimf} only holds when this is the case. But, as $Q(y)$ is a
polynomial of degree $|k_B|$, the final answer applies to all values
of $y$ by analytic continuation of the RHS.

Let us now take the 't Hooft large $N$ limit. In this limit
$ \log \left(1- y^{|\kappa_B|} (-1)^{N_B-1}\right)=0$ when $y < 1$ but
equals $|\kappa_B| \log w$ when $y > 1$.\footnote{We ignore the
  imaginary term in this logarithm arising from the factor
  $(-1)^{N_B}$ since it is of order unity and is small compared to the
  leading term which is order $|\kappa_B|$ in this limit.} It follows
that in the large $N$ limit
\begin{equation} \label{llargen}
  \log Q(y) =-\tr\log\left(1-y  U\right) + |\kappa_B| \Theta(y-1) \log y \ .
\end{equation} 
There is a similar formula for the antiparticle state with energy $E$
where $y$ above is replaced by $\tl{y} = \te^{-\beta(E + \mu)}$ and
$U$ is replaced by $U^\dag$:
\begin{equation}\label{llargenanti}
  \log \tl{Q}(\tl{y}) = -\tr\ln\left(1-\tl{y}  U^\dag \right) + |\kappa_B| \Theta(\tl{y}-1) \log \tl{y} \ .
\end{equation}
In the large $N$ limit, the trace over the fundamental gauge indices
on $U$ is replaced by an integral over the holonomy distribution
$\rho_B(\alpha)$:
\begin{equation}
  \tr \to N_B \int_{-\pi}^\pi d\alpha\, \rho_B(\alpha)\ .
\end{equation}
Taking these facts into account and integrating over the different
energies leads us to the following modified formula for the Fock space
trace:
\begin{align}\label{betbosdet}
-\frac{N_B T^3}{2 \pi} &\Bigg[\frac{\hat{c}_B^3}{3} - \int_{\hat{c}_B}^{\infty}  d\hat{\e}\ \hat{\e}  \int_{-\pi}^{\pi}  d\alpha\, \rho_B(\alpha)\left(\log\big(1-\te^{-\hat{\e}-\hat{\mu}-\i \alpha } \big) +\log\big(1-\te^{-\hat{\e}+ \hat{\mu}+\i \alpha }\right)\nonumber \\
&\ \ + \int_{\hat{c}_B}^{\infty}  d\hat{\e}\ \hat{\e} \left(\frac{\Theta({\hat \mu}-{\hat \epsilon})  ({\hat \mu}-{\hat \epsilon})}{|\lambda_B|}  
+ \frac{\Theta(-{\hat \mu}-{\hat \epsilon}) (-{\hat \mu}-{\hat \epsilon} ) }{|\lambda_B|}  \right) \Bigg] \ ,
\end{align}
where the $\tr\log(1 - y U)$ and $\tr\log(1-\tl{y}U^\dag)$ in
\eqref{llargen} and \eqref{llargenanti} give the first line of
\eqref{betbosdet} and the term with the $\Theta(y-1)$ and
$\Theta(\tl{y}-1)$ give the term in the second line of
\eqref{betbosdet}.

Performing the integral over ${\hat \epsilon}$ in the last lines of
\eqref{betbosdet}, we find that \eqref{betbosdet} reduces to
\begin{align}
&-\frac{N_B T^3 }{2 \pi}  \bigg[\frac{\hat{c}_B^3}{3} - \int_{\hat{c}_B}^{\infty}  d\hat{\e}\ \hat{\e}  \int_{-\pi}^{\pi}  d\alpha\, \rho_F(\alpha)\left(\log\big(1-\te^{-\hat{\e}-\hat{\mu}-\i \alpha }\big)+\log\big(1-\te^{-\hat{\e}+ \hat{\mu}+\i \alpha }\big)  \right)\nonumber  \\
&\qquad\qquad\ \ + \Theta(|{ \mu}|-{ c}_B ) \ \frac{(|\hat{\mu}| - {\hat c}_B)^2(|\hat{\mu}| + 2 {\hat c}_B)}{6 |\lambda_B| } \ \bigg] \ ,
\end{align}
which is the same as the quantity in \eqref{znsb}, as claimed.

\subsection{Discussion} \label{discoo} 

The starting point of the current paper is the previously obtained
results for the $S^2 \times S^1$ partition functions of Chern-Simons
matter theories computed only in the simultaneous large volume and
large $N$ limit. In Section \ref{inter} we argued that these known
results may be recast in the form
\begin{align}\label{extmatrbrep}
&\mc{Z}_{S^2 \times S^1} = \text{Ext}_{\{\varphi_{\rm aux}\}}\left[ \te^{-\mc{V}_2\beta F_{\rm int}[\varphi_{\rm aux}]}\ \wt{\mc{I}}_k(c)\right]\ ,\quad \wt{\mc{I}}_k(c) = \int [dU]_{\rm CS}\  \te^{-\mc{V}_2 \beta F_{\rm Fock}[c;\rho]}\ ,
\end{align}
In this section, we have identified a natural and completely precise
finite $N$ and finite volume generalization of $\wt{\mc{I}}_k(c)$ which
we denote by $\mc{I}_k(c)$. The precise quantity $\mc{I}_k(c)$ is a
genuine partition function in its own right (i.e. its expansion in
powers of $\te^{-\beta (E-\mu Q)}$ has integer coefficients). Moreover
it reduces to $\wt{\mc{I}}_k(c)$ at large $\mc{V}_2$ and large $N$.
Moreover we obtained the expression for $\mc{I}_k(c)$ using the
completely precise finite $N$ and $k$ Verlinde formula for the WZW
theory associated with the pure Chern-Simons theory under study. It is
thus natural to wonder if the following proposed generalization of
\eqref{extmatrbrep}
\begin{align}\label{extmatrbmod}
&\mc{Z}_{S^2 \times S^1} = \text{Ext}_{\{\varphi_{\rm aux}\}}\left[ \te^{-\mc{V}_2\beta F_{\rm int}[\varphi_{\rm aux}]}\ {\mc{I}}_k(c)\right]\ ,\quad 
\end{align}
where ${\mc{I}}_k(c)$ is ${\mc{I}_{B,k}}(c_B)$ in the case of bosons, or
${\mc{I}_{F,k}}(c_F)$ in the case of fermions, has a larger range of
validity than \eqref{extmatrbrep} to which it manifestly reduces in
the large $N$ and large $\mc{V}_2$ limit.

\subsubsection{Finite $N$, large $\mc{V}_2$ and large $N$, finite
  $\mc{V}_2$}

It seems extremely unlikely to us that \eqref{extmatrbmod} can apply
unmodified at generic temperatures at finite $N$ (see below for more
remarks about the low temperature limit). The non-planar Feynman
diagrams that contribute to the scattering of thermal excitations at
finite $N$ are generated by the path integral over auxiliary fields
like $\varphi_{\rm aux}$. Simply extremizing, as in
\eqref{extmatrbmod}, seems too simple to account for those
interactions that are mediated by non-planar graphs.

A similar objection to that outlined in the last paragraph does not
apply to the generalization to finite $\mc{V}_2$ while staying at
large $N$. Indeed as we have already explained in Section \ref{fici},
the partition function \eqref{ZRBungauged} (the analog of
\eqref{extmatrbmod} in the closest large $N$ ungauged version of the
gauged Chern-Simons matter theory), does in fact hold even at finite
$\mc{V}_2$. Motivated by these observations it is natural to wonder
whether an expression of the form \eqref{extmatrbmod} may be exact at
all values of $\mc{V}_2$ (and at every value of the 't Hooft coupling)
in the strict large $N$ limit. As the free spectrum of a free Fermion
of mass $c$ and free Boson at mass $c$ differ from each other on a
sphere of finite radius, it does not seem possible for the expressions
\eqref{extmatrbmod} to apply unmodified at finite volume (as the
fermionic and bosonic results would not map to each other violating
the conjectured duality between these theories). However it seems
entirely possible to us that an expression of the structural form
\eqref{extmatrbmod} - but with the quantity ${\mc{I}}_k(c)$ replaced
by the projection on to WZW singlets of a Fock space defined on a
suitably renormalized single particle energy spectrum\footnote{The
  idea here is that the renormalization would ensure that the single
  particle spectrum agrees across dual theories.} applies at every
value of the spherical radius; see the discussion in the last
paragraph of Section \ref{tpfv}. We leave a detailed investigation of
this suggestion to future work.

\subsubsection{Low temperatures, large volume} \label{ltlv}
There is one more limit in which we expect a simplified version of
\eqref{extmatrbmod} to correctly capture the  free energy of
Chern-Simons matter theories even at finite $N$ and $k$. The limit in
question is a simultaneous non-relativistic large volume
limit. Consider a limit in which the temperature much lower than the
thermal mass $T/c \ll 1$ but the volume of the sphere is
simultaneously scaled to be much larger than
$\frac{\te^{\frac{c}{T}}}{c^2}$. At these low temperatures the density
of fundamental and antifundamental excitations is of order
$c^2\te^{-\frac{c}{T}}$, and so is very small. It follows that the
chance that two particles will collide against each other is
negligible. Consequently we expect that the only effect of the part of
\eqref{extmatrbmod} which captures the effect of contact interactions
between particles - namely the extremization of $F_{\rm int}$ over
$\varphi_{\rm aux}$ - is to set the mass variable $c$ to the zero
temperature pole mass $c_0$ of the system. Recall that the
extremization over $\varphi_{\rm aux}$ was precisely the part of
\eqref{extmatrbmod} that seemed inextricably tied to the large $N$
limit; the remaining part of \eqref{extmatrbmod} (namely the factor
$\mc{I}_k(c)$) is precise even at finite $N$. In the limit described in
this paragraph, therefore, we expect the partition function of our
system to be given exactly by the extremely simple and completely
precise formula
\begin{align}\label{exm}
\mc{Z}_{S^2 \times S^1} = {\mc{I}}_k(c_0) ,\quad 
\end{align}
where $c_0$ is the zero temperature pole mass of our excitations. We
conjecture, in other words, that in the low temperature (and thus,
effectively non-relativistic) and simultaneously large
volume\footnote{The large volume limit is important to ensure that our
  system is not `empty' i.e. that it hosts a thermodynamically large
  number of particle excitations.} limit, the Hilbert space of our
Chern-Simons matter theory is precisely the space of quantum singlets
with the free Fock space Hamiltonian and no further corrections even
at finite values of $N$ and $k$. Once again we leave a careful study
of the conjecture \eqref{exm} to future work.

\section{The impact of the quantum singlet constraint in the limit
  $\mc{V}_2 \to \infty$} \label{winf}

In this section, we look at the quantum singlet constraint in the
limit of infinite spatial volume $\mc{V}_2$ with every other parameter
in the theory, including $N$ and $k$ fixed to finite values.  In this
limit the total number of (fundamental and antifundamental) particles
present on the sphere, at any nonzero temperature $T$, tends to
infinity. At the technical level the analysis in this section is
closely related to that in Section \ref{lwcoo}. Before turning to
technicalities, however, we first pause to motivate the study of the
large volume limit on physical grounds.

\subsection{Contrasting the `classical' and `quantum' Gauss laws at
  large volume}

We have discussed two kinds of gauge singlet conditions on gauged
matter theories. The first arises in the simple or `classical' problem
discussed in Section \ref{YMcase} which corresponds to Yang-Mills
gauged matter theories at zero gauge coupling. After rescaling so that
the vacuum energy on the sphere is set to zero, the partition function
takes the form
\begin{equation}\label{IYM}
  \mc{I}_{\rm cl} =  {\rm Tr}_{{\cal H}_{\rm cl}} \left(
    \te^{-\beta ( H - \mu Q)} \right)\ ,
\end{equation}
where the RHS of \eqref{IYM} was defined in \eqref{intou}. As
explained in Section \ref{YMcase}, ${\cal H}_{\rm cl}$ is the
projection of the Fock space onto the space of group theory singlets
which we call the space of `classical' singlets (this arises from the
usual Gauss law of Yang-Mills theory).

The second kind of singlet condition arises in the large $N$ limit of
Chern-Simons gauged matter theories which we have treated in detail in
the previous section. The analog of $\mc{I}_{\rm cl}$ \eqref{IYM} in
Chern-Simons gauged matter theories is
\begin{equation}\label{ICS}
  \mc{I}_k = \tr_{\mc{H}_k}\left( \te^{-\beta(H - \mu Q)}\right)\ .
\end{equation}
As we have discussed in the previous section, ${\mc{H}_k}$ is the
projection of the Fock space $\mathcal{H}_{\rm Fock}$ onto the Hilbert
space of \emph{quantum} singlets (which is obtained from the
Chern-Simons Gauss law or equivalently, which is the space of WZW
conformal blocks with representations corresponding to the various
states in Fock space).

In sectors of the Fock space where the particle occupation number is
less than or of order $N^2$, the removal of states due to the singlet
condition results in a very substantial reduction of entropy in both
the `classical' case \eqref{IYM} and the `quantum' case \eqref{ICS}.
On the other hand, when the particle occupation number is much greater
than $N^2$, the effect of the projection onto singlets is
qualitatively different in the `classical' and `quantum' situations,
as we now explain.

In sectors in which the particle occupation number and energy are each
large compared to $N^2$, the fractional reduction in the entropy of
states (resulting from the projection onto the space of singlets) is
negligible in the `classical' problem \cite{Aharony:2003sx}. This fact
may be qualitatively understood as follows.  The overall Gauss law
constraint consists of $N^2$ equations and crucially, is independent
of the particle number of the multi-particle state under
consideration. It follows that at occupation numbers much larger than
$N^2$, this constraint results in only a small fractional reduction of
the phase space volume of our system. In the limit
$\mc{V}_2 \to \infty$, it follows that the thermodynamics of the
singlet sector $\mc{H}_{\rm cl}$ becomes effectively identical to the
Fock space $\mc{H}_{\rm Fock}$ \cite{Aharony:2003sx} at fixed temperature
and chemical potential.

In the `quantum' problem, we demonstrate in this paper that the
projection onto the quantum singlet sector has a qualitatively
significant impact on the entropy of states even when the particle
number and energy of the states are large in units of $N^2$. This
reduction is a consequence of the fact that the singlet condition
follows from the fusion rules of the WZW model where the number of
degrees of freedom in the WZW primaries are drastically reduced after
fusion.

To understand this it is useful to consider a simple example. Recall
that the fusion of primaries of spin $j_1$ and $j_2$ in $SU(2)_k$ WZW
theory produces primaries in representations with spins ranging from
$|j_1-j_2|$ to ${\rm max}(j_1+j_2, k-j_1-j_2)$. The total number of
degrees of freedom in the primaries that go into this fusion is
$(2 j_1+1)(2j_2+1)$ while the total number of degrees of freedom in
the primaries that come out is
\begin{equation}\label{dofc}
\sum_{j=|j_1-j_2|}^{{\rm max}(j_1+j_2, k-j_1-j_2)} (2j+1) = \left\{\renewcommand{\arraystretch}{1.5}\begin{array}{cc} (2j_1+1)(2j_2+1) & {\rm if}\ j_1+j_2 \leq \frac{k}{2} \\ (k-2j_1+1)(k-2j_2+1) & {\rm if}\ j_1+j_2 \geq \frac{k}{2}\end{array}\right.
\end{equation} 
As $(k-2j_1+1)(k-2j_2+1) < (2j_1+1)(2j_2+1)$ when
$j_1+j_2> \frac{k}{2}$, it follows that fewer degrees of freedom come
out of the fusion than go into it in this second case\footnote{On
  the other hand when $j_1+j_2 \leq \frac{k}{2}$, the fusion rules
  coincide with that of the ordinary Lie group $SU(2)$ and the degrees
  of freedom in the primaries are conserved.}.  Repeated fusion of
representations generically encounters the second case in a finite
fraction of fusion processes. Suppose $L$ is the number of primaries
to be fused and $L$ is large. It follows that the ratio of the number
of degree of freedom that come out of the fusion process, to the
number of degrees of freedom put into it, is of order $\te^{-aL}$
where $a$ is an order one number whose precise value depends on
details. In the thermodynamic limit $L \to \infty$, it follows that
the entropy of the degrees of freedom that survive at then end of the
repeated fusion processes is an order one fraction of the entropy
entering the fusion.

In the rest of this section we explore this phenomenon quantitatively
by studying the integrals $\mc{I}_{F,k}$ \eqref{unimm},\eqref{IFtrace}
and $\mc{I}_{B,k}$ \eqref{uintebb},\eqref{IBtrace} in the limit
${\cal V}_2 / N \to \infty$ at fixed temperature and chemical
potential. As anticipated above, unlike in the `classical' case,
$\mc{I}_{F,k}$ and $\mc{I}_{B,k}$ do not simply reduce to the
partition function of the Fock space ${\cal H}_{F,\rm Fock}$ and
$\mc{H}_{B,\rm Fock}$ in this limit but do nonetheless take a simple
universal form, as we now explain.

\subsection{Simplification at large ${\cal V}_2$} \label{slv}

Let us study the limit of the partition functions $\mc{Z}_{F,k}$
\eqref{IFtrace} and $\mc{Z}_{B,k}$ \eqref{IBtrace} in the limit where
the spatial volume $\mc{V}_2$ is much larger than any other parameter
in the theory (recall that the $\mc{Z}_{F,k}$ and $\mc{Z}_{B,k}$
differ from their $\mc{I}_{F,k}$ and $\mc{I}_{B,k}$ counterparts by a
factor containing the vacuum energy).  The finite $N$ and $k$
partition functions $\mc{Z}_{F,k}$ and $\mc{Z}_{B,k}$ (and their large
$N$ counterparts \eqref{unimm} and \eqref{uintebb}) simplify in this
regime for the following technical reason.

The Fock space free energies $F_{F,\rm Fock}$ and $F_{B,\rm Fock}$
appear in $\mc{Z}_{F,k}$ and $\mc{Z}_{B,k}$ (up to the vacuum energy
term) as
\begin{equation}
  \mc{V}_2 \beta \times F_{\rm Fock}\ .
\end{equation}
The contribution from the measure over $U$ comes from the Vandermonde
factor
\begin{equation}\label{vandermonde}
  \prod_{1 \leq i < j \leq N} |w_i - w_j|^2\ ,
\end{equation}
which is crucially independent of the spatial volume $\mc{V}_2$. Thus,
in the large volume regime, the summation over the holonomies $U$ in
$\mc{Z}_{F,k}$ and $\mc{I}_{B,k}$ is dominated by the configuration
which minimizes the free energy factor\footnote{In the large $N$
  limit, where the volume scales as $N$ and no faster, the free energy
  factor and the measure over the holonomy $U$ are both of order $N^2$
  and there is an interesting interplay between the two factors in the
  evaluation of the partition function. However, in the limit where
  $\mc{V}_2$ scales to infinity faster than $N$, the contribution from
  the Vandermonde factor is sub-leading and the holonomy configuration
  that dominates in this limit is again the one which minimizes the
  free energy factor.}.

Now, it is well known that the holonomy configuration $U$ that
minimizes the free energy is the one in which all eigenvalues $w_i$ of
$U$ are as close as they can be to unity \cite{Jain:2013py}. In the
case of the `classical' problem \eqref{IYM}, the eigenvalues are free
to take any values, and so the configuration that dominates in the
limit ${\cal V}_2 \to \infty$ is $U= \mathbb{1}_N$ i.e.~all
eigenvalues are unity.

In the case of the `quantum' problem i.e.~the Chern-Simons matter
theory, the Verlinde formula described in Section \ref{verlinde} tells
us that the $N$ different eigenvalues are all forced to be distinct
(in particular, are separated by at least $2\pi / \kappa$) and so
cannot all be equal to unity. Instead, we expect the eigenvalue
configuration that dominates in this limit is one in which the
eigenvalues scatter themselves around unity in the closest possible
manner. This expectation has been rigorously demonstrated in the large
$N$ limit (see \cite{Aharony:2012ns, Jain:2013py}), and we have also
presented an argument for the same conclusion at finite $N$ and $k$ in
Section \ref{lwcoo}. As we have already explained in Section
\ref{lwcoo} the allowed diagonal unitary matrix that is closest to
unity is $U^{(0)}$:
\begin{equation}\label{configon}
  U^{(0)} = \diag\left\{\te^{-2\pi\i(N-1)/2\kappa}, \te^{-2\pi\i(N-3)/2\kappa},\ldots, \te^{2\pi\i(N-3)/2\kappa}, \te^{2\pi\i(N-1)/2\kappa}\right\}\ .
\end{equation}
It follows that, in the large volume limit, the summations over the
eigenvalue configurations in \eqref{finiteNsingF} and
\eqref{finiteNsingB} simplify to a single term; the term in which the
eigenvalue configuration is $U^{(0)}$. The simplified expressions for
$\mc{Z}_{F,k}$ and $\mc{Z}_{B,k}$ in the $\mc{V}_2 \to \infty$ limit
are given by
\begin{align}\label{largeV2IFIB}
  \mc{Z}_{F,k} &= C_F \times \prod_{a=1}^\infty \left( \sum_{n_a=0}^{N_F} \te^{-n_a \beta \left( E_a-\mu \right) } \ \chi_{\vgap A_{n_a}}(\ul{w}_{0}) \right) 
                 \left( \sum_{{\bar n}_a=0}^{N_F} \te^{-{\bar n}_a \beta \left( E_a+\mu \right) } \  \bar\chi_{\vgap A_{\bar{n}_a}}(\ul{w}_{0}) \right)\ ,\nonumber\\
  \mc{Z}_{B,k} &= C_B \times \prod_{a=1}^\infty \left( \sum_{n_a=0}^{|k_B|} \te^{-n_a \beta \left( E_a-\mu \right) } \ \chi_{\vgap S_{n_a}}(\ul{w}_{0}) \right) 
                 \left( \sum_{{\bar n}_a=0}^{|k_B|} \te^{-{\bar n}_a \beta \left( E_a+\mu \right) } \  \bar\chi_{\vgap S_{\bar{n}_a}}(\ul{w}_{0}) \right)\ ,
\end{align}
where the constants $C_F$ and $C_B$ are given by the Chern-Simons
corrected Haar measure
\begin{equation}
   \frac{1}{|\kappa|^{N-1}} \prod_{1\leq i < j \leq N} \left|w_{0i} - w_{0j}\right|^2 \ ,
\end{equation}
evaluated on the holonomy $U^{(0)}$ with eigenvalues $w_{0i}$,
$i=1,\ldots,N$. These constants can be ignored for thermodynamic
purposes since they are independent of the volume, temperature and
chemical potential.

At this stage, it is useful to introduce the symbol $q$ to denote the
primitive $|\kappa|^{\rm th}$ root of unity:
\begin{equation}\label{qdef}
  q = \te^{2\pi\i / |\kappa|}\ ,
\end{equation}
where $|\kappa| = |\kappa_B| = |\kappa_F|$. In terms of $q$, the
dominant holonomy matrix $U^{(0)}$ is
\begin{equation}
  U^{(0)} = \diag\left\{q^{-(N-1)/2}, q^{-(N-3)/2}, \ldots, q^{(N-3)/2}, q^{(N-1)/2}\right\}\ ,
\end{equation}
and the fermionic and bosonic partition functions can be written as
\begin{align}\label{qlargeV2IFIB}
  \mc{Z}_{F,k} &= C_F \times \prod_{a=1}^\infty \prod_{j=-(N_F-1)/2}^{(N_F-1)/2} \left(1 + \te^{-\beta(E_a-\mu)} q^j\right)\left(1 + \te^{-\beta(E_a+\mu)} {q}^j\right)\ ,\nonumber\\
  \mc{Z}_{B,k} &= C_B \times \prod_{a=1}^\infty  \prod_{j=-(N_B-1)/2}^{(N_B-1)/2} \left(\frac{1}{1 - \te^{-\beta(E_a-\mu)} q^j}\right)_{|k_B|} \left(\frac{1}{1 - \te^{-\beta(E_a+\mu)} {q}^j}\right)_{|k_B|}\ ,
\end{align}
where the subscript $|k_B|$ in the two factors above stands for
truncation at degree $|k_B|$ of the power series in
$\te^{-\beta(E_a - \mu)}$ and $\te^{-\beta(E_a + \mu)}$
respectively. We next discuss some identities involving the Fock space
factors above that will be useful in interpreting the above
expressions for the partition function.

\subsection{Some $q$-number identities}
We briefly review a few mathematical definitions and results involving
the parameter $q$ all of which are derived in Appendix \ref{qi}. First
we define the $q$-analog $[r]_q$ of a real real number $r$ as
\begin{equation}\label{defqnom}
[r]_q=\frac{q^{r/2}-q^{-r/2}}{q^{1/2}-q^{-1/2}} \ . 
\end{equation} 
For $q = \te^{2\pi\i/\kappa}$, it follows from the definition that
\begin{equation}\label{identsq}
  [m]_q= [2n\kappa +m]_q=-[2n\kappa -m]_q= -
[(2n+1)\kappa+m]_q= [(2n+1)\kappa-m]_q\ ,
\end{equation} 
where $n$ is any integer. In particular, we have
\begin{equation}\label{identsqsp}
  [m]_q=[\kappa-m]_q\  ,\quad[n\kappa]_q=0\ .
\end{equation} 
Note also that in the limit $\kappa \to \infty$, the parameter
$q \to 1$ and thus the $q$-analog of a number goes to the number
itself in this limit. That is,
\begin{equation}
  \lim_{q\to 1} \ [r]_q \ = \ r\ .
\end{equation}
The $q$-analog of the factorial function is defined by
\begin{equation}\label{qnfacdef} 
[m]_q!= [1]_q [2]_q \cdots [m]_q \ , 
\end{equation} 
and the $q$-analog of the binomial coefficient $\binom{n}{m}$ is the
$q$-binomial coefficient
\begin{equation} \label{defqchose}
\binom{n}{m}_{q} = \frac{[n]_{q}!}{[m]_{q}![n-m]_{q}!} \ . 
\end{equation}
We summarize a set of $q$-binomial identities which will turn out to
be useful in what follows (see Appendix \ref{qi} for proofs).
\begin{enumerate}
\item Using $|\kappa_B| = |\kappa_F| = N_B + |k_B| = N_F + |k_F|$, we have the following identity
\begin{equation}\label{chooseidentmain}  
\binom{|k_B|}{n}_{q}= \binom{N_B +n-1}{n}_q \ . 
\end{equation}
\item The product over the eigenvalues $q^j$ of $U^{(0)}$ in
  $\mc{I}_F$ in \eqref{qlargeV2IFIB} can be simplified as
  \begin{equation}\label{qbinfr}
    \prod_{j=-(N_F-1)/2}^{(N_F-1)/2} (1+q^{j} x) = \sum_{n=0}^{N_F}  \binom{N_F}{n}_{q} x^n \ . 
  \end{equation}
  where $x$ is either $\te^{-\beta(E_a-\mu)}$ or
  $\te^{-\beta(E_a+\mu)}$.
  
\item Similarly, the product over eigenvalues $q^j$ of $U^{(0)}$ in
  $\mc{I}_B$ in \eqref{qlargeV2IFIB} can be simplified as
  \begin{equation}\label{finbosansmain}
    \prod_{j=-(N_B-1)/2}^{(N_B-1)/2} \frac{1}{(1-q^{j} x)}\, \Bigg|_{|k_B|} = \sum_{n=0}^{|k_B|}
    \binom{N_B +n-1}{n}_q  x^n  \ . 
  \end{equation}
  where $x$ is either $\te^{-\beta(E_a-\mu)}$ or
  $\te^{-\beta(E_a+\mu)}$.
\end{enumerate}

\subsection{Interpolation between free bosons and free fermions} \label{interp}
Using the $q$-analog identities in the previous subsection, the
partition functions $\mc{Z}_{F,k}$ and $\mc{Z}_{B,k}$ in
\eqref{qlargeV2IFIB} can be written as (ignoring terms independent of
temperature and chemical potential)
\begin{equation}\label{largeV2form}
  \mc{Z}_{F,k} = \prod_{a=1}^\infty\mf{z}_F(y_a)\, \mf{z}_F(\tl{y}_a)\ ,\quad   \mc{Z}_{B,k} = \prod_{a=1}^\infty\mf{z}_B(y_a)\, \mf{z}_B(\tl{y}_a)\ ,
\end{equation}
where $y_a = \te^{-\beta(E_a - \mu)}$ and
$\tl{y}_a = \te^{-\beta(E_a + \mu)}$, and $\mf{z}_F$, $\mf{z}_B$ are
the single particle partition functions
\begin{align}\label{fbinlvo}
  &\mf{z}_F(y) = \sum_{r=0}^{N_F} \binom{N_F}{r}_{q} y^r = \sum_{r=0}^{N_F} \binom{|k_F|+r-1}{r}_{q} y^r = \prod_{j=-(N_F-1)/2}^{(N_F-1)/2} (1+q^{j} y)\ ,\nonumber\\
  &\mf{z}_B(y) = \sum_{r=0}^{|k_B|}  \binom{N_B+r-1}{r}_{q} y^r
    = \sum_{r=0}^{|k_B|}  \binom{|k_B|}{r}_{q} y^r =  \prod_{j=-(N_B-1)/2}^{(N_B-1)/2} \frac{1}{(1-q^{j} y)}\, \Bigg|_{|k_B|} \ .
\end{align}
Note that the second term in first line and the first term in second
line (and similarly, the first term in the first line and the second
term in the second line) map to each other under the Bose-Fermi
duality map
\begin{equation}\label{BFmap}
  N_B = |k_F|\ ,\quad |k_B| = N_F\ .
\end{equation}
Thus, provided that the energy spectra $E_a$ are the same for both the
fermionic and bosonic theories on $S^2$ (which we expect to be after
including possible energy renormalizations due to self-interactions of
the matter), the two partition functions $\mf{z}_F$ and $\mf{z}_B$
map to each other under Bose-Fermi duality.

For both the bosonic and fermionic cases, the formula for the
expectation value of the number of particles in a given energy
eigenstate is
\begin{equation}\label{nform}
  n= y\partial_{ y} \log \mf{z}(y)\ .
\end{equation} 
Using the expressions in \eqref{fbinlvo} we find 
\begin{align}\label{nexp}
n_F(y) &= \sum_{j=-(N_F-1)/2}^{(N_F-1)/2} \frac{1}{y^{-1}q^{-j}+1}\ ,\nonumber \\
n_B(y) &= \left( \sum_{j=-(N_B-1)/2}^{(N_B-1)/2} \frac{1}{ y^{-1} q^{-j}-1} \right)  -  \frac{|\kappa_B|}{1-y^{-|\kappa_B|}(-1)^{N_B-1} }\ ,
\end{align}
where we have used the identity \eqref{losfinbose} to obtain
$n_B$. The expressions \eqref{nexp} are generalizations of the famous
formulae of Bose and Fermi statistics
\begin{equation}\label{nexpfam}
  n^0_{F}(y)= \frac{N_F}{y^{-1}+1}\ ,\quad n^0_{B}(y) = \frac{N_B}{ y^{-1} -1}\ .
\end{equation}
Indeed the expressions \eqref{nexp} reduce to \eqref{nexpfam} in the limit $q \to 1$
i.e.~$|\kappa| \to \infty$ with $N$ fixed (at least for $y <1$ in the
case of bosons). In the large $N$ limit, the expressions in
\eqref{nexp} become
\begin{align}\label{nexpln}
  n_F(y) &= \frac{N_F}{2 \pi |\lambda_F|} \int_{-\pi |\lambda_F|}^{\pi |\lambda_F|} d \alpha 
           \frac{1}{y^{-1}\te^{\i \alpha}+1}\ ,\nonumber \\
  n_B(y) &= \frac{N_B}{2 \pi |\lambda_B|} \int_{-\pi |\lambda_B|}^{\pi |\lambda_B|} d \alpha 
           \frac{1}{y^{-1}\te^{\i \alpha}-1} - |\kappa_B|\Theta(y-1) \ ,
\end{align}
in agreement with Equations 1.1. and 1.3 of
\cite{Minwalla:2020ysu}\footnote{Note that $n$ in this paper refers to
  the total number of particles in a given energy eigenstate summing
  over all colours. On the other hand $n$ in \cite{Minwalla:2020ysu}
  refers to the average occupation per colour state. In other words
  $n^{\rm here}= N n^{\rm there}$ where $N=N_F$ for Fermions and $N_B$
  for bosons.}.

Just like the occupation numbers reducing to the standard Fermi and
Bose formulas in the $|\kappa| \to \infty$ limit, the partition
functions $\mf{z}_F$ and $\mf{z}_B$ reduce to the free fermion and
free boson partition functions respectively (at least for $y < 1$ for
the bosons):
\begin{equation}\label{freeferbos}
  \mf{z}^0_F(y) = (1 + y)^{N_F}\ ,\quad \mf{z}^0_{B}(y) = \frac{1}{(1 - y)^{N_B}}\ .
\end{equation}
The above fact holds more generally in the limit
$\frac{|k|}{N}\to \infty$ e.g.~in the limit $|k| \to \infty$ at fixed
$N$ or when $|\lambda|\to 0$ in the 't Hooft limit.

There is a sense in which the partition functions and occupation
numbers reported in this section interpolate between the formulae for
free fermions and free bosons. In order to see this consider, say, the
fermionic single particle partition function at a fixed but very large
value of $\kappa_F$. Then the partition function is parameterized by a
single integer $N_F$ and is given by
\begin{equation}
  \mf{z}_{F}(y; N_F, |k_F|) = \mf{z}_F(y; N_F, |\kappa_F| - N_F)\ ,
\end{equation}
where we have explicitly displayed that $\mf{z}_{F}$ is a function of
the rank $N_F$ and level $k_F$. This family of partition functions
interpolate between the free partition function of $N_F$ flavours of
free fermions in a single particle state, when $N_F$ is small (say of
order unity), to the free partition function of $|\kappa_F|-N_F$
flavours of free bosons in a single particle state when
$|\kappa_F|-N_F$ is small. In between these extremes, i.e. when
$\frac{N_F}{|\kappa_F|}$ is a fraction of order unity but is strictly
less than unity, our single particle partition functions capture
neither Bose nor Fermi but a new kind of statistics.

\subsection{A note on quantum dimensions}\label{qdnote}
The single particle partition functions $\mf{z}_F$ and $\mf{z}_B$ have the following expressions in terms of characters which can be read off easily from  \eqref{largeV2IFIB}:
\begin{equation}
  \mf{z}_F(y)= \sum_{r=0}^{N_F}  \chi_{\vgap A_{r}}(\ul{w}_{0}) y^r\ ,\quad \mf{z}_B(y)= \sum_{r=0}^{|k_B|}  \chi_{\vgap S_{r}}(\ul{w}_{0}) y^r\ .
\end{equation}
Comparing these expressions with \eqref{fbinlvo}, we get the following
expressions for the characters evaluated on the distinguished
eigenvalue configuration $U^{(0)}$:
\begin{equation}\label{qdimchar}
  \chi_{\vgap A_r}(\ul{w}_{0}) = \binom{N_F}{r}_q\ ,\quad \chi_{\vgap S_r}(\ul{w}_{0}) = \binom{N_B + r - 1}{r}_q\ .
\end{equation}
The right hand sides of the above are in fact the so-called
\emph{quantum dimensions} of the totally antisymmetric and totally
symmetric $r$-box representations of the $SU(N)_k$ WZW model. We
briefly discuss quantum dimensions in this subsection.

Recall that the quantum singlet condition was most conveniently phrased
in terms of the fusion rules of the WZW model: the number of quantum
singlets in the tensor product of some number of representations in
the Chern-Simons theory is given by the number of singlets in the
fusion of the corresponding primary operators of the WZW model.

The primary operators of a WZW model, say $SU(N)_k$ for concreteness,
are finite in number and correspond to the integrable representations
of the affine Lie algebra $\wh{su}(N)_k$ at level $k$. As for the
ordinary Lie algebra $su(N)$, the representations of $\wh{su}(N)_k$
are indexed in terms of Young tableaux. Integrable representations of
$\wh{su}(N)_k$ correspond to those Young tableaux whose first row
contains at most $k$ boxes.

\textbf{Note:} Only in this subsubsection, we use the notation $R$ to
denote the representation of an ordinary Lie algebra and $\hat{R}$ for
its counterpart for the affine Lie algebra.

Let us note a simple universal fact about representation theory of
affine Lie algebras. Let the WZW torus partition function (or
equivalently, the affine character) of the primary operator $\hat{R}$
(equivalently, the representation $\hat{R}$) be denoted by
$\chi^{\rm WZW}_{\hat{R}}(\mathsf{q})$ where
$\mathsf{q} = \te^{2\pi\i\tau}$ with $\tau$ is the complex structure
parameter of the torus. The character
$\chi^{\rm WZW}_{\hat{R}}(\mathsf{q})$ diverges in the limit
$\mathsf{q} \to 1$ for any $\hat{R}$, reflecting the fact that every
representation of the affine Lie algebra has an infinite number of
states. As this divergence is the same in every representation (they
arise from the action of the modes of the WZW currents on the highest
weight state) it follows that that the limit
\begin{equation}\label{ago} 
  \lim_{\mathsf{q} \to 1} \frac{\chi^{\rm WZW}_{\hat{R}}(\msf{q})}{\chi^{\rm WZW}_{\hat{0}}(\msf{q})} = \mc{D}(\hat{R})\ ,
\end{equation} 
is finite, where $\chi_{\hat{0}}^{\rm WZW}$ is the character of the
trivial representation. The quantity $\mc{D}(\hat{R})$ is called the
\emph{quantum} dimension of the representation $\hat{R}$ of the affine
Lie algebra or equivalently, of the primary operator corresponding to
$\hat{R}$ of the WZW model.

The quantum dimension of the affine representation $\hat{R}$ is in
general different from the classical dimension of the representation
$R$ of the ordinary Lie algebra. This can be understood roughly as
follows. In the absence of null states in the representation
$\hat{R}$, the character $\chi^{\rm WZW}_{\hat{R}}(\mathsf{q})$ would
have factorized as
$\chi_{\hat{R}}^{\rm WZW}(\mathsf{q})= \chi_{\vgap R}(\mathbb{1})
\chi_{\hat{0}}^{\rm WZW}(\mathsf{q})$ where
$\chi_{\vgap R}(\mathbb{1})$ is the ordinary Lie algebra character of
the representation $R$ evaluated on the identity matrix i.e.~the
classical dimension of the representation $R$ of the ordinary Lie
algebra. Had this naive analysis been correct the quantity
$\mc{D}(\hat{R})$ defined in \eqref{ago} would have simply been given
by the classical dimension of $R$, i.e.
$d(R) = \chi_{\vgap R}(\mathbb{1})$.

However, this is inaccurate since all WZW representations $\hat{R}$
have null states. It turns out that the naive, incorrect guess
$\mc{D}(\hat{R}) = \chi_{\vgap R}(\mathbb{1})$ is modified in a very beautiful
manner after accounting for the contribution of null states. The
correct formula for $\mc{D}(\hat{R})$ is
\begin{equation}\label{drago}
  \mc{D}(\hat{R}) = \chi_{\vgap R}(\ul{w}_{0})\ ,
\end{equation} 
where $U^{(0)}$ is the diagonal matrix \eqref{configon} that appeared
in the large number of insertions limit of the Verlinde formula
\eqref{configon1} in Section \ref{lwcoo}, which we reproduce here for
convenience:
\begin{equation}
  U^{(0)} = \diag\left\{\te^{-2\pi\i(N-1)/2\kappa}, \te^{-2\pi\i(N-3)/\kappa},\ldots, \te^{2\pi\i(N-3)/2\kappa}, \te^{2\pi\i(N-1)/2\kappa}\right\}\ .
\end{equation}
As we saw earlier in this section, this matrix also appears in the
large volume limit of Chern-Simons matter theories and is also the
eigenvalue configuration dual to the trivial representation that
appears in the sum over integrable representations in the original
form of Verlinde formula \eqref{verfor} (also, see footnote
\ref{verfoot}).

One way to derive this result is as follows. The high `temperature'
limit of the WZW character in \eqref{ago} is given in terms of the
Verlinde $\mc{S}$-matrix\footnote{ Recall that the Verlinde
  $\mc{S}$-matrix implements the $\tau \to -1/\tau$ transformation on
  the characters of integrable representations:
\begin{equation}
  \chi_{\vgap \hat{R}}(-1/\tau) = \sum_{\hat{R}'} \mc{S}_{\hat{R}\hat{R}'} \chi_{\vgap \hat{R}'}(\tau)\ .
\end{equation}
} (equation 14.238 in \cite{di1996conformal}):
\begin{equation}
  \lim_{\msf{q} \to 1} \chi_{\hat{R}}^{\rm WZW}(\msf{q}) = \mc{S}_{\hat{R} \hat{0}}\, \te^{\pi \i c / 12\tau}\ ,
\end{equation}
where $c$ is the central charge and $\msf{q} = \te^{2\pi\i\tau}$. It
follows from \eqref{ago} that
\begin{equation} \label{newformd}
\mc{D}(\hat{R}) = \frac{\mc{S}_{\hat{R} \hat{0}}}{\mc{S}_{\hat{0}\hat{0}}}  .
\end{equation}
We appeal to another striking result involving the Verlinde
$\mc{S}$-matrix which relates it to the characters of the ordinary Lie
algebra (see Section 14.6.3 in \cite{di1996conformal} for details or
Section \ref{verlindedetail} of this paper for a quick review):
\begin{equation}
  \frac{\mc{S}_{\hat{0} \hat{R}}}{\mc{S}_{\hat{0} \hat{0}}} = \chi_{\vgap R}(\ul{w}_{0})\ ,
\end{equation}
which immediately gives the formula \eqref{drago}.

The fact that $\chi_{\vgap R}(\mathbb{1}) \geq |\chi_{\vgap R}(U)|$
for any $U(N)$ matrix $U$ and the fact that the quantum dimension is a
real number implies in particular that the quantum dimension of an
integrable representation $\hat{R}$ is always smaller than the
classical dimension of the corresponding ordinary Lie algebra
representation $R$:
\begin{equation}\label{qdimineq}
  \mc{D}(\hat{R}) < d(R)\ .
\end{equation}
The quantum dimension $\mc{D}$ satisfies the fusion rules of the WZW
model:
\begin{equation}
  \mc{D}(\hat{R}_1) \mc{D}(\hat{R}_2) = \sum_{\hat{R}} \mc{N}_{\hat{R}_1 \hat{R}_2}^{\hat{R}} \mc{D}(\hat{R})\ ,
\end{equation}
where the fusion rule $\mc{N}_{\hat{R}_1 \hat{R}_2}^{\hat{R}}$ is a
positive integer that counts the number of times the representation
$\hat{R}$ appears in the fusion of $\hat{R}_1$ and $\hat{R}_2$. This
is analogous to the familiar statement that the classical dimensions
of representations satisfy the tensor product decomposition of
representations of an ordinary Lie algebra:
\begin{equation}
  d(R_1) d(R_2) = \sum_R (\mc{N}_{\rm cl})_{R_1R_2}^R d(R)\ ,
\end{equation}
where $\mc{N}_{\rm cl}$ counts the number of times $R$ appears in the
tensor product of the representations $R_1$ and $R_2$. See Appendix
\ref{proof} for another interesting property of the quantum dimension
which is extensively used in this section.

The classical dimension of a representation $R$ of $su(N)$ is given in
terms of its Young tableau as
\begin{equation}\label{classdimR}
  d(R) = \chi_{\vgap R}(\mbb{1}) = \prod_{1 \leq I < J \leq N} \frac{\ell_J - \ell_I + J - I}{J - I}\ ,
\end{equation}
where $\ell_I$, $I = 1,\ldots,N$ are the row lengths of the Young
tableau. The above can be inferred easily from the Weyl character
formula for ordinary Lie algebras. Given the formula \eqref{drago} for
the quantum dimension, it is then easy to see that
\begin{equation}\label{quandimR}
  \mc{D}(\hat{R}) = \chi_{\vgap R}({U}^{(0)}) = \prod_{1 \leq I < J \leq N} \frac{[\ell_J - \ell_I + J - I]_q}{[J - I]_q}\ ,
\end{equation}
where $[n]_q$ is the $q$-analog of the integer $n$, as defined in
\eqref{defqnom}. It is important to note that the classical dimension
\eqref{classdimR} is always a positive integer whereas the quantum
dimension \eqref{quandimR} is in general neither an integer nor a
positive real number. It is however possible to show that the quantum
dimensions of integrable representations are positive {\it real} numbers.

In particular, recall that the `classical' dimensions
of the $r$-box totally antisymmetric and $r$-box totally symmetric
representations of $su(N)$ are given by
\begin{equation}\label{classdim}
  \chi_{\vgap A_r}(\mbb{1}) = \binom{N}{r}\ ,\quad   \chi_{\vgap S_r}(\mbb{1}) = \binom{N +r - 1}{r}\ .
\end{equation}
The quantum dimensions of these representations are given in
\eqref{qdimchar} and clearly just the $q$-analogs of the above
classical dimensions.

\subsection{Universal distribution of representations} \label{udr} 

After the brief interlude on quantum dimensions, let us return to the
discussion of the large volume partition functions \eqref{largeV2form}
given by
\begin{equation}\label{largeV2form1}
  \mc{Z}_{F,k} = \prod_{a=1}^\infty \mf{z}_F(y_a) \mf{z}_F(\tl{y}_a)\ ,\quad   \mc{Z}_{B,k} = \prod_{a=1}^\infty \mf{z}_B(y_a) \mf{z}_B(\tl{y}_a)\ ,
\end{equation}
where, given our new notation for the quantum dimensions, the
partition functions $\mf{z}_F$ and $\mf{z}_B$ are simply written as
\begin{equation}\label{ZFZBqd}
\mf{z}_F(y) = \sum_{r=0}^{N_F} \mc{D}(A_r)\,y^r \ ,\quad \mf{z}_B(y) = \sum_{r=0}^{|k_B|} \mc{D}(S_r)\, y^r\ .
\end{equation}

The puzzling feature of \eqref{ZFZBqd} is the following. The partition
functions in \eqref{largeV2form1} take the form of a product of
partition functions, one for every single particle state. However each
individual factor $\mf{z}_F$ or $\mf{z}_B$ \eqref{ZFZBqd} {\it cannot}
be interpreted, by itself, as a `one particle partition function'
since $\mc{D}(A_r)$ and $\mc{D}(S_r)$ are not integers in general
(though they are positive real numbers). Recall that
$\mc{D}(A_r) < d(A_r)$ and $\mc{D}(S_r) < d(S_r)$, where
$d(A_r) = \binom{N_F}{r}$ and $d(S_r) = \binom{N_B - r +1}{r}$ are the
degeneracies of the states with $r$ fermions and $r$ bosons
respectively in the `classical' or weak-coupling limit
$\kappa \to \infty$. Thus, the non-integer `degeneracies' of the
$r$-filled states are reduced compared to the classical
situation\footnote{Everything is much simpler in the weak coupling
  limit $\kappa \to \infty$ (equivalently, $q \to 1$). The quantum
  dimensions $\mc{D}(A_n) = \binom{N_F}{n}_{q}$,
  $\mc{D}(S_n) = \binom{N + n - 1}{n}_{q}$ reduce to the classical
  dimensions $\binom{N_F}{n}$, $\binom{N_B+n-1}{n}$ respectively. In
  this limit \eqref{ZFZBqd} reduces to the ordinary single particle
  fermionic and bosonic partitions \eqref{freeferbos}. It follows that
  in the limit $\kappa \to \infty$ (and always first taking the limit
  ${\cal V}_2\to \infty$) the partition function $\mc{Z}_k$ over the
  Hilbert Space ${\cal H}_{k}$ reduces to the partition function over
  the Fock space ${\cal H}_{\rm Fock}$.}. How did this puzzling state
of affairs come about? In fact the answer to this question was already
provided in Section \ref{lwcoo} and we recall it below.

Consider any large collection of integrable representations (including
multiplicities) of $SU(N)_k$. Let the total number of these
representations be $n$. Let us now fuse all these representations and
determine the final $SU(N)_k$ content of this collection of
representations. In general, we will find $n(R)$ representations
transforming in the integrable representation $R$. What can we say
about the integers $n(R)$? For a finite number of representations,
there is nothing universal about the answer to this question since it
depends intricately on the details of the representations. On the
other hand, in the limit of large number of representations $n$
(i.e.~when $n$ is taken much larger than every parameter including $N$
and $k$) we argued in Section \ref{lwcoo} that the ratios of $n(R_1)$
and $n(R_2)$ take universal values \eqref{ratodim}
\begin{equation}\label{Nrlim}
  \frac{n(R_1)}{n(R_2)} = \frac{\mc{D}(R_1)}{\mc{D}(R_2)} \ .
\end{equation} 
We emphasize that \eqref{Nrlim} holds universally in the limit of
large number of representations, independent of the specific details
of the input representations.

The result \eqref{Nrlim} supplies an explanation of the factorized
form of \eqref{ZFZBqd}. In the infinite volume limit in which we work,
${\cal H}_{\rm Fock}$ has an infinite number of states that are
significantly occupied at any nonzero temperature $T$ and chemical
potential $\mu$. Let us set aside one particular multiparticle state
$(n_a, E_a)$ which is the energy eigenstate $E_a$ filled $n_a$ times,
and refer to the collection of all the other states in
$\mc{H}_{\rm Fock}$ as ``the sea'' (see under equation
\eqref{ratofsing} in Section \ref{lwcoo} for a very similar
discussion). As the total number of single particle sea states is very
large, the representation content of the sea is governed by the
distribution \eqref{Nrlim}.

Now the net state - that is, the state $(n_a, E_a)$ fused with the
state of the sea - must be a quantum singlet. It follows that the
total contribution to the partition function from the multiparticle
state in which $E_a$ is occupied $n_a$ times is proportional to the
number of times the conjugate of this representation ($A_{n_a}$ or
$S_{n_a}$ for fermions or bosons respectively) appears in the
sea. This number, however, is governed by \eqref{Nrlim}, explaining
the form of \eqref{ZFZBqd}.

The reduction of the dimension of the state occupied $r$ times in
$\mf{z}_F$ from $\binom{N_F}{r}$ in the genuine Fock Space
$\mathcal{H}_{\rm Fock}$ to $\binom{N_F}{r}_q$ in the final answer
(recall $\binom{N_F}{r}_{q} \leq \binom{N_F}{r}$, see
\eqref{qdimineq}) is entirely a consequence of the projection to
quantum singlets which involves an effective statistical `interaction'
between the state occupied $r$ times and the sea state (similar
statements apply to the bosons as well).

We emphasize that the factorized form \eqref{largeV2form1} is a
consequence of the universality of \eqref{Nrlim}, as a consequence of
which the effect of the sea on a particular multiparticle state does
not depend on the details of the sea (e.g.~the energy of the sea
state). This leads to a factorized partition function though the
Hilbert space itself does not factorize.

\section{Energy renormalization}\label{energy}

Recall the structure of the partition function of large $N$
Chern-Simons matter theories 
\begin{align}\label{extmatrb1}
  &\mc{Z}_{S^2 \times S^1} = \text{Ext}_{\{\varphi_{\rm aux}\}}\left[ \te^{-\mc{V}_2\beta F_{\rm int}[\varphi_{\rm aux}]}\ {\mc{I}_k}(c)\right]\ ,\quad \mc{I}_k(c) = \te^{-\beta \mc{V}_2 N E_{\rm vac}} \tr_{\mc{H}_k} \left(\te^{-\beta(H - \mu Q)}\right)\ ,
\end{align}
where $E_{\rm vac}$ is the vacuum energy density (see \eqref{IFtrace}
for the case of fermions and \eqref{IBtrace} for bosons) and
$\varphi_{\rm aux}$ are the auxiliary variables one of which is the
thermal mass $c$. Extremizing $F_{\rm int}(\varphi_{\rm aux})$ over
all auxiliary variables other than $c$ turns it into a function of
$c$, which we denote by $V_{\rm int}(c)$. Then, \eqref{extmatrb1} can
be recast as
\begin{align}\label{extmatrb2}
  &\mc{Z}_{S^2 \times S^1} = \text{Ext}_{c}\left[ \te^{-\mc{V}_2\beta V_{\rm int}(c)}\ \mc{I}_k(c)\right],
\end{align}
We see that in the large $N$ limit we obtain the full partition
function of Chern-Simons matter theories by extremizing over $c$ the
product of the completely precisely defined partition function
$\mc{I}_k(c)$ and the quantity $\te^{-\mc{V}_2\beta V_{\rm
    int}(c)}$. Recall that the structure of $F_{\rm int}$ (and hence
$V_{\rm int}(c)$) encodes the details of contact interactions in the
Chern-Simons matter theory. In this section we will give a Hamiltonian
interpretation of the partition function \eqref{extmatrb2}.

Throughout this section we work with the regular boson theory in its
unHiggsed phase but we expect all our final results to generalize to
all four classes of theories in each of their phases.

\subsection{Only forward scattering interactions contribute to the
  free energy at large $N$}
\label{fslN}

We begin the analysis of this section by reminding the reader of a
presumably well-known fact, namely that in the large $N$ limit of
vector-like theories only forward scattering interactions contribute
to the computation of the partition function in translationally
invariant situations e.g.~the thermal ensemble. In this subsection we
explain this fact in the context of the ungauged scalar theory \eqref{rblag}.
 As will be clear below,
the final result is general and applies to all the theories studied in
this section.

Consider the action \eqref{rblag}, which we rewrite here for convenience:  
\begin{align}\label{rblag1}
  &\mc{S}_{\text{RB}}[\phi] =  \int d^3 x\Bigg( (\partial_\mu \bar{\phi})(\partial^\mu \phi)+m_B^2 \bar{\phi}\phi + \frac{4\pi b_4}{\kappa_B}(\bar{\phi}\phi)^2+\frac{(2\pi)^2}{\kappa_B^2}(x_6^B+ \tfrac{4}{3}) (\bar{\phi}\phi)^3\Bigg)\ .
\end{align}
As we have explained in Section \ref{fici}, the action \eqref{rblag1} is exactly equivalent to \eqref{lagare} (which we once again rewrite for convenience)
\begin{align} \label{lagare1}
  &\mc{S}[c_B,\sigma_B, \phi] =\nonumber\\
  &\int d^3 x \ \Bigg[ \partial_\mu \bar{\phi} \partial^\mu \phi
    +c_B^2 \bar{\phi} \phi + \frac{N_B}{2 \pi} \left((m_B^2 -c_B^2)
    \sigma_B + 2 \lambda_B b_4 \sigma_B^2 + \lambda_B^2(x_6^B+
    \tfrac{4}{3}) \sigma_B^3\right) \Bigg] \ .
\end{align} 
($\sigma_B$ and $c_B^2$ are Lagrange multiplier type fields that are
path-integrated over the appropriate contours). The key simplification
in the large $N_B$ limit is that it suppresses fluctuations of the
Lagrange multipliers $c_B^2$ and $\sigma_B$. In other words, at
leading order in the large $N$ limit, $c_B^2$ and $\sigma_B$ can be
replaced by their constant saddle point values (in this step, we
assume that the thermal ensemble does not spontaneously break
translational invariance). We will now explain an important
consequence of this fact.

The original Lagrangian \eqref{rblag1} includes quartic and sextic
interaction terms. When re-expressed in momentum space, these terms
are respectively proportional to
\begin{equation}\label{scheme}
\frac{ 1}{\kappa_B}\int \prod_{i=1}^4 \frac{d^3 k_i}{(2 \pi)^3} \ (2 \pi)^3 \delta(k_1+k_2+k_3+k_4)   \   {\bar \phi}_m(k_1) \phi^m(k_2) {\bar \phi}_n(k_3) \phi^n(k_4) \ . 
\end{equation}
and 
\begin{equation}\label{sixintsch}\
  \frac{1}{\kappa_B^2} \int  \prod_{i=1}^6 \frac{d^3 k_i}{(2 \pi)^3} \ 
  (2 \pi)^3 \delta\Big(\sum_{\ell=1}^{6}k_\ell \Big)  \ {\bar \phi}_m(k_1) \phi^m(k_2) {\bar \phi}_n(k_3) \phi^n(k_4) {\bar \phi}_r(k_5) \phi^r(k_6) \ .
\end{equation}
The interaction term \eqref{scheme} allows any two particle state with
momenta $k_1$ and $k_2$ to mix (scatter) with any other two particle
state with momenta $k_3$ and $k_4$ as long as their net momentum is
conserved. Similarly the interaction \eqref{sixintsch} allows for
$3 \rightarrow 3$ scattering with arbitrary conserved momenta. This
mixing between momenta makes the interaction terms \eqref{scheme}
complicated and difficult to deal with.

In contrast, the fact that $c_B$ and $\sigma_B$ in the Lagrangian
\eqref{lagare1} are constants in the large $N$ limit tells us that the
part of the interaction terms that leads to mixing between momenta are
subleading in the large $N$ limit. At leading order in the large $N$
limit, the interaction structure of the Lagrangian \eqref{lagare1} is
actually very simple. The only interaction involving $\phi$ in this
Lagrangian is
\begin{equation}\label{schemo}
c_B^2\int \frac{d^3 k}{(2 \pi)^3}  {\bar \phi}_m(k) \phi^m(-k) \ . 
\end{equation}
\eqref{schemo} is an interaction only because the extremum value of
$c_B^2$ depends on ${\bar \phi} \phi$. Eliminating $c_B$ and $\sigma_B$
results in the interactions take the schematic form
\begin{equation}\label{scemformint}
\left( \int \frac{d^3 k}{(2 \pi)^3}  {\bar \phi}_m(k) \phi^m(-k) \right)^n \ . 
\end{equation}
From the viewpoint of diagrammatic perturbation theory, the
interaction terms \eqref{schemo} are generated by bubbles upon
bubbles.  \eqref{scemformint} describes the interaction of $2n$ bosons
in a forward scattering fashion since the $n$ momenta in the initial
state are identical to the $n$ momenta in the final state.

It follows that in the large $N$ limit (unlike at finite $N$), the
finite temperature free energy receives contributions only from
forward scattering interactions. Such interactions act momentum by
momentum; they shuffle around the energies and eigenstates
corresponding to different occupation numbers with a particular
momentum, but do not lead to mixing (or scattering) of particle states
with distinct momenta. These interactions are relatively simple to
handle.

The analysis of this subsection also applies to the study of a wider
class of configurations, namely all `states' in which $\sigma_B$ and
$c_B^2$ are constants (i.e.~translationally invariant) of order
unity. Note, however, that the analysis of this subsection does not
apply to non-translationally invariant configurations, for instance,
in the computation of local correlators or S-matrices.

\subsection{The forward scattering truncated effective Hamiltonian}

In this section we study the closest ungauged theory to the regular
boson theory \eqref{rblag1}, equivalently \eqref{lagare1}. In this
section, we construct an effective large $N$ Hamiltonian $H_{\rm eff}$
for this theory which is applicable to the study of translationally
invariant configurations like the thermal ensemble.

\subsubsection{Effective Hamiltonian from the Lagrangian written in
  terms of $c_B^2$ and $\sigma_B$}

The simplest way to obtain the Hamiltonian for the regular boson
theory with Lagrangian \eqref{rblag1} in the large $N_B$ limit is to
use the equivalent Lagrangian \eqref{lagare1} as the starting
point. This yields
\begin{align}\label{pfot} 
  H_{\rm eff}
  &= \sum_{i=1}^{N_B}  \int \frac{d^2k}{(2\pi)^2 }
    \Bigg[ {a}^{i\dag}({\vec k})   {a}_{i}({\vec k})  
    \left( {\omega}({\vec k})+   \mu \right)
    + {b}^{i}({\vec k}) { b}_i^{\dag}({\vec k})  
    \left( {\omega}({\vec k})  - \mu \right)  \Bigg]\nonumber  \\
  &\qquad\qquad\qquad +\frac{N_B {\cal V}_2}{2\pi} \left(2 \lambda_B {b}_4\sigma_B^2 + (x_6^B+ 
    \tfrac{4}{3}) \lambda_B^2\sigma_B^3 - (c_B^2-m_B^2)\sigma_B  \right)\ ,
\end{align}
with $\omega(\vec{k}) = \sqrt{c_B^2 + \vec{k}{}^2}$. Note that we have
included the contribution from the chemical potential as well to
$H_{\rm eff}$. The Hamiltonian \eqref{pfot} is a formal
(unrenormalized) Hamiltonian (note the oscillators are not normal
ordered). Adding in the contribution of the counterterms that are
effectively used in the dimensional regularization procedure turns the
Hamiltonian \eqref{pfot} into the well defined operator
\begin{align}\label{pfotnewfo} 
H_{\rm eff} &= N_B E_{B,\rm vac}(c_B)+ \sum_{i=1}^{N_B}  \int \frac{d^2k}{(2\pi)^2 }
\Bigg[ {a}^{i\dag}({\vec k})   {a}_{i}({\vec k})  
\left( {\omega}({\vec k})+   \mu \right)
+ { b}_i^{\dag}({\vec k}) {b}^{i}({\vec k})  
\left( {\omega}({\vec k})  - \mu \right)  \Bigg]\nonumber  \\
&\qquad\qquad\qquad +\frac{N_B {\cal V}_2}{2\pi} \left(2 \lambda_B {b}_4\sigma_B^2 + (x_6^B+ 
\tfrac{4}{3}) \lambda_B^2\sigma_B^3 - (c_B^2-m_B^2)\sigma_B  \right)\ ,
\end{align}
Here $\sigma_B$ and $c_B^2$ numbers rather than operators. These
numbers are determined by extremization, and so depend on the details
of the computation one performs using \eqref{pfotnewfo}; for instance
if we use \eqref{pfotnewfo} to evaluate
${\rm Tr}\, \te^{-\beta H_{\rm eff}}$ then the numbers $c_B^2$ and
$\sigma_B$ will depend on $\beta$ after extremization.

To see the dependence on the thermal state of $\sigma_B$ and $c_B^2$
in a little more detail, note that the saddle point equation for
$\sigma_B$ gives
\begin{equation}\label{cbeq}
  m_B^2 - c_B^2 + 4 \lambda_B b_4 \sigma_B + (3 x_6^B + 4)\lambda_B^2 \sigma_B^2 = 0\ ,
\end{equation}
while the saddle point equation for $c_B$ gives the following solution
for $\sigma_B$:
\begin{align}\label{valsig}
  \sigma_B& = \frac{2\pi}{\mc{V}_2 N_B} \int \frac{d^2 k }{(2\pi)^2 2\omega(\vec{k})} \sum_{i=1}^{N_B}\left\langle {a}^{i\dag}(\vec{k}) {a}_i(\vec{k}) + {b}^i(\vec{k}) {b}_i^\dag(\vec{k})\right\rangle\ ,\nonumber\\
  &= A_{\rm vac}^B(c_B) +
  \frac{2\pi}{\mc{V}_2 N_B} \int \frac{d^2 k }{(2\pi)^2 2\omega(\vec{k})} \sum_{i=1}^{N_B}\left\langle {a}^{i\dag}(\vec{k}) {a}_i(\vec{k}) + {b}^{i\dag}(\vec{k}) {b}_i(\vec{k})\right\rangle\ ,
\end{align}
which is nothing but the definition of $\sigma_B$ written in terms of
the oscillator modes of $\phi$:
\begin{equation}\label{sigphi} 
  \sigma_B = \frac{2\pi}{N_B} \langle \bar\phi\phi\rangle_0\ ,
\end{equation}
where the subscript $0$ indicates that it is the constant mode of
$\langle \bar\phi(x)\phi(x)\rangle$. The value of
$A^{B}_{\rm vac}(c_B)$ in \eqref{valsig} can be deduced by comparing
\eqref{valsig} with \eqref{sigphi} evaluated in the dimensional
regularization scheme. Assuming we are working in the limit of large
${\cal V}_2$ we find \footnote{The simplest way to see this is to use
  the fact that the value of $\sigma_B$ in the vacuum of the unHiggsed
  phase is $\frac{-c_B}{2}$, see Equation 5.6 of
  \cite{Dey:2018ykx}). }
\begin{equation} 
  A_{\rm vac}(c_B)= -\frac{c_B}{2}\ .
\end{equation} 
Equations \eqref{cbeq} and \eqref{valsig} make it immediately clear
that the effective value of $\sigma_B$ and $c_B^2$ depend on the
thermal state.
  
\subsubsection{Effective Hamiltonian from the Lagrangian written only in terms of $\phi$}

The Hamiltonian \eqref{pfotnewfo} is an unusual beast, as its
dependence on Lagrange multipliers makes it dependent on the thermal
state. The reader may wonder how this effective Hamiltonian emerges
from the original Lagrangian of the theory \eqref{rblag1} which is in
terms of the fields $\phi$ and ${\bar \phi}$ but has no Lagrange
multipliers. In this subsubsection we explain how this works.

As we will see below, the key step in obtaining the Hamiltonian
\eqref{pfotnewfo} out of the original Lagrangian \eqref{rblag1} is to
replace the interaction terms in the Lagrangian \eqref{rblag1} with
their forward scattering counterparts. Once that is done, it is not
difficult to demonstrate that the interactions in the original
Lagrangian \eqref{rblag1} lead to the Hamiltonian \eqref{pfotnewfo}
after making a Bogoliubov transformation on the creation /
annihilation operators.

In order to construct the effective Hamiltonian for the ungauged
regular boson theory, we first expand the scalar fields at $t=0$ in
terms of bare oscillators (assuming that the volume of the $S^2$ is
large, as we do throughout this section for simplicity, to allow the
use plane waves rather than spherical harmonics) as follows:
\begin{align} \label{expandfields} 
  \phi_i(x) &= \int \frac{d^2\vec{p}}{(2\pi)^2\sqrt{2{\tilde\omega}({\vec{p})}}} \ \Big( {{\tilde a}}_i(\vec{p}) \te^{\i {\vec p}\cdot {\vec x} } + {{\tilde b}}_i^\dag (\vec{p}) \te^{-\i {\vec p}\cdot {\vec x} } \Big)\ ,\nonumber\\
  { \dot \phi}_i(x) &= -\i \int  \frac{d^2\vec{p}  }{(2\pi)^2} \sqrt{\frac{{\tilde \omega}({\vec{p})}}{2}}  \ \Big( {{\tilde a}}_i(\vec{p}) \te^{\i {\vec p}\cdot {\vec x} } - {{\tilde b}}_i^\dag (\vec{p}) \te^{-\i {\vec p}\cdot {\vec x} } \Big)\ ,
\end{align}
and the complex conjugate expressions for $\bar\phi^i$ and
$\dot{\bar\phi}^i$, with
$\tl{\omega}(\vec{k}) = \sqrt{\vec{k}{}^2 + m_B^2}$.

As explained in Section \ref{fslN}, for computations performed in a
class of translationally invariant configurations, we are allowed to
replace all contact interaction terms in the Hamiltonian by their
forward scattering counterparts at leading order in the large $N$
limit.  It follows that the effective Hamiltonian of our system is
given by
\begin{align}\label{pfotl} 
  H_{\rm eff} &= \sum_{i=1}^{N_B} \int \frac{d^2k}{(2\pi)^2 } \left( {\tilde
      a}^{i\dag}({\vec k}) {\tilde a}_{i}({\vec k}) \left( {\tilde
      \omega}({\vec k}) + \mu \right) + {\tilde b}^{i}({\vec k})
      {\tilde b}_i^{\dag}({\vec k})
      \left( {\tilde \omega}({\vec k}) - \mu \right) \right)\nonumber  \\
    &\qquad\qquad +\frac{N_B {\cal V}_2}{2\pi}\left(2 \lambda_B {b}_4 \tl{\sigma}_B^2+(x_6^B+ \tfrac{4}{3})\lambda_B^2 \tl{\sigma}_B^3\right)\ ,\nonumber\\
     {\tilde \sigma_B} &=
      \frac{2\pi}{{\cal V}_2 N_B} \int \frac{d^2 {\vec k}}{(2 \pi)^2 \,2
      {\tilde \omega}({\vec k}) } \left( { \tilde a}^{i\dag}({\vec
      k}) {\tilde a}_i({\vec k}) + {\tilde b}^i({\vec k}) {\tilde
      b}_i^{\dag}({\vec k}) + { \tilde a}^{i\dag}({\vec k})
      {\tilde b^\dag}_i({\vec k}) + {\tilde b}^i({\vec k}) {\tilde
      a}_i({\vec k}) \right)\ .
\end{align}
The Hamiltonian \eqref{pfotl} can be further simplified by adding to
it the term
\begin{equation} \label{addedhampi}
  \frac{{\cal V}_2 N_B}{2\pi}(c_B^2-m_B^2)(\tl\sigma_B-{\sigma}_B )\ .
\end{equation} 
Here $c_B^2$ and $\sigma_B$ are Lagrange multipliers. The equation of
motion w.r.t.~$c_B^2$ tells us that $\sigma_B=\tilde{\sigma}_B$. This
has two consequences. The first is that the term in \eqref{addedhampi}
vanishes, and so its addition does not actually change the
Hamiltonian. The second is that every occurrence of $\tilde{\sigma}_B$
in \eqref{pfot} can be replaced by $\sigma_B$. Thus, the Hamiltonian
\eqref{pfot} can be rewritten as
\begin{align}\label{pfotn} 
 H &= \sum_{i=1}^{N_B} \int \frac{d^2k}{(2\pi)^2 } \bigg[ {\tilde
     a}^{i\dag}({\vec k}) {\tilde a}_{i}({\vec k}) \left( {\tilde\omega}({\vec k}) + \mu + \frac{c_B^2 - m_B^2}{2\tl{\omega}(\vec{k})}\right)\nonumber\\
   &\qquad\qquad + {\tilde b}^{i}({\vec k})
      {\tilde b}_i^{\dag}({\vec k})
     \left( {\tilde \omega}({\vec k}) + \frac{c_B^2 - m_B^2}{2\tl{\omega}(\vec{k})} - \mu \right) + \left(\tl{a}^{i\dag}(\vec{k}) \tl{b}^\dag_i(\vec{k}) + \tl{b}^i(\vec{k}) \tl{a}_i(\vec{k})\right) \frac{c_B^2 - m_B^2}{2\tl{\omega}(\vec{k})}\bigg]\nonumber  \\
   &\qquad\qquad +\frac{N_B {\cal V}_2}{2\pi}\left(2 \lambda_B {b}_4 {\sigma}_B^2 +(x_6^B+ \tfrac{4}{3})\lambda_B^2 {\sigma}_B^3 - (c_B^2 - m_B^2)\sigma_B\right)\ .
\end{align}
Let us now define
\begin{equation} \label{trueomega} 
\omega({\vec k})= \sqrt{ c_B^2+{\vec k}{}^2} \ ,
\end{equation} 
and perform the following Bogoliubov transformation (i.e.~a
transformation which preserves the canonical commutation relations)
that take us to the true oscillators\footnote{those appropriate to the
  expansion of a scalar field of squared mass $c_B^2$ rather than
  $m_B^2$.}
\begin{align}\label{strans}
  {\tilde a}_i({\vec k}) &=  F(\vec{k})\, { a}_i({\vec k}) + G(\vec{k})\, {b^\dagger}_i({\vec k})\ ,\quad {\tilde b}^i({\vec k}) = F(\vec{k})\, { b}^i({\vec k}) + G(\vec{k})\, {a^{i \dagger} }({\vec k})\ ,\nonumber \\
  {\tilde a^{i \dagger} }({\vec k}) &= F(\vec{k})\, { a^{i \dagger} }({\vec k}) + G(\vec{k})\, {b^I}({\vec k})\ ,\quad  {\tilde b^\dagger_i}({\vec k}) = F(\vec{k})\, { b^\dagger_i}({\vec k}) + G(\vec{k})\, {a_i}({\vec k})\ ,
\end{align}
where
\begin{equation}
  F(\vec{k}) = \frac{1}{2}\left( \sqrt{\frac{{\tilde \omega}({\vec k}) }{\omega({\vec k}) }}
    + \sqrt{\frac{{\omega}({\vec k}) }{{\tilde \omega}({\vec k}) }} \right)\ ,\quad G(\vec{k}) = \frac{1}{2}\left( \sqrt{\frac{{\tilde \omega}({\vec k}) }{\omega({\vec k}) }}
    - \sqrt{\frac{{\omega}({\vec k}) }{{\tilde \omega}({\vec k}) }} \right)\ .
\end{equation}
The inverse relations\footnote{This Bogoliubov transformation may be
  understood as follows. Instead of expanding the scalar fields of our
  problem as in \eqref{expandfields}, we are now choosing to expand
  the same fields instead as
	\begin{equation} 
	\phi_i(x) = \int \frac{d^2\vec{p}}{(2\pi)^2\sqrt{2{\omega}({\vec{p})}}} \ \Big( {{ a}}_i(\vec{p}) \te^{\i {\vec p}\cdot {\vec x} } + {{ b}}_i^\dagger (\vec{p}) \te^{-\i {\vec p}\cdot {\vec x} } \Big)\ ,\nonumber
      \end{equation}
      with $\omega(\vec{k}) = \sqrt{\vec{k}{}^2 + c_B^2}$. The
      difference between the mode expansion above and in
      \eqref{expandfields} is the value of the mass that goes into the
      definition of the frequency. The new expansion here is more
      useful as it accounts for the fact that interactions in our
      problem renormalize the bare mass $m_B^2$ to the renormalized
      mass $c_B^2$. Equating the mode expansion in
      \eqref{expandfields} and the mode expansion above leads to the
      transformations \eqref{strans} and \eqref{stransinv}.} are
\begin{align}\label{stransinv}
a_i(\vec{k}) &=  F(\vec{k})\, \tilde{a}_i(\vec{k}) - G(\vec{k})\, \tilde{b}_i^\dag(\vec{k})\ ,\quad b^i(\vec{k})= F(\vec{k})\, \tilde{b}^i(\vec{k}) - G(\vec{k})\, \tilde{a}^{i \dag}(\vec{k})\ ,\nonumber \\
a^{i \dagger}(\vec{k}) &= F(\vec{k})\,\tilde{a}^{i \dag}(\vec{k}) - G(\vec{k})\,  \tilde{b}^i(\vec{k})\ ,\quad b^\dagger_i(\vec{k})= F(\vec{k})\, \tilde{b}_i^\dagger(\vec{k}) - G(\vec{k})\, \tilde{a}_i(\vec{k})\ .
\end{align}
It is easy to verify that these transformations preserve the
commutation relations. Using \eqref{strans} and \eqref{stransinv} it
is easy to verify that the charge operator is the same in either the
tilde'd or the untilde'd basis of oscillators:
\begin{equation} \label{mutransf}
   {a}^{i\dagger}({\vec k})
  {a}_i({\vec k}) -{b}^i({\vec k}) {b}^\dag_i({\vec k})  =
  \tilde{a}^{i\dagger}({\vec k}) \tilde{a}_i({\vec k})
  - \tilde{b}^i({\vec k}) \tilde{b}_i^\dagger({\vec k}) \ ,
\end{equation}
while the Hamiltonians are related as
\begin{multline} \label{newen}
  \omega({\vec k}) \Big( {a}^{i\dagger}({\vec k}) {a}_i({\vec k}) +{b}^i({\vec k}) 
 {b}_i^\dagger({\vec k}) \Big) \\ = \left( {\tilde \omega}({\vec k}) + \frac{c_B^2-m_B^2}{2 {\tilde \omega}({\vec k})}
 \right)  \left( \tl{a}^{i\dagger}({\vec k}) \tl{a}_i({\vec k}) +  \tl{b}^i({\vec k}) \tl{b}_i^\dagger({\vec k})  \right) +\frac{c_B^2-m_B^2}{2{\tilde \omega}({\vec k})}\left(\tl{a}^{i\dagger}({\vec k}) \tl{b}^\dag_i({\vec k}) + \tl{b}^i({\vec k}) \tl{a}_i({\vec k}) \right)\ .
\end{multline}
The above two results \eqref{mutransf} and \eqref{newen} allow us to
re-express the Hamiltonian \eqref{pfotn} as
\begin{align}\label{pfotnew} 
  H &= \sum_{i=1}^{N_B}  \int \frac{d^2k}{(2\pi)^2 }
      \Bigg[ {a}^{i\dag}({\vec k})   {a}_{i}({\vec k})  
      \left( {\omega}({\vec k})+   \mu \right)
      + {b}^{i}({\vec k}) { b}_i^{\dag}({\vec k})  
      \left( {\omega}({\vec k})  - \mu \right)  \Bigg]\nonumber  \\
    &\qquad\qquad\qquad +\frac{N_B {\cal V}_2}{2\pi} \left(2 \lambda_B {b}_4\sigma_B^2 + (x_6^B+ 
      \tfrac{4}{3}) \lambda_B^2\sigma_B^3 - (c_B^2-m_B^2)\sigma_B  \right)\ ,
\end{align}
which is precisely the effective Hamiltonian $H_{\rm eff}$
\eqref{pfot}.

Since we started our discussion with a naive unrenormalized
Hamiltonian in \eqref{pfotl}, the Hamiltonian \eqref{pfot} we obtained
at the end of our procedure is also unrenormalized. Adding in the
contribution of counterterms and normal ordering (as discussed below
\eqref{pfot}) turns \eqref{pfot} into the renormalized Hamiltonian
\eqref{pfotnewfo}.

As we have emphasized above, the Hamiltonian \eqref{pfotnew} applies
in a large class of translationally invariant configurations, not just
the thermal ensemble. The formula \eqref{valsig} for the value of
$\sigma_B$ also applies in all such states.

To end this subsubsection let us reiterate its principal lesson:
$H_{\rm eff}$ follows from the canonical quantization of Lagrangian
\eqref{rblag1} only once the true canonical Hamiltonian is {\it
  truncated to its forward scattering sector}.

\subsection{The partition function of the RB theory as a trace over $H_{\rm eff}$} 

In the previous subsection we have explained that the effective
Hamiltonian, $H_{\rm eff}$, of the ungauged scalar theory `closest to
the regular boson theory' is given by \eqref{pfotnewfo} which we reproduce below for convenience:
\begin{align}\label{pfotnewfo1} 
H_{\rm eff} &= N_B E_{B,\rm vac}(c_B)+ \sum_{i=1}^{N_B}  \int \frac{d^2k}{(2\pi)^2 }
\Bigg[ {a}^{i\dag}({\vec k})   {a}_{i}({\vec k})  
\left( {\omega}({\vec k})+   \mu \right)
+ { b}_i^{\dag}({\vec k}) {b}^{i}({\vec k})  
\left( {\omega}({\vec k})  - \mu \right)  \Bigg]\nonumber  \\
&\qquad\qquad\qquad +\frac{N_B {\cal V}_2}{2\pi} \left(2 \lambda_B {b}_4\sigma_B^2 + (x_6^B+ 
\tfrac{4}{3}) \lambda_B^2\sigma_B^3 - (c_B^2-m_B^2)\sigma_B  \right)\ ,
\end{align}
The expression \eqref{pfotnewfo1} applies in every translationally
invariant state. Recall also that the chemical potential term $-\mu Q$
was also incorporated into the expression $H_{\rm eff}$ in
\eqref{pfotnewfo1}. We now demonstrate that using the above Hamiltonian
indeed reproduces the $S^2 \times S^1$ partition function
\eqref{extmatrbmod}:
\begin{equation}\label{extmatrbmod1}
  \mc{Z}_{S^2 \times S^1} = \text{Ext}_{\sigma_B,c_B}\left[ \te^{-\beta F_{\rm RB, int}} \mc{I}_{B,k}(c_B)\right]\ ,\quad \mc{I}_{B,k}(c_B) = \te^{-\beta\mc{V}_2 N_B E_{B,\rm vac}} \mc{Z}_{B,k}(c_B)\ .
\end{equation}
Let us evaluate
\begin{equation}\label{traceon}
  \tr_{{\cal H}_k} \te^{-\beta H_{\rm eff}}\ ,
\end{equation} 
where, as in previous sections, ${\cal H}_k$ denotes the projection of
the Fock space of bosons to the quantum singlet sector. Using the
explicit expression \eqref{pfotnewfo1} for $H_{\rm eff}$, this
computation is trivial to perform. The operator part of $H_{\rm eff}$
\begin{equation}
H_{\rm osc} = \sum_{I=1}^{N_B}  \int \frac{d^2k}{(2\pi)^2 }
\Bigg[ {a}^{I\dag}({\vec k})   {a}_{I}({\vec k})  
\left( {\omega}({\vec k})+   \mu \right)
+ { b}_I^{\dag}({\vec k}) {b}^{I}({\vec k})  
\left( {\omega}({\vec k})  - \mu \right)  \Bigg]\ ,
\end{equation}
gives\footnote{as usual, we perform the twisted trace
  $\tr\left(U\te^{-\beta H_{\rm osc}}\right)$ over the Fock space
  $\mc{H}_{B,\rm Fock}$ and impose the quantum singlet constraint by
  summing over the eigenvalue configurations in the Verlinde formula
  \eqref{verlindesu} or \eqref{verlindeu}.} the quantity
$\mc{Z}_{B,k}(c_B)$. Adding in the contribution of the term
$N_B E_{B,\rm vac}^B$ in \eqref{pfotnewfo1} turns this into
$\mc{I}_{B,k}(c_B)$. Finally, adding the contribution of the last line
of \eqref{pfotnewfo1} and extremizing over $\sigma_B$ and $c_B$ gives
\eqref{extmatrbmod1}.

We conclude, in other words, that the large $N$ thermal partition
function of the RB theory has the Hamiltonian interpretation presented
in \eqref{traceon}. In words, this partition function equals the trace
over the Hamiltonian of its closest ungauged scalar theory
\eqref{rblag1} on the Hilbert space obtained by restricting the Fock
Space of this theory to the quantum singlet sector.

In other words, at least as far as the large $N$ thermal partition
function can probe it, the Hilbert space of the regular boson theory
is simply that of its closest ungauged scalar cousin (which includes
forward scattering interactions) truncated to the quantum singlet
sector.

\subsection{Effect of $F_{\rm int}$ the expectation value of charge} 

Notice that partition function \eqref{extmatrbmod1} has two sources of
$\mu$ dependence. First the function $\mc{I}_{B,k}$ depends explicitly
on $\mu$. Second, the auxiliary parameters $\sigma_B$, $c_B$
(generally denoted as $\varphi_{\rm aux}$) also depend on $\mu$ after
extremization. As noted in \cite{Minwalla:2020ysu} (see Section 4), it
follows from the fact that $\varphi_{\rm aux}$ are are all extremized
on-shell that this second dependence drops out of the derivative of
the partition function \eqref{extmatrbmod1} w.r.t.~$\mu$. It follows
that the formula for the expectation value of charge in the thermal
ensemble are given by the extremely simple expressions listed in
Section 4 of \cite{Minwalla:2020ysu}. In particular, as explained in
\cite{Minwalla:2020ysu}, despite the fact that the Hamiltonian of the
system includes interactions, the total charge of the system is the
sum of the charges associated with each single particle energy state
of the system.

Unfortunately this simplicity does not generalize to other
thermodynamical quantities.  For example the second derivative of the
partition function w.r.t.~$\mu$ includes a contribution proportional
to the second derivative of the effective action
w.r.t.~$\varphi_{\rm aux}$. This second derivative is complicated; it
follows, in particular, that the compressibility of our system is not
simply the sum of compressibilities associated with single particle
states.

\section{Thermodynamics from an entropy functional} \label{entropy}

In this section, motivated by standard analyses of Fermi liquids, we
interpret our thermodynamical formulae of the previous sections in
terms of an entropy functional for the Chern-Simons matter theory.

As we have seen in the previous sections, the partition functions
$\mc{Z}_{F,k}$ \eqref{IFtrace} and $\mc{Z}_{B,k}$ \eqref{IBtrace}
share many features of the partition function corresponding to free
Fock spaces even though they are not quite based on free Fock
spaces. In particular, $\mc{Z}_{F,k}$ and $\mc{Z}_{B,k}$ factorize
into a product of single particle partition functions in the infinite
volume limit as we saw in Section \ref{winf}. We show that there is a
simple definition of an entropy functional for systems whose partition
functions factorize as a product of single partition functions and
apply it to our present case of interest i.e.~Chern-Simons matter
theories in the large volume regime.

\subsection{Thermodynamics of generalized `free' theories} 

Consider any partition function that is given by a product over single
particle partition functions
\begin{equation}\label{pfon}
  \mc{Z} = \prod_a \mf{z}(y_a) \mf{z}(\tl{y}_a) \ , 
\end{equation} 
where $y_a=\te^{- \beta(E_a-\mu)}$ and $\tl{y}_a=\te^{-\beta(E_a+\mu)}$
and
\begin{equation}\label{zabm}
  \mf{z}(y)= \sum_{n} B_n y^n\ .
\end{equation} 
If we think of $B_n$ as the effective degeneracy of the state in which
the single fundamental particle energy eigenstate $E_a$ is occupied
$n$ times \footnote{If $B_n$ is not an integer then it is not a true
  degeneracy as we have explained in Section \ref{udr}.}, it is
natural to define the entropy of our ensemble as
\begin{equation}\label{entgensys}
  S_a = - \tr\, \rho_a \log \rho_a\ ,
\end{equation}
where $\rho_a$ is the density matrix for the ensemble. It is a
diagonal matrix with entries $y_a^n$ repeated $B_n$ times,
successively for all $n$.  More explicitly, we have
\begin{equation}
S_a  = - \sum_{n=0}^{\infty} B_n  \frac{ y_a^n}{\mf{z}(y_a)} 
  \log \left( \frac{y_a^n}{\mf{z}(y_a)} \right) \ ,
\end{equation}
 It is also natural to define the average occupation number $n_a$ as
\begin{equation}  \label{numgensys} 
n_a(y_a) = y_a \frac{\partial}{\partial y_a}\log \mf{z}(y_a) = \sum_{n=0}^{\infty } \frac{n B_n y_a^n}{\mf{z}_a}  \ . 
\end{equation} 
The formula for the entropy associated with the state with energy
$E_a$ may then be manipulated as follows:
\begin{align}\label{relgo}
  S_a &= -\sum_{n=0}^{\infty } B_n  \frac{ y_a^n}{\mf{z}(y_a)} \big( n \log y_a - \log \mf{z}(y_a) \big)\ ,\nonumber\\
      & = -n_a(y_a) \log y_a + \log \mf{z}(y_a) \ . 
\end{align}
The above equation \eqref{relgo} asserts that the thermodynamical
relationship for the entropy of our subsystem (arbitrary number of
occupations of a fundamental particle in the eigenstate with energy
$E_a$) is \emph{exact}, even when there is no large `occupation'
parameter in the system\footnote{For instance, in a theory of a single
  free fermion, each state is either occupied or unoccupied and
  thermodynamics cannot be used in this two-state system, but
  \eqref{relgo} still applies.}.

Observe from \eqref{relgo} that $S_a$ is the Legendre transform of
$\log \mf{z}(y_a)$ w.r.t.~$\log y_a$, and thus it is natural of think
of $S_a$ as a function of $n_a$ rather than $\log y_a$. As usual, it
follows from \eqref{relgo} that the variation $\delta S_a$ is given by
\begin{align} \label{varent} \delta S_a
  &= -\delta n_a \log y_a - n_a \delta \left( \log y_a \right) + \delta \log \mf{z}(y_a)\ ,\nonumber\\
  &=  -\delta n_a \log y_a - n_a \delta \left( \log y_a  \right) +  n_a \delta \left( \log y_a  \right)\ ,\nonumber\\
  &= -\delta n_a \log y_a \ ,
\end{align}
which gives
\begin{equation}\label{abder} 
\frac{\partial}{\partial{n_a}}S_a= -\log y_a\ . 
\end{equation} 
In principle, this equation may be solved for $y_a$ as a function of
$n_a$; substituting this solution back into \eqref{relgo} gives $S_a$
as a function of $n_a$ which we denote as
\begin{equation}\label{sna}
 S_a = S(n_a) \ . 
\end{equation} 
Note that the detailed form of the function $S(n_a)$ depends on the
values of $B_n$.

\subsection{Specialization to Chern-Simons matter theories at large
  volume}
The discussion presented in this section so far applies to any system
in which partition function factorizes as a product of single particle
partition functions \eqref{pfon}, no matter what the values of the
numbers $B_n$ in \eqref{zabm} are. Specializing to the case of
Chern-Simons matter theories in the large volume limit (and ignoring
contact interactions for the moment), we define the single particle
entropy functions by the equations
\begin{equation}\label{spefs}
S_B(n)= {\rm Ext}_{y} 
\left(\log \mf{z}_B(y) - n \log y \right)\ ,\quad S_F(n)= {\rm Ext}_{y} \left(\log \mf{z}_F(y) - n \log y \right)\ ,
\end{equation} 
where $\mf{z}_F(y)$ and $\mf{z}_B(y)$ were defined in
\eqref{fbinlvo}. The extremization over $y$ in \eqref{spefs}
implements the Legendre transform described in \eqref{varent}. The
inverse relations to \eqref{spefs} are
\begin{equation}\label{spefs1}
\log \mf{z}_B(y)= {\rm Ext}_{n} \left(S_B(y) +n \log y \right)\ ,\quad \log \mf{z}_F(y)= {\rm Ext}_{n} \left(S_F(y) + n \log y \right)\ .
\end{equation} 
While \eqref{fbinlvo} gives explicit expressions for the functions
$\mf{z}_B(y)$ and $\mf{z}_F(y)$, we have not managed to find equally
explicit formulae for $S_B(n)$ and $S_F(n)$. The difficulty lies in
analytically performing the extremization over $y$ in \eqref{spefs}
(i.e. inverting the equations \eqref{nexp} to solve for $y$ in terms
of $n$). Even in the absence of completely explicit expressions, we
know certain things about these entropy functions on general
grounds. First, the Bose-Fermi duality of the single particle
partition functions immediately implies that
\begin{equation}\label{bdentfuo}
S_{B}(n)= S_{F}(n) \ ,\quad\text{with}\quad N_B = |k_F|\ ,\quad |k_B| = N_F\ .
\end{equation} 
We also know that in the limit $k_F\to \infty$ (see equation
\eqref{newent} in the Appendix \ref{thermo}),
\begin{equation}\label{saf}
S_{F}(n) \to -n\log n  -\big( N_F - n \big) \log \big( N_F - n \big)  + N_F \log N_F \ , 
\end{equation} 
and that in the limit $k_B \to \infty$ (see Appendix \ref{thermo}),
\begin{equation}\label{sab}
  S_{B}(n) \to -n \log n +\big( N_B + n \big) \log \big( N_B + n \big) - N_B \log N_B \ . 
\end{equation}

\subsection{The entropy functional}\label{entropyfun}

So far, we have discussed the entropy of the subsystem consisting of
any number of occupations by a fundamental particle of a single
particle energy eigenstate. The full system includes several energy
eigenstates and it also includes antifundamental particles. It follows
that the entropy $S$ of the full system is given by
\begin{equation}\label{totent}
S(\{n_a\}, \{\bar{n}_a\}) = \sum_a \Big( S(n_a) + S({\bar n}_a) \Big) \ ,
\end{equation} 
and is naturally thought of as a function of all $n_a$ and
${\bar n}_a$.

Suppose our starting point is the entropy $S(n_a)$ for each single
particle energy eigenstate rather than the partition function
$\mf{z}(y_a)$. In particular the `degeneracies' $B_n$ have to be
extracted from the formula for $S(n_a)$. Suppose that the total
entropy $S$ is then $S = \sum_a (S(n_a) +
S(\bar{n}_a))$. Let us extremize the entropy of our system at
fixed energy and charge, i.e.~the function
\begin{equation}\label{whatext}
 -\beta F=  S-\beta E +\beta \mu Q \ ,
\end{equation} 
where $\beta$ and $ \mu$ are Lagrange multipliers. The total energy
$E$ is $\sum_a(n_a +{\bar n}_a) E_a$, while the total charge $Q$ is
$\sum_{a}(n_a-{\bar n}_a)$ and so the function in \eqref{whatext} can
be rewritten as
\begin{equation}\label{whain}
\sum_a \Big( S(n_a)- n_a \beta ( E_a -\mu) \Big) +
\sum_a \Big(  S(\bar{n}_a)- {\bar n}_a \beta ( E_a +\mu) \Big) \ . 
\end{equation} 
Extremizing this function w.r.t.~$n_a$ we find the equation 
\begin{equation} \label{extent}
\frac{\partial}{\partial{n_a}} S(n_a)= \beta(E_a-\mu) \ . 
\end{equation} 
which we solve to obtain $n_a$. The above equation is the same as
\eqref{abder} once we identify $y_a$ with $\te^{-\beta( E_a-
  \mu)}$. Indeed once we use \eqref{extent} to solve for $n_a$ and
plug the result back into \eqref{whain}, this expression reduces
simply to (see \eqref{spefs})
\begin{equation}\label{expredu}
\sum_a \left( \log \mf{z}(y_a) +\log \mf{z}(\tl{y}_a) \right)\ ,\quad y_a = \te^{-\beta( E_a - \mu) }\ ,\quad \tl{y}_a = \te^{-\beta(E_a + \mu)}\ .
\end{equation} 
After extremizing w.r.t.~$n_a$ and ${\bar n}_a$, in other words, the
functional \eqref{whain} is simply the (log of the) partition function
of our system at inverse temperature $\beta$ and chemical potential
$\mu$. It follows that the information in the functional \eqref{whain}
is equivalent to that in the partition function. The main difference
is how the data is presented to us. When we are given the partition
function the `degeneracies' $B_n$ are given to us, and we deduce $n_a$
from \eqref{numgensys}. In order to compute the Legendre transform of
the partition function we have to solve for $y_a$ in terms of
$n_a$. On the other hand, when we are given the entropy function in
order to recover the partition function (and consequently the
degeneracies $B_n$), we need to solve \eqref{extent} for $n_a$ in
terms of $y_a$ and consequently, for the `degeneracies' $B_n$. In
other words the extremization of \eqref{whain} automatically produces
the average occupation numbers $n_a$ of every single particle energy
eigenstate $E_a$ and the degeneracy $B_n$ of the eigenstate filled $n$
times.

\subsection{The entropy functional for large volume Chern-Simons
  matter theories including interactions}
So far, the entropy functional $S_a(n_a)$ included only the effects of
imposing the quantum singlet constraint on the Fock space which
changed the degeneracies $B_n$ from positive integers to $k$-dependent
numbers which were not integers in general. We now include the energy
renormalization that results from the contact interactions of the
matter fields. We present our analysis for the large volume regular
boson theory in the unHiggsed phase, but a similar analysis applies to
all theories and in all phases.

Consider the free energy functional
\begin{multline}\label{entfn}
  -\beta F_{\rm RB}(n_{\vec k}, {\bar n}_{\vec k}, c_B^2, \sigma) =
  \int \frac{d^2\vec{k}}{(2 \pi)^2} \left( S_{B}(n_{{\vec k}}) +
    S_{B}( {\bar n}_{\vec k}) \right) \\ - \beta \left( \int
    \frac{d^2\vec{k}}{(2 \pi)^2}
    \bigg[ \Big( \sqrt{c_B^2+{\vec k}^2} - \mu \Big) n_{\vec{k}} + \Big( \sqrt{c_B^2+{\vec k}^2} +\mu \Big) \bar{n}_{\vec{k}} \bigg] + N_B E_{B,\rm vac} (c_B) \right)  \\
  - \frac{\beta}{2\pi} \left( 2 \lambda_B {b}_4\sigma_B^2 +
    \lambda_B^2(x_6^B + \tfrac{4}{3}) \sigma_B^3 - \left(c_B^2-m_B^2
    \right) \sigma_B \right)\ ,
\end{multline}
where the function $S_{B}(n)$ was defined in \eqref{spefs}. The first
two lines of \eqref{entfn} above is simply a special case of
\eqref{whain} with the discrete index $a$ replaced by the continuous
momentum $\vec{k}$ which indexes the energy spectrum of the free boson
in the infinite volume limit\footnote{Recall, in particular, that in
  the limit $k_B \to \infty$ we have
  \begin{multline}
    S_{B}(n_{\vec{k}},\bar{n}_{\vec{k}}) = N_B{\cal V}_2\int
    \frac{d^2\vec{k}}{(2 \pi)^2} \Big( -n_{\vec{k}} \log n_{\vec{k}} +
    \big(1 + n_{\vec{k}} \big) \log \big( 1+n_{\vec{k}} \big) \\
    -\bar{n}_{\vec{k}} \log \bar{n}_{\vec{k}} +\big(1 +
    \bar{n}_{\vec{k}} \big) \log \big( 1+ \bar{n}_{\vec{k}} \big)
    \Big)\ .
  \end{multline}}. The off-shell free energy functional $F_{\rm RB}$
is a functional of the occupation numbers $n_{\vec{k}}$ of the
particles, the occupation numbers $\bar{n}_{\vec{k}}$ of the
antiparticles as well as the auxiliary variables $c_B$ and
$\sigma_B$.

We will now demonstrate that the extremization of the free energy
functional defined in \eqref{entfn} simply yields the free energy for
the RB theory \eqref{RBoffshellfe} in the unHiggsed phase. This
demonstration is almost trivial. It follows immediately from
\eqref{spefs} that the extremization of \eqref{entfn} over
$n_{\vec k}$ and ${\bar n}_{\vec k}$ is
\begin{multline}\label{entfnext}
{\rm Ext}_{\{n_{\vec k}, {\bar n}_{\vec k} \} } \left( -\beta F_{\rm RB}(n_{\vec k}, {\bar n}_{\vec k}, c_B^2, \sigma) \right)  = \\ \int \frac{d^2\vec{k}}{(2 \pi)^2}  \left[ \log \mf{z}_{B}\left(\exp \left[ -\beta  \sqrt{c_B^2+{\vec k}^2} - \beta \mu  \right] \right)    + \log \mf{z}_{B}\left(\exp \left[ -\beta  \sqrt{c_B^2+{\vec k}^2} + \beta \mu  \right] \right) \right]
\\
-\beta N_B E_{B,\rm vac}(c_B)- \frac{1}{2\pi} \beta \left( 2 \lambda_B
  {b}_4\sigma_B^2 + \lambda_B^2(x_6^B + \tfrac{4}{3}) \sigma_B^3 -
  \left(c_B^2-m_B^2 \right) \sigma_B \right)\ ,
\end{multline}
from which the result follows\footnote{Note that the RHS of
  \eqref{entfnext} is precisely
$$\log \mc{I}_{B,k} - \frac{1}{2\pi} \beta \left( 2 \lambda_B {b}_4\sigma_B^2 + \lambda_B^2(x_6^B + \tfrac{4}{3}) \sigma_B^3 - \left(c_B^2-m_B^2 \right) \sigma_B  \right)$$
i.e. its exponential is precisely
$$ \te^{-\beta \mc{V}_2 V_{\rm int}(c_B)} \mc{I}_{B,k}\ .$$
}. It follows immediately that the extremization of the RHS of
\eqref{entfn} over all its variables reproduces the logarithm of the
partition function of the interacting large $N$ Chern-Simons matter
theory\footnote{ Although no more really needs to be said, the reader may find it useful to see how the extremization over all the variables in \eqref{entfn} works in a little more detail. 
Extremizing w.r.t.~$n_{\vec{k}}$ and $\bar{n}_{\vec{k}}$
determines $n_{\vec{k}}$ via the equation
\begin{equation} \label{varnk}
\frac{\delta}{\delta n_{\vec{k}}} S_B^{N_B, k_B}(n_{\vec{k}}) = N_B \mc{V}_2 \beta \left(\sqrt{c_B^2 + \vec{k}{}^2} - \mu\right)\ .
\end{equation}
It follows from \eqref{varnk} that $n_{{\vec k}}$ and ${\bar n}_{{\vec k}}$ are each given by \eqref{nform} (or more precisely by the second of \eqref{nexp}) with $y=\te^{-\beta \left(\sqrt{c_B^2 + \vec{k}{}^2} - \mu\right)}$ 
and  $y=\te^{-\beta \left(\sqrt{c_B^2 + \vec{k}{}^2} +\mu\right)}$ respectively. 
Extremizing w.r.t.~$\sigma_B$ yields \eqref{cbeq} while extremizing
w.r.t.~$c_B^2$ yields
\begin{equation}\label{nsinnew}
  \sigma_B = \partial_{c_B^2} E_B(c_B) + \int \frac{d^2 {\vec k}}{(2 \pi)^2\,2\omega(\vec{k}) }   \ 
  ( n_{\vec{k}} + \bar{n}_{\vec{k}} )  \ , 
\end{equation} 
in agreement with the equation for $\sigma_B$ in the previous section
\eqref{valsig}.  It follows that the entropy functional $S^B$ in
\eqref{entfn} reproduces every aspect of the thermodynamics of our
system.}.

In principle it would be possible to integrate $c_B^2$ and $\sigma_B$
out of the entropy functional \eqref{entfn}; we can do that by using
the saddle point equations \eqref{valsig} and \eqref{cbeq} to solve
for $\sigma_B$ and $c_B$ in terms of $n_{\vec{k}}$ and
$\bar{n}_{\vec{k}}$, and plug that solution back in
\eqref{entfn}. This resultant entropy functional, which is now a
functional only of the occupation numbers $n_{\vec k}$ and
${\bar n}_{\vec k}$, reproduces all aspects of the thermodynamics of
our system.
    
The free energy as a functional of occupation numbers - analogous to
that defined in \eqref{entfn} - is often used to good effect in the
analysis of Fermi liquids. Note that, just as in the case of
Chern-Simons matter theories, forward scattering dominates dynamics of
Fermi liquids for thermodynamical considerations. The equation
\eqref{entfn} (and its generalizations to the various other
Chern-Simons matter theories) may be thought of as a generalization of
the familiar Fermi liquid free energy functionals.

It would be interesting to use the functional \eqref{entfn} to repeat
computations that are standard for Fermi liquids in the current
context. The main deficiency in the current context is that we have
not managed to find a useful explicit formula for the entropy
functions $S_{B}(n)$ away from $\lambda_B=0$ and $\lambda_B=1$. It is
possible that a useful explicit expression of this nature exists but
has eluded our efforts to determine it and it is also possible that
one could perform computations to some extent using the implicit
expressions above. We leave further investigation in this direction to
future work.

\section{Counting quantum singlets}\label{counting}

In this section we present completely explicit formulae for the number
of quantum singlets in the product of representations
$R_1,\ldots,R_n$, both for $SU(N)_{k}$ as well as $U(N)_{k, k'}$
Chern-Simons theories at arbitrary values of the Chern-Simons
levels. As we have explained earlier, the number of \emph{quantum}
singlets in the tensor product of representations $R_1,\ldots,R_n$ is
the same as the dimension of the space of conformal blocks involving
the representations $R_1,\ldots,R_n$ in the corresponding WZW
$SU(N)_k$ or $U(N)_{k,k'}$ WZW model.

We first obtain our results using the well-established and completely
precise Verlinde formulae, and then re-obtain them by performing an
explicit evaluation of the Chern-Simons path integral on
$S^2 \times S^1$ with Wilson line insertions in representations
$R_1, \ldots, R_n$ following Blau and Thompson \cite{Blau:1993tv}. We
further obtain our formulae by evaluating the supersymmetric index of
pure $\mc{N}=2$ Chern-Simons theory in the presence of Wilson line
insertions. All methods give us the same answer and each has its own
advantage.

\subsection{Chern-Simons theories and WZW theories}\label{CSWZWconv}

We have described our notation for Chern-Simons theories in detail in
Section \ref{ntcs}. However, to keep the current section
self-contained we briefly review our notation.

The notation $SU(N)_k$ stands for Chern-Simons theory with $SU(N)$
gauge group with bare level $k$.\footnote{In the dimensional
  regularization scheme, the renormalized level
  $\kappa = k + \sgn(k) N$ appears in front of the Chern-Simons
  Lagrangian, while in the Yang-Mills regularization scheme the bare
  level $k$ appears in front of the Lagrangian. See
  \cite{Aharony:2015mjs} for details.} The $U(N)_{k, k'}$ Chern-Simons
theory is labelled by two levels: $k$ is the bare level for the $SU(N)$
part of the gauge group while $Nk'$ is the level for the $U(1)$ part
of the gauge group (when working in a normalization in which every
fundamental field carries overall $U(1)$ charge unity). As reviewed
e.g.~in Appendix A.3 of \cite{Minwalla:2020ysu}, consistency requires
that the level $k'$ be given by
\begin{equation}\label{kpdefm}
  k'= \sgn(k)(|\kappa| + s N) \ .
\end{equation}
where $s$ is an integer. In this paper, we will always work with
ranges of $\kappa$ such that, for a given $s$, $k'$ always has the
same sign as $\kappa$. In particular, we will be interested in two
values of $s$; the so called Type I $U(N)$ theory with $s=0$ (and so
$k'= \kappa$), and the so called Type II $U(N)$ theory with $s=-1$
(and so $k'=k$). See Appendix A of \cite{Minwalla:2020ysu} for a more
detailed discussion of these theories.

According to the Chern-Simons / WZW model correspondence, Chern-Simons
theory is dual to one chiral half of the WZW model and the chirality
of the WZW model is flipped when the sign of the Chern-Simons level is
flipped. In our conventions, a positive Chern-Simons level corresponds
to the holomorphic half of the WZW model. For instance, $SU(N)_k$
Chern-Simons theory is dual to the holomorphic half of the $SU(N)_k$
WZW model when $k > 0$ and to the antiholomorphic half of the
$SU(N)_{|k|}$ WZW model when $k < 0$. Since the sign of the $U(1)$
level $Nk'$ in a $U(N)_{k,k'}$ theory is the same as that of $k$ for
the cases of interest in this paper, the dual WZW theory will have the
same chirality for both the $SU(N)$ and $U(1)$ factors as well.

\textbf{Note:} Unless mentioned otherwise, in this section we take the
Chern-Simons level $k$ to be a positive integer (and hence, the
renormalized level $\kappa$ to be positive as well). In the case of
$U(N)_{k,k'}$ theories, we take both levels $k$, $k'$ to be positive.

We also briefly describe the spectrum of primaries of the chiral WZW
model. These are finite in number and are in one-to-one correspondence
with the integrable representations of the corresponding affine Lie
algebra. For instance, the primaries of the $SU(N)_k$ chiral WZW model
are described by the integrable representations of the $\wh{su}(N)_k$
affine Lie algebra (the affine Lie algebra arises as follows:
generators of the affine Lie algebra are the modes in the mode
expansion of the chiral WZW current $g^{-1} \partial g$.) The
integrable representations of $\wh{su}(N)_k$ are a finite subset of
the representations of the ordinary Lie algebra $su(N)$ are given by
those representations whose corresponding Young tableaux have at most
$k$ boxes in their first row. Since the same data goes into describing
the integrable representations of $\wh{su}(N)_k$ and the corresponding
representations of $su(N)$, we denote the two kinds of representations
by the same letter as long as there is no danger of confusion.

\subsection{The methods employed to evaluate the number of singlets}\label{methods}

In this subsection, we describe the various ways in which we derive
the number of quantum singlets in the tensor product of
representations in pure Chern-Simons theory.

\subsubsection{The Verlinde formula}
\label{sfvf} 

Let the highest weights of the primaries ${R}_1, {R}_2,\ldots,{R}_n$
of the $G_k$ WZW model be denoted ${\mu}_1, {\mu}_2,\ldots,{\mu}_n$
(recall that these correspond to integrable representations of the
affine Lie algebra $\wh{g}_k$). The Verlinde formula asserts that the
dimension $\mc{N}_{g,n}$ of the space of conformal blocks for the $n$
primaries ${R}_1, {R}_2,\ldots,{R}_n$ on a genus $g$ Riemann surface
is given by
\begin{equation}\label{ngtm} 
  \mc{N}_{g,n}({R}_1,\ldots,{R}_n) = \sum_{\lambda} (\mc{S}_{\lambda 0} )^{2-2g} \prod_{i=1}^n \frac{\mc{S}_{\lambda \mu_i}}{\mc{S}_{\lambda 0}}\ ,
\end{equation}
where the sum runs over the integrable representations $\lambda$ of
the affine algebra $\hat{g}_k$ corresponding to the WZW model,
$\mc{S}_{\lambda\mu}$ is the Verlinde $\mc{S}$-matrix and $0$ refers
to the trivial representation. See Appendix \ref{proofsub} for an
outline of the derivation of the formula \eqref{ngtm}. Later in this
subsection we will present completely explicit versions of the formula
\eqref{ngtm} for the special cases of $SU(N)_k$ and $U(N)_{k,k'}$
theories.

\subsubsection{Evaluation of the Chern-Simons path integral in the
  presence of Wilson lines}

Consider Chern-Simons theory on $\Sigma \times S^1$ where $\Sigma$ is
a genus $g$ Riemann surface and consider Wilson lines in the
representations $R_1,\ldots, R_n$ inserted at $n$ distinct points on
$\Sigma$ and winding along the $S^1$. In his path-breaking study of
Chern-Simons theory in the presence of Wilson loops
\cite{Witten:1988hf}, Witten demonstrated that the path integral of
Chern-Simons theory with the above Wilson line insertions equals the
dimension of the space of conformal blocks of the chiral WZW model
with primary operators in the representations $R_1\ldots, R_n$.

In the beautiful paper \cite{Blau:1993tv}, Blau and Thompson directly
evaluated the path integral of Chern-Simons theory described above,
providing another way to derive a formula for the number of conformal
blocks. Below we specialize the results of \cite{Blau:1993tv} to
$SU(N)_k$ theory where we find complete agreement with the Verlinde
formula, as anticipated and partially demonstrated already in
\cite{Blau:1993tv}. We also compute the path integral for the
$U(N)_{k, k'}$ Chern-Simons theory which we achieve with a minor
modification to the computation in \cite{Blau:1993tv} and here too we
find complete agreement with the Verlinde formula.

\subsubsection{Evaluation of the path integral using supersymmetric localization} 

The field content of ${\cal N}=2$ supersymmetric pure Chern-Simons
theory includes an adjoint gaugino in addition to the gauge
bosons. Like the gauge bosons, the superpartner gauginos are
non-dynamical: integrating them out simply shifts the level of
resulting pure Chern-Simons theory (see e.g.~equation D.4 of
\cite{Aharony:2018pjn} for the details of the shifts of levels).

We can thus obtain exact results for pure Chern-Simons theories by
evaluating the path integral for an ${\cal N}=2$ SUSY Chern-Simons
theory. The advantage of this manoeuvre lies in the fact that we can
use supersymmetric localization for evaluating path integrals in
supersymmetric Chern-Simons theories. Calculating the supersymmetric
index mentioned above indeed gives results that agree perfectly with
the Verlinde formula.

\subsection{$SU(N)_k$ Chern-Simons theory} \label{verlindedetail}

\subsubsection{A simple expression for the Verlinde $\cS$-matrix}

Recall that the Verlinde $\cS$-matrix $\mc{S}_{\hat\lambda\hat\mu}$ is
defined as the transformation on the space of affine characters (or
equivalently, WZW torus partition functions) that implements the
$\tau \to -1/\tau$ transformation:
\begin{equation}
  \chi_{\lambda}(-1/\tau) = \sum_{\mu} \mc{S}_{{\lambda}{\mu}}\, \chi_{ {\mu}}(\tau)\ ,
\end{equation}
where $\lambda$ is the highest-weight of an integrable representation
of the affine Lie algebra and $\mu$ in the summation runs over all
highest-weight representations of the affine Lie algebra.

We now give a simple formula for the $\cS$-matrix in terms of
quantities of the ordinary Lie algebra $g$ corresponding to the affine
Lie algebra $\wh{g}_k$ based on the discussion in \cite[Sections
14.6.2, 14.6.3]{di1996conformal}.

Let $r$ be the rank of the ordinary Lie algebra $g$ and let $|C|$ be
the size of the centre of the corresponding Lie group $G$. The
Verlinde $\cS$-matrix matrix $\mc{S}_{\lambda\mu}$ is given in terms
of $\chi_\mu$, the character of the representation $\mu$ of the
ordinary Lie algebra $g$, by
\begin{equation} \label{suverorm}
  \frac{\mc{S}_{\lambda\mu}}{\mc{S}_{\lambda 0}} =
  \chi_{ \mu}(\xi_\lambda)\ ,
\end{equation}
with
\begin{equation} \label{suverorm1}
  \xi_\lambda = -\frac{2\pi\i}{\kappa} (\lambda + \rho)\ ,\quad\text{and}\quad \mc{S}_{\lambda 0} = \frac{1}{\sqrt{|C| \kappa^{r}}} \prod_{\alpha \in \Delta_+} 2\sin\left(\frac{\pi}{\kappa} (\alpha,\lambda+\rho)\right)\ .
\end{equation}
(See also \cite{Zuber:1995ig} for a different point of view of the
above formulas.) An important point to note is that all quantities on
the right hand side are in terms of quantities defined for the
ordinary Lie algebra $g$. In detail, the quantity $\rho$ is the Weyl
vector of the ordinary Lie algebra $g$ and is given by half the sum of
all positive roots of $g$, $\lambda$ denotes the highest-weight of the
$g$ representation corresponding to an integrable representation of
$\wh{g}_k$, and $\xi_\lambda$ is a special $g$ weight associated to
$\lambda$ given by the formula in \eqref{suverorm1}. The product over
$\alpha$ runs over $\Delta_+$, the set of positive roots of the
ordinary Lie algebra $g$ and $(\mu,\nu)$ is the usual inner-product
induced on the weights of $g$ by the Killing form $K$ on the Lie
algebra\footnote{The Killing form is defined as
  \begin{equation} K(X, Y) = \frac{1}{2 h^\vee}
    \text{Tr}(\text{ad}\,X\,\text{ad}\,Y)\ ,\nonumber
  \end{equation} where $X$, $Y$
  are elements of the Lie algebra, $\text{ad}\,X$ is the adjoint
  representation of $X$ and $h^\vee$ is the dual Coxeter number of the
  Lie algebra ($h^\vee = N$ for $\text{su}(N)$). See Section 13.1.2 of
  \cite{di1996conformal} for more details.\label{killing}}.

We would like to specialize the above formula for $\wh{su}(N)_k$ and
eventually extend it to $\wh{u}(N)_{k,k'}$ as well. In this case,
there is a simple and down-to-earth basis for the weight space of
$su(N)$ in terms of which the above formulae become very explicit and
easy to compute.

\subsubsection{An orthonormal basis for $su(N)$ weights and a
  lightning review of the $su(N)$ Lie algebra}
\label{dteb}

In this subsubsection, we describe an orthonormal basis for the
$su(N)$ weights by embedding the $N-1$ dimensional weight space of
$su(N)$ into the $N$ dimensional weight space of $u(N)$.

The dual Cartan space of $u(N)$ has a natural basis $\epsilon_i$,
$i=1,2,\ldots,N$, dual to the $u(N)$ Cartan elements corresponding to
the $N \times N$ diagonal matrices $M^i$ such that the $i^{\rm th}$
diagonal element is 1 and the rest are zero\footnote{The pairing
  between the basis of the Cartan $M^i$ and the dual basis
  $\epsilon_i$ is given by $\epsilon_i(M^j) = \delta_i^j$.}. The basis
$\{\epsilon_i\}$ is defined to be orthonormal w.r.t.~a given inner
product $(\cdot,\cdot)$:
\begin{equation}\label{epnorm}
(\epsilon_i, \epsilon_j) = \delta_{ij}\ .
\end{equation} 
The $su(N)$ weights can be expanded in the orthonormal basis as
follows. A general $SU(N)$ weight $v$ is given by
\begin{equation}\label{mgwv}
v = \sum_{i=1}^N v_i \epsilon_i\ , \quad \textrm{with} \quad \sum_{i=1}^N v_i = 0\ .
\end{equation} 
The last condition in \eqref{mgwv} is a consequence of the condition
of tracelessness on $su(N)$ matrices and cuts down the dimensionality
of the $su(N)$ weight space from $N$ to $N-1$.\footnote{The inner
  product $(\cdot,\cdot)$ in \eqref{epnorm} restricted to this $N-1$
  dimensional plane coincides with the inner product induced on the
  dual Cartan space of $su(N)$ by the Killing form. The induced inner
  product is defined as follows. Every element $\alpha$ of the dual
  Cartan space is associated to an element $H^\alpha$ of the Cartan
  subalgebra via the Killing form (see footnote \ref{killing})
  \begin{equation}
    \alpha(H^a) = K(H^a, H^\alpha)\ ,\nonumber
  \end{equation}
  where $H^a$ is the basis of the Cartan subalgebra. The notation
  $\alpha(H^a)$ denotes the action of $\alpha$, an element of the dual
  Cartan space i.e.~space of linear functionals on the Cartan, on an
  element $H^a$ of the Cartan. Then, the inner product on the dual
  space is given by
  \begin{equation}
    (\alpha,\beta) = \alpha(H^\beta) = \alpha (\textstyle \sum_a \beta_a H^a) = \textstyle\sum_a \beta_a K(H^a, H^\alpha) = K(H^\beta, H^\alpha)\ .\nonumber
  \end{equation}
} In the orthonormal basis, the $N^2-N$ roots of $su(N)$ are given by
$\epsilon_i-\epsilon_j$ with $i \neq j$. Half of these roots are
positive. In our convention, the positive roots are given by
\begin{equation}\label{posroots} 
\alpha_{(ij)} =  \epsilon_i - \epsilon_j\ ,\quad\text{with}\quad 1\leq i < j \leq N\ .
\end{equation}
Our choice of simple roots is
$\alpha_1,\alpha_2, \ldots, \alpha_{N-1}$ given by
\begin{equation} \label{weightsofroots}
  \alpha_i = \epsilon_i - \epsilon_{i+1}\ ,\quad i = 1,\ldots, N-1\ .
\end{equation}
The positive roots are then the following linear combinations of the
simple roots
\begin{equation}
\alpha_{(ij)} = \alpha_i + \alpha_{i+1} + \cdots + \alpha_{j-1} = \epsilon_i - \epsilon_j\ ,\quad\text{with}\quad 1\leq i < j \leq N\ .
\end{equation}
The Weyl vector $\rho$ is  half the sum of all positive roots and is given by 
\begin{equation} \label{formrho} 
\rho = \sum_{i=1}^{N} \frac{N-2 i+1}{2}  \epsilon_i \ .
\end{equation} 
The highest-weight $\lambda$ of a given irrep of $su(N)$ is expanded
in the orthonormal basis in terms of $N-1$ positive integers
$\ell_i^\lambda$, $i=1,\ldots,N-1$ as
\begin{equation}\label{lambdaexp}
  \lambda = \sum_{i=1}^N (\ell^\lambda_i - q_\lambda) \epsilon_i\ ,\quad\text{where}\quad q_\lambda = \frac{1}{N} \sum_{i=1}^N \ell^\lambda_i\ .
\end{equation}
(In the formula above, $\ell^\lambda_N$ is understood to be zero). The
second condition in \eqref{mgwv} is automatically enforced on the
components of $\lambda$ in the orthonormal basis. Since all the
$\epsilon_i$ are on an equal footing, it is convenient to order the
$\ell^\lambda_i$ to satisfy
\begin{equation}\label{lord}
  \ell^\lambda_1 \geq   \ell^\lambda_2 \geq \cdots \geq   \ell^\lambda_{N-1} \geq \ell^\lambda_N = 0\ .
\end{equation}
The significance of the labels $\ell^\lambda_i$ ordered as in
\eqref{lord} is the following: $\ell^\lambda_i$ is the number of boxes
in the $i^{\rm th}$ row of the Young tableau corresponding to the
irrep with highest weight $\lambda$. Note that $q^\lambda$ is $1/N$
times the total number of boxes in the Young tableau corresponding to
$\lambda$.\footnote{The highest weight of a given irrep is usually
  expanded in terms of the basis of fundamental weights $\omega_i$,
  $i=1,\ldots,N-1$. For $su(N)$, the fundamental weights are defined
  by the relations $(\omega_i, \alpha_{j}) = \delta_{ij}$ where
  $\alpha_i$ is the basis of simple roots, and the $\omega_i$ are
  given explicitly in terms of the orthonormal basis as
  \begin{equation}
    \omega_i = \epsilon_1 + \cdots + \epsilon_i - \frac{i}{N}\sum_{i=1}^N \epsilon_i\ .\nonumber
  \end{equation}
  The components of $\lambda$ in the fundamental weight basis are
  called the Dynkin labels $d^\lambda_i$ and we have
  $\lambda = \sum_i d^\lambda_i \omega_i$. The relationship between
  $\ell^\lambda_i$ and the Dynkin labels $d^\lambda_i$ can be easily
  obtained to be
\begin{equation}
\ell^\lambda_i = d^\lambda_i + d^\lambda_{i+1} + \cdots +d^\lambda_{N-1}\ ,\quad\text{for}\quad i=1,\ldots,N-1\ ,\quad\text{and}\quad \ell^\lambda_N = 0\ .\nonumber
\end{equation} } The Weyl vector $\rho$ (see \eqref{formrho}) is given by the Young
tableau with rows
\begin{equation}\label{elrho}
\ell^\rho_i = N-i\ .
\end{equation} 
(one can check that inserting \eqref{elrho} into \eqref{lambdaexp}
yields \eqref{formrho}).

We finally come to the definition of the character of a representation
of $su(N)$. Given an element $\xi = \sum_i \xi_i \epsilon_i$ of the
dual Cartan space of $su(N)$, we can define an element $M_\xi$ of the
Cartan subalgebra as
\begin{equation}
  M_\xi = \sum_i \xi_i M^i\ ,
\end{equation}
where the $M^i$ are the diagonal $N \times N$ matrices defined at the
beginning of this subsubsection and the $\xi_i$ satisfy
$\sum_i \xi_i = 0$ so that $\tr\, M_\xi = 0$. Consider the exponential
$\te^{M_\xi}$ given by
\begin{equation}\label{suma}
  \te^{ M_\xi} = \begin{pmatrix} w_1(\xi) & 0 & \cdots & 0 \\ 0 & w_2(\xi) & \cdots & 0 \\ \vdots & \vdots & \ddots & \vdots \\ 0 & 0 & \cdots & w_N(\xi) \end{pmatrix}\ ,\quad\text{with}\quad w_i(\xi) = \te^{\xi_i} .
\end{equation}
Note that for the above matrix to be an element of the group $SU(N)$,
the $\xi_i$ must be imaginary i.e.~the matrix $M_\xi$ must be
antihermitian.

The Weyl character formula asserts that the character $\chi_\mu$ of an
irrep with highest weight $\mu$ evaluated on an element of the dual
Cartan space $\xi$ is given by
\begin{equation}\label{mucharmai}
  \chi_\mu(\xi) = \frac{\text{det}\ w_j(\xi)^{\ell^\mu_i + N - i}}{\text{det}\ w_j(\xi)^{N - i}} = \frac{\text{det}\ w_j(\xi)^{n^\mu_i}}{\text{det}\ w_j(\xi)^{N - i}}\ ,
\end{equation}
where the $w_i$ are given in terms of $\xi$ in \eqref{suma} and we
have defined, for future use,
\begin{equation}\label{nidef}
  n^\mu_i \equiv \ell_i^\mu + \ell^\rho_i = \ell_i^\mu + N-i , \quad\text{for}\quad i = 1,\ldots,N\ .
\end{equation}
Integrable representations of the affine Lie algebra $\wh{su}(N)_k$
(equivalently, the primaries of the $SU(N)_k$ WZW model) are also
parametrized by the weights of the ordinary Lie algebra $su(N)$ which
obey the additional \emph{level}-$k$ constraint
\begin{equation}\label{elamb}
  \ell^\lambda_1 \leq k\ .
\end{equation} 
In words, the integrable representations of $\wh{su}(N)_k$ correspond
to Young tableaux with at most $k$ boxes in their first rows. It
follows from the ordering \eqref{lord} and the level-$k$ constraint
\eqref{elamb} that
\begin{equation}\label{nord}
  \kappa-1 \geq n_1 > n_2 > \cdots > n_{N-1} \geq 1\ ,
\end{equation}
where recall that $\kappa = k + N$. It is easy to convince oneself
that the map between $\ell_i$ and $n_i$ is one-to-one: every choice of
$\ell_i$ obeying \eqref{lord} defines a choice of $\{n_i\}$ obeying
\eqref{nord} and vice versa. It follows that the total number of
integrable representations of $\wh{su}(N)_k$ is simply the number of
collections of $N-1$ distinct integers lying between (and including)
$1$ and $\kappa-1$ and so is given by
\begin{equation}\label{combno} 
\binom{\kappa - 1}{N-1} = \binom{\kappa-1}{k}\ .
\end{equation}

\subsubsection{Discretized $SU(N)$ eigenvalues}\label{disceigen}

According to equation \eqref{suverorm1}, the character $\chi_\mu$ is
evaluated on a diagonal matrix of the form \eqref{suma} whose
eigenvalues $w_i$ correspond to the special weight
$\xi_\lambda = -2\pi\i (\lambda + \rho)/\kappa$. That is
\begin{align}\label{whichmat} 
  w_i(\xi_\lambda) &= \exp\left(-\frac{2\pi\i}{\kappa} (\ell^\lambda_i - q_\lambda + \ell^\rho_i - q_\rho)\right)\ ,\nonumber\\
                   &= \exp\left(-\frac{2\pi\i n_i^\lambda}{\kappa} \right)\exp\left(\frac{2\pi\i}{N\kappa}\sum_i n_i^\lambda\right) \ ,
\end{align}
where, recall that $n^\lambda_i = \ell^\lambda_i + \ell^\rho_i$ and
$q_\lambda = \frac{1}{N}\sum_i \ell^\lambda_i$. We can then write
\begin{equation}\label{widef}
  w_i(\xi_\lambda) = z_i (\xi_\lambda)\,\te^{\i\theta_\lambda}\ ,
\end{equation}
where we have defined
\begin{equation}
  z_i(\xi_\lambda) = \exp\left(-\frac{2\pi\i n^\lambda_i}{\kappa}\right)\ ,\quad \theta_\lambda = \frac{2\pi}{\kappa} (q_\lambda + q_\rho) = \frac{2\pi}{N\kappa}\sum_{j=1}^{N-1} n^\lambda_j ,
\end{equation}
The diagonal $SU(N)$ matrix corresponding to an integrable
representation with highest weight $\lambda$ is then given by
\begin{equation}\label{won}
U^{(\lambda)} =  \begin{pmatrix} w_1(\xi_\lambda) & 0 & \cdots & 0 \\ 0 & w_2(\xi_\lambda) & \cdots & 0 \\ \vdots & \vdots & \ddots & \vdots \\ 0 & 0 & \cdots & w_N(\xi_\lambda) \end{pmatrix} = \te^{\frac{2\pi\i}{N\kappa} \sum_i n_i^\lambda} \begin{pmatrix} \te^{-\frac{2\pi\i n^\lambda_1}{\kappa}} & 0 & \cdots & 0 \\ 0 & \te^{-\frac{2\pi\i n^\lambda_2}{\kappa}} & \cdots & 0 \\ \vdots & \vdots & \ddots & \vdots \\ 0 & 0 & \cdots & 1 \end{pmatrix}\ ,
\end{equation} 
which indeed satisfies $\prod_i w_i(\xi_\lambda) = 1$.

Thus, for each integrable representation with highest weight
$\lambda$, there is a distinguished $SU(N)$ diagonal matrix with
eigenvalues
\begin{equation}
  \ul{w}_\lambda = \{w_1(\xi_\lambda),\ldots, w_N(\xi_\lambda)\}\ .
\end{equation}\label{disge}
with $w_i(\xi_\lambda)$ defined in \eqref{widef}. This map between
integrable representations and certain distinguished conjugacy classes
of $SU(N)$ has been described in \cite{Elitzur:1989nr,Moore:1989vd}
in the context of quantization of Chern-Simons theory on a cylinder
with a source (see \cite{Zuber:1995ig} for another (related) point of
view on the above map between integrable representations and
discretized $SU(N)$ matrices).

We call the set of these distinguished eigenvalue configurations
$\mc{P}_k$:
\begin{equation}\label{Pkdef}
  \mc{P}_k = \left\{\ul{w}_\lambda\ \bigg|\ \lambda\ \text{is the highest weight of an integrable representation of $\wh{su}(N)_k$}\right\}\ .
\end{equation}
The number elements of $\mc{P}_k$ is thus the number of integrable
representations of $\wh{su}(N)_k$ \eqref{combno}.

Observe that the eigenvalues
$\ul{w}_\lambda = \{w_1(\xi_\lambda),\ldots,w_N(\xi_\lambda)\}$ obey
the following properties for every choice of the highest weight
$\lambda$:
\begin{align}\label{propofwi}
  & w_i \neq w_j\ \text{for}\ i \neq j\ ,\quad w_i^\kappa =w_j^\kappa\ ,\quad \prod_{i=1}^N w_i=1\ .
\end{align}
In Appendix \ref{cpsu}, we argue that every solution
$\ul{w} = \{w_1,\ldots,w_N\}$ up to permutation of the above
constraints corresponds to an integrable highest weight of
$\wh{su}(N)_k$ and the solutions of the above constraints are in
one-to-one correspondence with highest weights of integrable
representations of $\wh{su}(N)_k$. Thus, the set $\mc{P}_k$ can be
alternatively characterized as the set of solutions up to permutation
of the constraints \eqref{propofwi}. The above constraints arise in
the path integral derivation of the Verlinde formula which we discuss
in Section \ref{suNpath} below.

\subsubsection{The Vandermonde factor} \label{tv}

Now let us turn to the evaluation of the quantity
$\mc{S}_{\hat\lambda 0}$ defined in \eqref{suverorm1}. For the case of
$su(N)$, the size of the centre is $|C| = N$, the rank is $r = N-1$,
the positive roots are $\alpha_{(ij)} = \epsilon_i - \epsilon_j$, and
the weights $\lambda$ and $\rho$ are given in the orthonormal basis as
\begin{equation}
  \lambda = \sum_{i=1}^N (\ell_i^\lambda - q_\lambda) \epsilon_i\ ,\quad   \rho = \sum_{i=1}^N (\ell_i^\rho - q_\rho) \epsilon_i\ ,
\end{equation}
with $\ell_i^\rho = N - i$. We thus obtain
\begin{equation} \label{alphadot}
  (\alpha_{(ij)}, \lambda+\rho) = \ell_i^\lambda + N - i - (\ell_j^\lambda + N - j) \equiv n^\lambda_i - n^\lambda_j\ ,
\end{equation} 
where $n^\lambda_i = \ell^\lambda_i + N - i$ was defined in
\eqref{nidef}. It follows that
\begin{align}\label{suNmeasure}
  (\mc{S}_{\lambda 0})^2
  &= \frac{1}{N \kappa^{N-1}} \prod_{1\leq i < j \leq N} \left| \te^{-\pi\i (n^\lambda_i-n^\lambda_j)/\kappa} - \te^{\pi\i (n^\lambda_i-n^\lambda_j)/\kappa}\right|^2\ ,\nonumber\\
  &= \frac{1}{N \kappa^{N-1}} \prod_{1\leq i < j \leq N} \left| w_i(\xi_\lambda) - w_j(\xi_\lambda)\right|^2\ .
\end{align}
In other words $(\mc{S}_{\hat\lambda 0})^2$ is an appropriately
normalized version of the $SU(N)$ Vandermonde factor evaluated on one
of the distinguished eigenvalue configurations $\{w_i(\xi_\lambda)\}$.

\subsubsection{The final formula} 
Putting all the results obtained above together, we conclude that the
dimension of the space of conformal blocks of the $SU(N)_k$ WZW model
on a genus $g$ surface \eqref{ngtm} is given by
\begin{equation}\label{ngtmnew} 
  \mc{N}_{g,n}({R}_1,\ldots,{R}_n) =    \frac{1}{(N \kappa^{N-1})^{1-g}} \sum_{\ul{w}_\lambda \in \mc{P}_k} \prod_{1 \leq i < j \leq N} |w_i(\xi_\lambda)-w_j(\xi_\lambda)|^{2-2g} \prod_{i=1}^n \chi_{\mu_i}(\xi_\lambda)\ ,
\end{equation}
where the sum over $\ul{w}_\lambda$ runs over the set of distinguished
eigenvalue configurations $\mc{P}_k$ with the explicit form of
$\ul{w}_\lambda = \{w_1(\xi_\lambda),\ldots,w_N(\xi_\lambda)\}$ is
given by \eqref{won}.

Since the eigenvalue configurations $\ul{w}_\lambda$ can be
alternatively characterized as solutions up to permutation of
\eqref{propofwi}, we sometimes drop all reference to the highest
weights $\lambda$ in \eqref{ngtmnew} and simply write
\begin{equation}\label{ngtmnewalt} 
  \mc{N}_{g,n}({R}_1,\ldots,{R}_n) =    \frac{1}{(N \kappa^{N-1})^{1-g}} \sum_{\ul{w} \in \mc{P}_k} \prod_{1 \leq i < j \leq N} |w_i - w_j|^{2-2g} \prod_{i=1}^n \chi_{\mu_i}(\ul{w})\ ,
\end{equation}
where we have used the same symbol $\mc{P}_k$ to denote the set of
solutions up to permutation of \eqref{propofwi}, and
$\chi_{\mu}(\ul{w})$ is the character of the representation $\mu$
evaluated on the eigenvalue configuration $\ul{w}$ as in
\eqref{mucharmai}.

\subsubsection{A special eigenvalue configuration} 
\label{sec} 

The trivial representation $\lambda=0$ is always an integrable
representation of $\wh{su}(N)_k$ for any $k$ and is defined by
$\ell^0_i=0$ which gives $n^0_i = N-i$. Plugging these into
\eqref{won} yields
\begin{equation}
  \ul{w}_0 = \left\{\te^{-\i\pi(N-1)/\kappa},\te^{-\i\pi(N-3)/\kappa},\ldots, \te^{\i\pi(N-1)/\kappa}\right\}\ ,
\end{equation}
that is, the diagonal $SU(N)$ matrix
\begin{equation}\label{evcI}
  U^{(0)} = \begin{pmatrix} \te^{-\i\pi (N-1)/\kappa} & 0 & \cdots & 0\\
            0 & \te^{-\i\pi (N-3)/\kappa} & \cdots & 0\\
            \vdots & \vdots & \ddots & \vdots\\
            0& 0 & \cdots & \te^{\i\pi (N-1)/\kappa}
          \end{pmatrix}\ .
\end{equation}
This matrix was encountered in Section \ref{lwcoo} in the situation
where there are a large number of character insertions in the Verlinde
formula \eqref{ngtmnew}. One way of thinking about $U^{(0)}$ is as
follows: it is the closest one can come to the identity matrix,
consistent with the conditions \eqref{propofwi}. Note that the
eigenvalues of \eqref{evcI} are as tightly packed around unity as
possible (with unity being one of the eigenvalues when $N$ is
odd). This fact was useful in the analysis in Section \ref{lwcoo}.

\subsubsection{Interpretation in terms of path integrals}\label{suNpath}

For this subsection we focus for convenience on the special case
$g=0$, i.e. the space of conformal blocks on the sphere (the
generalization to arbitrary genus $g$ surfaces is straightforward). In
this case, the final formula \eqref{ngtmnewalt} gives
\begin{equation}\label{ngtgeo} 
  \mc{N}_{0,n}({R}_1,\ldots,{R}_n) = \frac{1}{N \kappa^{N-1}}  \sum_{\ul{w} \in \mc{P}_k} \prod_{1 \leq i < j \leq N} |w_i-w_j|^2 \prod_{i=1}^n \chi_{\mu_i}(\ul{w})\ .
\end{equation}
The equation \eqref{ngtgeo} can be thought of as a very particular
discretization of the formula (see Appendix \ref{classcount}) for the
number of `classical' $SU(N)$ singlets i.e.~$SU(N)$ singlets in the
tensor product of $SU(N)$ representations $R_1,\ldots, R_n$ with
highest weights $\mu_1,\ldots, \mu_n$. Blau and Thompson, in their
beautiful paper \cite{Blau:1993tv}, explain why the formula for the
number of conformal blocks is a discretization of the formula for the
number of classical singlets.

The starting point of the analysis of \cite{Blau:1993tv} is the
observation by Witten \cite{Witten:1988hf} that the  number of
singlets in the fusion of WZW chiral primaries ${R}_1, \ldots, {R}_n$
can be obtained by computing the path integral on $S^2 \times S^1$ of
the $SU(N)_k$ Chern-Simons theory in the presence of Wilson loops in
the representations $R_1, \ldots, R_N$, each of which sit at a given
point on the $S^2$ and wind the $S^1$ once. Blau and Thompson
\cite{Blau:1993tv}) (see also \cite{Jain:2013py}) evaluated this path
integral and demonstrated that it could be reduced to the integral
over the (zero mode of the) $SU(N)$ holonomy matrix
\begin{equation}\label{sunmat}
  U(\ul{w}) = \begin{pmatrix} w_1 & 0  &\cdots &  0\\
    0 & w_2  &\cdots & 0\\
    \vdots &\vdots & \ddots  & \vdots \\
    0 & 0 & \cdots & w_N \end{pmatrix}\ ,
\end{equation} 
where the $w_i$ are all phases with $\prod_{i=1}^N w_i=1$. Following
\cite{Blau:1993tv} we obtain
\begin{multline}\label{nsingetsclnewel}
  \mc{N}_{0,n}({R}_1,\ldots,{R}_n) = \frac{1}{N!}  \oint
  \prod_{i=1}^N \frac{d w_i}{2 \pi \i w_i}\ 2\pi\delta(W-1) \prod_{1\leq
    i < j \leq N} |w_i-w_j|^2 \times \\ \times\left(\prod_{r=1}^{N-1}
    \sum_{m_r=-\infty}^{\infty} \left( \frac{w_{r}}{w_{r+1}}
    \right)^{-\kappa m_r} \right) \prod_{J=1}^m \chi_{\vgap
    R_J}(\ul{w}) \ ,
\end{multline}
where $W = \prod_i w_i$ and the integrand in the first line in
\eqref{nsingetsclnewel} above is the $SU(N)$ Haar measure\footnote{Our
  definition of the $\delta$-function $\delta(z)$ is given by
  \begin{equation}
    \oint \frac{dz}{\i z}\delta(z) = 1\ .\nonumber
  \end{equation}
  },
  $\chi_{\vgap R_J}(\ul{w})$ is the character of the representation
  $R_J$ evaluated on the matrix \eqref{sunmat}; the factor
\begin{equation}
  \prod_{r=1}^{N-1}\left( \frac{w_{r+1}}{w_r} \right)^{-\kappa m_r}\ ,
\end{equation}
is $\te^{-S_{\rm CS}}$ where $S_{\rm CS}$ is the Chern-Simons action
evaluated on the gauge field configuration with holonomy
\eqref{sunmat}. The integers $m_r$, $r=1,\ldots, N-1$, correspond to
magnetic fluxes through the $S^2$ w.r.t.~the $U(1)^{N-1}$ maximal
torus of $SU(N)$ generated by the following basis of the Cartan
generators:
\begin{equation}\label{sunkflux}
f_1= \begin{pmatrix}
	1 & 0 & 0 &\cdots & 0\\
	0 & -1 & 0 &\cdots & 0\\
	0 & 0 & 0 & \cdots & 0 \\
	\vdots &\vdots &\vdots & \ddots & \vdots \\
	0&0&0& \cdots & 0
      \end{pmatrix}\ ,\quad f_2 = \begin{pmatrix}
            0 & 0 & 0 &\cdots & 0\\
            0 & 1 & 0 &\cdots & 0\\
            0 & 0 & -1 & \cdots & 0 \\
            \vdots & \vdots & \vdots & \ddots & \vdots \\
            0&0&0& \cdots & 0\end{pmatrix}\ ,\ldots,\quad f_{N-1} =
          \begin{pmatrix}
                0 & \cdots & 0 &0 & 0\\
                \vdots  & \ddots  & \vdots & \vdots & \vdots \\
                0 & \cdots & 0 &0 & 0\\
                0& \cdots& 0 &1 &  0\\
                0 & \cdots & 0 & 0 & -1\end{pmatrix} \ .
\end{equation}
The sum over $m_r$ in \eqref{nsingetsclnewel} accounts for the fact
that one is required to sum over all flux sectors in the path integral
over $S^2$.
 
From the formula for the periodic delta function
\begin{equation}\label{discdfmt}
  \sum_{m=-\infty}^\infty 
  \te^{\i m \kappa \alpha} =  \frac{2 \pi}{\kappa} \sum_{n=-\infty}^\infty  \delta \bigg( \alpha -\frac{2 \pi n}{\kappa} \bigg) = 2\pi\sum_{n=-\infty}^\infty \delta( \kappa \alpha - 2\pi n)\ , 
\end{equation}  
it follows that
\begin{equation}\label{discdfmtn}
  \sum_{m=-\infty}^\infty 
  \left( \frac{w_{r}}{w_{r+1}} \right)^{-\kappa m_r} = 
  2 \pi \left( \frac{ w_{r}}{ w_{r+1}}\right)^{-\kappa} \delta\left( \left( \frac{ w_{r}}{ w_{r+1}}\right)^{-\kappa} -1\right)\ .
\end{equation}
We insert \eqref{discdfmtn} into \eqref{nsingetsclnewel} which gives
us $N-1$ $\delta$-functions of the form \eqref{discdfmtn}. These along
with $\delta(W-1)$ give a total $N$ $\delta$-functions which can be
integrated against the $N$ variables $w_i$. This step turns
\eqref{nsingetsclnewel} into a summation over all diagonal matrices
$U_{\ul{w}}$ with eigenvalues $w_i$ that satisfy
\begin{equation}
  w_i \neq w_j\ \text{for}\ i \neq j\ ,\quad  (w_i)^\kappa = (w_j)^\kappa\ ,\quad \prod_i w_i = 1\ .
\end{equation}
Note that these equations coincide exactly with the equations
\eqref{propofwi} that the distinguished eigenvalue configurations in
the Verlinde formula satisfy. It is not difficult to compute the
relevant Jacobian and convince oneself that
\begin{equation}\label{rrrr} 
\frac{1}{N!}  \int \prod_i  \frac{d w_i}{2 \pi \i w_i }\ 2\pi\delta(W-1)\ 
  \prod_{r=1}^{N-1} \left( \frac{ w_{r}}{ w_{r+1}}\right)^\kappa 2\pi\delta\left( \left( \frac{ w_{r}}{ w_{r+1}}\right)^\kappa -1\right) \rightarrow \frac{1}{N\kappa^{N-1}} \sum_{\ul{w} \in \mc{P}_k}\ ,
\end{equation} 
where the summation over $\ul{w} \in \mc{P}_k$ on the RHS of
\eqref{rrrr} is taken over the set $\mc{P}_k$ of eigenvalue
configurations up to permutation which obey the equations
\eqref{propofwi} (that the sum is over eigenvalue configurations up to
permutation is facilitated by the factor of $1/N!$ in \eqref{rrrr}). It
follows that \eqref{rrrr} reduces, on the nose, to \eqref{ngtmnew} at
$g=0$.

\subsubsection{Verification using supersymmetric localization} 

In Section \ref{pisl}, we verify using the method of supersymmetric
localization that the Chern-Simons path integral in the presence of
Wilson lines evaluates to \eqref{ngtmnew}.

\subsection{$U(N)_{k,k'}$ theories}

We now turn our attention to $U(N)_{k, k'}$ chiral WZW models which
correspond to the affine Lie algebra $\wh{u}(N)_{k,k'}$. Recall from
Section \ref{CSWZWconv} that $k'= \kappa+ s N$ (we have restricted
$k'$ to be positive in this subsection for brevity).

\subsubsection{$u(N)$ representation theory} \label{unrep}
Representations of $u(N)$, the Lie algebra corresponding to the group
$U(N)$, are specified by a highest weight $\hat\lambda$ of $su(N)$
along with a $u(1)$ charge $Q_\lambda$ which must satisfy
\begin{equation}
  Q_\lambda \equiv \hat{Q}_\lambda\ \text{mod}\ N\ ,
\end{equation}
where $\hat{Q}_\lambda$ is the number of boxes in the Young tableau
corresponding to the $su(N)$ representation $\hat\lambda$. Recall that
the row-lengths of the Young tableau are given by
$\hat\ell_i^\lambda$, $i =1,\ldots,N-1$ with
$\hat\ell^\lambda_1 \geq \cdots \geq \hat\ell^\lambda_{N-1} \geq 0$
and the number of boxes is
$\hat{Q}_\lambda = \sum_{i=1}^{N-1} \hat\ell_i^\lambda$. Thus, we
denote the weight of a $u(N)$ representation by
$\lambda = (\hat\lambda, Q_\lambda)$, where $\hat\lambda$ is the
$SU(N)$ weight and $Q_\lambda$ is the $U(1)$ charge.

Given the $su(N)$ Young tableau and the $u(1)$ charge $Q_\lambda$, we can
define an `extended' Young tableau which incorporates both the $su(N)$
and $u(1)$ charges as follows. Let us write the $u(1)$ charge as
$Q_\lambda = \hat{Q}_\lambda + r N$ with $r$ being a integer (could be
positive or negative, depending on the value of $Q_\lambda$). Then,
the $u(N)$ Young tableau has row-lengths
\begin{equation}\label{extendyd}
  \ell^\lambda_i = \hat\ell^\lambda_i + r \ ,\quad\text{for all}\quad i=1,\ldots, N\ .
\end{equation}
The $u(1)$ charge of a $u(N)$ Young tableau is then simply
\begin{equation}
  Q_\lambda = \sum_{i=1}^N \ell^\lambda_i\ .
\end{equation}
In particular, the `extended' Young tableau \eqref{extendyd} typically
has a non-zero $N^{\rm th}$ row $\ell^\lambda_N = r$ in contrast to
an $su(N)$ Young tableau. The row-lengths $\ell_i$ of a $u(N)$
Young tableau then satisfies
\begin{equation}
  \ell_1 \geq \ell_2 \geq \cdots \geq \ell_{N-1} \geq \ell_N\ .
\end{equation}
Since the row-lengths can be zero or negative, we need to include the
so-called \emph{antiboxes} in a pictorial depiction of the Young
tableau which fill up the rows with negative lengths. As usual,
ordinary boxes fill up rows with positive lengths and rows with zero
length have no boxes.

As in the case of $su(N)$, we use the weight vectors $\epsilon_i$
\eqref{epnorm} as a basis for $u(N)$ weights. The $u(N)$
representation with weight $\lambda = (\hat\lambda, Q_\lambda)$ is
expanded as
\begin{equation}\label{unweight1}
  \lambda = \sum_{i=1}^N (\hat\ell^\lambda_i - \hat{q}_\lambda) \epsilon_i + \frac{Q_\lambda}{N} \sum_{i=1}^N \epsilon_i\ ,
\end{equation}
where that $\hat{q}_\lambda = \hat{Q}_\lambda / N$. Using the
definition of the row-lengths $\ell^\lambda_i$ of the $u(N)$ Young
tableau \eqref{extendyd}, we can simply write the above decomposition
as
\begin{equation}\label{uNweight}
  \lambda = \sum_{i=1}^N \ell^\lambda_i\, \epsilon_i\ .
\end{equation}
The $N^2-N$ roots of $u(N)$ continue to be given by
$\epsilon_i - \epsilon_j$, $i \neq j$. We continue to use the $su(N)$
Weyl vector which we denote $\hat\rho$ and is half sum over positive
roots \eqref{formrho}. The character of a $u(N)$ representation
$\mu = (\hat\mu, Q)$ is the product of the $su(N)$ character
$\chi_{\hat\mu}(\hat\xi)$ and the $u(1)$ character
$\chi_Q = \te^{\i Q \theta}$:
\begin{equation}
  \chi_{\mu}(\hat\xi,\theta) = \chi_{\hat\mu}(\hat\xi) \times \chi_Q(\theta)\ .
\end{equation}
Using the Weyl character formula, we see that the above character is
given by
\begin{equation}\label{unchar}
  \chi_{\mu}(\hat\xi,\theta) = \te^{\i Q\theta}\, \frac{\text{det}\ w_j(\hat\xi)^{\hat\ell^\mu_i + N - i}}{\text{det}\ w_j(\hat\xi)^{N - i}}\ ,
\end{equation}
where the phases $w_j(\hat\xi)$ are the elements of the diagonal
$SU(N)$ matrix \eqref{suma}. Recall that
$Q_\lambda = \hat{Q}_\lambda + rN$; using this, we see that the $u(1)$
character can be absorbed into the determinants by redefining
$w_j(\hat\xi)$ and $\hat\ell^\mu_i$:
\begin{equation}
  w_j(\xi) = w_j(\hat\xi) \te^{\i\theta}\ ,\quad \ell^\mu_i = \hat\ell^\mu_i + r\ ,
\end{equation}
where we have already encountered the row-lengths $\ell^\mu_i$ of
the $u(N)$ Young tableau \eqref{extendyd}. The $\theta$ can be
absorbed in a redefinition of $\hat\xi$ to get
\begin{equation}\label{tlxi}
  \xi = \sum_{i=1}^N \hat\xi_i \epsilon_i + \theta \sum_{i=1}^N \epsilon_i = \sum_{i=1}^N (\hat\xi_i + \theta) \epsilon_i\ ,\quad\text{with}\quad \sum_{i=1}^{N} \hat\xi_i = 0\ .
\end{equation}
Thus, the $u(N)$ character $\chi_{\mu}(\xi)$ is given by
\begin{equation}\label{charmaiun}
  \chi_{\mu}(\xi) = \frac{\text{det}\ w_j(\xi)^{\ell^\mu_i + N - i}}{\text{det}\ w_j(\xi)^{N - i}} = \frac{\text{det}\ w_j(\xi)^{{n}^\mu_i}}{\text{det}\ w_j(\xi)^{N - i}}\ ,
\end{equation}
with ${n}^\mu_i \equiv \ell_i^\mu + N - i$ for $i = 1,\ldots,N$.

Note that the character formula for $su(N)$ that was discussed in
Section \ref{dteb} carries through for $u(N)$ representations as well,
when we interpret the weight $\mu$ as a $u(N)$ weight
\eqref{unweight1}, the row-lengths $\ell^\mu_i$ as $u(N)$ Young
tableau row-lengths \eqref{extendyd} and the dual Cartan element $\xi$
as a $u(N)$ dual Cartan element \eqref{tlxi}.

\subsubsection{Integrable representations for the $U(N)_{k, k'}$ WZW
  model}\label{unintrep}

The group $U(N)$ of $N \times N$ unitary matrices can be defined as
\begin{equation}\label{UNdef}
  \frac{SU(N) \times U(1)}{\mbb{Z}_N}\ ,
\end{equation}
where the `diagonal' $\mbb{Z}_N$ group in the denominator is generated
by the $SU(N)$ central element $\te^{-\frac{2 \pi \i}{N}}$ together
with a simultaneous rotation $\te^{\frac{2 \pi \i}{N}}$ in $U(1)$. The
$U(N)_{k,k'}$ WZW model also possesses this $\mbb{Z}_N$ gauge symmetry
which acts on its primaries. A $U(N)_{k,k'}$ primary is labelled by an
$SU(N)_k$ integrable representation $\hat\sigma$ and a $U(1)_{Nk'}$
integrable representation with charge $Q_\sigma$ which must satisfy
$0\leq Q_\sigma < Nk'$. The $U(1)$ charge $Q_\sigma$ must further
satisfy
\begin{equation}\label{modN}
  Q_\sigma - \hat{Q}_\sigma = rN\ ,
\end{equation}
where $\hat{Q}_\sigma$ is the number of boxes in the Young tableau of
the $SU(N)$ representation $\hat\sigma$ and $r$ is an integer. The
above constraint arises due to the $\mbb{Z}_N$ gauge symmetry as
follows. The $U(N)$ representation $\sigma$ transforms under the
generator of the diagonal $\mbb{Z}_N$ in the denominator of
\eqref{UNdef} with the phase
\begin{equation}\label{diagact}
  \exp\left(\frac{2\pi\i \hat{Q}_\sigma}{N}\right) \times \exp\left(-\frac{2\pi\i Q_\sigma}{N}\right)\ ,
\end{equation}
where the first factor arises from the diagonal $\mbb{Z}_N$ acting as
the centre symmetry $\mbb{Z}_N$ on the $SU(N)$ representation with
$\hat{Q}_\sigma$ boxes, and the second factor arises from the action
of the simultaneous $\mbb{Z}_N$ as a $U(1)$ transformation on the
$U(1)$ representation with charge $Q_\sigma$. Since the diagonal
$\mbb{Z}_N$ is supposed to be not part of the $U(N)$ group (it is
`gauged' away), the phase \eqref{diagact} must be trivial. This gives
the condition \eqref{modN}.

In the absence of the $\mbb{Z}_N$ gauge symmetry, the total number of
$U(N)_{k, k'}$ primaries would have been the number of integrable
representations of $SU(N)_k$ \eqref{combno} times the number of
integrable representations of $U(1)_{Nk'}$, i.e. would have been
\begin{equation}\label{naiveest} 
  \binom{\kappa - 1}{N-1} \times Nk'\ .
\end{equation}
This number is reduced by a factor of $N^2$ due to the $\mbb{Z}_N$
gauge symmetry in \eqref{UNdef}. First, the condition \eqref{modN}
cuts down the number of allowed $U(1)$ charges by a factor of $N$. The
reduction by a second factor of $N$ results from the fact that the
centre $\mbb{Z}_N$ acts on the integrable representations $\hat\sigma$
of the $SU(N)_k$ WZW model $\hat\sigma \to R(\hat\sigma)$ (see
\eqref{znmove} for a description of this action). There is a similar
action on the $U(1)$ representations which shifts the $U(1)$ charge as
$Q\to Q + k'$.\footnote{Note that this is a genuine $\mbb{Z}_N$ action
  since shifting $N$ times gives $Q + Nk'$ which is equivalent to $Q$
  as an integrable representation of $U(1)_{Nk'}$.}. Thus, the entire
$\mbb{Z}_N$ orbit generated by the simultaneous $\mbb{Z}_N$ action
$(\hat\sigma, Q_\sigma) \to (R(\hat\sigma), Q_\sigma + k')$ must
correspond to one $U(N)_{k,k'}$ representation. It follows that the
number of $U(N)_{k,k'}$ representations equals
\begin{equation}\label{actest} 
\binom{\kappa - 1}{N-1} \times \frac{k'}{N}\ .
\end{equation} 
Recall that integrable representations of $SU(N)_k$ are labelled by
$SU(N)$ Young tableaux with no more than $k$ columns. In Appendix
\ref{expir} we find a similar description of integrable
representations of $U(N)_{k, k'}$ for $s \leq 0$ (recall that
$k' = \kappa + sN)$. We present strong evidence that these may be
taken to be labelled by $U(N)$ Young tableaux with all row lengths
positive, and further subject to the following restrictions:
\begin{itemize} 
\item When $s=0$ or $s=-1$, the Young tableaux have no more than $k+s$
  columns. This simple rule encompasses the case of the Type I and Type
  II theories of principal interest to this paper.

\item When $s \leq -2$, the length of the $i^{\rm th}$ row of the
  Young tableau is restricted to be \emph{strictly} smaller than
  $k+(s+1)i$. In other words the length of the first row of the Young
  tableau is $\leq k+s$, the length of the second row is
  $\leq k+2(s+1)-1$, and so on, with the length of the last
  i.e.~$N^{\rm th}$ row $\leq k+N(s+1) - 1 = k'-1$.
\end{itemize} 
Although the extension to $s>0$ may not be difficult, we do not have a
well motivated conjecture for a `fundamental domain' of integrable
representations in this case.

\subsubsection{Verlinde $\mc{S}$-matrices} 

The Verlinde $\mc{S}$-matrix for the $U(N)_{k,\kappa}$ i.e.~the Type I
$U(N)$ WZW model was worked out by Naculich and Schnitzer
\cite{Naculich:2007nc}. A slight generalization of their discussion
allows us to demonstrate that the Verlinde $\mc{S}$-matrix for the
family of theories $U(N)_{k,k'}$ is given by
\begin{equation}
  \mc{S}_{\lambda\mu} = \sqrt{\frac{N}{k'}}  \,\mc{S}_{\hat\lambda\hat\mu}\,\exp\left(-\frac{2\pi\i}{Nk'} {Q}_\lambda {Q}_\mu\right)\ ,
\end{equation}
where $\lambda = (\hat\lambda, Q_\lambda)$, $\mu = (\hat\mu, Q_\mu)$
are $u(N)$ integrable weights with $\hat\lambda$ and $\hat\mu$ being
$su(N)$ highest weights, and $\mc{S}_{\hat\lambda\hat\mu}$ on the
right hand side is the $SU(N)_k$ Verlinde $\mc{S}$-matrix. Thus, the
analog for $U(N)_{k,k'}$ of the relation between $\mc{S}$-matrices and
classical characters for $SU(N)_k$ \eqref{suverorm} becomes
\begin{equation}\label{suverormu} 
  \frac{\mc{S}_{\lambda\mu}}{\mc{S}_{\lambda 0}}  =  \te^{-2\pi\i {Q}_\lambda {Q}_\mu/Nk'}\ \chi_{\hat\mu}(\xi_{\hat\lambda})\ ,
\end{equation}
where $\mc{S}_{\lambda 0}$ is given by
\begin{equation}\label{sl0u}
  \mc{S}_{\lambda 0} =  \sqrt{\frac{N}{k'}} \mc{S}_{\hat\lambda 0} = \frac{1}{\sqrt{k' \kappa^{N-1}}} \prod_{\alpha \in \Delta_+} 2\sin\left(\frac{\pi}{\kappa} (\alpha,  \hat\lambda+\hat\rho)\right)\ ,
\end{equation}
where all quantities on the right hand side are $su(N)$ quantities as
in the $su(N)$ case \eqref{suverorm1}. The character
$\chi_{\hat\mu}(\xi_{\hat\lambda})$ is the $su(N)$ character for
highest weight $\hat\mu$ evaluated on the special weight
$\xi_{\hat\lambda}$ as defined in \eqref{suverorm1} which we reproduce
here for convenience:
\begin{equation} 
  \xi_{\hat\lambda} = -\frac{2\pi\i}{\kappa} ({\hat\lambda} + \hat\rho)\ .
\end{equation}
(We have an additional hat on the $su(N)$ weights $\hat\lambda$ and
$\hat\rho$ compared to \eqref{suverorm1} since unhatted quantities
correspond to $u(N)$ weights in this subsection.) The right hand side
of the relation \eqref{suverormu} can be reinterpreted as a $u(N)$
character $\chi_{\mu}$ (see discussion after \eqref{unchar}) by
treating the phase as the $u(1)$ character $\te^{\i {Q}_\mu\theta}$ of
the representation $\mu = (\hat\mu, Q_\mu)$ evaluated on the special
element
\begin{equation}
  \theta = -\frac{2\pi}{Nk'}{Q}_\lambda\ .
\end{equation}
This leads us to define the special $u(N)$ weight (see \eqref{tlxi}
for the definition of a $u(N)$ weight)
\begin{align}\label{tlxiu}
  \xi_{\lambda} &= -\frac{2\pi\i}{\kappa} (\hat\lambda + \rho) - \frac{2\pi\i {Q}_\lambda}{Nk'} \sum_{i=1}^N \epsilon_i\ ,\nonumber\\
                &= -\frac{2\pi\i}{\kappa} \sum_{i=1}^N(\hat\ell^\lambda_i + N - i - q_\lambda - q_\rho)\epsilon_i - \frac{2\pi\i}{k'}\frac{{Q}_\lambda}{N} \sum_{i=1}^N \epsilon_i\ ,\nonumber\\
                &= -\frac{2\pi\i}{\kappa} \sum_{i=1}^N(\ell^\lambda_i + N - i - q_\rho)\epsilon_i + \left(\frac{1}{\kappa} - \frac{1}{k'}\right) \frac{2\pi\i{Q}_\lambda}{N} \sum_{i=1}^N \epsilon_i\ ,\nonumber\\
                &= -\frac{2\pi\i}{\kappa} \sum_{i=1}^N(\ell^\lambda_i + N - i - q_\rho)\epsilon_i +  \frac{2\pi\i s {Q}_\lambda}{\kappa k'} \sum_{i=1}^N \epsilon_i\ ,\nonumber\\
                &= -\frac{2\pi\i}{\kappa} \sum_{i=1}^N \left(\ell^\lambda_i + N - i - q_\rho - \frac{s {Q}_\lambda}{k'}\right)\epsilon_i\ .
\end{align}
Thus, the right hand side of \eqref{suverormu} becomes
\begin{equation}\label{suverormu1} 
  \frac{\mc{S}_{\lambda\mu}}{\mc{S}_{\lambda 0}} = \chi_{\mu}(\xi_{\lambda})\ ,
\end{equation}
where $\chi_\mu$ is the $u(N)$ character of the $u(N)$ representation
$\mu$, the special $u(N)$ weight $\xi_\lambda$ is given in
\eqref{tlxiu}, and the quantity $\mc{S}_{\lambda 0}$ is given in
\eqref{sl0u}.

\subsubsection{Discretized $U(N)$ eigenvalues}\label{undisc}
The special $u(N)$ weight $\xi_\lambda$ in \eqref{tlxiu} gives
eigenvalues $w_i(\xi_\lambda) = \te^{\xi_{\lambda i}}$ of a diagonal
$U(N)$ matrix:
\begin{equation}\label{wonuo}
  w_i(\xi_{\lambda}) = \exp\left(-\frac{2\pi\i  n^{\lambda}_i}{\kappa}\right) \exp\left(\frac{2\pi\i q_\rho}{\kappa} + \frac{2\pi \i s {Q}_\lambda}{\kappa k'}\right)\ ,
\end{equation}
where $n_i^{\lambda} = \ell^\lambda_i + N - i$. The formula
\eqref{wonuo} can be rewritten using
${Q}_\lambda = \sum_i {\ell}^\lambda_i = \sum_i (n^\lambda_i - (N-i))$
as
\begin{equation}\label{wonuo1}
  w_i(\xi_{\lambda}) = \exp\left(-\frac{2\pi\i  n^{\lambda}_i}{\kappa}\right) \exp\left(\frac{2\pi\i q_\rho}{k'} + \frac{2\pi\i s}{\kappa k'}\sum_{i=1}^N n^{\lambda}_i\right)\ .
\end{equation}
We obtain an invariant characterization of the eigenvalues
\eqref{wonuo} for comparison with the $U(N)_{k,k'}$ Chern-Simons path
integral later. From \eqref{wonuo}, we note that each of the
eigenvalues $w_i$ obeys
\begin{equation}\label{wilamb} 
  w_i^\kappa = \exp\left(2\pi\i q^\rho + \frac{2\pi\i s {Q}_\lambda}{k'}\right)\ ,
\end{equation} 
and the product is given by
\begin{equation}\label{prodw}
  \prod_{i=1}^N w_i = \exp\left(-\frac{2\pi\i {Q}_\lambda}{\kappa} + \frac{2\pi\i sN {Q}_\lambda}{\kappa k'}\right) = \exp\left(-\frac{2\pi\i {Q}_\lambda}{k'} \right)\ .
\end{equation} 
where we have used $\sum_{i=1}^N n^\lambda_i= Q_\lambda + N
q_\rho$. It follows that
\begin{equation}\label{prodw1}
  w_i^\kappa  \left( \prod_{j=1}^N w_j \right)^s = \te^{2 \pi \i q_\rho} = (-1)^{N-1}\ .
\end{equation} 
In Appendix \ref{cpu}, we show that every solution of the above
equations modulo permutation corresponds to an integrable
representation of the affine algebra $\wh{u}(N)_{k,k'}$ i.e. a primary
operator of the $U(N)_{k,k'}$ WZW model discussed in Section
\ref{unintrep}.
 
\subsubsection{The Vandermonde} \label{tvu} 

The Vandermonde-like factor in the Verlinde formula for the
$U(N)_{k,k'}$ WZW model is given by
\begin{align}\label{uNmeasure}
  (\mc{S}_{\lambda 0})^2
  &= \frac{1}{k' \kappa^{N-1}} \prod_{1\leq i < j \leq N} \left| \te^{-\pi\i (\hat{n}^\lambda_i-\hat{n}^\lambda_j)/\kappa} - \te^{\pi\i (\hat{n}^\lambda_i-\hat{n}^\lambda_j)/\kappa}\right|^2\ ,\nonumber\\
    &= \frac{1}{k' \kappa^{N-1}} \prod_{1\leq i < j \leq N} \left| \te^{-\pi\i (n^{\lambda}_i-n^{\lambda}_j)/\kappa} - \te^{\pi\i (n^{\lambda}_i-n^{\lambda}_j)/\kappa}\right|^2\ ,\nonumber\\
  &= \frac{1}{k' \kappa^{N-1}} \prod_{1\leq i < j \leq N} \left| w_i(\xi_\lambda) - w_j(\xi_\lambda)\right|^2\ .
\end{align}
In going to the second step, we have used that
$n^{\lambda}_i = \hat{n}^{\lambda}_i + \frac{{Q}_\lambda -
  \hat{Q}_\lambda}{N}$ which follows from
$n^{\lambda}_i = \ell^\lambda_i + N - i$ and the definition of
$\ell^\lambda_i$ \eqref{extendyd}.

\subsubsection{The final formula} \label{tffu}

We conclude that the Verlinde formula for the dimension of the space
of conformal blocks $\mc{N}_{g,n}$ for the $U(N)_{k,k'}$ WZW theory
for the $n$ primaries $\mu_1, \mu_2, \ldots, \mu_n$ on a genus
$g$ surface may be recast in the following simple explicit form
\begin{equation}\label{ngtmnewu} 
  \mc{N}_{g,n} = \sum_{\ul{w} \in \mc{P}_{k,k'}}   \frac{1}{(k' \kappa^{N-1})^{1-g}} \prod_{1 \leq i < j \leq N} |w_i-w_j|^{2-2g} \prod_{i=1}^n \chi_{\mu_i}(\ul{w})\ ,
\end{equation}
where the sum over $\ul{w}$ runs over the set of distinguished
eigenvalue configurations $\mc{P}_{k,k'}$ indexed by the integrable
representations of $\wh{u}(N)_{k,k'}$ or equivalently, the set
of solutions modulo permutation of the equations \eqref{prodw}. The
eigenvalues $\ul{w} = \{w_1(\xi_\lambda),\ldots,w_N(\xi_\lambda)\}$
are given by \eqref{wonuo}.

In Appendix \ref{cpu}, we have explicitly given the one-to-one
correspondence between integrable representations of
$\wh{u}(N)_{k,k'}$ and the solutions of \eqref{prodw} up to
permutation. This shows the equivalence between the Chern-Simons path
integral and the Verlinde formula for the $U(N)_{k,k'}$
theories.

\subsubsection{Eigenvalues corresponding to the trivial representation} 
\label{secu}

As in the case of the $SU(N)_{k}$ theory, it is interesting to ask
what eigenvalue configuration corresponds to the identity
representation in the case of the $U(N)_{k, k'}$ theory. The answer to
this question is given by plugging $\ell^0_i=0$ i.e.~$n^0_i=N-i$ into
\eqref{wonuo}. Interestingly enough we find that the corresponding
eigenvalue configuration, $U^{(0)}$ is once again given by
\eqref{evcI} as for the $SU(N)_k$ theory for every value of $k'$. Once
again this is the eigenvalue configuration nearest to the identity
that is allowed by the quantization conditions \eqref{prodw}. It is
interesting that though the detailed quantization conditions
\eqref{prodw} depend on $k'$, the special eigenvalue configuration
\eqref{evcI} obeys these conditions for every value of $k'$.

\subsubsection{A path integral derivation}\label{UNpath}
Recall that the level of the $U(1)$ part of the $U(N)_{k,k'}$ (working
in a normalization in which $\theta$ is $2 \pi$ periodic, i.e. the
charges under this $U(1)$ are all integers) is $N k'$ where, $k'$ is
defined by
\begin{equation}\label{kpdefalt}
  k'= \kappa + s N \ ,
\end{equation}
where $s$ is an integer. Consider the diagonal $N \times N$ $U(N)$
gauge field $x$ which is the result of gauge fixing to the maximal
abelian gauge (see Appendix \ref{oneloop} for details). The Euclidean
Chern-Simons action on this configuration takes the form
\begin{equation}\label{rcsa}
S_{\rm CS} = \frac{\i \kappa}{4\pi} \int {\rm Tr} \left( x dx \right)  
+ \frac{\i s}{4 \pi} \int {\rm Tr} (x)\, d {\rm Tr} (x)  \ , 
\end{equation} 
(see equation A.14 of \cite{Minwalla:2020ysu}).

Now consider a diagonal $U(N)$ gauge field configuration in which each
of the $N$ $U(1)$ factors carries magnetic flux $2 \pi m_i$,
$i=1,\ldots,N$, and the holonomies over the time circle are given by
the constant $U(N)$ matrix $U_{\ul{w}} = \diag\{w_1,\ldots,
w_N\}$. The factor $\te^{-S_{\rm CS}}$ in the path integral
($S_{\rm CS}$ is the Euclidean action \eqref{rcsa}) evaluates on this
configuration to
\begin{equation}\label{expse}
  \prod_{i=1}^N w_i^{-\kappa m_i -s \sum_{i} m_i }\ .
\end{equation}
In addition to the above, there is a contribution from the one-loop
determinants that arise from integrating out the off-diagonal modes of
the gauge field:
\begin{equation}\label{oneloopsign}
  (-1)^{(N-1)\sum_i m_i} \prod_{1 \leq i < j \leq N} |w_i - w_j|^2\ .
\end{equation}
The sign above was absent in the $SU(N)$ case since it couples to the
total magnetic flux $\sum_i m_i$ which was zero for $SU(N)$. We
describe the calculation of the one-loop determinants in Appendix
\ref{oneloop}, closely following \cite{Blau:1993tv} and
\cite{Jain:2013py}.

It follows that the path integral on $S^2 \times S^1$ for
$U(N)_{k,k'}$ Chern-Simons theory with Wilson lines along $S^1$ in the
representations $R_1,\ldots,R_n$ is
\begin{multline}\label{csmf} 
\mc{N}_{0,n}(R_1,\ldots,R_n) = \frac{1}{N!}
\oint \prod_i  \frac{d w_i}{2 \pi i w_i} \prod_{1 \leq i < j \leq N}|w_i-w_j|^2  \times \\ \times \left( \sum_{m_i=-\infty}^{\infty} (-1)^{(N-1)\sum_i m_i} \prod_{r=1}^{N} 
w_i^{- \kappa m_i - s \sum_{i} m_i}  \right)
 \prod_{J=1}^n \chi_{\vgap R_J}(\ul{w})\ .
\end{multline}
The terms in \eqref{csmf} involving $m_i$ for a fixed $i$ are
\begin{equation}
  \left[(-1)^{N-1} w_i^{-\kappa} 
    (w_1 w_2 \ldots w_N)^{-s} \right]^{m_i}\ .
\end{equation}
It follows that the sum over $m_i \in \mbb{Z}$ for a fixed $i$
evaluates to
\begin{equation}
  2 \pi (-1)^{(N-1)} w_i^{-\kappa} 
    \left(w_1 w_2 \ldots w_N\right)^{-s}
  \delta \left( (-1)^{N-1} w_i^{-\kappa} 
    \left(w_1 w_2 \ldots w_N\right)^{-s}-1 \right)\ .
\end{equation}
It follows that \eqref{csmf} equals 
\begin{multline}\label{csmfn} 
  \mc{N}_{0,n}(R_1,\ldots,R_n)\\ = \frac{1}{N!}  \oint \prod_i \frac{d
    w_i}{2 \pi \i w_i} \prod_{1 \leq i < j \leq N}|w_i-w_j|^2
  \prod_{j=1}^N W_j \delta(W_j-1)  \prod_{J=1}^n
  \chi_{\vgap R_J}(\ul{z})\ ,
\end{multline}
with
\begin{equation}
  W_j = (-1)^{N-1} w_j^{-\kappa } \left( \prod_{i=1}^N w_i \right)^{-s}\ ,\quad\text{for}\quad j=1,\ldots,N\ .
\end{equation}  
It follows immediately from the $\delta$-functions in \eqref{csmfn}
that the integral over the eigenvalues $w_i$ reduces to a sum over
eigenvalue configurations that obey
\begin{equation}\label{obey}
w_j^{\kappa } \left( \prod_{i=1}^N w_i \right)^s = (-1)^{N-1}\ ,\quad\text{for}\quad j = 1,\ldots,N\ .
\end{equation} 
Note that the above equations are exactly those satisfied by the
distinguished eigenvalue configurations of Section \ref{undisc}. Due
to the $1/N!$ in the path integral \eqref{csmfn}, we have to count the
solutions to the above equations modulo $\mc{S}_N$ permutations. We
label the set of such solutions modulo permutations as
$\mc{P}_{k,k'}$.

It is not difficult to evaluate the relevant Jacobian factors to
verify that
\begin{equation}\label{rrrrun} 
  \frac{1}{N!}  \oint \prod_i  \frac{d w_i}{2 \pi \i w_i } 
  \prod_{j=1}^{N}  W_j \delta(W_j)
  \rightarrow \frac{1}{k'\kappa^{N-1}} \sum_{\ul{w} \in \mc{P}_{k,k'}}
\end{equation}
where the summation on the RHS runs over the set $\mc{P}_{k,k'}$ of
all solutions modulo permutations of \eqref{obey}. Using \eqref{rrrr}
in the formula for $\mc{N}_{0,n}$ in \eqref{csmfn} yields the Verlinde
formula \eqref{ngtmnewu}.

\subsection{Supersymmetric localization and the path
  integral} \label{pisl}

\subsubsection{{\cal N}=2 Chern-Simons matter theories}

In this subsection, we consider the ${\cal N}=2$ supersymmetric
generalizations of the $SU(N)_k$ and $U(N)_{k,k'}$ Chern-Simons matter
theories on $S^2 \times S^1$ described in Section \ref{dsr}. We
present the superconformal index for these theories in the presence of
supersymmetric Wilson lines and relate them to the Wilson line
correlation functions of the non-supersymmetric Chern-Simons theories
discussed earlier in this paper\footnote{See
  \cite{Benini:2015noa,Benini:2016hjo,Closset:2016arn} for similar,
  more general results that are obtained by computing a supersymmetric
  index for the topologically twisted $\mc{N}=2$ Chern-Simons matter
  theories on $\Sigma_g \times S^1$.}. In fact, we will actually only
study the pure ${\cal N}=2$ Chern-Simons theory (i.e. the theory with
no matter multiplets. Equivalently, the $N_f=0$ case of \eqref{n2act}
below).  However, with an eye to future generalizations, we first
discuss and present the superconformal index for a more general class
of theories than we actually need, namely $\mc{N}=2$ Chern-Simons
matter theories with $N_f$ chiral multiplets. Later in this section we
will specialize our analysis to the simple case ${ N}_f=0$. For the
purposes of the current paper, the reader will lose nothing if she
simply sets all occurrences of ${ N}_f$ to zero in all formulae of
this subsection.

The off-shell field content of the theory is in two supersymmetry
multiplets. The $SU(N)$ or $U(N)$ gauge field $A_\mu$ sits in the
vector multiplet $V \equiv (A_\mu, \lambda_\alpha, \sigma, D)$ which
also consists of the gaugino $\lambda_\alpha$, a real scalar $\sigma$
and a real auxiliary field $D$. All of these fields transform in the
adjoint representation of the gauge group. The matter fields
transforming in the fundamental representation of gauge group belong
to $N_f$ chiral multiplets
$\Phi^j \equiv (\phi^j, \psi_\alpha^j, F^j)$ where $F^j$ are auxiliary
complex scalars. The $U(N)_{\tl{k}, \tl{k}'}$ theory is the described
by the following action
\begin{align}\label{n2act}
  S & = \int d^3x \bigg[ \frac{\i}{4\pi} \left( \tl{k} - \frac{N_f}{2}\sgn(\tl{k}) \right) 
      \textrm{Tr} \left( \varepsilon^{\mu\nu\rho} (A_\mu \partial_\nu A_\rho - \tfrac{2\i}{3}A_\mu A_\nu A_\rho) - 
      \bar\lambda \lambda + 2\sigma D \right)\nonumber  \\
    &\qquad + \frac{\i}{4\pi} N \left( \tl{k}' - \tl{k} - \frac{N_f}{2}\sgn(\tl{k}) \right) \left(\varepsilon^{\mu\nu\rho} (\tr A_\mu) \partial_\nu (\tr A_\rho) - (\tr \bar\lambda) (\tr\lambda) + 2 (\tr\sigma) (\tr D)\right)\nonumber\\
    & \qquad  + \sum_{j=1}^{N_f} \left( D_\mu \bar\phi_j D^\mu \phi^j 
      + i \bar\psi_j \gamma^\mu D_\mu \psi^j + \bar\phi^j (- \sigma^2 + D) \phi_j - \bar\psi_j \sigma \psi^j + 
      i \bar\phi_j \bar\lambda \psi^j - i \bar\psi_j \lambda \phi^j \right) \bigg]\ .
\end{align}
For consistency, the level $\tl{k}'$ of the $U(1)$ part of the gauge
group must be quantized as $\tl{k}'=\tl{k} + sN$ with $s$ an integer
(e.g. see \cite{Minwalla:2020ysu}). The shift by
$-\tfrac{N_f}{2}\sgn(\tl{k})$ is a consequence of the parity anomaly
of the fermions in the chiral multiplet i.e.~integrating out $N_f$
fermions with mass $m_F$ in the fundamental representation generates
an additional Chern-Simons term with level $\tfrac{N_f}{2}\sgn(m_F)$
so that the level of the infrared Chern-Simons theory is
$\tl{k} - N_f$ or $\tl{k}$ depending on whether
$\sgn(m_F) = \pm \sgn(\tl{k})$ (see e.g.~Eq (D.5) of
\cite{Aharony:2018pjn}). We will see below that the shift by
$\tfrac{N_f}{2} \sgn(\tl{k})$ is also required for the consistency of
the superconformal index.

In the 't Hooft large $N$ limit (while keeping
$N_f \sim {\cal O}(N^0)$), the $\mc{N}=2$ supersymmetric theory (and
massive deformations thereof) is also solvable using techniques
similar to those used to solve the bosonic and fermionic
theories. All-orders-exact (in 't Hooft coupling) results for the
thermal partition function \cite{Jain:2012qi,Jain:2013py}, correlation
functions \cite{Inbasekar:2019wdw} and scattering amplitudes
\cite{Inbasekar:2015tsa} are known for this theory.

The auxiliary fields $\lambda_\alpha, \sigma, D$ and $F$ can easily be
integrated out since they appear quadratically in the action
\eqref{n2act} and the action can be recast as a non-supersymmetric
Chern-Simons matter theory purely in terms of the dynamical fields
$A_\mu, \phi^j, \psi_\alpha^j$. However, the levels in the Chern-Simons
action may be shifted in the process of integrating out the gauginos
$\lambda_\alpha$. Before moving on, let us discuss this possible shift
in the Chern-Simons levels.

\subsubsection{Relationship between supersymmetric and
  non-supersymmetric Chern-Simons levels}
As discussed in detail \cite[Appendix D]{Aharony:2018pjn}, the level
in front of the $SU(N)$ Chern-Simons action depends on the
regularization scheme used to regulate the divergences, the two most
common regularization schemes used in the literature being Yang-Mills
regularization (YM-reg) and dimensional regularization
(dim-reg). Integrating out the gaugino in YM-reg scheme shifts the
level in front of the $SU(N)$ Chern-Simons action by $-\sgn(\tl{k})N$
whereas it does not shift the $U(1)$ level since the gauginos are
neutral under the $U(1)$. That is,
\begin{align}\label{levelrel1}
  \mc{N}=2\  U(N)_{\tl{k},\tl{k}'} ~ \textrm{ in YM-reg} 
  & \quad \longrightarrow \quad \textrm{non-susy}\ U(N)_{\tl{k}-\sgn(\tl{k})N,\tl{k}'} ~ \textrm{in YM-reg}\ .
\end{align}
Additionally, we also know that the results of a non-supersymmetric
pure Chern-Simons theory in the YM-reg scheme and dim-reg scheme agree
only if their levels are matched as
\begin{equation}\label{levelrel2}
  \textrm{non-susy}\ U(N)_{\tl{k}-\sgn(\tl{k})N,\tl{k}'} ~ \textrm{in YM-reg} \quad \equiv \quad \textrm{non-susy}\ U(N)_{\tl{k},\tl{k}'} ~ \textrm{in dim-reg}\ .
\end{equation}
\footnote{Only in this subsection, the subscripts on the group of a
  Chern-Simons theory e.g.~$SU(N)_{k}$ will indicate the level in
  front of the corresponding Chern-Simons action to reflect the
  dependence on regularization scheme. In the rest of the paper,
  irrespective of the regularization scheme, the subscript always
  indicates the `bare level' $k$. Recall that we also exclusively use
  the dim-reg scheme in the rest of the paper in which the level in
  front of the action is $\kappa$.} Thus, to match the results of the
dimensionally regularized non-supersymmetric $U(N)_{\kappa,k'}$
Chern-Simons theory with the YM-regularized supersymmetric theory, the
corresponding levels should be related as
\begin{equation} 
  (\tl{k},\tl{k}') = (\kappa,k')\ .
\end{equation} 
We will replace $\tl{k}$ and $\tl{k}'$ by $\kappa$ and $k'$ in the
rest of this subsection.

\subsubsection{The superconformal index}
The theory \eqref{n2act} is also $\mc{N}=2$ superconformal in addition
to being $\mc{N}=2$ supersymmetric. The presence of superconformal
symmetry allows one to define the superconformal index
\cite{Bhattacharya:2008zy,Bhattacharya:2008bja} as the path integral
of the theory on $S^2 \times S^1$ with supersymmetry preserving
boundary conditions along the $S^1$ direction (i.e.~periodic for both
bosons and fermions and possibly twisted by supersymmetry preserving
global charges).  In Hamiltonian language the index can be defined as
\begin{equation}\label{indexexp}
  {\cal I}= {\rm Tr}_{\mc{H}} \left( (-1)^{F} x^{2j_3+R} \te^{-\beta(\epsilon - j_3-\mc{R})} \right) \ ,
\end{equation} 	 
where $\mc{H}$ is the Hilbert space of the theory on
$S^2 \times \mbb{R}$, $F$ is the fermion number, $\epsilon$ is the
energy, $j_3$ is the $z$-component of the angular momentum on $S^2$
and $\mc{R}$ is the $R$-charge of the superconformal algebra\footnote{In
  our (standard) convention, all bosonic fields in the vector
  multiplet carry zero $R$-charge whereas the gaugino carries
  $R$-charge $+\tfrac{1}{2}$.}  and $x = \te^{-\xi}$ is the (in
general complex) fugacity for the charge $2j_3 + \mc{R}$. There exists a
supercharge $\mc{Q}$ in the superconformal algebra such that
\begin{equation}
  \{\mc{Q}, \mc{Q}^\dag\} = \epsilon - j_3 - \mc{R}\ ,
\end{equation}
where, recall that $\mc{Q}^\dag$ is the hermitian conjugate of $\mc{Q}$ in
radial quantization. By standard arguments, the index is then
independent of the parameter $\beta$ and receives contributions only
from BPS states on $S^2$ i.e.~states satisfying
\cite{Bhattacharya:2008zy}
\begin{equation}\label{BPScond}
 \epsilon - j_3 - \mc{R} = 0\ .
\end{equation}
Since the index is independent of $\beta$, we can choose the value
$\beta = 2\xi$ to simplify calculations \cite{Benini:2013yva}. As a
consequence of \eqref{BPScond}, the \eqref{indexexp} can equally well
be written as
\begin{equation}\label{indexexpbps} 
{\cal I}= {\rm Tr}_{\mc{H}_{\rm BPS}} \left( (-1)^{F} x^{\epsilon + j_3} \right)\ ,
\end{equation}
where $\mc{H}_{\rm BPS}$ is the subspace of states which satisfy
\eqref{BPScond}. The path integral that evaluates the index can be
computed exactly for any finite rank and level of the Chern-Simons
theory using the method of supersymmetric localization
\cite{Kim:2009wb,Imamura:2011su,Benini:2013yva,Fujitsuka:2013fga}
which reduces the full path integral to a finite dimensional
integral. Very briefly, one adds a $\mc{Q}$-exact deformation i.e.~a term
of the form $t\mc{Q}V$ with $t$ a parameter and $V$ a functional of the
fields such that the asymptotic behaviour of the potential is
unchanged after the addition of the deformation. Since the index is
invariant under such $\mc{Q}$-exact deformations, one can take the
parameter $t$ to be very large so that the path integral localizes on
the saddle point locus $\mc{Q}V = 0$.

With the standard choice of $\mc{Q}$-exact deformation term used in
\cite{Kim:2009wb,Imamura:2011su}\footnote{
  \cite{Benini:2013yva,Fujitsuka:2013fga} considers a more general
  class of deformations and refers to \eqref{loclocus} as the
  Coulomb-like solutions.}  the saddle point locus when the $R$-charge
of the chiral multiplet is positive is given by
\begin{equation}\label{loclocus}
  A_\mu dx^\mu  = \left[ \frac{\bf{a}}{2\xi R} d\tau + \frac{\bf{m}}{2}(\eta - \cos\theta) d\phi \right]\ , \quad \sigma = - \frac{\bf{m}}{2R}\ , \quad \phi = 0\ ,\quad \text{all fermions} = 0\ ,
\end{equation}
where $R$ is the radius of $S^2$, $\eta = \pm 1$ in north and south
hemisphere on $S^2$ respectively. The quantities $\bf{a}$ and $\bf{m}$
are holonomy and quantized magnetic flux matrices both lying along the
Cartan subalgebra of the gauge group. For the $U(N)$ and $SU(N)$ gauge
groups which are of interest in this paper, they are simply given by
\begin{equation}
{\bf a} = \text{diag}(a_1,a_2,\ldots, a_N)\ , \quad {\bf m} = \text{diag}(m_1, m_2, \ldots, m_N)\ ,
\end{equation}
where $\{ a_i \} \in (0,2\pi)$ and $\{ m_i \}$ take integer values,
with the further constraints $\tr(\bf{a}) = \tr(\bf{m}) = 0$ for the
$SU(N)$ case.

Taking into account the extra contribution to the fermion number from
the `spin' of magnetic flux configurations $\{ m_i \}$
\cite{Aharony:2013dha,Dimofte:2011py,Hwang:2012jh} which effectively
replaces the fermion number operator $F$ by $2j_3$ (more on this in
Section \ref{spinmono} below), the superconformal index for
${\cal N}=2$ $U(N)_{\kappa,\kappa+sN}$ Chern-Simons theory coupled to
$N_f$ chiral multiplets in the fundamental representation with the
same $R$-charge $r$ for all the $N_f$ chiral multiplets is
\begin{align}\label{index1cm1}
  & {\cal I}(N,\kappa,k';x) = \textrm{Tr}_{\mc{H}_{\rm BPS}} \left[(-1)^F x^{\epsilon + j_3}\right]\nonumber \\
  & = \quad \sum_{\{m_i\} \in {\mathbb Z}^N/\mc{S}_N} \frac{1}{sym}
  \oint \left( \prod_{i=1}^N \frac{dz_i}{2\pi \i z_i}  ((-1)^{m_i} z_i)^{-(\kappa - \frac{1}{2}N_f \sgn(\kappa)) m_i} \right) \left( (-1)^M z \right)^{-s M}\nonumber \\
  & \hspace{3.5cm}\times \prod_{i \neq j} \left[x^{-\frac{1}{2} |m_i - m_j|}\left(1- \frac{z_i}{z_j} x^{|m_i - m_j|} \right) \right]\nonumber \\
  & \hspace{3.5cm} \times \prod_{i=1}^{N} \left( \left( x^{1-r}
      ((-1)^{m_i} z_i)^{-1} \right)^{|m_i|/2}
    \prod_{j=0}^{\infty} \frac{ 1- z_i^{-1} x^{|m_i|+2-r+2j} }{1- z_i x^{|m_i|+r+2j}} \right)^{N_f}  \ ,
\end{align}
where $z_i = \te^{\i a_i}$, $z = \prod_{i=1}^N z_i$ and $ M = \sum_{i=1}^N m_i$ and
$\mc{S}_N$ is the Weyl group of $SU(N)$. The summation over fluxes
$\{ m_i \}$ and the integral over holonomies $\{ z_i \}$ above
represents the ``sum'' over the saddle points i.e.~zero locus of the
$\mc{Q}$-exact deformation term \cite{Kim:2009wb}. The rest of the terms in
the first line come from the Chern-Simons action evaluated at the
saddle points \eqref{loclocus}. The terms in the second line comes
from the one-loop determinants of the fields in the vector multiplet
around \eqref{loclocus}\footnote{Note that this vector multiplet
  contribution reduces to the $U(N)$ Vandermonde determinant in the
  zero flux sector at general $x$ and also for all flux sectors at
  $x=\pm 1$ (see Sections \ref{xm1} and \ref{spinmono} below).}. The
third line is the one-loop determinant of the fields in the $N_f$ chiral
multiplets around the locus \eqref{loclocus}. The integral over each
holonomy $z_i$ runs counterclockwise over a unit circle in the complex
plane.

The summation over flux configurations $\{m_i\}$ is a restricted sum
over integer values modulo permutation equivalence while the ``$sym$''
factor is the order of subgroup of the Weyl group $\mc{S}_N$ of
$SU(N)$ preserved by the flux configuration $\{m_i\}$ and is given by
\begin{equation}\label{wsf}
  sym = \prod_{i=1}^{N} \left( \sum_{j=i}^N \delta_{m_i,m_j} \right) \ .
\end{equation}
More explicitly, for a configuration with $s$ distinct values for
fluxes appearing $n_1, n_2, \ldots n_s$ times such that
$n_1+n_2+\ldots n_s = N$, we have $sym = n_1! n_2! \cdots n_s!$.
Since the integrand is permutation invariant, it is convenient to
rewrite the permutation-restricted summation over the flux
configurations into a free un-restricted sum over all $\{m_i\}$ and
dividing by the number of equivalent permutations
$\frac{N!}{n_1! n_2! \ldots n_s!}$.  Thus, we can rewrite the flux
summation as \cite{Hwang:2012jh}
\begin{align}\label{index1cm2}
& {\cal I}(N,\kappa,k';x) =\nonumber \\
& \quad \frac{1}{N!}  \sum_{\{m_i\} \in {\mathbb Z}^N}
         \oint \left( \prod_{i=1}^N \frac{dz_i}{2\pi \i z_i}  ((-1)^{m_i} z_i)^{-(\kappa - \frac{1}{2}N_f \sgn(\kappa)) m_i} \right) \left( (-1)^M z \right)^{-s M}\nonumber \\
   & \hspace{3cm} \times \prod_{i \neq j} \left[ x^{-\frac{1}{2} |m_i - m_j|}\left(1- \frac{z_i}{z_j} x^{|m_i - m_j|} \right) \right]\nonumber \\
   & \hspace{3cm} \times \prod_{i=1}^{N} \left( \left( x^{1-r} ((-1)^{m_i} z_i)^{-1} \right)^{|m_i|/2} 
          \prod_{j=0}^{\infty} \frac{ 1- z_i^{-1} x^{|m_i|+2-r+2j} }{1- z_i x^{|m_i|+r+2j}} \right)^{N_f}  \ .
\end{align}
Note that for odd values of $N_f$ the Chern-Simons level in front of the action
$\kappa - \tfrac{1}{2}N_f \sgn(\kappa)$ takes half-integral values
(i.e.~values in $\mbb{Z} + \tfrac{1}{2}$) due to the parity anomaly in
three dimensions; the factor $((-1)^{m_i} z_i)^{-{|m_i|/2}}$ from the
third line of \eqref{index1cm2} combines with terms in the
first line of this equation, effectively `renormalizing' this
half-integral value $\kappa - \tfrac{1}{2}N_f \sgn(\kappa)$ back to an
integer (see discussion after \eqref{n2act}).

In the following subsection we will adapt the localization computation
of the index to include supersymmetric Wilson loop operators in
arbitrary representation of the gauge group.

\subsubsection{Supersymmetric Wilson loops on $S^2 \times S^1$}
Consider the following generalized Wilson loop labelled by a
representation $K$ and loop $\mc{C}$
\begin{equation}\label{gswl}
  W_K(\mc{C}) = \textrm{Tr}_K {\cal P}\exp\left( \oint_{\cal C} d\lambda \left( \i A_\mu \dot x^\mu 
      + \zeta \sigma |\dot x| \right) \right)\ ,
\end{equation}
where $\tau$ is the parameter along the curve ${\cal C}$,
$\dot x^\mu = \partial_\tau x^\mu$ and
$|\dot x| = \sqrt{\dot x^\mu \dot x_\mu}$. $\zeta$ is an arbitrary
real parameter which will fix by requiring invariance under
supersymmetry. Thus, the loop operator \eqref{gswl} with the
appropriate value of $\zeta$ is a supersymmetric generalization of a
Wilson loop\footnote{Supersymmetric Wilson loops (with $\zeta=1$)
  were considered in ${\cal N}=2$ theories on $S^3$ in
  \cite{Kapustin:2009kz} where it was shown that these are
  supersymmetric if the curve ${\cal C}$ is along the $S^1$ Hopf
  fibres.}.

For the discussion of supersymmetry transformations and localization
on $S^2 \times S^1$, we will follow the conventions of
\cite{Benini:2013yva}\footnote{The superconformal index was first computed for the
  ${\cal N}=6$ ABJM theory in \cite{Kim:2009wb} and was later
  generalized for arbitrary ${\cal N}=2$ theories in
  \cite{Imamura:2011su}.}. For convenience we summarize these in
appendix \ref{susyS2conv}. The supersymmetry transformations for
$A_\mu$ and $\sigma$ are
\begin{equation}\label{susyvar}
  \quad \delta A_\mu = - \frac{\i}{2} (\bar\epsilon \gamma_\mu
  \lambda - \bar\lambda \gamma_\mu \epsilon) \quad, \quad \delta
  \sigma = \frac{1}{2} (\bar\epsilon \lambda - \bar\lambda \epsilon)\ ,
\end{equation}
where $\epsilon, \bar\epsilon$ are independent spinors and not complex
conjugates of each other, and satisfy the Killing spinor equation on
$S^2$:
\begin{equation}\label{ksps1}
 D_\mu \epsilon = \frac{1}{2R} \gamma_\mu \gamma_3 \epsilon \quad , 
 \quad D_\mu \bar\epsilon = -\frac{1}{2R} \gamma_\mu \gamma_3 \bar\epsilon\ ,
\end{equation}
whose solution is
\begin{equation}\label{kspin}
 \epsilon = \te^{\tau/2R} \te^{-\i \frac{\theta}{2} \sigma_2} \te^{\i \frac{\phi}{2} \sigma_3 } \epsilon_0 \ , \quad \bar\epsilon = \te^{-\tau/2R} \te^{\i \frac{\theta}{2} \sigma_2} \te^{\i \frac{\phi}{2} \sigma_3 } \bar\epsilon_0 \ ,
\end{equation}
with $\epsilon_0$, $\bar\epsilon_0$ being arbitrary constant
spinors. The Killing spinors used for supersymmetric localization
correspond to the choice
$\epsilon_0 = \begin{pmatrix}\textstyle 1 \\ 0\end{pmatrix}$ and
$\bar\epsilon_0 = \begin{pmatrix}\textstyle 0 \\ 1\end{pmatrix}$, giving
\begin{equation}\label{ksps2}
\epsilon = \te^{\tau/2R} \te^{\i\frac{\phi}{2}} \begin{pmatrix} \cos\tfrac{\theta}{2} \\ 
\sin\tfrac{\theta}{2} \end{pmatrix}\ , \quad \bar\epsilon = \te^{-\tau/2R} \te^{-\i\frac{\phi}{2}} 
\begin{pmatrix} \sin\tfrac{\theta}{2} \\ \cos\tfrac{\theta}{2} \end{pmatrix}\ .
\end{equation}
The supersymmetry variation of the loop operator \eqref{gswl} is
proportional to
\begin{equation}\label{wlsusyvar}
  \delta \left(\i A_\mu \dot x^\mu + \zeta \sigma |\dot x| \right) = 
  \frac{1}{2} \left[ \bar\epsilon (\zeta |\dot x| \mathbb{1}_2 +\gamma^\mu \dot x^\mu)\lambda - \bar\lambda (\zeta |\dot x| \mathbb{1}_2 
    +\gamma^\mu \dot x^\mu) \epsilon \right]. 
\end{equation}
where $\mathbb{1}_2$ is the $2\times 2$ identity matrix. 

In this paper, we are interested in Wilson loops along the $S^1$
direction on $S^2\times S^1$. For the above generalized Wilson loop to
be invariant under the supersymmetries generated by $\epsilon$ and
$\bar\epsilon$ requires the coefficient of $\lambda$ and $\bar\lambda$
to vanish.  Working with the standard coordinates and metric on
$S^2\times S^1$ (see Appendix \ref{susyS2conv}) gives
\begin{align}\label{WLsolns}
\lambda\  &: \quad 0 =  \bar\epsilon (\zeta \mbb{1}_2 + \sigma_3) = \i \te^{-\frac{\tau}{2R}-\frac{\i\phi}{2}}  \begin{pmatrix} -(\zeta+1)\cos \tfrac{\theta}{2} & (\zeta-1)\sin\tfrac{\theta}{2} \end{pmatrix}\ ,\nonumber \\
\bar\lambda\ &: \quad 0 =  (\zeta \mbb{1}_2 + \sigma_3)\epsilon = \te^{-\frac{\tau}{2R} + \frac{\i\phi}{2}} \begin{pmatrix} (\zeta+1) \cos\tfrac{\theta}{2} \\ (\zeta-1) \sin\tfrac{\theta}{2} \end{pmatrix}\ .
\end{align}
This gives two solutions, a supersymmetric Wilson loop located at the
north pole ($W^N$) and another supersymmetric Wilson loop located at
the south pole ($W^S$):
\begin{align}\label{susywlns}
 (1) \quad (\zeta = -1, \theta = 0)\ &: \quad W^N_K = 
                 \textrm{Tr}_K {\cal P}\exp\left( \int_0^{\beta R} d\tau (\i A_\tau - \sigma) \right)\ ,\nonumber\\
 (2) \quad (\zeta = +1, \theta = \pi)\ &: \quad W^S_K = 
                 \textrm{Tr}_K {\cal P}\exp\left( \int_0^{\beta R} d\tau (\i A_\tau + \sigma) \right)\ ,
\end{align}
where recall from Appendix \ref{susyS2conv} that $\beta R$ is the perimeter of temporal $S^1$ (so
that the dilatation operator in the conformal algebra is the
translation operator along $S^1$) and $R$ is the radius of the spatial
$S^2$.

To compute the expectation values of a product of these supersymmetric
Wilson loops, we insert them in the index path integral. The path
integral with any number of insertions of the two kinds of
supersymmetric Wilson loops, $W^{N,S}_R$, simply reduce to the
insertions evaluated on the saddle points \eqref{loclocus}. The value
of the supersymmetric Wilson loops at the saddle points
\eqref{loclocus} give
\begin{align}\label{nswlmchar}
  W^N_K &\rightarrow  \tr_K \exp \left( \te^{\i{\bf a}} x^{{\bf -m}} \right) = \chi_{\vgap K}(\{z_i x^{-m_i}\})\nonumber \\
  W^S_K &\rightarrow  \tr_K \exp \left( \te^{\i{\bf a}} x^{{\bf m}} \right) = \chi_{\vgap K}(\{z_i x^{m_i}\})\ .
\end{align}
Here the $z_i$ factors arise from the gauge field term in
\eqref{susywlns} whereas the $x^{m_i}$ come from the $\sigma$ term
\eqref{susywlns}.

The final result for the superconformal index path integral in
presence of arbitrary insertions of north pole Wilson loops
$\prod_{\alpha} W^N_{R_\alpha}$ and south pole Wilson loops
$\prod_{\alpha} W^S_{R_\beta}$ in an $U(N)_{\kappa,\kappa+sN}$
Chern-Simons theory with $N_f$ chiral multiplets in the fundamental
representation is simply
\begin{align}\label{IWLwm}
& {\cal I}\left( N,\kappa,k';x;\{R_\alpha\},\{ R_\beta \} \right)\nonumber  \\
  & = \frac{1}{N!} \sum_{\{m_i\} \in {\mathbb Z}^N} 
    \oint \left( \prod_{i=1}^N \frac{dz_i}{2\pi \i z_i}  ((-1)^{m_i} z_i)^{-(\kappa -\frac{1}{2}N_f \sgn(\kappa)) m_i} \right) \left( (-1)^M z \right)^{-s M}\nonumber \\
  & \hspace{3cm} \times  \prod_{i \neq j} \left[ x^{-\frac{1}{2} |m_i - m_j|}\left(1- \frac{z_i}{z_j} 
    x^{|m_i - m_j|} \right) \right]\nonumber \\
  & \hspace{3cm} \times \prod_{i=1}^{N} \left( \left( x^{1-r} ((-1)^{m_i} z_i)^{-1} \right)^{|m_i|/2} 
    \prod_{j=0}^{\infty} \frac{ 1- z_i^{-1} x^{|m_i|+2-r+2j} }{1- z_i x^{|m_i|+r+2j}} \right)^{N_f}  \nonumber \\
  & \hspace{3cm} \times  \prod_{\alpha,\beta} \chi_{\vgap R_\alpha}(\{z_i x^{-m_i}\}) 
    \chi_{\vgap R_\beta}(\{z_i x^{m_i}\}) \ .
\end{align}

\paragraph{Pure ${\cal N}=2$ Chern-Simons theory} In the rest of this section below we will restrict our attention to the special case of Wilson loops in
pure ${\cal N}=2$ $U(N)_{\kappa,\kappa+sN}$ Chern-Simons theory and
analyse it consequences for the algebra of supersymmetric Wilson
loops. Dropping the contribution of chiral multiplet from
\eqref{IWLwm}, i.e. setting $N_f=0$, we get the much simpler
expression
\begin{align}\label{PCSwl1}
& {\cal I}\left( N,\kappa,k';x;\{R_\alpha\},\{ R_\beta \} \right)\nonumber \\
       & = \frac{1}{N!} \sum_{\{m_i\} \in {\mathbb Z}^N}  
         \oint \left( \prod_{i=1}^N \frac{dz_i}{2\pi \i z_i}  ((-1)^{m_i} z_i)^{-\kappa m_i} \right) \left( (-1)^M z \right)^{-s M}\nonumber \\
       & \hspace{3cm} \times \prod_{i \neq j} \left[ x^{-\frac{1}{2} |m_i - m_j|} \left(1- \frac{z_i}{z_j} 
              x^{|m_i - m_j|} \right) \right]\nonumber \\
       & \hspace{3cm} \times  \prod_{\alpha,\beta} \chi_{\vgap R_\alpha}(\{z_i x^{-m_i}\}) 
             \chi_{\vgap R_\beta}(\{z_i x^{m_i}\})\ .
\end{align}
The flux dependent phase factor can be factored out and written
explicitly as
\begin{align}\label{PCSwl2}
& {\cal I}\left( N,\kappa,k';x;\{R_\alpha\},\{ R_\beta \} \right)\nonumber \\
       & =  \sum_{\{m_i\} \in {\cal Z}^N}   \frac{(-1)^{(\kappa + s)M}}{N!}
         \oint \left( \prod_{i=1}^N \frac{dz_i}{2\pi \i z_i}  z_i^{-\kappa m_i} \right)  z^{-s M}\nonumber \\
       & \hspace{3cm} \times \prod_{i \neq j} \left[ x^{-\frac{1}{2} |m_i - m_j|} \left(1- \frac{z_i}{z_j} 
              x^{|m_i - m_j|} \right) \right]\nonumber \\
       & \hspace{3cm} \times  \prod_{\alpha,\beta} \chi_{\vgap R_\alpha}(\{z_i x^{-m_i}\}) 
             \chi_{\vgap R_\beta}(\{z_i x^{m_i}\})\ .
\end{align}

\subsubsection{The fugacity $x$ in the index of pure $\mc{N} = 2$
  Chern-Simons theory with Wilson loops}

Another observation about pure $\mc{N}=2$ Chern-Simons theory that we
will use heavily below is the following.  In general, the fugacity $x$
in the index \eqref{index1cm1} measures the charge $2j_3 + R$ of the
supersymmetric states contributing to the index (see
\eqref{indexexp}). However, for pure $\mc{N}=2$ Chern-Simons theory,
the only dynamical field is $A_\mu$ which carries zero $R$
charge. Thus, in the index for pure Chern-Simons theory with Wilson
loop insertions \eqref{PCSwl1} above, the fugacity $x$ just measures
$2j_3$. This observation along with equivalence of the pure
supersymmetric and pure non-supersymmetric Chern-Simons theory tells
us that in this special case, the index with Wilson loop insertions
simply counts the states on $S^2$ weighted by $x^{2j_3}$. Thus, upon
setting $x=-1$, we expect to get the usual thermal partition function
i.e.~the partition function of pure Chern-Simons theory on
$S^2 \times S^1$!  We will now show that this is in fact true and
confirm that the result correctly reproduces the quantization of the
holonomies and hence the algebra of Wilson loops as demonstrated
previously in this section from the path integral and from the
Verlinde formula.

\subsubsection{Partition function from index at $x=-1$}\label{xm1}
Motivated by the discussion of the previous subsubsection we set
$x=-1$ in the result for the index with Wilson loop insertions of pure
$\mc{N} =2$ Chern-Simons theory \eqref{PCSwl1}:
\begin{align}\label{PCSwl3}
& {\cal I}\left(N,\kappa,k';x=-1;\{R_\alpha\},\{ R_\beta \} \right)\nonumber \\ 
& \qquad = \frac{1}{N!} \sum_{\{m_i\} \in {\mathbb Z}^N}  
         \oint \left( \prod_{i=1}^N \frac{dz_i}{2\pi \i z_i}  ((-1)^{m_i} z_i)^{-\kappa m_i} \right) \left( (-1)^M z \right)^{-s M}\nonumber \\
& \hspace{4cm} \times \prod_{i \neq j} \left[ (-1)^{-\frac{1}{2} |m_i - m_j|}\left(1- \frac{z_i}{z_j} (-1)^{|m_i - m_j|} \right) \right]\nonumber \\
&  \hspace{4cm} \times  \prod_{\alpha,\beta} \chi_{\vgap R_\alpha}(\{z_i (-1)^{-m_i}\}) 
             \chi_{\vgap R_\beta}(\{z_i (-1)^{m_i}\})  \ .
\end{align}
Now, using $(-1)^{|m|} = (-1)^{m} = (-1)^{-m} = (-1)^{m^2}$ and
performing the change of variables $z_i \rightarrow (-1)^{m_i} z_i$, we
get
\begin{align}
  1.\ & ((-1)^{m_i} z_i)^{-\kappa m_i} \rightarrow  z_i^{-\kappa m_i}\ ,\nonumber \\ 
  2.\ & ((-1)^M z)^{-s M} \rightarrow  = z^{-s M}\nonumber \\
  3.\ & \prod_{i\neq j} (-1)^{|m_i-m_j|/2} = \prod_{i>j} (-1)^{|m_i-m_j|} = \prod_{i>j} (-1)^{m_i-m_j} = (-1)^{(N-1)M}\ . 
\end{align}
Using these, the index \eqref{PCSwl3} with $x = -1$ becomes
\begin{align}\label{PCSwl4}
  & {\cal I}\left( N,\kappa,k';x=-1;\{R_\alpha\},\{ R_\beta \} \right)\nonumber \\    
  & =  \sum_{\{m_i\} \in {\mathbb Z}^N} \frac{(-1)^{(N-1)M} }{N!}
    \oint \left( \prod_{i=1}^N \frac{dz_i}{2\pi \i z_i}  z_i^{-\kappa m_i} \right) z^{-s M}
 \prod_{i < j} | z_i- z_j|^2 \prod_{\alpha,\beta} \chi_{\vgap R_\alpha}(\{z_i\}) \chi_{\vgap R_\beta}(\{z_i \}) \ ,
\end{align}
in perfect agreement with \eqref{csmf}. 
Note that at $x=-1$, the rescaling $z_i \rightarrow (-1)^{m_i} z_i$, converts both the north and the 
south pole Wilson loop insertions to the usual character of the corresponding representation. 

The fact that the result of the localization computation perfectly
matches the phase corrected non supersymmetric path integral evaluated
around \eqref{csmf} is one of the principal results of this
subsection. As explained around \eqref{csmf}, in this limit the flux
$\{m_i\}$ dependence of the index integrand simplifies significantly and
the summation over fluxes reduces to
\begin{equation}
  \sum_{\{m_i\}} \left[ \prod_{i=1}^N \left( (-1)^{(N-1)} z_i^{-\kappa} z^{-s} \right)^{m_i} \right] 
\end{equation}
The summation over $\{m_i\}$ constrains each the $U(N)$ holonomies
$\{z_i\}$ according to
\begin{equation}\label{holquant1}
z^{s} z_i^{\kappa} = (-1)^{(N-1)}\ ,  
\end{equation} 
which is the quantization condition \eqref{obey}. We have verified
that the index of pure $\mc{N}=2$ Chern-Simons theory with Wilson loop
insertions \eqref{PCSwl1} evaluated at $x=-1$ reproduces the pure
non-supersymmetric Chern-Simons partition function with Wilson loop
insertions. In particular, it follows that the index evaluated at
$x=-1$ is a positive integer. At general values of $x$, on the other
hand, the index is a polynomial in $x$. The coefficients of this
polynomial are integers that could be either positive or negative. In
Appendix \ref{susywlresults} we have explicitly computed the index
\eqref{PCSwl4} for small values of $\kappa$ and $N$ with insertions of
Wilson loops in certain integrable representations (and, in
particular, have verified in every example that the index is a
positive integer at $x=-1$).  The physical interpretation of the more
fine grained integers we obtain as coefficients of the different
powers of $x$ is more mysterious; we discuss this further in the
Section \ref{disc} below.

\subsubsection{The index at $x=1$ and the Bose-Fermi nature of states
  on $S^2$}\label{spinmono}
Let us now look at another special value of the parameter $x$
viz.~$x=1$ in some detail and compare it with the result at
$x=-1$. Setting $x=1$, we get
\begin{align}\label{indexx1}
& {\cal I}\left( N,k,k';x=1;\{R_\alpha\},\{ R_\beta \} \right)\nonumber \\ 
& = \sum_{\{m_i\} \in {\mathbb Z}^N}  \frac{(-1)^{(\kappa + s)M}}{N!}
         \oint \left( \prod_{i=1}^N \frac{dz_i}{2\pi \i z_i}  z_i^{-\kappa m_i} \right) z^{-s M}           \times \prod_{i < j} |z_i- z_j|^2  \prod_{\alpha,\beta} \chi_{\vgap R_\alpha}(\{z_i \}) 
             \chi_{\vgap R_\beta}(\{z_i \}) \ .
\end{align}
The above expression of index evaluated at $x=1$ simply counts the
state on $S^2$ with a relative minus sign for fermionic states
compared to the partition function \eqref{PCSwl4} which counts the
same states without any relative sign. Note the only difference
between \eqref{indexx1} and \eqref{PCSwl4} is in the flux-dependent
phase factor which again only depends on the overall $U(1)$ flux
$M=\sum_{i=1}^N m_i$ and not on the individual $m_i$. The
relative difference between these two phase factors thus carries
information about the Bose/Fermi nature of the corresponding states.

It is clear from \eqref{PCSwl4} and \eqref{indexx1} that in the
absence of any Wilson loops, only the trivial flux sector $\{m_i\}=0$
contributes and the result is simply unity, corresponding to a unique
bosonic ground state.  In the presence of Wilson loops in some
representations ${ R}_1,{ R}_2, \ldots, { R}_n$ one gets a
non-vanishing result only when the sum total of the $U(1)$ charges
$\sum_i |{R}_i|$ of the Wilson loop insertions is an integer
multiple of $k'$ i.e.~$\sum_i |{R}_i| = k' Q$, reflecting the fact
that the $U(1)$ charge is conserved modulo $k'$ in the $U(1)_{k, k'}$
theory. The expectation value of such a collection of Wilson loops
receives contributions from those flux sectors for which the total
$U(1)$ flux is $M = \frac{1}{k'}\sum_i |{R}_i|$.\footnote{The
  relationship between the overall $U(1)$ charge of insertions and the
  total $U(1)$ flux is a manifestation of the $U(1)$ Chern-Simons
  equation of motion $ 2 \pi j^0 = k' F_{ij}$. Integrating both sides
  over the sphere, we conclude that configurations with a net $U(1)$
  flux $M$ only contribute when the total charge of all insertions is
  $M k'$. }

The sign of such flux sectors is $(-1)^{(\kappa+s)M}$ in the index
\eqref{indexx1} while it is $(-1)^{(N-1)M}$ in the partition function
\eqref{PCSwl4}.  The relative sign is then
\begin{equation}\label{mpspin}
  (-1)^{(\kappa+s-N+1)M} = (-1)^{(k+s+1)M}\ .
\end{equation}
Thus the corresponding state on $S^2$ is fermionic if $(k+s+1)M$ is
odd and bosonic when it is even. In particular, it also implies that
for $SU(N)$ theories, all the states are bosonic since $M=0$. This
result matches precisely with the results discussed in Appendix A.6 of
\cite{Minwalla:2020ysu} where it was argued that, for positive $SU(N)$
level $k$, the basic monopole operator of the Type I theory has spin
$-\frac{k+1}{2}$ (see Eq.~A.64 of \cite{Minwalla:2020ysu}) while the
basic monopole of the Type II theory has spin $-\frac{k}{2}$ (see
Eq.~A.61 of \cite{Minwalla:2020ysu}). This contributes
resp.~$(-1)^{k+1}$ or $(-1)^k$ to the fermion number in the case of
Type I and Type II theories which precisely matches with
\eqref{mpspin} for Type I ($s=0$) or Type II ($s=-1$) respectively.

It is easy to extend the computation in Appendix A.6 of
\cite{Minwalla:2020ysu} to see that the spin of the basic monopole of
the $U(N)_{k, \kappa +sN}$ theory with $k > 0$ equals
$-\frac{k+1+s}{2}$ and hence contributes $(-1)^{k+s+1}$ to the fermion
number.


\subsection{A Check of level-rank duality in Type I $U(N)$ theory}\label{typeIdual}

Specializing the Verlinde formula \eqref{ngtmnewu} for the
$U(N)_{k,k'}$ WZW model to the case $k' =\kappa$ i.e.~$s=0$, we
conclude that the number of singlets in the product of the integrable
representations $R_1, \ldots, R_n$ in type I $U(N)$ WZW theory is
given by the formula
\begin{equation}\label{NUgenmtne}
  \mc{N}_{g,n} = \frac{1}{\kappa^{N(1-g)}} \sum_{ \ul{w} \in \mc{P}_{k,\kappa}} \ \prod_{1 \leq i < j \leq N} |w_i-w_j|^{2-2g} \prod_{p=1}^{n} \chi_{\vgap R_p}( \ul{w} )\ .
\end{equation}
The summation over $\ul{w}$ is taken over the set of solutions modulo
permutation $\mc{P}_{k,\kappa}$ of the equations
\begin{equation}\label{loie}
w_i^\kappa  =(-1)^{N-1}\ ,\quad\text{for}\quad i=1, \ldots, N\ ,\quad\text{with}\quad w_i \neq w_j\ \text{when}\ i \neq j\ .
\end{equation} 
Restated, the complex numbers $w_i$ in \eqref{NUgenmtne} are each
$\kappa^{\rm th}$ roots of unity when $N$ is odd and are
$\kappa^{\rm th}$ roots of $-1$ when $N$ is even. Thus, in each term
in the sum in \eqref{NUgenmtne} we pick $N$ distinct roots among the
$\kappa$ $\kappa^{\rm th}$ roots of $(-1)^{N-1}$. The sum in
\eqref{NUgenmtne} ranges over the $\binom{\kappa}{N}$ possibilities
for this choice.

We will now demonstrate that the formula \eqref{NUgenmtne} respects
the Type I - Type I level-rank duality (see e.g. Appendix A of
\cite{Minwalla:2020ysu} for a precise description of this duality).

First, we observe that the complement of $\{w_i\}$, the $N$ distinct
$\kappa^{\rm th}$ roots of $(-1)^{N-1}$, defines a choice of
$\kappa-N=k$ distinct $\kappa^{\rm th}$ roots of $(-1)^{N-1}$. Let
these $k$ excluded eigenvalues be denoted by $\{w'_J\}$, $J=1,\ldots,k$. Note that
\begin{equation}\label{sumoverev}
  \sum_{i=1}^{N} w_i + \sum_{J=1}^{k} w'_J = 0\ ,
\end{equation} 
which follows from the familiar fact that the sum of all $\kappa$
roots of unity vanishes.

In Appendix \ref{vander} we demonstrate that 
\begin{equation}\label{sumoverpairs}
  \frac{1}{\kappa^N} \prod_{1 \leq i < j \leq N} |w_i-w_j|^2
  = \frac{1}{\kappa^k} \prod_{1 \leq I < J \leq k} |w'_I-w'_J|^2\ ,
\end{equation}
for any choice of the $N$ eigenvalues $\{w_i\}$ (and consequently, the
complement $\{w'_J\}$).

Let us define
\begin{equation}\label{wdef}
  z_J = - w'_J\ .
\end{equation}
Since $(w'_J)^\kappa = (-1)^{N-1}$, it follows that
\begin{equation}\label{neree}
  (z_J)^\kappa = (-1)^{\kappa - N - 1} = (-1)^{k-1}\ ,\quad z_I \neq z_J\quad\text{for}\quad I \neq J\ .
\end{equation} 
In other words $z_J$, $J=1,\ldots,k$, satisfy the conditions to be
distinguished eigenvalues of the form \eqref{prodw} for the Type I
$U(k)_{N, \kappa}$ antichiral WZW model (which is dual to the
$U(k)_{-N,-\kappa}$ Chern-Simons theory in three
dimensions)\footnote{The additional minus sign in the levels of the
  Chern-Simons theory is due a parity inversion that appears to be
  required for level-rank duality of Chern-Simons theories. This
  corresponds to flipping the chirality of the dual WZW
  model. However, since the formula for singlets is the same for the
  chiral half or the antichiral half of the $U(N)_{N,N}$ WZW model and
  we ignore this distinction henceforth.}.  That is, the $U(k)$
eigenvalues $\ul{z} = \{z_I\}$ belong to the set of solutions modulo
permutation $\mc{P}_{N,\kappa}$ of the equations \eqref{neree}.

It follows from \eqref{sumoverev} and
\eqref{sumoverpairs} that
\begin{equation}\label{sumoverpairsn}
  \frac{1}{\kappa^N} \prod_{1 \leq i < j \leq N} |w_i-w_j|^2
  = \frac{1}{\kappa^k} \prod_{1 \leq I < J \leq k} |z_I - z_J|^2\ ,
\end{equation} 
and
\begin{equation}\label{sumoverevn}
  \sum_{i=1}^{N} w_i = \sum_{J=1}^{k} z_J\ ,\quad  \sum_{i=1}^{N} w_i^* =\sum_{J=1}^{k} z_J^*\ .
\end{equation} 
The left hand sides of the two formulas in \eqref{sumoverevn} are
simply the characters of the $U(N)$ fundamental and antifundamental
representations respectively whereas the right hand sides of
\eqref{sumoverevn} are the characters of the $U(k)$ fundamental and
anti-fundamental representations respectively. In other words,
\eqref{sumoverev} can be rewritten as
\begin{equation}\label{charequiv}
  \chi_{\vgap F}(\ul{w}) =\chi_{\vgap F}(\ul{z})\ ,\quad  \chi_{\vgap A}(\ul{w}) = \chi_{\vgap A}(\ul{z})\ ,
\end{equation} 
where $F$ and $A$ are the fundamental and antifundamental
representations of $U(N)$ on the left hand sides, and of $U(k)$ on the
right hand sides of the above equations.

It follows from \eqref{sumoverpairsn} and \eqref{charequiv} that for
any given choice of $\ul{w} \in \mc{P}_{k,\kappa}$ and the resultant
choice for $\ul{z} \in \mc{P}_{N,\kappa}$, we have
\begin{equation}\label{themo}
  \frac{1}{\kappa^N} \prod_{1 \leq i < j \leq N} |w_i-w_j|^{2-2g} \chi_{\vgap F}(\ul{w})^{n_1} \chi_{\vgap A}(\ul{w})^{n_2}
= \frac{1}{\kappa^k} \prod_{1 \leq I < J \leq k} |z_i-z_j|^{2-2g}
\chi_{\vgap F}(\ul{z})^{n_1} \chi_{\vgap A}(\ul{z})^{n_2}\ .
\end{equation}
Summing the LHS of \eqref{themo} over all
$\ul{w} \in \mc{P}_{k,\kappa}$ (and thus summing the RHS over all
resultant choices for $\ul{z} \in \mc{P}_{N,\kappa}$) it follows that
the number of singlets in the product of $n_1$ fundamentals and $n_2$
anti-fundamentals is the same for level-rank dual theories.

Notice that this test of level-rank duality in works in an extremely
simple manner. It works at the level of the integrand - i.e.~the sums
on the two sides match because each term in the sum matches another
term in the dual sum. The agreement happens at the level of each
individual eigenvalue configuration, and not after we sum over all
such configurations.

Using the obvious identities
$\chi_{\vgap F}^*(\ul{z}^*)= \chi_{\vgap F}(\ul{z})$ and
$\chi_{\vgap A}^*(\ul{z}^*)= \chi_{\vgap A}(\ul{z})$, \eqref{themo}
can be rewritten in the (at first sight more complicated) form
\begin{equation}\label{themocomp}
\frac{1}{\kappa^N} \prod_{1 \leq i < j \leq N} |w_i-w_j|^{2-2g} \chi_{\vgap F}(\ul{w})^{n_1} \chi_{\vgap A}(\ul{w})^{n_2}
= \frac{1}{\kappa^k} \prod_{1 \leq I < J \leq k} |z_i-z_j|^{2-2g}
 \chi_{\vgap F}^*(\ul{z}^*)^{n_1}  \chi_{\vgap A}^*(\ul{z}^*)^{n_2}\ .
\end{equation}
In this section, so far, we have studied the Verlinde formula with
insertions of only fundamentals and antifundamentals.  In the next
subsection we explain, however, that \eqref{themo} and
\eqref{themocomp} admit a simple generalization to the Verlinde
formula with insertions of characters in arbitrary representations. In
particular we will argue that, for any collection of representations
$R_m$
\begin{equation}\label{themogenrep}
\frac{1}{\kappa^N} \prod_{1 \leq i < j \leq N} |w_i-w_j|^{2-2g} \prod_m \chi_{\vgap R_m}(\ul{w}) 
= \frac{1}{\kappa^k} \prod_{1 \leq I < J \leq k} |z_I-z_J|^{2-2g}
\prod_m\chi^*_{R_m^T}(\vgap \ul{z}^*)\ .
\end{equation}
In \eqref{themogenrep}, the representation
$R_m^T$ is defined as the representation whose Young tableau is obtained from that of $R_m$ by a `transposition', i.e. an interchange of rows and columns.

\subsubsection{Type I-Type I duality: map between
  representations} \label{typeonedualmap}

In the previous section we have found a simple map between eigenvalues
of the two participating theories in
$U(N)_{k, (N+|k|)} \leftrightarrow U(k)_{-N,-(N+|k|)}$ level-rank
duality; any given collection of eigenvalues $\{w_i\}$ of the $U(N)$
theory defines a collection of `excluded' eigenvalues $\{w_i'\}$; we
defined the collection $\{-(w_I')^*\}$ as the $U(k)$ eigenvalues
associated with the collection of $U(N)$ eigenvalues $\{w_i\}$.

However earlier in this section we have independently described the
well-established map between distinguished eigenvalues and integrable
representations of the theory \cite{Elitzur:1989nr}. It follows that
the map between eigenvalues, described above, effectively determines a
map between the integrable representations of the level-rank dual
theories. In this subsection we work of the details of this map and
investigate its implications.

The equation \eqref{wonuo1} establishes the following map between
integrable representations of the Type I theory and elements of
${\cal P}_{k, \kappa}$ (i.e.  the distinguished $U(N)$ eigenvalues
that appear in the sum over eigenvalues \eqref{verlindeu} at $s=0$):
\begin{equation}\label{TOeigento}
w_i(\xi_\lambda) =\te^{ \i\pi  (N-1)/\kappa} \exp\left(-\frac{2\pi\i n^\lambda_i}{\kappa}\right) 
\end{equation}
Recall that the integers $n^\lambda_i$ are ordered as
\begin{equation}\label{orderto}
\kappa > n^\lambda_1 > n^\lambda_2 > \cdots > n^\lambda_{N} \geq  0\ .
\end{equation}
As we have explained above, \eqref{TOeigento} are the eigenvalues that
appear in the sum \eqref{NUgenmtne}. Let us label the complement of
the set of $n^\lambda_i$ in the range $[0,\kappa-1]$ as $\hat{n}_I$ with
$I = 1,\ldots, k$.  We then order the integers so that
\begin{equation}\label{orderingto}
  \kappa > \hat{n}_1 > \hat{n}_2 > \cdots > \hat{n}_k \geq 0\ ,
\end{equation}
and define the integers $m_I$ as
\begin{equation} \label{defmto} 
{\hat m}_I = \kappa - \hat{n}_{k-I+1}\ .
\end{equation}
Notice that the $\hat{m}_I$ satisfy the ordering
\begin{equation}
  \kappa \geq  {\hat m}_1 > \cdots > {\hat m}_k > 0\ .
\end{equation}
With this labelling, the set of excluded eigenvalues $\{w'_I\}$ (see
\eqref{wdef})is given by
\begin{equation}\label{toputative}
  w_I'=\te^{ \i\pi (N-1)/\kappa} \exp\left(-\frac{2\pi\i \hat{n}_I}{\kappa}\right)=\te^{\i\pi (N-1)/\kappa} \exp\left(\frac{2\pi\i \hat{m}_{k-I+1}}{\kappa}\right)\ .
\end{equation}
for $I=1 \ldots k$. The collection of eigenvalues
$\{z_J^* \}=\{-w'^*_{k-J+1} \}$ is thus given by
\begin{equation}\label{toputative1}
  z_I^*=-\te^{ -\i\pi(N-1)/\kappa} \exp\left(-\frac{2\pi\i \hat{m}_I}{\kappa}\right)\ =\te^{\i\pi (k-1)/\kappa} \exp\left(-\frac{2\pi\i (\hat{m}_I-1)}{\kappa}\right)
\end{equation}
where we have used
$-\te^{-\i\pi (k+N)/\kappa}=-\te^{-\i\pi \kappa/\kappa}=1$. Let us now
define
\begin{equation}\label{midef}
m_I={\hat m}_I -1\ .
\end{equation} 
It follows that $m_I$ obey
\begin{equation}
\kappa > {m}_1 > \cdots > {m}_k \geq  0.
\end{equation}
and that 
\begin{equation}\label{toputativen}
  z_I^* =\te^{ \i\pi (k-1)/\kappa} \exp\left(-\frac{2\pi\i {m}_I}{\kappa}\right)\ .
\end{equation}
The collection of $\{z_i^*\}$ take the form \eqref{TOeigento}, but
with $k$ playing the role of $N$ and with $m_I$ playing the role of
$n^\lambda_I$ i.e.~the configurations appropriate for Type I
$U(k)_{N,\kappa}$ theory.

Let us define the representation $\lambda^T$ by the equation
$m_I= n^{\lambda^T}_I$. In other words, $\lambda^T$ is the $U(k)$
representation the collection of eigenvalues $\{z_I^* \}$ corresponds
to. It is not difficult to find the relationship between $\lambda^T$
and $\lambda$.  The Young tableau corresponding to $\lambda^T$ has row
lengths given by
\begin{equation}\label{mapbetrep}
  \ell'_I = m_I- k + I\ =N +I-{\hat n}_{k-I+1}-1\ .
\end{equation}
It follows from \eqref{orderto} that 
\begin{equation}\label{nordtab}
I-1 < {\hat n}_{k-I+1} \leq  N+I-1\ ,
\end{equation} 
so that $$ 0 \leq \ell_I' \leq N\ ,$$
as expected.

We will now demonstrate that $\lambda^T$ is the representation whose
Young tableau is the transpose (rows and columns interchanged) of the
Young tableau corresponding to $\lambda$.  Consider a $U(N)$ Young
tableau whose $i^{\rm th}$ column has length $c_i$. Since the first
column has length $c_1$, we have
$\ell_{c_1+1}, \ell_{c_1+2} \ldots \ell_{N}=0$ so that
$n_{N}=0, n_{N-1}=1, \ldots, n_{c_1+1}=N-c_1-1$ and
$n_{c_1} = \ell_{c_1} + N - c_1$ which means that $n_{c_1}$ is
strictly greater than $N-c_1$. It follows that the smallest value
among the excluded integers ${\hat n}_I$ is $N-c_1$
i.e.~${\hat n}_{k}=N-c_1$.

In a similar manner we see that the second lowest unoccupied $n$ value
is $N-c_2+1$, the third highest unoccupied value is $N-c_3+2$ and so
on. It follows that
\begin{equation}
  {\hat n}_{k-I+1}=N-c_I +I-1\ .
\end{equation}
Plugging this result into \eqref{mapbetrep} we conclude that
\begin{equation}
  \ell'_i= c_i\ ,
\end{equation}
establishing that the Young tableau for the representation $\lambda^T$
is the transpose of that for $\lambda$, as we sought to prove.

It may be useful consider a few examples. To start with, consider the
identity representation defined by $n_I=N-I$. For this representation
it follows that ${\hat n}_I= \kappa-I$ so that
${\hat n}_{k-I+1}=N+I-1$, and so $l_I'=0$ for all $I$. In other words
the identity representation maps to itself. Next consider a
representation with $m$ boxes in the first row, i.e. $n_1=N-1+m$,
$n_I=N-I$ (for $I=2 \ldots N$). It follows that
$${\hat n}_{k-I+1}= N+I, ~~~I\leq m; ~~~~~~{\hat n}_{m-I+1}=N+I-1, ~~~~I >m$$
so that $\ell'_I=1$ for $I \leq m$, and $\ell'_I=0$ for $I>m$. In
other words the representation with $m$ boxes in the first row maps to
the representation with $m$ boxes in the first column, in agreement
with our general result.

\subsubsection{Implications}

Given the results of the previous subsubsection,
it is natural to suspect that the result
\begin{equation}\label{charequivn}
\chi_{\vgap F}(\ul{w}) =\chi_{\vgap F}^*(\ul{z}^*)\ ,\quad  \chi_{\vgap A}(\ul{w}) = \chi_{\vgap A}^*(\ul{z}^*)\ ,
\end{equation} 
(which follows immediately from \eqref{charequiv}) has the following generalization to arbitrary integrable representations
\begin{equation}\label{charequivngen}
  \chi_{\vgap R}(\ul{w}) =\chi_{\vgap R^T}^*(\ul{z}^*)\ .
\end{equation} 
While we have not managed to prove \eqref{charequivngen} from a `three
dimensional' viewpoint, that is prove it using the Weyl character
formula evaluated on the distinguished eigenvalue configurations (see
Appendix \ref{levelranktoev} for a crude, partially successful first
attempt), this result follows directly from the relationship, well
known in the study of the level-rank duality of WZW theory (see
\cite[Chapter 16.6]{di1996conformal}),
\begin{equation}\label{wellknown}
  \mc{S}_{\lambda \mu} = \sqrt{\frac{k}{N}} \te^{2\pi\i \hat{Q}_\lambda \hat{Q}_\mu/Nk}\ \mc{S}^*_{\lambda^T \mu^T}\ .
\end{equation} 
That \eqref{wellknown} implies \eqref{charequivngen} follows
immediately from \eqref{suverormu1} and the results of the previous
subsubsection. It follows from this discussion that
\eqref{themogenrep} holds.

\subsection{Check of Level-Rank duality between $SU(N)_k$ and Type II
  $U(k)_{-N,-N}$  theory}\label{sundual}

The discussion of TypeI-TypeI level rank duality presented in
subsection \ref{typeIdual}, has a simple generalization to the study
of $SU(N)$-Type II duality.

Recall from \eqref{ngtmnew} that the number of singlets in $SU(N)_k$
chiral WZW theory on a genus $g$ Riemann surface with primary
operators in representations $R_1,\ldots, R_n$ (see around
\eqref{ngtmnew} for details) is given by
\begin{equation}\label{ngtmnewer} 
  \mc{N}_{g,n} = \frac{1}{(N \kappa^{N-1})^{1-g}}  \sum_{\ul{w} \in \mc{P}_{k}} \prod_{1 \leq i < j \leq N} |w_i-w_j|^{2-2g} \prod_{p=1}^n \chi_{\vgap R_p}(\ul{w})\ , 
\end{equation}
where the summation is taken over the set $\mc{P}_{k}$ of solutions
modulo permutations of the equations
\begin{align}\label{sueq}
  &1.\quad (w_i)^\kappa = (w_j)^\kappa\ ,\quad i, j=1, \ldots, N\ ,\quad\text{with}\quad w_i \neq w_j\quad\text{for}\quad i\neq j\ ,\nonumber\\
  &2.\quad  \prod_{i=1}^N w_i = 1\ .
\end{align}
On the other hand the equivalent formula for the $U(k)_{N,N}$ chiral
WZW model is
\begin{equation}\label{NUgenmtnett}
    \mc{N}_{g,n} = \frac{1}{(N\kappa^{N-1})^{1-g}} \sum_{\ul{z} \in
      \mc{P}_{N,N}} \ \prod_{1 \leq I < J \leq k} |z_i-z_j|^{2-2g}
    \prod_{p=1}^{n} \chi_{\vgap R_p}( \ul{z}) \ .
\end{equation}
The summation in \eqref{NUgenmtnett} is taken over the set of
solutions modulo permutation $\mc{P}_{N,N}$ of the equations
\begin{equation}\label{loiett}
  \frac{(z_I)^\kappa}{\prod_{J=1}^k z_J}=(-1)^{k-1}\
  ,\quad\text{for}\quad I=1,\ldots, k\ ,\quad\text{with}\quad z_I\neq
  z_J\ \text{when}\ I \neq J\ .
\end{equation}
In this subsection we will demonstrate that the formulae
\eqref{ngtmnewer} and \eqref{NUgenmtnett} map to each other under
level-rank duality. Consider any particular solution $\ul{w}$ of the
equation \eqref{sueq} and define the collection of $k$ distinct phases
$w_I'$ as follows. Start with one of the eigenvalues in $\ul{w}$, say
$w_N$ (any eigenvalue works equally well). Multiplying this by all
$\kappa^{\rm th}$ roots of unity gives a collection of $\kappa$
phases. The eigenvalues $\ul{w}$ are $N$ of these $\kappa$ phases. We
define $w_I'$ to be the remaining $\kappa - N = k$ phases in this
collection. We can summarize the definition of $w'_I$ as follows.
\begin{align}\label{wip}
  &(w_I')^\kappa= (w_i)^\kappa\quad\text{with}\quad w'_I \neq w_i\ ,\quad\text{for all}\quad I=1,\ldots, k\ ,\quad i=1,\ldots, N\ ,\nonumber\\
  &w_I' \neq w_J'\ ,\quad\text{for}\quad I \neq J\ .
\end{align}
Note that the union of $\{w_i\}$ and $\{w'_I\}$ is the collection of
$\kappa^{\rm th}$ roots of unity rotated by a common phase (which can
be taken to be one of the eigenvalues among the $\{w_i\}$). Using the
fact that the sum of all $\kappa^{\rm th}$ roots of unity vanishes, it
follows that
\begin{equation}\label{sumoverw}
  \sum_{i=1}^N w_i + \sum_{J=1}^k w'_J = 0\ .
\end{equation} 
Using the fact that every $w_i$ or $w'_J$ is a product of a
$\kappa^{\rm th}$ root of unity with $w_N$, and that the product of
all $\kappa^{\rm th}$ roots of unity equals $(-1)^{\kappa-1}$, we get
\begin{equation}\label{prodev}
  \prod_{j=1}^N w_j \prod_{J=1}^k w'_J = (-1)^{\kappa-1} (w_N)^\kappa = (-1)^{\kappa+1} (w'_I)^\kappa\ ,\text{for any}\ I = 1,\ldots,k\ ,
\end{equation}
where the second equality was arrived at from the first equation in
\eqref{wip}. Plugging the second condition in \eqref{sueq} into
\eqref{prodev} yields
\begin{equation}\label{prodevn}
\prod_{J=1}^k w'_J = (-1)^{\kappa-1} (w_J)^\kappa\ .
\end{equation}
Now define 
\begin{equation} \label{choiceofeig} 
z_J = - w'_J\ ,\quad\text{for}\quad J = 1,\ldots,k\ .
\end{equation}
Plugging into \eqref{prodevn} yields 
\begin{equation}\label{prodevnac}
\prod_{J=1}^k z_J = (-1)^{\kappa+1} (-1)^k (-z_I)^{\kappa} = (-1)^{k-1} (z_I)^\kappa\ ,\quad\text{for every}\quad I = 1,\ldots,k\ .
\end{equation}
Thus, equation \eqref{prodevnac} tells us that the collection of $k$
eigenvalues $\{z_J\}$ obey \eqref{loiett} for $U(k)_{N,N}$ WZW
theory. In other words we have produced a map between solutions of the
$SU(N)_k$ theory \eqref{sueq} and those of the $U(k)_{N,N}$ theory
\eqref{loiett}. The algebra above can be reversed, demonstrating that
this map between solutions is one-to-one. It follows from
\eqref{sumoverw} that
\begin{equation}\label{sumovernn}
  \sum_{i=1}^N w_i = \sum_{J=1}^k z_J\ ,
\end{equation}  
so that 
\begin{equation}\label{charactermap} 
\chi_{\vgap F}(\ul{w}) =\chi_{\vgap F}(\ul{z})\ ,\quad \chi_{\vgap A}(\ul{w}) =\chi_{\vgap A}(\ul{z})\ ,
\end{equation}
where $F$ and $A$ are resp.~fundamental and antifundamental
representations of $SU(N)$ on the left hand sides and of $U(k)$ on the
right hand sides.

As phases cancel out upon taking the absolute value, it follows as in
\eqref{sumoverpairs} that
\begin{equation}\label{sumoverpairsh}
\frac{1}{\kappa^N} \prod_{1 \leq i < j \leq N} |w_i-w_j|^2
= \frac{1}{\kappa^k} \prod_{1 \leq I < J \leq k} |z_I-z_J|^2\ .
\end{equation}
Combining \eqref{sumoverpairsh} and \eqref{charactermap} we see that
for any collection $\{w_i\}$ and its corresponding $\{z_I\}$,
\begin{multline}\label{themot}
\frac{1}{(N \kappa^{N-1})^{1-g}} \prod_{1 \leq i < j \leq N} |w_i-w_j|^{2-2g} \chi_{\vgap F}(\ul{w})^{n_1} \chi_{\vgap A}(\ul{w})^{n_2}
\\ =\frac{1}{(N \kappa^{k-1})^{1-g}} \prod_{1 \leq I < J \leq k} |z_I- z_J|^{2-2g}
\chi_{\vgap F}(\ul{z})^{n_1} \chi_{\vgap A}(\ul{z})^{n_2}\ ,
\end{multline}
for any choice of the positive integers $n_1$ and $n_2$. Summing the
LHS of \eqref{themot} over all solutions $\ul{w} \in \mc{P}_k$ (and
consequently summing the RHS over all solutions
$\ul{z} \in \mc{P}_{N,N}$), we conclude that the number of singlets in
the product of any number of fundamentals and any number of
antifundamentals matches across the $SU(N)_k$ and Type II $U(k)_{N,N}$
duality. As in the case of the Type I - Type I duality, the matching
across the $SU$ - Type II duality occurs already at the level of the integrand.

In the next subsection we will demonstrate that - as in the case of
Type I-Type I duality - \eqref{themot} has a simple generalization to
insertions of more general representations
\begin{multline}\label{themotgen}
  \frac{1}{(N \kappa^{N-1})^{1-g}} \prod_{1 \leq i < j \leq N}
  |w_i-w_j|^{2-2g} \prod_m \chi_{\vgap R_m}(\ul{w}) \\ =\frac{1}{(N
    \kappa^{k-1})^{1-g}} \prod_{1 \leq I < J \leq k} |z_I- z_J|^{2-2g}
  \prod_{m} \chi^*_{R_m^T}(\ul{z}^*) \ .
\end{multline}

\subsubsection{$SU(N)$-Type II duality: map between representations}

As in the case of Type I-Type I duality, the level rank map between eigenvalues in the context of   $SU(N)_k \leftrightarrow U(k)_{-N,-N}$ level-rank duality, induces a map between the integrable representations of these theories. In this section we work out this map.

The distinguished eigenvalue configurations of $SU(N)_k$ are given in
terms of integrable representations $\lambda$ of the $SU(N)_k$ WZW model by
\eqref{whichmat}:
\begin{equation}\label{SUeigen}
w_i(\xi_\lambda) = \exp\left(-\frac{2\pi\i n^\lambda_i}{\kappa}\right) \left(\frac{2\pi\i}{N\kappa}\sum_{i=1}^{N-1} n^\lambda_i\right)\ .
\end{equation}
Recall that the integers $n^\lambda_i$ are ordered as
\begin{equation}
\kappa > n^\lambda_1 > n^\lambda_2 > \cdots > n^\lambda_{N-1} > 0\ .
\end{equation}
Let us label the complement of the set of $n^\lambda_i$ in the range
$[1,\kappa-1]$ as $\hat{n}_I$ with $I = 1,\ldots, k$ where
\begin{equation}
\hat{n}_1 > \hat{n}_2 > \cdots > \hat{n}_k\ ,
\end{equation}
and define the integers $m_I$ as
\begin{equation}
{\hat m}_I = \kappa - \hat{n}_{k-I}\ .
\end{equation}
Then, it is clear that the $m_I$ also satisfy
\begin{equation}
\kappa >  {\hat m}_1 > \cdots > {\hat m}_k>0\ .
\end{equation}
Let us now look at the $k$ $\kappa^{\rm th}$ roots of unity multiplied
by the common phase as in \eqref{SUeigen}
\begin{equation}\label{IIputative}
w_I'=\exp\left(-\frac{2\pi\i \hat{n}_I}{\kappa}\right) \exp\left(\frac{2\pi\i}{N\kappa}\sum_{i=1}^{N-1} n_i^\lambda\right)\ .
\end{equation}
Since the $n^\lambda_i$ and $\hat{n}_I$ together give the integers
$1,\ldots,\kappa-1$, we have
\begin{equation}
\sum_i n^\lambda_i + \sum_I \hat{n}_I = \frac{\kappa(\kappa-1)}{2}\ .
\end{equation}
Using this and that $\hat{n}_I = \kappa - {\hat m}_{k-I}$, the formula \eqref{IIputative} becomes
\begin{align}
w_I'=&\exp\left(\frac{2\pi\i {\hat m}_{k-I}}{\kappa}\right) \exp\left(\frac{\pi\i(\kappa-1)}{N} -\frac{2\pi\i}{N\kappa}\sum_{I=1}^{k} (\kappa - {\hat m}_I)\right)\ ,\nonumber\\
&\quad=-\exp\left(\frac{2\pi\i {\hat m}_{k-I}}{\kappa}\right) \exp\left(\frac{\pi\i(k-1)}{N} -\frac{2\pi\i}{N\kappa}\sum_{I=1}^{k} (\kappa - {\hat m}_I)\right)\ ,\nonumber\\
&\quad=-\exp\left(\frac{2\pi\i ({\hat m}_{k-I}-1)}{\kappa}\right) \exp\left(-\frac{\pi\i(k-1)}{N} + \frac{2\pi\i}{N\kappa}\sum_{I=1}^{k} ({\hat m}_I-1)\right)\ .
\end{align}
As in Section \ref{typeonedualmap}, we define the new collection of
integers $m_I$ by
\begin{equation}\label{newmi}
m_I={\hat m}_I-1\ ,
\end{equation}
in terms of which 
\begin{align}\label{zstarsu}
z_I^*=-(w_I')^*&=\exp\left(-\frac{2\pi\i { m}_{k-I}}{\kappa}\right) \exp\left(\frac{\pi\i(k-1)}{N} - \frac{2\pi\i}{N\kappa}\sum_{I=1}^{k}  {m}_I\right)\ .
\end{align}
Note that $m_I$ lie in the range
\begin{equation}
\kappa - 1 > m_1 > \cdots > m_k  \geq 0\ .
\end{equation}
Comparing with \eqref{wonuo1}, we see that the eigenvalues
\eqref{zstarsu} are precisely those of the Type II $U(|k|)_{-N,-N}$
theory in the representation $\lambda^T$ whose Young tableau has rows
\begin{equation}\label{mapbetrepsu}
\ell'_I = m_I - k + I= {\hat n} =N +I-{\hat n}_{k-I+1}-1\ .
\end{equation}
As \eqref{mapbetrep} is identical to \eqref{mapbetrepsu}, the
arguments of subsection \ref{typeonedualmap} apply equally well here,
and we conclude that the Young tableau for $\lambda^T$ is the
transpose of the Young tableau for $\lambda$. The remaining analysis
of subsection \ref{typeonedualmap} also carries through for this case;
once again we find that \eqref{charequivngen} holds and that
\eqref{themotgen} also holds.

\section{Discussion} \label{disc} 

In this paper we have argued that the partition function of large $N$
Chern-Simons matter theories is that of a free Fock space projected
onto the space of quantum singlets along with all states subject to a
well understood energy renormalization.

A key qualitative consequence of this observation is the `bosonic
exclusion principle' (first encountered in
\cite{Minwalla:2020ysu}). As we have explained in Section \ref{bep},
this principle, which asserts that no bosonic single particle state
can be occupied more than $k_B$ times, follows from the fact that
conformal blocks with insertions of primary operators in
non-integrable representations vanish which in turn imply that
occupations that violate the bosonic exclusion principle do not
contribute to the number of quantum singlets. It would be very useful
to find a direct path integral explanation of this phenomenon. As we
have already discussed in detail in Section \ref{nonint} below, there
is a seemingly related issue already in pure Chern-Simons theory,
whose resolution could prove the first step in fully understanding the
bosonic exclusion principle.

If it is indeed the case that integrable and non-integrable representations decouple from each other within a Chern-Simons path integral, it would also be interesting to understand the
implications of this decoupling for the dynamics of Chern-Simons theories coupled to
matter in both integrable as well as non-integrable
representations. Gauge theories with matter in large representations
are hard to solve even at large $N$ but it might be possible to
explore this question in the corresponding supersymmetric theories
(provided supersymmetry can be preserved in theories with matter in
large representations) via the supersymmetric index which can be
written down for arbitrary gauge groups and matter representations.

Returning to the Bosonic Exclusion Principle, we believe that this
principle is part of a richer and more detailed statistics that, as of
yet, remains to be completely uncovered. As an analogy, recall the
Pauli exclusion principle, which may be taken to be the statement that
every Fermionic creation operator squares to zero. This statement,
taken by itself, may seem to be in conflict with the demand of basis
independence. Let $\alpha^\dagger$, $\beta^\dagger$ and
$\gamma^\dagger$ denote three Fermionic creation operators, related as
$\gamma^\dagger = \cos \theta \alpha^\dagger+ \sin \theta
\beta^\dagger$. It follows that
$$ \left( \gamma^\dagger \right)^2= \cos^2 \theta  \left( \alpha^\dagger \right)^2 + \sin^2 \theta  \left( \beta^\dagger \right)^2 + \cos \theta 
\sin \theta \left( \alpha^\dagger \beta^\dagger + \beta^\dagger
  \alpha^\dagger \right) $$ The principle that fermionic operators
square to zero tells us that the LHS, and first two terms on the RHS
of the above equation vanish. But what of the last term on the RHS?
Have we arrived at a contradiction? Of course, the reader knows that
this is not the case; the last term on the RHS also vanishes, because
the Pauli principle is a particular case of a broader fact that
asserts that all Fermionic creation operators (equal or not)
anticommute with each other. In a similar manner to the discussion
above, the consistency of the bosonic exclusion principle with a
requirement of basis independence presumably requires that this
principle be embedded into a larger, more detailed statistics that
effectively changes commutation relations between creation
operators. Note that this fact is already suggested by the nontrivial
Aharonov-Bohm phase between two fundamental operators (this phase is
small, of order $1/\kappa$, e.g.~see equation 2.46 of
\cite{Jain:2014nza}, but it builds up in states with order $k$
insertions so cannot be ignored even in the large $N$ limit). We
suspect that this broader statistics could connect with the
representation theory of quantum groups. It is possible that this
discussion will also connect with the interesting recent observations
\cite{Jain:2021wyn, Jain:2021vrv, Gandhi:2021gwn} that the three point
functions of spin $s$ current operators all take a form that looks
free (up to dressing by some phases) when viewed in momentum space
using spinor helicity variables. We hope to return to this topic in
the near future.

Another (less pressing) puzzle that awaits resolution is the
interpretation of the detailed results of Wilson Line expectations
values obtained using the method of supersymmetric localization. In
Section \eqref{pisl} we evaluated
\begin{equation}\label{trouble}
  {\rm Tr}_{\mc{H}} \left( (-1)^F x^{2J+\mc{R}} \right)\ ,
\end{equation} 
over the Hilbert space $\mc{H}$ of pure $\mc{N} = 2$ Chern-Simons
theory on $S^2$ with supersymmetric Wilson line insertions at the
North pole of $S^2$. Here $J=J_z$ is the generator of rotations around
the $z$ axis and $\mc{R}$ is the $R$-charge. As we have explained in
Section \ref{pisl}, the two choices $x=-1$ and $x=1$ respectively
reproduce ${\rm Tr}$ and ${\rm Tr}(-1)^F$ over this Hilbert space
(i.e. space of conformal blocks). The results of our computation at
$x=-1$ are in perfect agreement with the Verlinde formula, while the
results at $x=1$ allows us to grade conformal blocks into bosons and
fermions in a physically satisfying manner.  However our results at
generic values of $x$ are more puzzling. Our final answer includes
terms like $x^2-x^4$, which vanish at $x= \pm 1$, but are puzzling to
interpret at generic $x$. The relative minus suggests that the two
terms above capture the contributions of states with opposite (Bose or
Fermi) statistics.  However the fact that the ratio of the two terms
terms is an even power of $x$ suggests that the angular momentum of
the corresponding states differs by an integer. The tension between
these two statements and the spin-statistics connection prevents us
from interpreting our results for \eqref{trouble} at general values of
$x$ as a trace over a physically acceptable Hilbert space.  The
resolution to this puzzle may lie in the fact that the charge $J$ is
well defined only when all Wilson lines are inserted at precisely the
same point (the North Pole). Perhaps it is not really possible to
weight the trace in \eqref{trouble} with $x^{2 J}$ as the
point-splitting that is needed to give meaning to coincident operators
breaks the rotational symmetry generated by $J$. It would be good to
understand this better.

In this paper we have presented a Hilbert space interpretation of the
already known formulae for the partition function of large $N$
Chern-Simons matter theories in the large volume limit. However there
are some indications that the final formulae of this paper have a
larger range of validity than the large $N$ and large volume limit
(see section \ref{discoo}). This question deserves further
investigation.  In particular, it would be interesting to investigate
the effectively non-relativistic limit of Section \ref{ltlv} from the
viewpoint of the relevant Schrodinger equations.

It would be very interesting - and may prove possible - to extend the
analysis of this paper to provide a precise Hilbert space
interpretation for the (exactly known expressions for the)
superconformal index \cite{Bhattacharya:2008zy} in Chern-Simons matter
theories at finite values of $N$ and $k$. The fact that superconformal
indices receive contributions only from protected states makes it
plausible that this exercise has a tractable solution. If it does
prove possible to carry this exercise through, it would allow us to
test the ideas of this paper (or an appropriate deformation thereof)
in a completely precise finite $N$ context.

It would be very interesting - and may well prove possible - to
directly derive some of the results of this paper by analysing the
relevant Schrodinger equations in the non-relativistic limit (see the
last paragraph of Section \ref{mainres} for additional related
discussion).

As we have explained, we think it is possible that the new universal
infinite volume (effectively) free particle statistics, described in
detail in Section \ref{slv}, apply not only in the large $N$ limit but
also in the dilute and non-relativistic limit of Section \ref{ltlv}
even at finite $N$ and $k$. 

As explained in Section \ref{slv} there is
a sense in which the new effectively free statistics presented in this
paper interpolate between the familiar formulae of Bose-Einstein and
Fermi-Dirac statistics. Going out on a limb, one might hope that, one
day, one may find experimental realizations of this new statistics of
this paper, perhaps in gases of quasi-particles in quantum Hall
systems.

It would be interesting to better understand the thermodynamical
formulae \eqref{spefs} for the entropy as a function of occupation
number that follows from the new `free' statistics presented in this
paper, and perhaps obtain a explicit analytic formula for the entropy
functions presented in these formulae.

As we have explained in the main text, $\mc{I}_B$ and $\mc{I}_F$ are
Fock space partition functions projected down to the space of
quantum singlets. As the quantum group plays such an important role
the  thermodynamics of these systems, it is natural to suspect that
the same structure also governs aspects of the detailed
statistics\footnote{The suggestions that it should be possible to
  understand several of the properties of large $N$ matter
  Chern-Simons theories via an alternative `anyonic' quantization of
  the basic fields has been emphasized to us, on several occasions, by
  S. Wadia. } (e.g.~commutation relations between creation operators
as mentioned above) and dynamics in these theories. It would be
interesting to investigate this possibility.

We hope to return to several of these issues in future work.

\section*{Acknowledgements}
We would like to thank S.~Jain, I.~Halder and L.~Janagal for several
discussions, over a period of several years, on the issues discussed
in this paper We would also like to thank S.~Jain for collaboration
during the initial stages of this paper. We would like especially to
thank A.~Gadde and S.~Mukhopadhyay for extremely useful discussions on
the inter relationship between the path integral and Verlinde formulae
in Section 3, O. Parrikar for helping us to better understand the
Verlinde formula in the limit of a large number of insertions and
A. Nair for assistance with group theory. We would also like to thank
O. Aharony,  K. Damle,   A. Gadde, R. Gopakumar, G. Mandal, G. Moore,
O. Parrikar, N. Seiberg, R. Sensharma, V. Tripathy, S. Trivedi,
S. Wadia and E. Witten for useful discussions.  We would further like to
thank O. Aharony,T. Dimofte,  D. Gaiotto, S. Giombi, O. Parrikar, D. Radicevic,  N. Seiberg, D. Tong and 
S. Wadia for comments on a preliminary draft of this manuscript. The
work of S.~M., A.~M. and N.~P. was supported by the Infosys Endowment
for the study of the Quantum Structure of Spacetime and a J. C. Bose
Fellowship. Finally, we would like to acknowledge our debt to the
steady support of the people of India for research in the basic
sciences.

\appendix

\section{Enumerating classical $SU(N)$ and 
$U(N)$ singlets} \label{classcount} 

The number of singlets in the tensor product of irreducible
representations $R_1$, $R_2$,\ldots, $R_n$ of a group $G$ is given by
\begin{equation} \label{wif}
  \mc{N}_{\rm cl}(R_1,\ldots,R_n) = \int_G dg\ \chi_{\vgap R_1}(g) \chi_{\vgap R_2}(g) \cdots \chi_{\vgap R_n}(g)\ .
\end{equation}
where $dg$ is the normalized Haar measure on $G$ and $\chi_{\vgap R}$
is the character of the representation $R$. Since the characters
$\chi_{\vgap R}$ are class functions i.e.~are invariant under
conjugation $g \to h^{-1} g h$ for $h \in G$, we can use the Weyl
integral formula to simplify the above integral:
\begin{equation}\label{weylint}
  \int_G dg = \frac{1}{|W|} \int_T dt\, D(t)\ ,\quad\text{where}\quad D(t) = \prod_{\alpha \in \Delta_+} \left|\te^{\alpha/2}(t) - \te^{-\alpha/2}(t)\right|^2\ .
\end{equation}
Here, $W$ is the Weyl group of $G$, $T$ is a maximal torus of $G$ and
$dt$ is the normalized Haar measure on $T$. The product runs over the
set $\Delta_+$ of positive roots $\alpha$ of the Lie algebra of
$G$. The notation $\te^{\alpha/2}(t)$ is interpreted as follows. We
can write $t = \te^{\i\phi}$ for an angular variable $\phi$ that
belongs to the Cartan subalgebra which can be expanded as
$\phi = \phi^I H_I$ where $H_I$ are the Cartan generators. The root
$\alpha$ has a corresponding decomposition
$\alpha = \alpha_I \epsilon^I$ where $\{\epsilon^I\}$ are a basis for
the dual of the Cartan subalgebra. Then,
$\te^{\alpha/2}(t) = \te^{\i \alpha_I \phi^I}$.

The number of singlets is then given by
\begin{equation}\label{NsingSUN}
  \mc{N}_{\rm cl}(R_1,\ldots,R_n) = \frac{1}{|W|} \int_T dt\, D(t) \prod_{I=1}^n \chi_{\vgap R_I}(t)\ .
\end{equation}
In this appendix, we give explicit forms of the above integral for the
groups $SU(N)$ and $U(N)$.

\subsection{$SU(N)$}
The maximal torus of $SU(N)$ can be described in terms of the defining
or fundamental representation of $SU(N)$ which is defined as the set
of $N \times N$ unitary matrices $M$ with $\det\,M = 1$. A given
$SU(N)$ matrix can always be diagonalized; let the diagonal matrix be
\begin{equation}\label{sutowmat}
  t_\theta = \begin{pmatrix} \te^{\i \theta_1} & 0  &\cdots & 0\\
    0 & \te^{\i \theta_2}  &\cdots & 0\\
    \vdots & \vdots & \ddots & \vdots\\
    0 & 0 & \cdots & \te^{\i \theta_N} \end{pmatrix} \ ,\quad\text{with}\quad \sum_{i=1}^N \theta_i = 2 \pi m \ , 
\end{equation}
where $m$ is an integer. The maximal torus $T$ is described by the
angular coordinates $\theta_i$ satisfying the constraint above with
appropriate ranges for the angles which we will describe
presently. The Weyl integral formula thus \eqref{weylint} instructs us
to integrate over all diagonal matrices $t$ \eqref{sutowmat}. We now
obtain an explicit form of that formula in terms of the angles
$\theta_i$. Firstly, the quantity $D(t)$ in \eqref{weylint} is given by
\begin{equation}
  D(t_\theta) = \prod_{1 \leq i < j \leq N} \left|\te^{\i(\theta_i-\theta_j)/2} - \te^{-\i(\theta_i-\theta_j)/2}\right|^2 = \prod_{1 \leq i < j \leq N} 4\sin^2\left(\frac{\theta_i-\theta_j}{2}\right)\ .
\end{equation}
Let us next determine the ranges of the angular variables
$\theta_i$. It is clear that the shifts
$\theta_i \to \theta_i + 2\pi m_i$ for integer $m_i$ do not change the
matrix $t$. We use this freedom to restrict the ranges of
$\theta_1$,\ldots,$\theta_{N-1}$ to
\begin{equation}\label{rangef}
0 \leq  \theta_i < 2\pi \ ,\quad i=1, \ldots, N-1\ ,
\end{equation}
and set $m = 0$ in \eqref{sutowmat}\footnote{For instance, this can be
  done by shifting the $\theta_1$,\ldots,$\theta_{N-1}$ by the
  appropriate integers $2 \pi m_1$,\ldots,$2\pi m_{N-1}$ to bring them
  to the range $[0,2\pi)$ and then shifting $\theta_N$ by
  $-m-m_1-\ldots-m_{N-1}$.}. In this choice of ranges of the
$\theta_i$, we then have
\begin{equation}\label{alphai}
  \wt\theta_N = -\sum_{i=1}^{N-1} \theta_i  \ ,\quad\text{with}\quad -2\pi(N-1) < \wt\theta_N \leq 0\ . 
\end{equation} 
We have put a tilde on $\theta_N$ to indicate that its range is
different from the other $\theta_i$. Thus, we have achieved our goal
of obtaining an explicit parametrization of the matrices
$t$.
However, this parametrization breaks the manifest $\mc{S}_N$ symmetry
by treating $\wt\theta_N$ differently from the other $\theta_i$. We
fix this issue by dividing the integration range of $\wt\theta_N$ into
$N-1$ pieces
\begin{equation}
  \wt\theta_N \in (-2\pi(N-1), -2\pi(N-2)] \cup (-2\pi(N-2), -2\pi(N-3)] \cup \cdots \cup (-2\pi, 0]\ ,
\end{equation}
and in each interval of the form $(-2\pi s, -2\pi(s-1)]$ we redefine
$\wt\theta_N$ as
\begin{equation}
  \wt\theta_N = \theta_{s,N} - 2\pi s\ ,\quad\text{with}\quad 0 < \theta_{s,N} \leq 2\pi\ .
\end{equation}
The new variable $\theta_{s,N}$ then satisfies
\begin{equation}\label{thetasN}
  \theta_{s,N} = - \sum_{i=1}^{N-1} \theta_i + 2\pi s\ ,\quad\text{for each}\quad s = 1,\ldots,N-1\ .
\end{equation}
We have thus constructed a one-to-one map from the variables
$\{\theta_1,\ldots,\theta_{N-1},\wt\theta_N\}$ to the $N-1$ sets of
variables $\{\theta_1,\ldots,\theta_{N-1},\theta_{s,N}\}$ with the
advantage that all the variables take values $(0,2\pi]$ and hence
respect the $\mc{S}_N$ symmetry. We impose the constraint
\eqref{thetasN} through a $\delta$-function in the integrand. We thus
get the following explicit formula for the number of singlets
$\mc{N}_{\rm cl}(R_1,\ldots,R_n)$:
\begin{align}\label{nsingetsclnew}
  &\mc{N}_{\rm cl}(R_1,\ldots,R_n) =\nonumber\\
  &\frac{1}{N!}\sum_{s=1}^{N-1}\left( \prod_{i=1}^{N}\int_0^{2\pi} \frac{d\theta_i}{2\pi}\right) \delta\left(\frac{1}{2\pi}\sum_{i=1}^N \theta_i - s\right) \prod_{1 \leq i < j \leq N} 4\sin^2\left(\frac{\theta_i-\theta_j}{2}\right)\ \prod_{i=1}^n \chi_{\vgap R_i}(t_\theta)\ ,
\end{align}
where $\chi_{\vgap R}(t_\theta)$ is the character in the
representation $R$ evaluated on the element $t_\theta$ in
\eqref{sutowmat}.

The formula \eqref{nsingetsclnew} can be rewritten as a contour
integral over eigenvalues
\begin{equation}\label{eigenval} 
  z_i = \te^{\i \theta_i}\ .
\end{equation} 
Using $dz_i = \i z_i d\theta_i$ and
\begin{equation}
  \delta(z_1 z_2 \ldots z_N - 1) =  \sum_{s=1}^{N-1} \frac{ \delta \left( 
      \frac{1}{2 \pi} \sum_{i=1}^N \theta_i  -s \right) }{2\pi z_1 z_2 \cdots z_N}\ ,
\end{equation}
we can rewrite \eqref{nsingetsclnew} in the elegant form
\begin{align}\label{nsingetsclnewelt}
  &\mc{N}_{\rm cl}(R_1,\ldots,R_n) = \frac{1}{N!} \left(\prod_{i=1}^N \oint \frac{d z_i}{2 \pi \i z_i} \right)  2\pi\delta(Z-1) \prod_{1 \leq i < j \leq N} |z_i-z_j|^2 \prod_{i=1}^n \chi_{\vgap R_i}(t_\theta)  \ ,
\end{align}
where $Z = z_1 z_2\cdots z_N$ and the contour of integration for each
$z_i$ runs anticlockwise along the unit circle.

\subsubsection{Weyl character formula for $SU(N)$}

Consider an $SU(N)$ matrix $U$ whose eigenvalues are given by $w_1$,
$w_2 \ldots w_N$. Consider a representation $R$ of $SU(N)$ whose
corresponding Young tableau has $\ell_i$ boxes in the $i^{\rm th}$ row
(of course, $\ell_N=0$). Define the quantity
\begin{equation}\label{detmat}
  \Big[f_1(w), \ f_2(w), \ \ldots, \ f_N(w) \Big] = \det\, f_i(w_j)\ , 
\end{equation} 
i.e.~the determinant of the $N\times N$ matrix whose $ij^{\rm th}$
element is $f_i(w_j)$. Then, the Weyl character formula asserts that
the character of the representation $R$ evaluated on $U$ is given by
the functions $f_i(w) = w^{N-i + \ell_i}$ and thus
\begin{equation}\label{weylchar}
\chi_{\vgap R}(U) = \frac{\Big[w^{N-1+\ell_1}, w^{N-2+\ell_2}, \ldots,  w^{\ell_N} \Big]}{\Big[w^{N-1}, w^{N-2}, \ldots 1 \Big]} = \frac{\det\, w_j^{N-i+\ell_i}}{\det\, w_j^{N-i}} \ . 
\end{equation}

\subsubsection{$SU(N)$ character polynomials}

Another useful formula for the character evaluated on $U$ in the
representation $R$ is
\begin{equation}\label{chirchid}
  \chi_{\vgap R}(U)= \frac{1}{n!} \sum_{g\in \mc{S}_n} \chi^{\mc{S}}_{\vgap R}(g) \prod_m (\tr\, U^m)^{C_{g,m}} \ , 
\end{equation} 
where $n$ is the number of boxes in the Young tableau $Y_R$
corresponding to $R$, $\mc{S}_n$ is the permutation group on $n$
objects, $\chi^{\mc{S}}_{\vgap Y(R)}(g)$ is the $\mc{S}_n$ character
of the element $g$ in the representation $Y(R)$ and $C_{g,m}$ is the
number of cycles of length $m$ in the permutation $g$.

\subsubsection{The case of $SU(2)$}
When $N=2$, the matrix \eqref{sutowmat} reduces to
\begin{equation}\label{sutwo}
  t_\theta = \begin{pmatrix} \te^{\i \theta} & 0 \\ 0 & \te^{-\i \theta}  \end{pmatrix}\ ,
\end{equation} 
and the range of $\theta$ is $0 \leq \theta < 2 \pi$. In this simplest
case, the Weyl integral formula \eqref{nsingetsclnew} reduces to
\begin{equation}\label{classsu2}
  \mc{N}_{\rm cl}(R_1,\ldots,R_n) = \frac{1}{2} \int_{0}^{2\pi} \frac{d\theta}{2\pi}  4\sin^2 \theta  \prod_{i=1}^n \chi_{\vgap R_i}(t_\theta)\ ,
\end{equation}
where the character of a representation $R$ with spin $j$ is
\begin{equation}\label{chja}
  \chi_{\vgap R}(t_\theta)= \sum_{m=0}^{2j}\te^{2 \i(j-m) \alpha} = 
  \frac{\sin\big( (2j+1) \alpha \big)}{\sin \alpha} \ , 
\end{equation}
The representation with spin $j$ has $2j$ symmetrized boxes in the
first and only row of its Young tableau. The allowed range of values
for $j$ is $j=0, \frac{1}{2}, 1, \frac{3}{2},\ldots$. The
dimensionality of the representation with spin $j$ is $2j+1$.  The
$j=\frac{1}{2}$ representation $R_{1/2}$ is two-dimensional and its
character is given by
\begin{equation}\label{chjhalf}
  \chi_{\vgap R_{1/2}}(t_\theta) = \te^{-\i \theta}+ \te^{\i \theta}  \ . 
\end{equation}
The usual $J_z$ operator is implemented by the operator
\begin{equation}\label{formulaforjz}
  J_z = -\frac{\i}{2}\partial_\theta  \ . 
\end{equation} 
The extra factor of $\frac{1}{2}$ reflects the fact that $\theta$ is
half of the $SO(3)$ rotational angle around the $z$-axis in three
dimensional space. In particular a $\pi$ rotation in $\theta$
corresponds to a $2 \pi$ rotation around the $z$-axis.

It is easy to directly verify that 
\begin{equation}\label{charorthog}
  \int_{0}^{2\pi} \frac{d\theta}{2\pi}   2\sin^2 \theta  
  \chi_j(\theta) \chi_{j'}(\theta) = \delta_{j j'} \ . 
\end{equation}

The classical $SU(2)$ tensor product decomposition of representations
translates to the following algebraic identity for characters
\begin{equation}\label{chardec}
\chi_{j_1} \chi_{j_2} = \sum_{i=|j_1-j_2|}^{j_1+j_2} \chi_i \ . 
\end{equation} 
This identity is true when the characters are evaluated at any
$t_\theta$.

\subsection{$U(N)$}

Any $U(N)$ matrix can be diagonalized to the form
\begin{equation}\label{utowmat}
  t_\theta = \begin{pmatrix} \te^{\i \theta_1} & 0  &\cdots & 0\\
    0 & \te^{\i \theta_2}  &\cdots & 0\\
    \vdots & \vdots & \ddots & \vdots\\
    0 & 0 & \cdots & \te^{\i \theta_N} \end{pmatrix} \ ,
\end{equation}
with no constraint on the determinant of $t_\theta$. As a consequence,
all $U(N)$ formulae are simpler than their $SU(N)$ counterparts. The
ranges of the $\theta_i$ are simply
\begin{equation}\label{rangefun} 
  0 \leq  \theta_i < 2\pi \ ,\quad i=1, \ldots, N \ , 
\end{equation} 
and the Weyl integral formula for the total number of singlets in the
product of the representations $R_1, R_2, \cdots, R_{n}$ is
\begin{equation}\label{nsingets}
\mc{N}_{\rm cl}(R_1,\ldots,R_n) = \frac{1}{N!} \prod_{i=1}^N \int_{0 }^{2\pi}  \frac{d\theta_i}{2\pi}  \prod_{1 \leq i < j \leq N} 4 \sin^2 \left(\frac{\theta_i - \theta_j}{2}\right) \prod_{I=1}^n \chi_{\vgap R_I}(t_\theta) \ . 
\end{equation}
Rewritten as a contour integral over the eigenvalues
$z_i = \te^{\i\theta_i}$, we have
\begin{equation}\label{unclasssing}
\mc{N}_{\rm cl}(R_1,\ldots,R_n) = \frac{1}{N!} \left( \prod_i \oint \frac{d z_i}{2 \pi \i z_i } \right) \left( \prod_{1 \leq i < j \leq N} |z_i-z_j|^2\right) \prod_{I=1}^n \chi_{\vgap R_I}(t_\theta)\ .
\end{equation}

\subsubsection{$U(N)$ representations and the Weyl character
  formula} \label{lwc}

Recall that irreducible representations of $SU(N)$ correspond to Young
tableaux with row lengths $\ell_i$ satisfying
$\ell_1 \geq \ell_2 \geq \ldots \geq \ell_{N-1} \geq 0$. In the same
way, representations of $U(N)$ correspond to Young tableaux with $N$
rows with ``row lengths'' satisfying
$\ell_1 \geq \ell_2 \geq \cdots \geq \ell_N$, where the integers $\ell_i$
can now be either positive or negative. This representation descends
to an $SU(N)$ representation with
\begin{equation}\label{lmapusu}
  \ell^{SU(N)}_i = \ell^{U(N)}_i -\ell^{U(N)}_N \ .
\end{equation}
The $U(1)$ charge of a $U(1)$ representation is $\sum_i \ell_i$. Note
that any two representations which descend to the same $SU(N)$
representation must differ in $U(1)$ charge by a multiple of $N$
reflecting the fact that adding a column of $N$ boxes to an $SU(N)$
Young tableau does not change the $SU(N)$ representation since a
column of $N$ boxes corresponds to the trivial representation;
however, the same operation increases the $U(1)$ charge by $N$.

For example, the antifundamental representation of $U(N)$ has $U(1)$
charge $-1$ and corresponds to the Young tableau
$\ell_1=\ell_2=\cdots=\ell_{N-1}=0$ and $\ell_N=-1$. The
antifundamental of $SU(N)$ corresponds to the Young tableau with
$\ell_1=\ell_2=\cdots=\ell_{N-1}=1$ (and $\ell_N=0$). Both these Young
tableaux denote the antifundamental of $SU(N)$ but with $U(1)$ charges
$-1$ and $N-1$ respectively and the latter is obtained from the former
by adding a column of $N$ boxes.

With all these conventions in place the Weyl character formula
\eqref{weylchar} applies without change to $U(N)$ characters:
\begin{equation}\label{weylcharag}
  \chi_{\vgap R}(U) = \frac{\Big[w^{N-1+\ell_1}, w^{N-2+\ell_2}, \ldots,  w^{\ell_N} \Big]}{\big[w^{N-1}, w^{N-2}, \ldots, 1 \big]} \ , 
\end{equation}
See also the discussion in Section \ref{unrep}.

\subsubsection{Conjugating $U(N)$ representations} 

Recalling that $w$ is a phase (and so $w^*=w^{-1}$), it follows from
\eqref{weylcharag} that
\begin{align}\label{weylcharag1}
  \chi^*_{\vgap \{\ell_1, \ell_2, \ldots, \ell_N\}}
  &= \frac{\Big[w^{-N+1-\ell_1}, w^{-N+2-\ell_2}, \ldots,  w^{-\ell_N} \Big]}{\big[w^{-N+1}, w^{-N+2}, \ldots, 1 \big]}\ ,\nonumber\\
  &=\frac{w^{-N+1} \Big[w^{-\ell_1}, w^{-\ell_2+1}, \ldots,  w^{-\ell_N + N - 1} \Big]}{w^{-N+1}\big[1, w, \ldots, w^{N-1}\big]}  \ ,\nonumber  \\
  &= \frac{\Big[w^{N-1-\ell_N}, w^{N-2-\ell_{N-1}}, \ldots,  w^{-\ell_1} \Big]}{\big[w^{N-1}, w^{N-2}, \ldots, 1 \big]}  = \chi_{\vgap \{-\ell_N, -\ell_{N-1}, \ldots, -\ell_1\}}\ .
\end{align}

\section{Solutions of the $SU(N)_k$ and $U(N)_{k,k'}$ eigenvalue equations} 
\label{cpsuto}

\subsection{$SU(N)_k$} \label{cpsu}
Consider the following equations for $N$ distinct phases
$\ul{f} = \{f_1,\ldots,f_N\}$:
\begin{equation}\label{progen}
  (f_i)^\kappa =(f_j)^\kappa\ ,\quad \prod_{i=1}^N f_i = 1\ .
\end{equation} 
In this section, we solve these equations and show that solutions of
\eqref{progen} modulo permutation are in one-to-one correspondence with
the $w_i(\xi_\lambda)$ discussed in Section \ref{disceigen} and hence
in one-to-one correspondence with the integrable representations of
$\wh{su}(N)_k$.

From the first of \eqref{progen} we can write
\begin{equation}\label{mou}
  \frac{f_{i}}{f_N}= \exp\left(-\frac{2 \pi \i p_i }{\kappa}\right)\ ,\quad\text{for}\quad i=1,2,\ldots,N\ ,
\end{equation} 
for some collection of integers $p_i$ (with $p_N = 0$) which are
distinct modulo $\kappa$ i.e.~$p_i \neq p_j\ \text{mod}\ \kappa$ for
$i \neq j$. Since the integers $p_i$ are only defined modulo $\kappa$
we choose to fix this ambiguity by shifting each $p_i$ by appropriate
multiples of $\kappa$ so that they satisfy
\begin{equation}\label{pirange}
  0 < p_i < \kappa\ ,\quad\text{for}\quad i=1,\ldots,N-1\ .
\end{equation}
($p_i = 0$ is not allowed since in that case $f_i = f_N$ which is
ruled out by the requirement that the $f_i$ are distinct.) The second
of \eqref{progen} then gives
\begin{equation}\label{eqforfo}
1 = \prod_{i=1}^N f_i = (f_N)^N \exp\left(-\frac{2\pi\i}{\kappa}\sum_{j=1}^{N-1} 
    p_j\right)\ ,
\end{equation}
from which we conclude that 
\begin{equation}\label{foeq}
  f_N= \exp\left(\frac{2\pi\i}{N\kappa}\sum_{j=1}^{N-1} 
    p_j\right) \times \exp\left(\frac{2 \pi \i r}{N}\right)\ ,
\end{equation}
where $r$ is some fixed integer which we choose to lie the range
\begin{equation}\label{lrange}
  0 \leq r \leq N-1\ .
\end{equation}
Thus, the general solution of \eqref{progen} is
\begin{equation}\label{ninteg} 
  \left\{ f_1, f_2, \ldots, f_N \right\} = \te^{2 \pi \i r/N} \te^{\frac{2\pi\i}{N\kappa}\sum_j p_j} \left\{ \te^{-2 \pi \i p_1/\kappa}, \ldots, \te^{-2 \pi \i p_{N-1}/\kappa}, 1 \right\}\ ,
\end{equation} 
where the $p_i$ lie in the range $(0,\kappa)$ \eqref{pirange}.

Since the equations \eqref{progen} are invariant under permutations of
the $f_i$, an $\mc{S}_N$ permutation of the solution \eqref{ninteg} is
also a solution of the equations \eqref{progen}. Thus, solutions of
\eqref{progen} modulo $\mc{S}_N$ permutations can be described by
ordering the integers $p_i$ as follows:
\begin{equation}\label{rop}
\kappa -1 \geq p_1 > p_2 > \cdots > p_{N-1} \geq 1\ .
\end{equation}
This corresponds to ordering $\{f_1,f_2,\ldots,f_N\}$ counterclockwise
on the unit circle by arbitrarily picking one of the $N$ eigenvalues
to be the `first' eigenvalue $f_1$ and then labelling the eigenvalues
we subsequently encounter by moving counterclockwise along the unit
circle as $f_2,\ldots$.

Let us now look at the above solutions for different values of $r$.

\subsubsection{$r=0$}
When $r = 0$, the solution takes the form
\begin{equation}\label{req0}
  \left\{ f_1, f_2, \ldots, f_N \right\} =  \te^{\frac{2\pi\i}{N\kappa}\sum_j p_j}
  \left\{ \te^{-2 \pi \i p_1/\kappa}, \ldots, \te^{-2 \pi \i p_{N-1}/\kappa}, 1 \right\}\ .
\end{equation} 
Observe that the $p_i$ satisfy the same ordering \eqref{rop} as the
integers $n^\lambda_i$ \eqref{nord} in Section \ref{verlindedetail}
corresponding some highest weight $\lambda$. In addition, the form of
the above $f_i$ is exactly the same as $w_i(\xi_\lambda)$ when we
identify $n_i^\lambda = p_i$. In particular, the rows of the Young
tableau corresponding to $p_i$ can be extracted by inverting the
equation \eqref{nidef} in Section \ref{verlindedetail}:
\begin{equation}\label{r0yd}
  \ell^\lambda_i = p_i - (N-i)\ .
\end{equation}

\subsubsection{$r \neq 0$}
Let us look at some non-zero value of $r$ (recall that the range of
$r$ is $0 \leq r \leq N-1$ \eqref{lrange}). In this case, since we
have the additional phase $\te^{2\pi\i r/ N}$, the solution
\eqref{ninteg} does not take the form of $w_i(\xi_\lambda)$. However,
we show that it is possible to reinterpret \eqref{ninteg} so that it
does take the form of $w_i(\xi_\lambda)$ for some $\lambda$. 

Let us apply a particular $\mbb{Z}_N$ cyclic permutation on the $f_i$
that shifts them by $r$ units counterclockwise:
\begin{equation}\label{rcyclic}
  \{f_1,\ldots, f_{N-r}, f_{N-r+1},\ldots, f_{N}\} \to \{f_{N-r+1},\ldots, f_{N}, f_{1},\ldots, f_{N-r}\}\ .
\end{equation}
We first bring \eqref{rcyclic} to the form \eqref{req0} by scaling out
$f_{N-r}$ out of every $f_i$ on the right hand side of
\eqref{rcyclic}:
\begin{multline}\label{fNrscale}
  \{f_{N-r+1},\ldots, f_{N}, f_{1},\ldots,f_{N-r-1}, f_{N-r}\} \\ = f_{N-r}   \left\{\frac{f_{N-r+1}}{f_{N-r}},\ldots, \frac{f_{N}}{f_{N-r}}, \frac{f_{1}}{f_{N-r}},\ldots,\frac{f_{N-r-1}}{f_{N-r}}, 1\right\}\ .
\end{multline}
From the definition of the $p_i$ in \eqref{mou}, we have
\begin{align}
  \frac{f_{N-r+1}}{f_{N-r}} &= \exp\left(-\frac{2\pi\i}{\kappa} (p_{N-r+1} - p_{N-r})\right)\ ,\quad \ldots\ ,\quad   \frac{f_{N}}{f_{N-r}} = \exp\left(-\frac{2\pi\i}{\kappa}(-p_{N-r})\right) \nonumber\\
  \frac{f_{1}}{f_{N-r}} &= \exp\left(-\frac{2\pi\i}{\kappa}(p_1-p_{N-r})\right)\ ,\quad  \ldots\ , \quad \frac{f_{N-r-1}}{f_{N-r}} = \exp\left(-\frac{2\pi\i}{\kappa}(p_{N-r-1}-p_{N-r})\right)\ .  
\end{align}
Since the differences between the $p_i$ in the first line above are
negative (due to the ordering \eqref{rop}), we can shift the
difference by $\kappa$ without changing the ratio
$f_{N-i+1}/f_{N-r}$. It is then clear that the integers $p'_i$
defined by
\begin{align}\label{ptldef}
  &p'_1 =  \kappa + p_{N-r+1} - p_{N-r}\ ,\quad\ldots,\quad p'_{r-1} = \kappa + p_{N-1} -  p_{N-r}\ ,\quad p'_r = \kappa - p_{N-r}\ ,\nonumber\\
  & p'_{r+1} = p_1 - p_{N-r}\ ,\quad\ldots, \quad p'_{N-1} = p_{N-r-1} - p_{N-r}\ ,
\end{align}
satisfy the inequalities \eqref{rop}:
\begin{equation}
  \kappa > p'_1 > p'_2 > \cdots > p'_{N-1} > 0\ .
\end{equation}
Further, the common factor $f_{N-r}$ simplifies as
\begin{align}
  f_{N-r} &= f_N \exp\left(-\frac{2\pi\i p_{N-r}}{\kappa}\right) = \exp\left(\frac{2\pi\i}{N\kappa}\sum_{j=1}^N p_i\right) \exp\left(\frac{2\pi\i r}{N}\right)\ ,\nonumber\\
          &=\exp\left(\frac{2\pi\i}{N\kappa} \sum_{j=1}^{N} (p_i - p_{N-r})\right) \exp\left(\frac{2\pi\i r}{N}\right)\ ,\nonumber\\
          &=\exp\left(\frac{2\pi\i}{N\kappa} \sum_{j=1}^{N-r} (p_i - p_{N-r}) + \frac{2\pi\i}{N\kappa} \sum_{j=N-r+1}^{N} (\kappa + p_i - p_{N-r}) \right)\ ,\nonumber\\
          &= \exp\left(\frac{2\pi\i}{N\kappa} \sum_{j=1}^{N} p'_i\right)\ ,
\end{align}
where, in the last step, we have absorbed the $r / N$ into the last
$r$ differences $p_i - p_{N-r}$, $i= N-r+1,\ldots,N$, by shifting each
of them by $\kappa$. Thus, we have recast the cyclically permuted
$f_i$ \eqref{fNrscale} into the form \eqref{req0}:
\begin{equation}
  \te^{\frac{2\pi\i}{N\kappa}\sum_j p'_j} \left\{ \te^{-2 \pi \i p'_1/\kappa}, \ldots, \te^{-2 \pi \i p'_{N-1}/\kappa}, 1 \right\}\ ,
\end{equation}
where, in particular, there is no longer a non-trivial $\mbb{Z}_N$
phase. The above configuration is indeed of the form of
$w_i(\xi_{\lambda'})$ for some $\lambda'$. The rows of the Young
tableau corresponding to $\lambda'$ are then given by
\begin{equation}\label{rn0yd}
  \ell^{\lambda'}_i = p'_i - (N-i)\ .
\end{equation}
Thus, in summary, solutions $\ul{f}$ of \eqref{progen} modulo
permutation are in one-to-one correspondence with the integrable
representations $\lambda$ of $\wh{su}(N)_k$. When there is no
non-trivial $\mbb{Z}_N$ phase in the solution, it directly matches the
form of $w_i(\xi_\lambda)$ for an integrable highest weight
$\lambda$. If there is a non-trivial $\mbb{Z}_N$ phase
$\te^{2\pi\i r/N}$, then the above discussion tells us that our choice
of what we call the first eigenvalue of the collection $\ul{f}$ was
incorrect and that the correct choice is $f_{N-r+1}$. With this choice,
the expression for $\ul{f}$ again corresponds to an integrable highest
weight.

\subsubsection{Relation between integrable highest weights under the
  $\mbb{Z}_N$ action}\label{monopole}

In this subsection we ask the following question. Say we start with legal configuration of $SU(N)$ eigenvalues
\begin{equation}\label{collection}
\te^{\frac{2\pi\i}{N\kappa}\sum_j p_j} \left\{ \te^{-2 \pi \i
	p_1/\kappa}, \ldots, \te^{-2 \pi \i p_{N-1}/\kappa}, 1 \right\}\ ,
\end{equation}
which corresponds to a Young tableau with row-lengths
$\ell^\lambda_i = p_i - (N-i)$ (see \eqref{r0yd}). We know from the discussion around \eqref{rephase} that the eigenvalues obtained by multiplying \eqref{collection} by the constant phase $\te^{\frac{2 \pi i m}{N}}$ also gives us a legal $SU(N)$ eigenvalue configuration. Question: to what representation does this new set of eigenvalues correspond?

The discussion earlier in this subsection gives us an immediate answer. We have demonstrated above that we can `get rid of an overall phase that is a power of 
$\te^{\frac{2 \pi i}{N}}$' (i.e. put the collection of eigenvalues in the standard Verlinde form) by changing our choice of `origin' (which eigenvalue we associate with $p_1$). Above we have worked out the transformation the integers $p_m$ under such a move and demonstrated that the configuration which includes the phase $\te^{2\pi\i r/N}$
corresponds to a Young tableau with row-lengths
$\ell^{\lambda'}_i = p'_i - (N-i)$ (see \eqref{rn0yd}) with the integers $p_i'$  given in \eqref{ptldef}. 

Focusing on  the simplest non-trivial central element $\te^{2\pi\i/N}$ corresponding to $r=1$, (i.e. the relations between
$p'_i$ and $p_i$ in \eqref{ptldef} with $r=1$) we find that the row lengths for the Young tableau corresponding to the new representation are given by 
\begin{align}\label{znmove}
\ell^{\lambda'}_1 &= p'_1 - (N-1) = \kappa - p_{N-1} - (N-1) = k - \ell^\lambda_{N-1}\ ,\nonumber\\
\ell^{\lambda'}_i &=  p'_i  - (N-i) = p_{i-1} - p_{N-1} - (N-i) = \ell^\lambda_{i-1} - \ell^\lambda_{N-1}\ ,\quad\text{for}\quad i=2,\ldots,N\ .
\end{align}
Pictorially, the above operation corresponds to adding a row of $k$
boxes to the top of the Young tableau $\lambda$ and then removing
columns of length $N$ since they correspond to the trivial
representation. Since the number of columns of length $N$ is given by
$\ell^\lambda_{N-1}$ we have to subtract this quantity from every row
of the new tableau. This gives the formulae \eqref{znmove}.

\subsection{$U(N)_{k,k'}$} \label{cpu}

\subsubsection{Counting solutions} \label{csu}

As in Section \ref{cpsu} will now argue that the converse of
\eqref{prodw} is also true; i.e.~that every collection of $N$ distinct
eigenvalues $\ul{f} = \{f_1,\ldots,f_N\}$ modulo permutation that obey
\begin{equation}\label{progenu}
  f_i^\kappa \left( \prod_{i=1}^N f_i \right)^s = (-1)^{N-1}\ ,
\end{equation} 
can be cast in the form $\ul{w}_{\lambda}$ for some $\wh{u}(N)_{k,k'}$
highest weight $\lambda$. The argument that this is the case can be made in a surprisingly simple manner. 

First, we observe that the equations \eqref{progenu} also ensure that
\begin{equation}\label{ratiok}
  (f_i)^\kappa = (f_j)^\kappa\ ,\quad\text{for all}\quad i,j=1,\ldots, N\ .
\end{equation}
We next make use of the analysis of the $SU(N)$ case as follows. We
write the diagonal matrix of $U(N)$ eigenvalues as a $SU(N)$ matrix times a
$U(1)$ phase:
\begin{equation}\label{unsun}
  \begin{pmatrix} f_1 & & & \\ & f_2 & & \\ & & \ddots & \\ & & & f_N \end{pmatrix} = \te^{\i\hat\phi} \times \begin{pmatrix} \hat{f}_1 & & & \\ & \hat{f}_2 & & \\ & & \ddots & \\ & & & \hat{f}_N \end{pmatrix}\ ,
\end{equation}
where $\{\hat{f}_1,\ldots, \hat{f}_N\}$ are the diagonal elements of
an $SU(N)$ matrix and, following \eqref{ratiok}, they satisfy
$(\hat{f}_i)^\kappa = (\hat{f}_j)^\kappa$ for all
$i,j=1,\ldots,N$. 

We already know that the number of inequivalent choices for the
collection of eigenvalues, ${\hat f}_i$ (up to permutations) is the
same as the number of inequivalent $SU(N)$ Young tableaux with at most
$k$ columns, and so is given by $\binom{N+k-1}{N-1}$. For any given
choice of $\{\hat f_i\}$ we must now choose all allowed values of
$\te^{\i{\hat \phi}}$. Plugging \eqref{unsun} into \eqref{progenu}
uniquely determines the variable
$$\te^{\i (\kappa +Ns){\hat \phi}}=\te^{\i k'{\hat \phi}}\ .$$
Taking the $k'^{\rm th}$ root, we find $k'$ different values of
${\hat \phi}$ for every choice of $\{\hat f_i\}$.  Thus, naively, the
number of solutions up to permutation of the equations \eqref{progenu}
is $k'$ times the number of $SU(N)$ eigenvalue configurations:
\begin{equation} \label{naivecount} 
k' \times \binom{N+k-1}{N-1}\ .
\end{equation}
\eqref{naivecount} is, however, an overcount.  We know that the
solutions $\{ {\hat f}_n \}$ appear in $N$ dimensional orbits of the
centre symmetry group $\mbb{Z}_N$. As ${\hat \phi}$ in \eqref{unsun}
is determined from a condition on the final eigenvalues $\{f_i\}$, it
is clear that our procedure above will output each distinct collection
$\{f_i\}$ $N$ times\footnote{ The $\mbb{Z}_N$ phase cancels between
  the $SU(N)$ factor and the $U(1)$ phase in \eqref{unsol} and
  different values of $r$ all correspond to the same $U(N)$
  solution. Completely explicitly, if a set $\{{\hat f}_i\}$ and the
  choice $\te^{\i{\hat \phi}}=\alpha$ outputs the legal eigenvalue set
  $\{f_i\}$, then the set $\{\te^{{2 \pi i r}{N}}{\hat f}_i\}$ and the
  choice $\te^{\i\hat \phi}=\alpha \te^{-{2 \pi \i r}/{N}}$ also
  outputs $\{f_i\}$.}. Thus the true number of solutions is
\begin{equation}\label{unsolnum}
  k' \times  \frac{1}{N} \times \binom{N+k-1}{N-1}\ .
\end{equation}
Since \eqref{unsolnum} matches the number of integrable
representations of the $U(N)$ theory, and since each integrable
representation certainly yields an inequivalent solution (via
Verlinde's formula) of \eqref{progenu}, it follows that the distinct
solutions (up to permutations) of \eqref{progenu} are in one to one
correspondence with integrable representations of $U(N)$ via the
Verlinde formula.

\subsubsection{Explicit parametrization of integrable representations}\label{expir}

We will now flesh out the idea outlined in Section \ref{csu} in more
detail, and use this construction to obtain an explicit
parametrization of the integrable representations of the $U(N)_{k,k'}$
theory.

In the $SU(N)$ case, the $\mbb{Z}_N$ phase was responsible for
cyclically permuting the $SU(N)$ eigenvalues, resulting in a different
solution \eqref{ptldef} of the $SU(N)$ equations \eqref{progen}. It is
useful to study the effect of the $\mbb{Z}_N$ phase quantities that (redundantly)
parametrize the $U(N)$ solution.

Following the analysis of the previous
subsection, the $SU(N)$ eigenvalues can be parametrized as
\eqref{req0}:
\begin{equation}\label{usol}
 \hat{f}_i = \hat{f}_N \exp\left(-\frac{2\pi\i \hat{p}_i}{\kappa}\right)\ ,\quad \text{for}\quad i=1,\ldots,N\ ,\quad\text{and}\quad \hat{f}_N = \te^{2\pi\i r/N}\exp\left( \frac{2\pi\i}{N\kappa}\sum_{i=1}^{N-1} \hat{p}_i\right)\ .
\end{equation}
where $\hat{p}_N = 0$ by definition, the integer $r$ is in the range
$0 \leq r < N$ and the $N-1$ integers $\hat{p}_i$ are ordered as
\begin{equation}\label{hatpord}
  \kappa > \hat{p}_1 > \cdots > \hat{p}_{N-1} > 0\ ,
\end{equation}
We then determine the phase $\te^{\i\hat\phi}$ by plugging back the
expression \eqref{unsun} for $f_i$ in the equations
\eqref{progenu}. We get
\begin{equation}
  (-1)^{N-1} = \te^{\i \kappa \hat\phi} \te^{\i Ns \hat\phi} \exp\left(\frac{2\pi\i}{N}\sum_{i=1}^{N-1} \hat{p}_i\right) \te^{2\pi\i (\kappa + Ns)r/N}\ ,
\end{equation}
which, upon using $k' = \kappa + Ns$, gives
\begin{equation}\label{phival}
  \te^{\i\hat\phi} = \te^{\i\pi (N-1)/k'} \te^{-2\pi\i p_N/k'}  \te^{-2\pi\i r/N} \exp\left(-\frac{2\pi\i}{N k'}\sum_{i=1}^{N-1} \hat{p}_i\right)\ ,
\end{equation}
where $p_N$ is an integer whose range we are free to choose to be a
contiguous interval of length $k'$ i.e.~$-m \leq p_N < k'-m$ where $m$
is an arbitrary integer. We will often make the choice $m=0$, so $p_N$
is chosen to be non-negative.

Thus, a solution to the equations \eqref{progenu} constitutes of the
$SU(N)$ eigenvalues \eqref{usol} along with the $k'$ choices of the
phase $\te^{\i\hat\phi}$ where the $k'$ choices correspond to the
different values of the integer $p_N$.

We can recast the solution described above elegantly in terms of
variables that are more suited for $U(N)$ matrices. Let us define
\begin{equation}\label{phatdef}
  p_i = \hat{p}_i + p_N\ ,\quad\text{for}\quad i=1,\ldots,N\ ,
\end{equation}
where, recall that $\hat{p}_N =0$ by definition. Then, the solution is given by
\begin{align}\label{unsol}
  f_i &= \te^{\i\hat\phi} \hat{f}_i\ ,\quad\text{where}\ ,\nonumber\\
    \hat{f}_i &= \te^{2\pi\i r/N} \exp\left(\frac{2\pi\i}{N\kappa} \sum_{i=1}^N p_i\right) \exp\left(-\frac{2\pi\i p_i}{\kappa}\right)\ ,\nonumber\\
  \te^{\i\hat\phi} &= \te^{\i\pi(N-1)/k'} \te^{-2\pi\i r/N}  \exp\left(-\frac{2\pi\i}{Nk'} \sum_{i=1}^N p_i\right)\ .
\end{align}
Simplifying, we obtain
\begin{align}\label{unsolsimp}
  f_i &= 
        \exp\left(\i\pi\frac{N-1}{k'} + \frac{2\pi\i s}{k'\kappa} \sum_{i=1}^N p_i\right) \exp\left(-\frac{2\pi\i p_i}{\kappa}\right)\ .
\end{align}
From the ordering of the $\hat{p}_i$ in \eqref{hatpord} and the
definitions \eqref{phatdef}, we see that the $p_i$ satisfy
\begin{equation}\label{pord}
p_1 > p_2 > \cdots > p_{N-1} > p_N\ ,\quad\text{and}\quad \kappa + p_N > p_1\ .
\end{equation}
\eqref{unsolsimp} is in the Verlinde form \eqref{wonuo1}, except for
the fact that the $\mbb{Z}_N$ gauge symmetry described in Section
\eqref{csu} has not been taken into account, and thus there are
$N$-fold redundant descriptions of the same solution in \eqref{pord}.

\subsubsection{The $\mbb{Z}_N$ `gauge' symmetry}\label{gaugezn}
We now focus on the $\mbb{Z}_N$ phase $\te^{2\pi\i r/N}$ in
\eqref{unsol}. As noted above, this phase cancels out between
$\hat{f}_i$ and $\te^{\i\hat\phi}$ in \eqref{unsol}. This reflects the
fact the group $U(N)$ is defined as
\begin{equation}\label{undef}
  U(N) = \frac{SU(N) \times U(1)}{\mbb{Z}_N}\ ,
\end{equation}
where the $\mbb{Z}_N$ in the denominator acts simultaneously as the
centre of the $SU(N)$ and a $\mbb{Z}_N$ subgroup of $U(1)$. We see
this simultaneous action precisely in the solution \eqref{unsol} where
the two $\mbb{Z}_N$ phases in $\hat{f}_i$ and $\te^{\i\hat\phi}$
respectively are both controlled by the same integer $r$.

Next, instead of cancelling the opposite $\mbb{Z}_N$ phases between
$\hat{f}_i$ and $\te^{\i\hat\phi}$ in \eqref{unsol}, we can absorb
them into redefinitions of the eigenvalues. First, from the analysis
in the $SU(N)$ subsection (see equation \eqref{ptldef}), we know that
the $\mbb{Z}_N$ phase $\te^{2\pi\i r/N}$ can be compensated for by
reordering the eigenvalues such that they are shifted by $r$ units in
the counterclockwise direction. For instance, when $r=1$, we have
\begin{multline}
\te^{2\pi\i /N} \exp\left(\frac{2\pi\i}{N\kappa}\sum_{i=1}^{N-1} \hat{p}_i \right) \begin{pmatrix} \te^{-2\pi\i \hat{p}_1/\kappa} &  &  &  \\  & \te^{-2\pi\i \hat{p}_2/\kappa} &  &  \\  &  & \ddots &  \\  &  &  & 1 \end{pmatrix} \\ =  \exp\left(\frac{2\pi\i}{N\kappa}\sum_{i=1}^{N-1} \hat{p}'_i\right) 
\begin{pmatrix} \te^{-2\pi\i \hat{p}'_2/\kappa} &  &  &  &  \\  & \te^{-2\pi\i \hat{p}'_3/\kappa} &  &  &  \\  &  & \ddots &  &  \\  &  &  & 1 &  \\  &  &  &  & \te^{-2\pi\i \hat{p}'_1/\kappa} \end{pmatrix}\ ,
\end{multline}
where the $\hat{p}'_i$ are given by \eqref{ptldef} (with $r=1$):
\begin{equation}\label{ptldefn}
  \hat{p}'_1 = \kappa - \hat{p}_{N-1}\ ,\quad \hat{p}'_2 = \hat{p}_1 - \hat{p}_{N-1}\ ,\quad\ldots\ ,\quad \hat{p}'_{N-1} = \hat{p}_{N-2} - \hat{p}_{N-1}\ .
\end{equation}
Next, the phase $\te^{\i\hat\phi}$ \eqref{phival} (with $r=1$) can be
rewritten in terms of the $\hat{p}'_i$ as
\begin{align}
  \te^{\i\hat\phi} &= \te^{\i\pi (N-1)/k'} \te^{-2\pi\i p_N/k'}  \te^{-2\pi\i /N} \exp\left(-\frac{2\pi\i}{N k'}\sum_{i=1}^{N-1} \hat{p}'_i + \frac{2\pi\i\kappa}{Nk'} - \frac{2\pi\i \hat{p}_{N-1}}{k'}\right)\ ,\nonumber\\
  &=  \te^{\i\pi (N-1)/k'} \te^{-2\pi\i p_N/k'} \exp\left(-\frac{2\pi\i}{N k'}\sum_{i=1}^{N-1} \hat{p}'_i - \frac{2\pi\i s}{k'} - \frac{2\pi\i \hat{p}_{N-1}}{k'}\right)\ ,
\end{align}
where we have used $\kappa = k' + Ns$ in going to the second line. It
is easy to see that the above expression can again be recast into the
form of \eqref{phival} by defining
\begin{equation}\label{pndef}
  p'_N = p_N + \hat{p}_{N-1} + s = p_{N-1} + s\ ,
\end{equation}
where we have used the definition of $p_{N-1}$ from
\eqref{phatdef}. Defining $p'_i = \hat{p}'_i + p'_N$, the
transformation \eqref{ptldef} along with \eqref{pndef} can be written
succinctly as
\begin{align}\label{znact}
  p'_1 = \kappa + p_N + s \ ,\quad   p'_2 = p_{1} + s\ ,\quad p'_3 = p_2 + s\ ,\quad \ldots\ ,\quad p'_N = p_{N-1} + s \ .
\end{align}

Let us summarize. We first explained that the eigenvalues $\{f_i\}$
given in \eqref{unsun} - with $\{{\hat f}_i\}$ given in \eqref{usol}
and $\te^{\i {\hat \phi}}$ given in \eqref{phival} - can be recast in
the canonical $U(N)$ Verlinde form \eqref{unsolsimp} with $p_i$ given
in \eqref{phatdef}. We then pointed out that the same collection of
$\{f_i\}$ can equally well be described by multiplying each of the
${\hat f}_i$ by $\te^{\frac{2 \pi \i}{N} }$ and multiplying
$\te^{\i {\hat \phi}}$ by $\te^{-\frac{2 \pi \i}{N} }$. Then using the
fact that we can absorb the factor of $\te^{\frac{2 \pi \i}{N} }$ into
a cyclical relabelling of ${\hat f}_n$ we produced an alternate
representation of the same set of eigenvalues $\{ f_i\}$, also in the
form \eqref{unsol}, but with the numbers $p_m$ replaced by $p_m'$
listed in \eqref{znact}.

Thus, the generator of the simultaneous $\mbb{Z}_N$ in \eqref{undef}
acts on the integers $p_i$ as above. The entire $\mbb{Z}_N$ orbit
generated from a given configuration $p_i$ by the transformation
\eqref{znact} corresponds to a single $U(N)$ eigenvalue
configuration. This is responsible for the factor $1/N$ in the formula
for the number of solutions up to permutation in \eqref{unsolnum}.

\subsubsection{A conjectured fundamental domain for the $\mbb{Z}_N$ action}

In the previous subsubsection we have explained that sets of $\{p_i\}$
that are related by $\mbb{Z}_N$ moves (\eqref{znact} and iterations)
all describe the same eigenvalues. In order to uniquely parametrize
distinct eigenvalues, it is convenient to `gauge fix' this $\mbb{Z}_N$
symmetry.

In this subsubsection we present a conjecture for a `fundamental
domain' for the integers $p_i$ that results from such a gauge
fixing. We conjecture that in the case that $s \leq 0$, the set of
distinct legal $U(N)_{k,k'}$ eigenvalues are in one-to-one
correspondence with elements in the fundamental domain that we now
describe.

Recall that $p_i$ satisfy the ordering \eqref{pord}:
\begin{equation}\label{pordn}
\kappa + p_N >  p_1 > p_2 > \ldots > p_{N-1} > p_N\ .
\end{equation}
Since the $p_i$ appear in $\kappa^{\rm th}$ roots of unity, we
can further restrict them to the range
\begin{equation}\label{prestr}
  \kappa > p_1 > p_2 > \cdots > p_N \geq 0\ .
\end{equation}
Our conjectured fundamental domain is given by the quantities
$\{ p_i\}$ that, in addition, obey the inequalities
\begin{equation}\label{fundomainalt}
  N-j \leq p_j < \kappa + js\ ,\quad\text{for}\quad j=1,\ldots,N\ ,
\end{equation}
or equivalently, using $\kappa = k' - Ns$,
\begin{equation}\label{fundomain}
  N-j \leq p_j < k' - (N-j)s\ ,\quad\text{for}\quad j=1,\ldots,N\ .
\end{equation}
That is,
\begin{align}\label{fundomainexpl}
  N-1 &\leq p_1 < k' - (N-1)s\ ,\nonumber\\ N-2 &\leq p_2 < k' - (N-2)s\ ,\nonumber\\ &\phantom{\leq}\ \ \vdots\nonumber\\ 1 &\leq p_{N-1} < k' -s \ ,\nonumber\\ 0 &\leq p_N < k'\ .
\end{align}

The bound on $p_N$ is the same as the choice $m=0$ (see under \eqref{phival}). Recall $p_N$ was defined only modulo
$k'$ and we have effectively chosen the interval $[0,k')$ for its
range. The ordering \eqref{pordn} then gives the lower bounds on
$p_1,\ldots,p_{N-1}$. 

While we do not have a computation that deduces the upper bounds in
\eqref{fundomainexpl}, it is easy to see that our conjecture at least
obeys the following consistency check. If the domain
\eqref{fundomainexpl} is indeed correct then it must follow that if we
start with any collection of $\{p_i\}$ that obey
\eqref{fundomainexpl}, then the action of any nontrivial $\mbb{Z}_N$
element takes us out of this domain (i.e. maps us to a collection of
$\{p_i\}$ that does not obey \eqref{fundomainexpl}).

To see this is the case, note that the action of the generating
$\mbb{Z}_N$ action \eqref{znact} gives $p'_1 = \kappa + s + p_N$. As
$p_N>0$ it follows immediately that
$p'_1 \geq \kappa + s= k' - (N-1)s$, and so $p_1'$ violates the first
of \eqref{fundomainexpl}. Similarly, if we act again with the
$\mbb{Z}_N$ generator \eqref{znact} we find
$p''_2 = \kappa + 2s + p_N=k'-(N-2)s$ which violates the second of
\eqref{fundomainexpl}. In a similar way it follows that the
configuration obtained by acting with the $\mbb{Z}_N$ generator $m$
times violates the bound on $p_m$ in \eqref{fundomainexpl}.

Another bit of evidence for the correctness of the fundamental domain
above comes from the observation that it correctly reproduces the
correct fundamental domains for the Type I and Type II theories (in
Appendices \ref{sucount}, \ref{to}, \ref{tt} we determine these
fundamental domains using the so-called `quantum identities' on
characters). In the $s = 0$ (Type I) and $s = -1$ (Type II) our
conjectured fundamental domains reduce to
\begin{enumerate}
\item \textbf{Type I}: For $s = 0$, we have $k' = \kappa$ and
  $N - i \leq p_i < \kappa$ which corresponds to the following bounds
  on the Young tableau rows $\ell_i = p_i - (N-i)$:
\begin{equation}
  0 \leq \ell_i < k + i\ .
\end{equation}
When we take into account the ordering
$\ell_1 \geq \ell_2 \geq \cdots \geq \ell_N$, the above inequalities
are automatically enforced by this ordering plus the following bounds:
\begin{equation}\label{fundomainI}
  0 \leq \ell_1 \leq k\ ,\quad 0 \leq \ell_N\ .
\end{equation}
The elements in the fundamental domain corresponds to the Young
tableaux that fit in a $k \times N$ box which is given by
\begin{equation}
  \binom{k+N}{N}\ .
\end{equation}
Note that this is the same as the number of integrable representations
of the $U(N)_{k,\kappa}$ WZW model.

\item \textbf{Type II}: For the type II theory, i.e.~$s = -1$ and
  $k' = k$, we have $ N - i \leq p_i < k + N-i$ which gives which
  gives $0 \leq \ell_i < k$ i.e.~$0 \leq \ell_i \leq k-1$. Thus, the
  elements of the fundamental domain are all possible Young tableaux
  that fit in a $(k-1) \times N$ box which is given by
  \begin{equation}
    \binom{k+N-1}{N}\ .
  \end{equation}
  Again, note that the above answer matches the number of integrable
  representations of the $U(N)_{k,k}$ WZW model.
\end{enumerate}
The final piece of evidence we present for the correctness of the
fundamental domain \eqref{fundomain} comes from counting the distinct
$\{p_i\}$ that obey \eqref{fundomainexpl} and demonstrating that the
result of this counting problem agrees with the independently known
number of integrable representations of $U(N)_{k, k'}$ listed in
\eqref{actest}. We now turn to this problem.

As for the Type I and Type II cases, it is easier to perform the
counting in terms of Young tableau rows $\ell_i = p_i - N + i$ rather
than the $p_i$ themselves. From \eqref{fundomain}, it is easy to see
that $\ell_i$ satisfy $0 \leq \ell_i < k' - (N-i)(s+1)$. Using
$k' = \kappa + Ns = k + N(s+1)$, we can write the bounds on $\ell_i$
as
\begin{equation}\label{ellfund}
  0 \leq \ell_i < k + i(s+1)\ .
\end{equation}
Since $s = 0$ corresponds to the Type I case and has been dealt with
previously, let us restrict the range of $s$ to $s \leq -1$ and define
$t = -(s+1)$. The integer $t$ satisfies $t \geq 0$ for $s \leq -1$. In
terms of $t$, \eqref{ellfund} becomes
\begin{equation}\label{ellfundt}
  0 \leq \ell_i < k - it\ .
\end{equation}
Recall that in our physical problem $k' = k - Nt > 0$. In physically sensible cases, consequently, the upper bound on $\ell_N$ is positive. In the mathematics below, however, we will also encounter unphysical values of 
$k' \leq Nt$. In such situations there are no values of $\ell_N$ that obey the inequalities 
\eqref{ellfundt}, and so the number of Young tableaux obeying these conditions vanishes. 
Finally, there is only one tableau (the empty one) at $N=0$.
 
Let $Q(k, N)$ denote the number of Young tableaux with (at most) $N$
rows, obeying the inequalities \eqref{ellfundt}. Using the fact that
every tableau has either an empty $N^{\rm th}$ row (in which case,
effectively, $N\rightarrow N-1$) or a full first column (in which case
the rest of the tableaux obeys the inequalities \eqref{ellfundt} with
$k \rightarrow k-1$) it follows immediately that
\begin{equation}\label{recrel}
  Q(k, N)=Q(k,N-1)+ Q(k-1, N)\ .
\end{equation}
From the discussion of the previous paragraph it also follows that
\begin{equation}\label{dipp}
  Q(k, 0)=1\ , \quad\text{for}\quad k>0\ ,\qquad Q(Nt, N)= 0\ ,\quad\text{for}\quad N>0\ .
\end{equation}
The recursion relations \eqref{recrel}, together with the boundary
conditions \eqref{dipp}, completely determine the solution $Q(k,
N)$. To see how this works note that it follows from \eqref{recrel}
and \eqref{dipp} that
\begin{equation}
  Q(t+1, 1)=Q(t, 1)+Q(t+1,0)=0+1=1\ .
\end{equation}
Using \eqref{recrel} again we find 
\begin{equation}
  Q(t+2, 1)=Q(t+1,1)+Q(t+2,0)=1+1=2\ .
\end{equation}
Continuing this way we conclude that 
\begin{equation}
  Q(t+m, 1)=m\ ,\quad \text{or}\quad Q(k, 1)=k-t\ .
\end{equation}
Clearly this answer that is output by the recursion relations is the
correct answer for the number of tableaux with one row and less than
$k-t$ boxes, see \eqref{ellfundt}.

Using the value of $Q(k,1)$ obtained above, we may now find $Q(k, 2)$
as follows. We first note that from \eqref{recrel}
\begin{equation}
  Q(2t+1, 2)=Q(2t+1, 1)+Q(2t, 2)=(t+1)+0=t+1\ ,
\end{equation}
so that
\begin{equation}
  Q(2t+2,2)=Q(2t+2,1) + Q(2t+1,2)= \left(t+2 \right) + \left( t+1 \right) =2t+3\ .
\end{equation}
Continuing, we get
\begin{equation}
  Q(2t+m, 2)=\sum_{j=1}^m (t+m)=\frac{m(2t+m+1)}{2}\ ,\quad\text{or}\quad Q(k, 2)=\frac{(k-2t)(k+1)}{2}\ .
\end{equation}
In a similar way we can use the known values of $Q(k, 2)$ to determine
$Q(k, 3)$ and so on. It is thus clear that the recursion relation
\eqref{recrel}, together with the boundary conditions \eqref{dipp},
completely determine $Q(k, N)$ for every value of $k$ and $N$
s.t. $k-Nt>0$.

Now consider the function
\begin{equation}\label{Qguess}
  Q(k, N)  = \frac{k-Nt}{k+N} \binom{k+N}{N}\ .
\end{equation}
It is obvious that \eqref{Qguess} obeys the boundary conditions
\eqref{dipp}. Moreover, it is easy to verify that \eqref{Qguess} also
obeys the recursion relation \eqref{recrel}. To see this we plug
\eqref{Qguess} into the right hand side of \eqref{recrel}:
\begin{align}
  \text{RHS of \eqref{recrel}}
  &= \frac{k-(N-1)t}{k+N-1}\binom{k+N-1}{N-1} + \frac{k-1-Nt}{k+N-1}\binom{k+N-1}{N}\ ,\nonumber\\
  &= \frac{k-1-Nt}{k+N-1}\left[\binom{k+N-1}{N-1} + \binom{k+N-1}{N}\right] + \frac{t+1}{k+N-1}\binom{k+N-1}{N-1}\ ,\nonumber\\
  &= \frac{k-1-Nt}{k+N-1}\binom{k+N}{N} + \frac{t+1}{k+N-1}\frac{N}{k+N}\binom{k+N}{N}\ ,\nonumber\\
  &= \frac{(k-1-Nt)(k+N) + N(t+1)}{(k+N-1)(k+N)} \binom{k+N}{N}\ ,\nonumber\\
  &= \frac{k-Nt}{k+N} \binom{k+N}{N}\ ,
\end{align}
which is the left-hand side of \eqref{recrel} according to the ansatz
\eqref{Qguess}.

It follows that the number of elements in the fundamental domain
\eqref{fundomain} is given by \eqref{Qguess}. However \eqref{Qguess}
independently equals \eqref{actest}, the number of integrable
representations of $U(N)_{k,k-Nt}$. We view this match as rather
nontrivial evidence in favour of our fundamental domain
\eqref{fundomainexpl}. We leave a full deductive proof of
\eqref{fundomainexpl}, together with its extension to positive values
of $s$, to future work.

\subsubsection{A $\mbb{Z}_{k'}$ global symmetry}
Recall the general form of the solution \eqref{unsol}:
\begin{align}\label{unsoln}
  f_i &= \te^{\i\hat\phi} \hat{f}_i\ ,\quad\text{where}\ ,\nonumber\\
    \hat{f}_i &=  \exp\left(\frac{2\pi\i}{N\kappa} \sum_{i=1}^N p_i\right) \exp\left(-\frac{2\pi\i p_i}{\kappa}\right)\ ,\nonumber\\
  \te^{\i\hat\phi} &= \te^{\i\pi(N-1)/k'} \exp\left(-\frac{2\pi\i}{Nk'} \sum_{i=1}^N p_i\right)\ ,
\end{align}
where we have gotten rid of the $\mbb{Z}_N$ phase in \eqref{unsol}
since it is a gauge symmetry as discussed in Section \ref{gaugezn}.

Consider the transformation
\begin{equation}\label{zkptr}
  f_i \to f'_i = \te^{2\pi\i/k'} f_i\ ,\quad\text{for}\quad i = 1,\ldots,N\ .
\end{equation}
From the formulas in \eqref{unsoln}, it is easy to see that $f'_i$ can
also be written in the same form as \eqref{unsoln} with the
corresponding integers $p'_i$ given by
\begin{equation}\label{pzkptr}
  p'_i = p_i + 1\ ,\quad\text{for}\quad i=1,\ldots,N\ .
\end{equation}
\footnote{To wit, the $SU(N)$ part $\hat{f}'_i$ remains unchanged and is equal
to $\hat{f}_i$ whereas the phase $\te^{\i\hat\phi'}$ becomes
\begin{equation}
 \te^{\i\hat\phi'} = \te^{\i\pi(N-1)/k'} \exp\left(-\frac{2\pi\i}{Nk'} \sum_{i=1}^N p'_i\right) =  \te^{\i\pi(N-1)/k'} \exp\left(-\frac{2\pi\i}{Nk'} \sum_{i=1}^N p'_i\right) = \te^{2\pi\i/k'} \te^{\i\hat\phi}\ .\nonumber
\end{equation}}
If the $\mbb{Z}_{k'}$ generated by \eqref{zkptr} (or equivalently,
\eqref{pzkptr}) takes the $p_i$ out of the fundamental domain
\eqref{fundomain}, we may have to shift the $p_i$ by appropriate
multiples of $\kappa$ and then perform a compensating $\mbb{Z}_N$
transformation to bring it back inside the fundamental domain (see
Appendices \ref{to} and \ref{tt} for a plethora of such manipulations
in the Type I and Type II cases).

This `global' $\mbb{Z}_{k'}$ symmetry is the analog of the global
centre $\mbb{Z}_N$ symmetry for $SU(N)_k$ Chern-Simons theories. Under
Type I-Type I level-rank duality, $k' = \kappa$ so that the two global
symmetries match. Similarly, under $SU(N)_k$-Type II $U(k)_{-N,-N}$
duality, the global symmetry is again $\mbb{Z}_N$ on both sides.

\section{$SU(N)_k$ quantum identities} \label{sucount}

Integrable representations of $SU(N)_k$ are those whose Young tableaux
have no more than $k$ columns. Recall the set of solutions modulo
permutation $\mc{P}_k$ of the equations \eqref{ngtmnew} on eigenvalue
configurations $\ul{w} = \{w_1,\ldots, w_N\}$:
\begin{equation}\label{ngtmnewapp}
  (w_i)^\kappa = (w_j)^\kappa\ ,\quad\text{for all}\quad i,j = 1,\ldots,N\ ;\quad w_i \neq w_j\quad\text{for}\quad i \neq j\ ;\quad \prod_{i=1}^N w_i = 1\ ,
\end{equation}
In this appendix, we show that characters evaluated on the eigenvalue
configurations in $\mc{P}_k$ either vanish or reduce to the character
of a related integrable representation (up to a possible minus sign).
 
In other words in this section will find a class of relations between
characters that are evaluated on solutions (up to permutation) of the
equations \eqref{ngtmnewapp}. These equations take the schematic form
\begin{equation}\label{qidef}
  \chi_{\vgap R}(\ul{w}) = \pm \chi_{\vgap R'}(\ul{w})\ ,\quad\text{or}\quad \chi_{\vgap R}(\ul{w}) = 0\ ,\quad\text{for every}\quad \ul{w} \in \mc{P}_{k}\ ,
\end{equation} 
where $R$ and $R'$ are different representations. To reiterate, the
above relations are valid only when the characters are evaluated on
eigenvalue configurations in $\mc{P}_k$ and not generally on arbitrary
eigenvalue configurations. However, we must emphasize that they are
satisfied for \textbf{any} eigenvalue configuration in
$\mc{P}_k$. Throughout this paper we refer to characters evaluated on
solutions of \eqref{ngtmnewapp} as \emph{quantum characters} and the
identities of the form \eqref{qidef} as \emph{quantum
  identities}.

In order to derive quantum identities we make crucial use of the
following facts.
\begin{enumerate} 
\item The eigenvalues $w_i$ satisfy $\prod_{i=1}^N w_i = 1$,

\item The eigenvalues $w_i$ in \eqref{ngtmnew} are of the form
  $w_i= \te^{\i \alpha} z_i$ where $z_i$ are $\kappa^{\rm th}$ roots
  of unity with $\alpha$ is independent of $i$. Since
  $w_1\cdots w_N = 1$, the phase $\te^{\i\alpha}$ is constrained as
  \begin{equation}\label{alphaval}
    \te^{\i N\alpha} = \frac{1}{\prod_{i=1}^N z_i}\ .
  \end{equation}

\item The Weyl character formula for an $SU(N)$ representation with
  Young tableau row-lengths
  $\{\ell_i\} = \{\ell_1,\ldots,\ell_{N-1}\}$ is
\begin{equation}\label{wcpca}
  \chi_{\vgap \{\ell\}}(\ul{w}) = \frac{\left[w^{n_1}, w^{n_2}, \ldots, w^{n_{N-1}}, 1 \right]}{\left[w^{N-1}, w^{N-2}, \ldots, 1 \right]} = \frac{D_{\{\ell\}}}{D_{\{0\}}}\ , 
\end{equation}
where we have defined $D_{\{\ell\}}$ to be the determinant
\begin{equation}\label{detdef}
  D_{\{\ell\}} = [w^{n_1},\ldots, w^{n_{N-1}}, 1]\ ,
\end{equation}
with $n_i = \ell_i+N-i$.    
\end{enumerate}
Note crucially that since $(w_i)^\kappa = \te^{\i\kappa\alpha}$ for
every $i$, whenever $n_i$, for any $i$, is a multiple of $\kappa$, the
above determinant \eqref{detdef} vanishes since the $i^{\rm th}$ row
is proportional to the last row. Similarly, if the difference of $n_i$
and $n_j$ is a multiple of $\kappa$ for any $i \neq j$, the
determinant again vanishes. This will be the method we will use to
prove identities of the form of the second type listed in
\eqref{qidef}. On the other hand we will prove identities of the first
type listed in \eqref{qidef} using observation 2 above together with
row rearrangement identities for determinants.

To see how all this works, we begin our discussion of quantum
identities for characters of representations with a single non-zero
row in their Young tableaux, and then build up to the general case.

\subsection{Representations whose Young tableaux have a single
  row} \label{sr}
 
To start with, let us focus on the representation with $m$ boxes in
the first row and no boxes in any other row i.e.~the character
$\chi_{\vgap \{m, \vec{0}\}}$. For this representation, the $n_i$ become
\begin{equation}\label{nor} 
n_1=m+N-1\ ,\quad n_2=N-2\ ,\quad n_3=N-3\ ,\quad \ldots\ ,\quad n_{N-1}=1\ ,\quad n_N=0\ .
\end{equation} 
Suppose $n_1 = p\kappa + n$ where $0\leq n < \kappa$ (this corresponds
to $m = p\kappa + n - N + 1$). Let us focus on the numerator
determinant in the Weyl character formula \eqref{wcpca}. We have
\begin{equation}\label{generalonerow}
  D_{\{m,\vec{0}\}} = [w^{p\kappa + n},w^{N-2},\ldots,w,1] = \te^{\i p\kappa\alpha} [w^n,w^{N-2},\ldots,w,1]\ .
\end{equation}
The phase $\te^{\i p\kappa \alpha}$ is $1$ when $p$ is a multiple on
$N$; this follows from the fact \eqref{alphaval} which says that
$\te^{\i N\alpha}$ is a $\kappa^{\rm th}$ root of unity whence
$\te^{\i N \kappa \alpha} = 1$. Thus, we can restrict our attention to the range
\begin{equation}
  0 \leq p \leq N-1\ .
\end{equation}
Now, if $n$ is one of $N-2,N-3,\ldots,1,0$ then the determinant is
zero since two rows become identical. Thus, we restrict our attention
to the cases
\begin{equation}
  N-1 \leq n < \kappa\ .
\end{equation}
We thus have the following ordering:
\begin{equation}
  0 \leq p \leq N-1 \leq n < \kappa \ .
\end{equation}
We first treat the borderline case $p = N-1$. In this case, we can
absorb the phase $\te^{\i(N-1)\kappa\alpha}$ into the last $N-1$ rows
of the determinant:
\begin{equation}
  \te^{\i p\kappa\alpha} [w^n,w^{N-2},\ldots,w,1] = [w^n,w^{\kappa + N-2},\ldots,w^{\kappa + 1},w^\kappa]\ ,
\end{equation}
further take out the factor $w_1^n \cdots w_N^n = 1$ and reshuffle the rows to get
\begin{equation}
  [w^n,w^{\kappa + N-2},\ldots,w^{\kappa + 1},w^\kappa] = (-1)^{N-1} [w^{\kappa + N-2 - n},\ldots,w^{\kappa + 1-n},w^{\kappa-n}, 1]\ .
\end{equation}
This corresponds to a Young tableau with rows $\kappa - n -1$,
$\kappa - n-1$,\ldots, $\kappa-n-1$. Thus, we have
\begin{equation}\label{redu1}
  \chi_{\vgap \{m,\vec{0}\}}(\ul{w}) = (-1)^{N-1} \chi_{\vgap \{\kappa - n -1,\ldots,\kappa-n-1\}}(\ul{w}) \ .
\end{equation}
Since $N-1 \leq n$, the largest row-length that is possible above is
$\kappa - (N-1)-1 = k$.

We next look at the cases $p < N-1$. As above, we absorb the phase
$\te^{\i p\kappa \alpha}$ into the last $p$ rows of the
determinant by using the property
$(w_i)^\kappa = \te^{\i\kappa\alpha}$:
\begin{equation}
  \te^{\i p\kappa\alpha} [w^n,w^{N-2},\ldots,w,1] = [w^n,w^{N-2},\ldots,w^{p}, w^{\kappa + p - 1}, \ldots,w^{\kappa + 1},w^\kappa]\ ,
\end{equation}
and take out the factor $w_1^p w_2^p \cdots w_N^p = 1$ from the
determinant to get
\begin{multline}
  [w^n,w^{N-2},\ldots,w^{p}, w^{\kappa + p - 1}, \ldots,w^{\kappa + 1},w^\kappa] \\ = [w^{n-p}, w^{N-2-p},\ldots, 1, w^{\kappa-1},\ldots, w^{\kappa -(p-1)}, w^{\kappa - p}]\ .
\end{multline}
Since $N-1 \leq n < \kappa$ and $p < N-1$, we have
$0 < n - p < \kappa - p$. Thus, we can move the last $p$ rows to the
top to get
\begin{multline}
[w^{n-p}, w^{N-2-p},\ldots, 1, w^{\kappa-1},\ldots, w^{\kappa -(p-1)}, w^{\kappa - p}] \\ = (-1)^{p(N-p)} [w^{\kappa-1},\ldots, w^{\kappa -(p-1)}, w^{\kappa - p}, w^{n-p}, w^{N-2-p},\ldots, 1]\ .
\end{multline}
The Young tableau that corresponds to the determinant on the right
hand side is obtained by subtracting $N-i$ from the $i^{\rm th}$
exponent:
\begin{equation}
  \ell'_1 = \cdots = \ell'_{p} = k\ ,\quad \ell'_{p+1} = n - N + 1\ ,\quad\ell'_{p+2} = \cdots = \ell'_{N-1} = 0\ .
\end{equation}
That is, the new Young tableau has $k$ boxes in the first $p$ rows,
$n - (N-1)$ boxes in the $(p+1)^{\rm th}$ row and zero in the rest. The characters are then related as
\begin{equation}\label{redu2}
  \chi_{\vgap \{m,\vec{0}\}}(\ul{w}) = (-1)^{p(N-p)} \chi_{\vgap \{\underbrace{\scriptstyle k,\ldots,k}_{\text{$p$ times}}, n-N+1,\vec{0}\}}(\ul{w}) \ .
\end{equation}
Thus, the identities \eqref{redu1} and \eqref{redu2} tell us that the
quantum character of any representation can be reduced to that of a
representation whose $n_i$ satisfy
\begin{equation}
  \kappa > n_1 > n_2 > \cdots > n_{N-1} > 0 \ ,
\end{equation}
or equivalently, whose row-lengths satisfy
\begin{equation}
  k \geq \ell_1 \geq \cdots \geq \ell_{N-1} \geq 0\ .
\end{equation}
These representations are precisely the integrable representations of
the $SU(N)_k$ WZW model. In the next subsection, we will show a
similar identity for general representations i.e.~representations with
more than one row with non-zero length.

\subsection{General representations}

Consider a general representation with Young tableau $\{\ell_i\}$ and
the corresponding $n_i = \ell_i + N - i$. Suppose
$n_i = p_i\kappa + \nu_i$ where $0 \leq \nu_i < \kappa$. The case
$\nu_i = 0$ is trivial since this implies that the corresponding $n_i$
is a multiple of $\kappa$ which in turn implies that the quantum
character vanishes (see discussion after equation \eqref{detdef}). So,
we restrict the integers $\nu_i$ to satisfy $0 < \nu_i < \kappa$.

The integers $\nu_i$ resulting from shifting $n_i$ by multiples of
$\kappa$ need not be decreasing in general though the $n_i$ themselves
were strictly decreasing. However, there is a non-trivial permutation
$\sigma$ which rearranges the $\nu_i$ so that they are indeed strictly
decreasing. Let us call these integers $n'_i$ which indeed satisfy
\begin{equation}
  \kappa > n'_1 > n'_2 > \cdots > n'_{N-1} > 0\ ,
\end{equation}
and the corresponding Young tableau row-lengths
$\ell'_i = n'_i - (N-i)$. Thus, we have
\begin{equation}\label{prequan1}
  \chi_{\vgap \{\ell_i\}}(\ul{w}) = \te^{\i P\kappa\alpha}\, \sgn(\sigma)\,  \chi_{\vgap \{\ell'_i\}}(\ul{w})\ .
\end{equation}
where $P = \sum_i p_i$. This is the analog of the integer $p$ that we
encountered in our analysis of single row Young tableaux in the
previous subsection. As in that subsection, we can reabsorb the phase
$\te^{\i P\kappa\alpha}$ into the determinant of the character
formula. We proceed by studying different ranges of $P$ separately.

First, we look at the case when $P$ is a multiple of $N$. Since
$\te^{\i N\kappa\alpha} = 1$ (see the discussion after
\eqref{generalonerow}). This gives
\begin{equation}\label{PmultN}
  \chi_{\vgap \{\ell_i\}}(\ul{w}) = \sgn(\sigma)\,  \chi_{\vgap \{\ell'_i\}}(\ul{w})\ ,\quad\text{when $P$ is a multiple of $N$}.
\end{equation}
Thus, we can restrict our analysis to the range
\begin{equation}
  0 < P < N\ .
\end{equation}
The numerator determinant $D_{\{\ell'\}}$ for the integrable
representation $\{\ell'_i\}$ is
\begin{equation}
  D_{\{\ell'_i\}} = [w^{n'_1},\ldots,w^{n'_{N-1}},1]\ .
\end{equation}
We perform the following manipulations on
$\te^{\i P\kappa\alpha} D_{\{\ell'_i\}}$:
\begin{align}\label{detman1}
  &\te^{\i P\kappa \alpha}  D_{\{\ell'_i\}} \nonumber\\
  &= [w^{n'_1},\ldots,w^{n'_{N-P}},w^{\kappa + n'_{N-P+1}},\ldots,w^{\kappa+n'_{N-1}},w^\kappa]\ ,\nonumber\\
  &= [w^{n'_1-n'_{N-P}},\ldots,w^{n'_{N-P-1}-n'_{N-P}},1,w^{\kappa + n'_{N-P+1} - n'_{N-P}},\ldots,w^{\kappa-n'_{N-P}}]\ ,\nonumber\\
  &= (-1)^{P(N-P)} [w^{\kappa + n'_{N-P+1} -n'_{N-P}},\ldots,w^{\kappa-n'_{N-P}},w^{n'_1-n'_{N-P}},\ldots,w^{n'_{N-P-1}-n'_{N-P}},1]\ .
\end{align}
In the second step, using $(w_i)^\kappa = \te^{\i \kappa\alpha}$, we
have absorbed $\kappa$ into the last $P$ rows of the $N$ rows in the
determinant (this is procedure is valid since we have $P < N$). In
going to the third line we have taken out a factor of
$1 = w_1^{n'_{N-P}} w_2^{n'_{N-P}}\cdots w_N^{n'_{N-P}}$, and in going
to the fourth line, we have performed a cyclic permutation which
shifts the rows by $P$ units in the determinant.

Let us look at the row-lengths $\{\ell''_i\}$ of the Young tableau
that we get from the exponents $w$ in the determinant in the last line
of \eqref{detman1}:
\begin{align}
  \ell''_1 &= k + \ell'_{N-P+1} - \ell'_{N-P}\ ,\nonumber\\
  \ell''_2 &= k + \ell'_{N-P+2} - \ell'_{N-P}\ ,\nonumber\\
           &\ \,\vdots\nonumber\\
  \ell''_P &= k - \ell'_{N-P}\ ,\nonumber\\
  \ell''_{P+1} &= \ell'_1 - \ell'_{N-P}\ ,\nonumber\\
           &\ \,\vdots\nonumber\\
  \ell''_{N-1} &= \ell'_{N-P-1} - \ell'_{N-P}\ .
\end{align}
We recognize the last permutation as the $\mbb{Z}_N$ action
\eqref{znmove} discussed in Appendix \ref{monopole} repeated $P$
times.

\section{Type I $U(N)$ quantum identities} \label{to}

The distinguished eigenvalue configurations
$\ul{w} = \{w_1,\ldots,w_N\}$ that arise from the Type I $U(N)$
Chern-Simons path integral satisfy equation \eqref{obey} (with $s=0$):
\begin{equation}\label{obeyIapp}
  (w_i)^\kappa = (-1)^{N-1}\ ,\quad\text{for}\quad i=1,\ldots,N\ ,\quad\text{with $w_i$ distinct.}
\end{equation}
Recall that the set of solutions of the above equations modulo
permutations is called $\mc{P}_{k,\kappa}$.

As in the $SU(N)_k$ case, the characters of representations of $U(N)$
evaluated on the above eigenvalue configurations will be referred to
as \emph{quantum characters} of the Type I theory. In this appendix,
we will establish various identities between quantum characters of
different $U(N)$ representations using the Weyl character formula for
$U(N)$ representations. Let the rows of the Young tableau
corresponding to a $U(N)$ representation be $\{\ell_1,\ldots,\ell_N\}$
and define the integers $n_i = \ell_i + N - i$. Then the character of
this representation evaluated on an eigenvalue configuration
$\ul{w} = \{w_1,\ldots,w_N\}$ is given by
\begin{equation}\label{UNwcf}
  \chi_{\vgap \{\ell\}}(\ul{w}) = \frac{\left[w^{n_1}, w^{n_2}, \ldots, w^{n_{N-1}}, w^{n_N} \right]}{\left[w^{N-1}, w^{N-2}, \ldots, 1 \right]}\ .
\end{equation}
We also recall that the $n_i = \ell_i + N - i$ satisfy the ordering
\begin{equation}
  n_1 > n_2 > \cdots > n_N\ ,
\end{equation}
since the $\ell_i$ satisfy
$\ell_1 \geq \ell_2 \geq \cdots \geq \ell_N$.

\subsection{Periodicity}

Since all eigenvalues in Type I theory are $\kappa^{\rm th}$ roots of
$(-1)^{N-1}$, it follows immediately that the operation
\begin{equation}\label{trivident}
  n_i \rightarrow n_i + p_i  \kappa \ ,\quad\text{with}\quad  p_i \in \mbb{Z}\ ,\quad i=1,\ldots,N\ ,
\end{equation} 
simply multiplies the original character by the phase $(-1)^{P(N-1)}$
where $P = \sum_i p_i$. Of course, if we start with a legal Young
tableau i.e.~a Young tableau which satisfies $n_1 > n_2 \cdots > n_N$,
the operation \eqref{trivident} also yields a legal Young tableau only
when
\begin{equation}\label{inequal}
  p_1 \geq p_2 \geq \cdots \geq p_N  \ . 
\end{equation}

\subsection{Conditions for vanishing of characters}

It follows from the Weyl character formula \eqref{UNwcf} that the
character evaluated on an eigenvalue configuration in
$\mc{P}_{k,\kappa}$ vanishes whenever there exist two rows $i$ and $j$
such that
\begin{equation}\label{condforvan}
  n_i = n_j + Q \kappa\ , 
\end{equation} 
where $Q$ is an integer. When this happens, the determinant in the
numerator of \eqref{UNwcf} vanishes because the $i^{\rm th}$
and $j^{\rm th}$ rows of the numerator become proportional to each
other.

\subsection{Young tableaux with a single row}

Consider representations $R = \{m,\vec{0}\}$ i.e.~Young tableaux with
$\ell_1=m$ and $\ell_i=0$ for $i>1$. The corresponding $n_i$ are given by
\begin{equation}
  n_1 = m+N-1\ ,\quad n_2 = N-2\ ,\quad \ldots\ ,\quad n_{N-1} = 1\ ,\quad n_N = 0\ .
\end{equation}
From \eqref{condforvan}, we see that the character vanishes for
$m=k+1, k+2 \ldots, \kappa-1$ since then
$n_1 = \kappa + n_N, \kappa + n_{N-1}, \ldots, \kappa + n_2$
respectively. This result, together with
\begin{equation}\label{qkap}
  \chi_{\vgap \{m+Q\kappa, \vec{0}\}}(\ul{w}) = (-1)^{Q(N-1)} \chi_{\vgap \{m, \vec{0}\}}(\ul{w})\ ,
\end{equation} 
determines the quantum character of all representations with a general
number of boxes in only the first row of the Young tableau in terms of
representations with up to $k$ boxes in the first row.

\subsection{Adding a `monopole'} 

A slightly less trivial operation is the following. Consider a
representation in which $\ell_1 -\ell_N \leq k$. Then define
\begin{align} \label{trividentm}
&\ell_1' =\ell_N+1+k \ ,\nonumber \\
&\ell_2'=\ell_1+1 \ ,\nonumber \\
&\vdots\nonumber \\
&\ell_N'=\ell_{N-1}+1 \ .
\end{align}
We will now demonstrate that
$\chi_{\vgap \{\ell'\}}(\ul{w}) = \chi_{\vgap \{\ell\}}(\ul{w})$. To
see this note that $\chi_{\vgap \{\ell'\}}$ is given by
\begin{align} \label{charkmanip} 
  \chi_{\vgap \{\ell'\}}(\ul{w}) &= \frac{\left[w^{\ell_1'+N-1}, w^{\ell_2'+N-2}, \ldots,  w^{\ell_{N}'} \right]}{\left[w^{N-1}, w^{N-2}, \ldots,  1 \right]} \nonumber \\
  &= \frac{\left[w^{\kappa+ \ell_N}, w^{N-1+\ell_1}, \ldots,  w^{1+\ell_{N-1}} \right]}{\left[w^{N-1}, w^{N-2}, \ldots,  1 \right]}  \nonumber \\
  &=(-1)^{N-1} \times (-1)^{N-1}~\frac{\left[w^{N-1+\ell_1}, w^{N-2+\ell_2}, \ldots, w^{\ell_{N}} \right]}{\left[w^{N-1}, w^{N-2}, \ldots, 1 \right]} \nonumber \\
  &=\chi_{\vgap \{\ell\}}(\ul{w}) \ .
\end{align}
In the second last line, the first factor of $(-1)^{N-1}$ comes from
$\omega_i^\kappa=(-1)^{N-1}$ and the second factor of $(-1)^{N-1}$
from moving the first row of the determinant to the end.

In words, the operation \eqref{trividentm} can be understood as
follows. The most general $U(N)$ Young tableau can be understood as
$\ell_N$ columns of length $N$ appended on the left of an $SU(N)$
tableau with lengths $\ell_1 - \ell_N,\ldots,
\ell_{N-1}-\ell_N$. Since we consider only representations with
$\ell_1 - \ell_N \leq k$, the $SU(N)$ tableau has $\leq k$ boxes in
its first row. The operation above consists of adding an extra row of
$k$ boxes on top of the $SU(N)$ Young tableau and appending
$\ell_N + 1$ columns of $N$ boxes to the resulting tableau. The
resulting $U(N)$ tableau has $SU(N)$ part with row lengths
$k - (\ell_{N-1}-\ell_N), \ell_1 - \ell_{N-1},\ldots, \ell_{N-2} -
\ell_{N-1}$. Note that this is precisely the result of the $\mbb{Z}_N$
action on an $SU(N)$ tableau discussed in Appendix \ref{monopole}. In
the Type I theory, the centre $\mbb{Z}_N$ is a gauge symmetry as
discussed in Appendix \ref{gaugezn}. Thus, different $U(N)$ tableaux
related by this $\mbb{Z}_N$ symmetry must be equivalent and they are
indeed since they have the same characters according
\eqref{charkmanip}.

\subsection{Reducing the general character to a `fundamental domain'} \label{toint}

We will now argue that representations with Young tableaux whose
row-lengths $\ell_i$ satisfy
\begin{equation}\label{rangeoftab}
  0 \leq \ell_i \leq k\ ,\quad\text{that is}\quad 0 \leq n_i < \kappa\ ,
\end{equation} 
constitute a `fundamental domain' in the space of all
representations. The first observation is that the shift operation
$n_i \to n_i + p_i \kappa$ \eqref{trivident} clearly allows one to
move all $n_i$ to the range
\begin{equation}
  0 \leq n_i < \kappa\ .
\end{equation}
However, as we have commented before, once we have accomplished this,
we have obtained the character of a Young tableau that is in general
illegal as it does not obey the inequality \eqref{inequal} in
general. However the character of this illegal tableau is easily
rewritten in terms of the character of a legal tableau as
follows. Consider the integers $\nu_i$ defined by
$n_i = p_i \kappa + \nu_i$ with $0 \leq \nu_i < \kappa$. If two of the
$\nu_i$ are equal then the character vanishes according to
\eqref{condforvan}. When all $\nu_i$ are all distinct then there
exists a unique element $\sigma$ of the permutation group $\mc{S}_N$ that
rearranges these integers so that they are now in decreasing
order. Let this permutation be $\sigma$ and let the integers obtained after
reordering be
\begin{equation}\label{reshuffleP}
  \{n_1', n_2',\ldots, n_N'\}\ ,\quad\text{with}\quad \kappa > n_1' > n_2' > \cdots > n_N' \geq 0\ ,
\end{equation}
and the corresponding Young tableau row-lengths
$\ell'_i = n'_i - (N-i)$ which satisfy
\begin{equation}
  k \geq \ell'_1 \geq \ell'_2 \geq \cdots \geq \ell'_N \geq 0\ .
\end{equation}
It then follows that
\begin{equation}\label{chareeident}
  \chi_{\vgap \{ \ell_i\}} 
  =  {\rm sgn}(\sigma)\, \chi_{\vgap \{\ell'_i\}}\ ,
\end{equation}
where the $\sgn(\sigma)$ arises from reshuffling the rows of the
numerator determinant in the Weyl character formula according to
\eqref{reshuffleP}. Thus, a general Young tableau can be reduced to
one that has at most $k$ columns.

\section{Type II $U(N)$ quantum identities} \label{tt}

The distinguished eigenvalue configurations for the Type II $U(N)$
Chern-Simons theory are the solutions $\ul{w}$ of \eqref{obey} with
$s = -1$:
\begin{equation}\label{obeyII}
  \frac{  (w_i)^\kappa }{\prod_{j=1}^N w_j} = (-1)^{N-1}\ ,\quad\text{for}\quad i=1,\ldots,N\ ,\quad\text{with $w_i$ distinct.}
\end{equation}
In particular, we have $(w_i)^\kappa = (w_j)^\kappa$ for all
$i,j$. This means that $w_i$ is of the form $w_i = \te^{\i\alpha} z_i$
where $z_i$, $i=1,\ldots,N$, are distinct $\kappa^{\rm th}$ roots of
unity. The common phase $\te^{\i\alpha}$ satisfies
\begin{equation}\label{commonconst}
  \te^{\i k\alpha} = (-1)^{N-1} \prod_{i=1}^N z_i\ ,\quad\Longrightarrow \te^{\i k\kappa\alpha} = (-1)^{\kappa(N-1)} = (-1)^{k(N-1)}\ .
\end{equation}
We will also make use of the Weyl character formula for $U(N)$
representations:
\begin{equation}\label{UNwcfrepeat}
  \chi_{\vgap \{\ell\}}(\ul{w}) = \frac{\left[w^{n_1}, w^{n_2}, \ldots, w^{n_{N-1}}, w^{n_N} \right]}{\left[w^{N-1}, w^{N-2}, \ldots, 1 \right]}\ ,
\end{equation}
where $\ell_i$ are the row-lengths of the Young tableau for a $U(N)$
representation and the integers $n_i$ are given by
$n_i = \ell_i + N - i$.

As in the Appendices \ref{sucount} and \ref{to}, we start by deriving
quantum identities for the simple case of representations with only
one non-zero row in their Young tableaux and then treat the general
case.

\subsection{Representations with a single non-zero row}

Let us look at the $U(N)$ Young tableau $\{m,\vec{0}\}$ i.e.~one which
has $m$ boxes in the first row and none in the other $N-1$ rows. The
integers $n_i$ are given by
\begin{equation}
  n_1 = m + N -1\ ,\quad n_2 = N-2\ ,\quad \ldots\ ,\quad n_{N-1} = 1\ ,\quad n_N = 0\ .
\end{equation}
The numerator determinant in the Weyl character formula for
$\chi_{\vgap \{m,\vec{0}\}}$ is given by
\begin{equation}\label{onerowdet}
D_{\{m,\vec{0}\}} = [w^{n_1}, w^{n_2},\ldots, w, 1] =  [w^{m+N-1}, w^{N-2},\ldots, w, 1]\ .
\end{equation}
We study various simple examples for the number of boxes $m$ and then
give a general result.

\subsubsection{$m = k$}
The determinant on the right hand side of \eqref{onerowdet} becomes
\begin{align}
  [w^{\kappa-1}, w^{N-2},\ldots, w, 1]
  &= \frac{\te^{\i\kappa\alpha} }{w_1\cdots w_N} [1, w^{N-1},\ldots,w^2, w]\ ,\nonumber\\
  &= (-1)^{N-1} \times (-1)^{N-1} [w^{N-1},\ldots,w^2, w,1]\ ,
\end{align}
where we have used $(w_i)^\kappa = \te^{\i\kappa\alpha}$ and removed a
factor of $w^{-1}$ from each row. In going to the second line, we have
used \eqref{obeyII} which gives one factor of $(-1)^{N-1}$ and shifted
the first row to the end which gives another factor of
$(-1)^{N-1}$. This implies
\begin{equation}
  \chi_{\vgap \{k,\vec{0}\}}(\ul{w}) = \chi_{\vgap \{\vec{0}\}}(\ul{w}) = 1 \ .
\end{equation}

\subsubsection{$ k+1 \leq m < \kappa$}

Let $m = k + \mu$ with $1 \leq \mu < N$. The determinant
$D_{\{m,\vec{0}\}}$ vanishes:
\begin{align}\label{vanishII}
  D_{\{m,\vec{0}\}} &= \frac{\te^{\i\kappa\alpha} }{w_1\cdots w_N} [w^{\mu}, w^{N-1},\ldots,w^2, w] = 0\ .
\end{align}
This is because the first row becomes identical to one of the other
$N-1$ rows precisely for $\mu = 1,\ldots,N-1$.

\subsubsection{$m = p\kappa$}
The determinant \eqref{onerowdet} can be written as
\begin{align}\label{detmultkappa}
  D_{\{m,\vec{0}\}}
  &= \te^{\i p\kappa\alpha} [w^{N-1}, w^{N-2}, \ldots, w, 1]\ ,\nonumber\\
  &= (-1)^{p(N-1)} (w_1\cdots w_N)^p [w^{N-1}, w^{N-2}, \ldots, w, 1]\ ,\nonumber\\
  &= (-1)^{p(N-1)} [w^{p+N-1}, w^{p+N-2}, \ldots, w^{p+1}, w^p]\ .
\end{align}
In going to the second line, we have used \eqref{obeyII} and in the
third line we have absorbed $w_i^p$ into the $i^{\rm th}$ of the
determinant. This gives
\begin{equation}\label{charmultkappa}
  \chi_{\vgap \{p\kappa,\vec{0}\}}(\ul{w}) = (-1)^{p(N-1)} \chi_{\vgap \{p,p,\ldots,p\}}(\ul{w})\ .
\end{equation}
The Young tableau for the representation on the right hand side has
$N$ rows with $p$ boxes each.

For $p = p' k$, we can simplify the quantum identity
\eqref{charmultkappa} further. We have
$\te^{\i p\kappa \alpha} = \te^{\i p'k\kappa\alpha} = (-1)^{p'k(N-1)}$
from \eqref{commonconst}. Thus, we get
\begin{equation}
  \chi_{\vgap \{p'k\kappa,\vec{0}\}}(\ul{w}) = (-1)^{p'k(N-1)} \chi_{\vgap \{\vec{0}\}}(\ul{w})\ .
\end{equation}
Then, It is not hard to see from \eqref{charmultkappa} that
\begin{equation}
  \chi_{\vgap \{p'k,\ldots,p'k\}}(\ul{w}) =  \chi_{\vgap \{\vec{0}\}}(\ul{w})\ .
\end{equation}
Thus, we can restrict the range of $p$ to be $0 \leq p < k$.

\subsubsection{$m = p\kappa + m'$, $0\leq p < k$,
  $0 < m' < \kappa$}\label{genm}
The determinant \eqref{onerowdet} can be written as
\begin{align}
  D_{\{m,\vec{0}\}} &= \te^{\i p\kappa\alpha} [w^{m'+N-1}, w^{N-2}, \ldots, w, 1]\ .
\end{align}
For $m' = k + 1, k+2,\ldots,\kappa-1$, the determinant on the right
hand side again vanishes for the same reason as in equation
\eqref{vanishII}. We thus focus on the non-vanishing cases
$0 < m' \leq k$. For these cases, it is convenient to rewrite the
determinant using \eqref{obeyII} as
\begin{align}
  D_{\{m,\vec{0}\}}
  &= (-1)^{p(N-1)} (w_1\cdots w_N)^p [w^{m'+N-1}, w^{N-2}, \ldots, w, 1]\ ,\nonumber\\
  &= (-1)^{p(N-1)}  [w^{p+m'+N-1}, w^{p+N-2}, \ldots, w^{p+1}, w^p]\ .
\end{align}
This gives
\begin{equation}
  \chi_{\vgap \{p\kappa+m',\vec{0}\}}(\ul{w}) = (-1)^{p(N-1)} \chi_{\vgap \{p+m',p,\ldots,p\}}(\ul{w})\ .
\end{equation}
The Young tableau on the right hand side has $p+m'$ boxes in the first
row and $p$ boxes in the remaining rows. Note that when
$p + m' < k$, the Young tableau has at most $k-1$ columns.

When $p + m' \geq k$ i.e.~$p + n_1 \geq \kappa - 1$, we can simplify
the above formula further. Then we have
\begin{align}
  D_{\{m,\vec{0}\}}
  &= (-1)^{p(N-1)}  [w^{p + n_1}, w^{p+n_2}, \ldots, w^{p+n_{N-1}}, w^{p+n_N}]\ ,\nonumber\\
  &= (-1)^{p(N-1)}  [w^{p + n_1 - \kappa + \kappa}, w^{p+n_2}, \ldots, w^{p+n_{N-1}}, w^{p+n_N}]\ ,\nonumber\\
  &= (-1)^{p(N-1)} \te^{\i\kappa \alpha}  [w^{p + n_1 - \kappa}, w^{p+n_2}, \ldots, w^{p+n_{N-1}}, w^{p+n_N}]\ .
\end{align}
We have $\kappa > n_1 > \cdots > n_N$ with $\kappa + n_N > n_1$ (the
last inequality is easy to see because $n_N = 0$ here, whence
$\kappa + n_N = \kappa > n_1$ is obviously true). This implies that
\begin{equation}
  \kappa + n_2 > \kappa + n_3 > \cdots > \kappa + n_N > n_1\ ,
\end{equation}
which in turn gives
\begin{equation}
p + n_2 > p+ n_3 > \cdots > p + n_N > p + n_1 - \kappa\ .
\end{equation}
Using this, we can further manipulate the determinant above:
\begin{align}
  D_{\{m,\vec{0}\}}
  &= (-1)^{(p+1)(N-1)} \te^{\i\kappa \alpha}  [ w^{p+n_2}, \ldots, w^{p+n_{N-1}}, w^{p+n_N},w^{p + n_1 - \kappa}]\ ,\nonumber\\
  &= (-1)^{(p+1)(N-1)} \te^{\i\kappa \alpha}  [ w^{p+n_2+1 - 1}, \ldots, w^{p+n_{N-1}+1-1}, w^{p+n_N+1-1},w^{p + n_1 - \kappa+1-1}]\ ,\nonumber\\
  &= (-1)^{(p+1)(N-1)} \frac{\te^{\i\kappa \alpha}}{(w_1\cdots w_N)} [ w^{p+n_2+1}, \ldots, w^{p+n_{N-1}+1}, w^{p+n_N+1},w^{p + n_1 - \kappa+1}]\ ,\nonumber\\
  &= (-1)^{p(N-1)} [ w^{p+n_2+1}, \ldots, w^{p+n_{N-1}+1}, w^{p+n_N+1},w^{p + n_1 - \kappa+1}]\ ,\nonumber\\
  &= (-1)^{p(N-1)} D_{\{p, p, \ldots, p, p + m' - k\}}\ .
\end{align}
Since $n_2 = N-2$ and $p < k$, we have
$p + n_2 +1 = p + N - 1 < \kappa - 1$. Thus, we have
\begin{equation}
  \kappa - 1 > p + n_2 + 1 > \cdots > p + n_N + 1 > p + n_1 -\kappa + 1\ .
\end{equation}
Thus, we have succeeded in relating the quantum character of a
representation with a general number of boxes in the first row and
none in any other row to a representation with at most $k-1$ columns.

\subsection{Adding a monopole}\label{IImono}
Consider a representation $\{\ell_i\}$ in which
$\ell_1 -\ell_N \leq k$. Then define a new Young tableau with rows
\begin{align} \label{ltrividenttypeII}
  &\ell_1' =\ell_N+k\ ,\quad \ell_2'=\ell_1\ ,\quad\ldots\ ,\quad \ell_N'=\ell_{N-1}\ .
\end{align}
\begin{equation}\label{IIZN}
  n'_1 = \kappa + n_N - 1\ ,\quad n'_2 = n_1 - 1\ ,\quad \ldots\ ,\quad n'_N = n_{N-1} - 1\ .
\end{equation}
The character of the new representation is 
\begin{equation}\label{chilprw}
\chi_{\vgap \{\ell'\}}(\ul{w}) = \frac{[w^{n'_1},w^{n'_2},\ldots,w^{n'_N}]}{[w^{N-1},w^{N-2},\ldots, 1]}\frac{[w^{\kappa + n_N - 1},w^{n_1 - 1},\ldots, w^{n_{N-1}-1}]}{[w^{N-1},w^{N-2},\ldots, 1]}\ .
\end{equation}
Let us process the numerator determinant in \eqref{chilprw}.
\begin{align}\label{chilprrw}
  D_{\{\ell'\}} &= [w^{\kappa + n_N - 1},w^{n_1 - 1},\ldots, w^{n_{N-1}-1}]\ ,\nonumber\\
                &= \frac{\te^{\i \kappa \alpha}}{w_1\cdots w_N}  [w^{n_N},w^{n_1},\ldots, w^{n_{N-1}}]\ ,\nonumber\\
                &=  (-1)^{N-1} [w^{n_N},w^{n_1},\ldots, w^{n_{N-1}}]\ ,\nonumber\\
                &=  (-1)^{N-1} \times (-1)^{N-1} [w^{n_1},\ldots, w^{n_{N-1}},w^{n_N}] \ ,\nonumber\\
                &= D_{\{\ell_i\}}\ .
\end{align}
The factor of $(-1)^{N-1}$ comes from \eqref{obeyII} and the
additional factor of $(-1)^{N-1}$ in the last line comes from pushing
the first row to the end. Thus, we have
\begin{equation}\label{chil}
\chi_{\vgap\{\ell'\}}(\ul{w}) = \chi_{\vgap \{\ell\}}(\ul{w})\ .
\end{equation}

\subsection{General representations}
Let $n_i = p_i\kappa + \nu_i$ with $0 \leq \nu_i < \kappa$. Since the
$\nu_i$ may not be ordered in decreasing order, we reorder them by
hand with the permutation $\sigma$ and call the resulting integers
$n'_i$ and the corresponding Young tableau rows
$\ell'_i = n'_i - N + i$. The $n'_i$ satisfy the ordering
\begin{equation}\label{nnaiverange}
  \kappa > n'_1 > \cdots > n'_N \geq 0\ .
\end{equation}
It follows that $\kappa \geq \kappa - n'_N > n'_1 - n'_N > \cdots$,
which gives, upon adding $n'_N$ to each node in the inequality,
\begin{equation}\label{nordII}
  \kappa + n'_N > \kappa > n'_1 > \cdots > n'_N \geq 0\ .
\end{equation}
The above ordering \eqref{nordII} will be useful later. Let
$P = \sum_{i=1}^N p_i$. We then have
\begin{align}\label{detIIgen}
  D_{\{\ell_i\}} &= [w^{n_1}, w^{n_2},\ldots, w^{n_N}]\ ,\nonumber\\
                 &= \te^{\i P\kappa\alpha} [w^{\nu_1},w^{\nu_2},\ldots, w^{\nu_N}]\ ,\nonumber\\
                 &= \sgn(\sigma)\te^{\i P\kappa\alpha}[w^{n'_1},w^{n'_2},\ldots, w^{n'_N}]\ ,\nonumber\\
                    &= \sgn(\sigma) (-1)^{P(N-1)} (w_1\cdots w_N)^P [w^{n'_1},w^{n'_2},\ldots, w^{n'_N}]\ ,\nonumber\\
                  &= \sgn(\sigma) (-1)^{P(N-1)}  [w^{P+n'_1},w^{P+n'_2},\ldots, w^{P+n'_N}]\ .
\end{align}
When $P = k$, we have $\te^{\i k \kappa\alpha} =
(-1)^{k(N-1)}$. Substituting the above into the third line in
\eqref{detIIgen}, we get
\begin{equation}\label{P=k}
D_{\{\ell_i\}} = \sgn(\sigma) (-1)^{k(N-1)} D_{\{\ell'_i\}}\ .
\end{equation}
Similarly substituting $P = k$ in the last line of \eqref{detIIgen},
we get
\begin{equation}\label{P=k1}
  D_{\{\ell_i\}} = \sgn(\sigma) (-1)^{k(N-1)}  [w^{k+n'_1},w^{k+n'_2},\ldots, w^{k+n'_N}]\ .
\end{equation}
Comparing \eqref{P=k} and \eqref{P=k1}, we get
\begin{equation}\label{P=k2}
  D_{\{\ell'_i\}} = D_{\{k+\ell'_i\}}\ ,
\end{equation}
which in particular tells us that
\begin{equation}
  D_{\{k,\ldots,k\}} = D_{\{\vec{0}\}}\ .
\end{equation}
Now, let us focus on the last line of \eqref{detIIgen}:
\begin{equation}
  D_{\{\ell_i\}} = \sgn(\sigma) (-1)^{P(N-1)} [w^{P+n'_1},w^{P+n'_2},\ldots, w^{P+n'_N}]\ .
\end{equation}
Since the original $n'_i$ are arbitrary integers, the $p_i$ can be
either negative or positive and consequently, $P$ can also be of
either sign. Let us define $P = p k + P'$ where $0 \leq P' < k$. Then,
from \eqref{P=k2}, we have
\begin{equation}
  D_{\{\ell_i\}} = \sgn(\sigma) (-1)^{P(N-1)} [w^{P'+n'_1}, w^{P'+n'_2},\ldots, w^{P'+n'_N}]\ .
\end{equation}
Suppose there is an $I$ with $1\leq I \leq N$ such that
\begin{equation}\label{assumpII}
  P' + n'_i \geq \kappa-1\quad\text{for}\quad i = 1,\ldots, I\ ,\quad\text{and}\quad P' + n'_i < \kappa-1\quad\text{for}\quad i = I+1,\ldots,N\ .
\end{equation}
We can apply the $\mbb{Z}_N$ transformation \eqref{IIZN} $-I$ times to
get
\begin{align}\label{ZNminusI}
  &[w^{P'+n'_1}, w^{P'+n'_2},\ldots, w^{P'+n'_N}]\nonumber\\
  &= [w^{P'+ n'_{I+1} + I}, w^{P' + n'_{I+2} + I}, \ldots, w^{P' + n'_N + I}, w^{P' + n'_1 - \kappa + I}, \ldots, w^{P' + n'_{I} - \kappa + I}]\ .
\end{align}
Note that every exponent in the second line above is positive by our
assumptions \eqref{assumpII} on the $n'_i$ and since $I \geq 1$. Let
us look at the sum of exponents before and after the $\mbb{Z}_N$
action.
\begin{align}
  \text{Before the  $\mbb{Z}_N$ action}:&\quad NP' + \sum_{i=1}^N n'_i\ ,\nonumber\\
  \text{After the  $\mbb{Z}_N$ action}:&\quad NP' + \sum_{i=1}^N n'_i + I N - I\kappa = NP' + \sum_{i=1}^N n'_i - Ik\ .
\end{align} 
We see that the sum of exponents has reduced by $I k$, or
equivalently, the number of boxes in the corresponding Young tableau
has reduced by $I k$. Suppose $P' + n'_{I+1} + I < \kappa - 1$. Then,
we have nothing more to do. In case
$P' + n'_{I+1} + I > \cdots > P' + n'_{I+J} + I \geq \kappa - 1$ for
some $J$ with $1 \leq J \leq N$, then we apply the $\mbb{Z}_N$
transformation $-J$ times to get a new determinant whose sum of
exponents reduces by $Jk$. This procedure has to terminate since the
sum of exponents is positive at each step (see the explanation after
\eqref{ZNminusI}). At the end, we will have exponents $\tl{n}_i$ which satisfy
\begin{equation}
  \kappa - 1 > \tl{n}_1 > \cdots > \tl{n}_N \geq 0\ .
\end{equation}
Note that the demarcating value $\kappa - 1$ in \eqref{assumpII}
cannot be replaced by a different value $\kappa - t$ for some $t > 1$
since the exponents need not be positive at each stage.

\section{$SU(2)_k$ as an example}\label{sutq}

In the case of the $SU(2)_k$ theory, the holonomy eigenvalues that
obey the equation \eqref{progen} are given by
\begin{equation}\label{eivr}
  \ul{w}_p = \{w_{1,p},w_{2,p}\} = \{ \te^{\frac{-\i p \pi}{\kappa}}, \te^{\frac{\i p \pi}{\kappa}}\}\ ,
\end{equation}
which corresponds to the diagonal $SU(2)$ matrix
\begin{equation}
  U^{(p)} = \begin{pmatrix} \te^{\frac{-\i p \pi}{\kappa}} & 0 \\
    0 & \te^{\frac{\i p \pi}{\kappa}}\end{pmatrix} \ ,\quad p = 1, \ldots, \kappa-1=k+1 \ .
\end{equation}
Note that the number of distinct eigenvalue configurations match the
number of integrable primaries as expected. The formula
\eqref{ngtmnewalt} for the number of $SU(2)_k$ quantum singlets in the
product of representations with spins $j_1,\ldots, j_n$ reduces to
\begin{equation}\label{Neq2discsum} 
  \mc{N}_{0,n} = \frac{1}{2 \kappa} \sum_{p=1}^{\kappa-1} 4 \sin^2
  \Big( \frac{p \pi}{\kappa}\Big)
  \prod_{i=1}^n \chi_{ j_i}(\ul{w}_p) \ .
\end{equation}  
Recall that the spin $j$ that labels an irreducible representation of
$SU(2)$ is quantized in half-integer units \footnote{The number of
  boxes in the first row of the Young tableau corresponding to a spin
  $j$ representation is $2j$.}. Note that the RHS of
\eqref{Neq2discsum} must evaluate to unity in the case that no
characters are inserted. Indeed it is not difficult to directly check
that
\begin{equation}\label{directcheck}
\sum_{p=1}^{ \kappa-1}  4 \sin^2 \Big( \frac{p \pi}{\kappa} \Big) = 2\kappa \ . 
\end{equation} 
providing a consistency check on the normalization of
\eqref{Neq2discsum}.

For the case of $SU(2)_k$, it is easy to use the fact that eigenvalues
inside the path integral are all quantized as in \eqref{eivr},
together with the explicit character formulae \eqref{chja} to
independently derive the following quantum character identities. For
for every integer $m$,
\begin{equation} \begin{split} \label{charprop}
&\chi_{\vgap \frac{\kappa-1 + m}{2}} = -\chi_{\vgap \frac{\kappa-1 -m}{2}} \ , \\
& \chi_{\vgap \kappa+m} = \chi_{\vgap m} \ . \\
\end{split}
\end{equation}

The second identity is obvious from \eqref{chja} and \eqref{eivr} while the first follows from \eqref{chja} upon using the 
identity 
\begin{equation}\label{identiyo}
\sin \big( x +n\pi \big)= -\sin\big(-x + n \pi \big)  \ . 
\end{equation} 
which holds for any integer $n$. This identity 
can be seen as a special case of the results of 
subsection \ref{sr} as follows. The analysis of 
subsection \ref{sr} tells us that - using the notation of that subsection - that  
$\chi_{\vgap \kappa+m, 0}=-\chi_{\vgap k, m}$. In other words the character of the representation with $\kappa+m$ boxes in the first row is the same as the character with $k$ boxes in the first row and $m$ in the second row. In the case of $SU(2)$ the last gadget should be interpreted by dropping filled columns, i.e. as the character of the representation with $k-m$ boxes in the first row.
This is the same as the identity in the first line of \eqref{charprop}. 

It follows in particular from the first of \eqref{charprop} 
that
\begin{equation}\label{spcchar}
\chi_{\vgap \frac{k+1}{2}}=0 \ . 
\end{equation}
In other words, the representation with $k+1$ boxes 
in the first row of the Young tableau vanishes, in agreement with the general analysis of subsection \ref{sr}.

From the second of \eqref{charprop} it then follows that  $\chi_{\vgap -\frac{1}{2}}$ also vanishes, and more generally  
\begin{equation}\label{whichzer}
\chi_{\vgap \frac{n \kappa}{2}-\frac{1}{2}}=0 \ . 
\end{equation}
Again this agrees with the general results of 
subsection \ref{sr}. 
Note also that 
\begin{equation} \label{chiup}
\chi_{\vgap \kappa-1}= \chi_{\vgap \frac{\kappa-1}{2} + \frac{\kappa-1}{2}}
=-  \chi_{\vgap \frac{\kappa-1}{2} - \frac{\kappa-1}{2}} = 
- \chi_{\vgap 0} =- 1 \ . 
\end{equation} 
It follows also from  \eqref{charprop} and \eqref{spcchar} that characters at any value of $j$ 
are equal, up to a possible minus sign, to a character within the `fundamental domain', i.e. in the range  
\begin{equation}\label{range}
j= 0,  \frac{1}{2}, 1, \frac{3}{2},  \ldots ,  \frac{k}{2} \ . 
\end{equation} 
These are the $k+1$ integrable representations in which the tableau has no more than $k$ boxes in the first (and only) row.

\eqref{chardec} and  
\eqref{charprop} imply the following fusion rules 
\begin{equation}\label{frulessu}
\chi_{j_1} \chi_{j_2} = \sum_{n=|j_1-j_2|}^{j_1+j_2} \chi_n
=  \sum_{n=|j_1-j_2|}^{{\rm min}(k-j_1-j_2, ~ j_1+j_2)}\chi_n \ , 
\end{equation}
where we have used the second of \eqref{charprop} to cancel terms on the 
RHS of \eqref{frulessu}. This is the famous fusion algebra for 
primaries in $SU(2)$ level $k$ WZW theory. 

In the case of the $SU(2)_k$ the lowest action holonomy (see Section
\ref{sec}) is
\begin{equation}\label{sutowmatmin}
U= \begin{pmatrix}\te^{\frac{\i \pi}{\kappa}} & 0 \\
	0 &\te^{-\frac{\i \pi}{\kappa}} \end{pmatrix} \ . 
\end{equation}

\section{The quantum dimension is the largest eigenvalue of a fusion
  matrix }\label{proof}
In this appendix\footnote{We thank O. Parrikar for the suggestion that
  led to this Appendix.}, we revisit the proof of the fact that the
quantum dimension is the largest among the absolute values of
eigenvalues of a fusion matrix. This has been derived using the
positivity of the fusion matrix and the Perron-Frobenius theorem in
many different but related contexts e.g.~non-abelian quantum Hall
states (see \cite[Section 4]{tong2016lectures}, \cite[Chapter
2]{chamon2017topological} for a review); see also
\cite{dong2012quantum, etingof2017fusion} for a proof of this fact. We
then use this fact to prove the theorem stated in Section \ref{lwcoo}.

Recall that the fusion coefficients $\mc{N}_{\mu\nu}^\lambda$ are
defined as
\begin{equation}
  R_\mu \times R_\nu = \sum_{\lambda} \mc{N}_{\mu\nu}^\lambda R_\lambda\ ,
\end{equation}
where $R_\mu$, $R_\nu$ and $R_\lambda$ are integrable representations
with $su(N)$ (or $u(N)$) highest weights $\mu$,$\nu$,$\lambda$
respectively, and the notation $R \times R'$ indicates the fusion of
primaries associated to the integrable representations $R$ and
$R'$. Note that the fusion coefficient $\mc{N}_{\mu\nu}^\lambda$ is
symmetric in $\mu \nu$ since the fusion $R_\mu \times R_\nu$ is a
commutative operation. Alternatively, the fusion coefficient can also
be defined as the number of singlets in the fusion of $R_\mu$, $R_\nu$
and $R_{\lambda^*}$ where $\lambda^*$ is the conjugate of $\lambda$:
\begin{equation}\label{Ntotsym}
  \mc{N}_{\mu\nu}^\lambda \equiv \mc{N}_{\mu\nu\lambda^*} = \mc{N}_{0,3}(R_\mu, R_\nu,R_{\lambda^*})\ ,
\end{equation}
where we have defined the quantity $\mc{N}_{\mu\nu\lambda^*}$ with
three lower indices by the above equation and the quantity
$\mc{N}_{0,3}$ is defined in \eqref{ngtm} and is the number of
singlets in the fusion of three primary insertions on the sphere. Note
that $\mc{N}_{\mu\nu\lambda}$ is symmetric in all three indices (see
\cite{di1996conformal} for a detailed discussion of these properties).

The basic identity, \emph{the Verlinde formula}, between the fusion
coefficients and the Verlinde $\mc{S}$-matrix is the following
\begin{equation}\label{basiciden}
\sum_{\lambda'}  \mc{N}_{\mu\nu}^{\lambda'} \mc{S}_{\lambda'\lambda} =   \frac{\mc{S}_{\mu\lambda}}{\mc{S}_{\lambda 0}} \mc{S}_{\nu\lambda}\ ,
\end{equation}
where $0$ corresponds to the trivial representation. This identity can
be derived in various ways: (1) by comparing the monodromies of the
one-point function of the primary $\mu$ on the torus around the $a$-
and $b$-cycles \cite{Moore:1988uz}, \cite[Exercise 3.6]{Moore:1989vd},
(2) by surgery of an appropriate choice of Wilson loops on $S^3$
\cite{Witten:1988hf}, (3) by looking at the modular properties of the
partition function on an annulus \cite{Cardy:1989ir}. There are
several things one can infer from the above identity.

Recall that the Verlinde $\mc{S}$ matrix enjoys the following
properties: $\mc{S}$ is a symmetric, unitary matrix, and, moreover,
$\mc{S}_{\mu\nu}= \mc{S}^*_{\nu\mu^*}$, where $\mu^*$ is the complex
conjugate of the representation $\mu$ (e.g.~see
\cite{di1996conformal}). Multiplying the formula \eqref{basiciden} by
$(S^\dag)_{\lambda \rho}$ and summing over $\lambda$ gives the
following formula for fusion coefficients:
\begin{equation}\label{Nverlin}
  \mc{N}_{\mu\nu}^\rho = \sum_{\lambda} \frac{\mc{S}_{\mu\lambda} \mc{S}_{\nu\lambda} \mc{S}^*_{\rho\lambda}}{\mc{S}_{\lambda 0}} = \sum_{\lambda} \frac{\mc{S}_{\mu\lambda} \mc{S}_{\nu\lambda} \mc{S}_{\lambda\rho^*}}{\mc{S}_{\lambda 0}}\ .
\end{equation}
The formula \eqref{ngtm} for the number of singlets $\mc{N}_{g,n}$ for
$n$ insertions on a genus $g$ surface specialized to the sphere
($g = 0$) and three representations $\mu$, $\nu$ and $\rho^*$ gives
\begin{equation}\label{spcg}
\mc{N}_{\mu\nu\rho^*} = \mc{N}_{0,3}(R_\mu, R_\nu, R_{\rho^*}) = \sum_\lambda \frac{\mc{S}_{\lambda \mu }\mc{S}_{\lambda \nu}\mc{S}_{\lambda \rho^*}}{\mc{S}_{\lambda 0}}\ ,
\end{equation}
which agrees with the formula \eqref{Nverlin}. Later, in Appendix
\ref{proofsub}, we outline the derivation of the general formula
\eqref{ngtm} based on the results of this section.

Next, consider the set of \emph{fusion matrices} $N_\mu$ defined as
\begin{equation}
  (N_\mu)_\nu^\lambda = \mc{N}_{\mu\nu}^\lambda\ .
\end{equation}
The fusion matrices mutually commute as a result of the associativity
of the fusion product:
\begin{equation} \label{fcom} 
  (R_\mu \times R_\nu) \times R_\rho = R_\mu \times (R_\nu \times R_\rho)\ .
\end{equation}
\footnote{The algebra proceeds as follows. The number of copies of the
  representation $R_\theta$ that is obtained by fusing representations
  as on the LHS of \eqref{fcom} is
  $\mc{N}_{\mu\nu}^\alpha \mc{N}_{\alpha \rho}^\theta$. On the other
  hand the fusion on the RHS of \eqref{fcom} gives
  $\mc{N}_{\mu \alpha}^\theta \mc{N}^\alpha_{\nu \rho}$ copies of the
  representation $R_\theta$. The fact that these two quantities are
  equal tells us that
  \begin{equation}
    \mc{N}_{\mu\nu}^\alpha \mc{N}_{\alpha \rho}^\theta = \mc{N}^\alpha_{\nu \rho} \mc{N}_{\mu \alpha}^\theta\ .\nonumber
  \end{equation} 
  In other words, the fusion matrices $N_\mu$ and $N_\rho$ commute
  with each other.} As a result the matrices $N_\mu$ for all choices
of $\mu$ are simultaneously diagonalizable. This fact can be seen more
explicitly, as we now explain. The identity \eqref{basiciden} can be
recast as
\begin{equation} \label{recastas}
  \mc{S}^{-1} N_\mu \mc{S} = \mc{V}_\mu\ ,\quad\text{with}\quad (\mc{V}_\mu)_\nu^\lambda = \frac{\mc{S}_{\mu\nu}}{\mc{S}_{\nu 0}} \delta_\nu^\lambda\ ,
\end{equation}
where recall that $0$ stands for the trivial representation. We see
from \eqref{recastas} that a universal similarity transformation - in
particular the similarity transformation with the Verlinde
$\mc{S}$-matrix - does indeed simultaneously diagonalize all the
fusion matrices, and that the eigenvalues of the matrix $N_\mu$ are
\begin{equation} \label{mfrep} 
\frac{\mc{S}_{\mu\nu}}{\mc{S}_{\nu 0}} = \chi_\mu(\xi_\nu)\ ,
\end{equation}
(where $\xi_\nu$ is the distinguished dual Cartan element
$\xi_\nu = -\frac{2\pi\i}{\kappa}(\nu + \rho)$ and we have used the magic formula \eqref{suverorm}). 

In summary, we see that fusion matrices $N_\mu$ are all simultaneously
diagonalized by the Verlinde $\mc{S}$-matrix and the eigenvalues of
$N_\mu$ (equivalently the diagonal elements of $\mc{V}_\mu$) are the
independent quantum characters $\chi_\mu(\xi_\nu)$, i.e. the ordinary
Lie algebra characters evaluated successively on the distinguished
group elements associated with the various integrable representations
of the theory (see e.g.~Section \eqref{disge} and also
\cite{Zuber:1995ig} for an explicit formula for these special group
elements in the case of $SU(N)_k$ theory).

Note that one of the eigenvalues of the matrix $N_\mu$ - namely the
eigenvalue corresponding to $\nu=0$ in \eqref{mfrep} - is simply the
quantum dimension $\mc{D}_\mu = \chi_\mu (\xi_0)$ of the
representation $\mu$ \footnote{See Section \ref{qdnote} for the
  definition of the quantum dimension and some further discussion of
  this quantity. In that section, the notation for the quantum
  dimension of a representation $R$ is $\mc{D}(R)$ whereas in this
  appendix, we use $\mc{D}_\mu$ to avoid clutter.} Now setting
$\lambda= 0$ in \eqref{basiciden}, we get
\begin{equation}
  \sum_{\lambda'} \mc{N}_{\mu\nu}^{\lambda'} \mc{S}_{\lambda'0} = \mc{D}_\mu \mc{S}_{\nu 0}\ .
\end{equation}
Dividing both sides by $\mc{S}_{00}$, we get
\begin{equation} \label{spcvec} 
  \sum_{\lambda'}  \mc{N}_{\mu\nu}^{\lambda'} \mc{D}_{\lambda'} = \mc{D}_\mu \mc{D}_{\nu}\ .
\end{equation}
\eqref{spcvec} is the matrix equation
\begin{equation} \label{spcvecrw} 
 \mc{N}_{\mu} v_{\mc{D}} = \mc{D}_\mu v_{\mc{D}} \ ,
\end{equation}
where the vector $v_{\mc{D}}$ is given by
\begin{equation}\label{vdgb} 
  v_{\mc{D}} = 
  \begin{pmatrix}
    \mc{D}_{\mu_1}\\ \mc{D}_{\mu_2}\\ \vdots \\ \mc{D}_{\mu_r}
  \end{pmatrix} \ ,
\end{equation}
where $r$ is the number of integrable generators of the given WZW
theory and $\mc{D}_{\mu_1},\ldots,\mc{D}_{\mu_r}$ are the quantum
dimensions of these representations. The equation \eqref{spcvecrw} is
the assertion that the eigenvector corresponding to the eigenvalue
$\mc{D}_\mu$ is given by the vector \eqref{vdgb}, i.e. the vector of
quantum dimensions of all integrable representations! Note that every
entry of the eigenvector is positive since the quantum dimension of
every integrable representation is a positive real number.

We now prove that the modulus of every eigenvalue of the fusion matrix
$N_\mu$ is bounded above by the quantum dimension $\mc{D}_\mu$.
Consider the matrix $M$ defined by 
\begin{equation}\label{Mdef}
  M=\sum_\mu N_\mu\ ,
\end{equation} 
where the summation runs over all integrable representations of the
theory. As the quantities $(N_\mu)_\nu^\lambda$ are all non negative
integers, it is clear that every element of the matrix $M$ is greater
than or equal to zero. In fact more is true: every element of the
matrix $M$ is actually a positive integer so that
\begin{equation} \label{posf}
M_\nu^\lambda = \sum_\mu (N_\mu)_\nu^\lambda > 0\ ,\quad\text{for every $\nu$ and $\lambda$}.
\end{equation}
In order to establish \eqref{posf} we need to show that given any two
integrable representations $\nu$ and $\lambda$, it is always possible
to find at least one other integrable representation $\mu$ so that
$N_{\mu\nu}^\lambda$ is nonzero. Clearly the fusion of $\nu$ and
$\lambda^*$ produces at least one representation; let us denote any
one such any representation by $\mu^*$. We now have
\begin{equation}\label{exire}
  \mc{N}_{\lambda^* \nu}^{\mu^*} = \mc{N}_{\lambda^*\nu\mu} = \mc{N}_{\mu\nu\lambda^*} = \mc{N}^{\lambda}_{\mu \nu}\ .
\end{equation} 
(see equation \eqref{Ntotsym} and the subsequent text for the
properties of the fusion coefficients that enable the above
manipulation.) It follows from \eqref{exire} that
$\mc{N}_{\mu\nu}^\lambda$ is nonzero and so \eqref{posf} follows.

To summarize, we have shown that every element of the matrix
$M = \sum_\mu N_\mu$ is a strictly positive integer.

Recall that the Perron-Frobenius theorem asserts that if all the
elements of a square matrix are strictly positive then its largest (in
modulus) eigenvalue is a positive real number and has a unique
eigenvector. It also asserts that all elements of the eigenvector
corresponding to this largest eigenvalue are non-zero real numbers of
the same sign (and so, by a choice of normalization, can be chosen to
be strictly positive). Finally it assures us that no other eigenvector
of this matrix - other than the one corresponding to the largest
eigenvalues - has this equal sign property (i.e. that every other
eigenvector of the matrix has elements of differing sign).

A slight generalization of the same theorem (by Frobenius) establishes
similar, but weaker, results, for a slightly larger class of
matrices. Consider a square matrix all of whose elements of a square
matrix are non-negative (note that some elements are allowed to
vanish). In this case the eigenspace of the largest eigenvalue of the
matrix may or may not be unique. In either case, the generalization
assures us that there exists an eigenvector in the eigenspace of the
largest eigenvalue, which has all real elements of the same sign (in
this case some elements are allowed to be zero). In this case,
however, there is no claim to uniqueness; other eigenvectors of the
matrix may also have the same non-negativity property.
 
We will now apply these theorems to the matrices $N_\mu$ and $M$.  Let
us first consider $N_\mu$ for any given value of $\mu$.  As each
element $\mc{N}_{\mu\nu}^\lambda$ of this matrix is a non-negative
integer, the conditions of the generalization of the Perron-Frobenius
theorem apply to $N_\mu$ and we conclude that the largest eigenvalue
of $N_\mu$ must have an eigenvector all of whose elements have the
same sign (and so can, by choice of normalization, be chosen to be
either positive or zero). In other words, $N_\mu$ has a non-negative
eigenvalue $\Lambda_{(\mu)}$ such that
\begin{equation} \label{hiev}
  \Lambda_{(\mu)} \geq |\Lambda|\ ,
\end{equation} 
for every eigenvalue $\Lambda$ of $N_\mu$, and that the corresponding
eigenvector $v_{(\mu)}$ is non-negative, i.e.
\begin{equation}\label{evelempo}
  \left( v_{(\mu)} \right)_\nu \geq 0\ ,\quad\text{for all $\nu$}.
\end{equation} 
As all $N_\mu$ are simultaneously diagonalizable, and as $M$ is simply
a sum of the $N_\mu$, it follows that $v_{(\mu)}$ is also an
eigenvector of $M$. In other words
\begin{equation}\label{mvev}
  \sum_\rho  M_\lambda^\rho \left( v_{(\mu)} \right)_\rho = m_{(\mu)} \left( v_{(\mu)} \right)_\lambda\ .
\end{equation} 
Since each $M_\lambda^\rho$ is a positive integer, it follows
immediately from \eqref{evelempo} that the LHS of \eqref{mvev} is
strictly positive for every value of $\lambda$ (we have used the fact
that $v_{(\mu)}$ is a nontrivial eigenvector, and so at least one
$(v_{(\mu)})_\lambda$ is nonzero). It follows that the RHS of
\eqref{mvev} must also be strictly positive for every value of
$\lambda$. We have thus established that all the elements
$\left( v_{(\mu)} \right)_\lambda$ are nonzero, and so $v_{(\mu)}$ is
a strictly positive eigenvector.

However, as all elements of $M$ are strictly positive, the first
(stricter) Perron-Frobenius theorem applies to $M$, and so it is
guaranteed that $M$ has only one strictly positive eigenvector
$v_\mu$. It follows that the eigenvectors $v_\mu$, for the different
possible values of $\mu$ - are all the same (upto scalar multiples).

We independently already know of one eigenvector of each $N_\mu$ - and
therefore of $M$ - that is strictly positive. This is simply the
eigenvector $v_{\mc{D}}$ \eqref{vdgb} whose elements are the quantum
dimensions of the various integrable representations. The uniqueness
of positive eigenvectors of $M$ allow us to conclude the
$v_{(\mu)}= v_{\mc{D}}$ for all $\mu$. It thus follows from
\eqref{spcvecrw} that $\Lambda_{(\mu)} = \mc{D}_\mu$, the quantum
dimension of the representation $\mu$. Recalling the definition of
$\Lambda_\mu$ (see around \eqref{hiev}) we conclude that the largest
eigenvalue of the matrix $N_\mu$ is $\mc{D}_\mu$.

Recall from \eqref{mfrep} that the collection of eigenvalues of
$\mc{D}_\mu$ are the characters evaluated on the distinguished
eigenvalues corresponding to the various integrable representations. It
follows that
\begin{equation}
  \mc{D}_\mu = \chi_\mu(\xi_0) \geq |\chi_\mu(\xi_\nu)|\quad\text{for all $\nu$}.
\end{equation}
as we set out to show. We have thus established the theorem stated in
Section \ref{lwcoo}.

\subsection{Derivation of the dimension of space of conformal
  blocks in terms of Verlinde $\mc{S}$-matrices}\label{proofsub}
The dimension of the space of conformal blocks on $S^2$ with
representations $\mu_1$, $\mu_2$,\ldots, $\mu_n$ is given in terms of
the fusion coefficients by
\begin{equation}
  \mc{N}_{0,n}(\mu_1,\ldots,\mu_n) = \mc{N}_{\mu_2\mu_1}^{\nu_1} \mc{N}_{\mu_3\nu_1}^{\nu_2} \cdots \mc{N}_{\mu_{n-1}\nu_{n-2}}^{\mu^*_n}\ .
\end{equation}
This is the same as the following matrix element in the appropriate
product of fusion matrices
\begin{equation}\label{numsing}
  (N_{\mu_2} N_{\mu_3} \cdots N_{\mu_{n-1}})_{\mu_1}^{\mu^*_n}\ .
\end{equation}
Plugging $N_\mu = \mc{S} \mc{V}_\mu \mc{S}^{-1}$ and using the
explicit form the diagonal matrix $\mc{V}_\mu$ \eqref{recastas}, we
have
\begin{align}\label{genus0}
  \mc{N}_{0,n}(\mu_1,\ldots,\mu_n) &= (\mc{S} \mc{V}_{\mu_2} \mc{V}_{\mu_3} \cdots \mc{V}_{\mu_{n-1}} \mc{S}^{-1})_{\mu_1}^{\mu^*_n}\nonumber\\
                                   & = \sum_{\nu_1,\ldots,\nu_{n-1}} \mc{S}_{\mu_1\nu_1} \frac{\mc{S}_{\mu_2\nu_1}}{\mc{S}_{\nu_1 0}} \delta_{\nu_1 \nu_2} \frac{\mc{S}_{\mu_3\nu_2}}{\mc{S}_{\nu_2 0}} \delta_{\nu_2 \nu_3} \cdots \frac{\mc{S}_{\mu_{n-1}\nu_{n-2}}}{\mc{S}_{\nu_{n-2} 0}} \delta_{\nu_{n-2} \nu_{n-1}} (\mc{S}^{-1})^{\nu_{n-1} \mu^*_n}\nonumber\\
                                   & = \sum_{\nu_1} (\mc{S}_{\nu_1 0})^2 \frac{\mc{S}_{\mu_1\nu_1}}{\mc{S}_{\nu_1 0}} \frac{\mc{S}_{\mu_2\nu_1}}{\mc{S}_{\nu_1 0}} \frac{\mc{S}_{\mu_3\nu_1}}{\mc{S}_{\nu_1 0}}  \cdots \frac{\mc{S}_{\mu_{n-1}\nu_{1}}}{\mc{S}_{\nu_{1} 0}}  \frac{S_{\mu_n \nu_{1}}}{\mc{S}_{\nu_1 0}}\ ,
\end{align}
where in the last step we have used $\mc{S}^{-1} = \mc{S}^\dag$,
$(\mc{S}^*)^{\mu\nu} = \mc{S}_{\mu\nu^*} = \mc{S}_{\mu^*\nu}$ and
$\mc{S}^T = \mc{S}$.

For the dimension of the space of $n$-point conformal blocks on a
genus $g > 0$ surface, we start with $2g + n$ insertions
$\lambda_1,
\lambda_1^*,\ldots,\lambda_g,\lambda_g^*,\mu_1,\ldots,\mu_n$ on $S^2$
and then sum over the $\lambda_i$, $i=1,\ldots,g$:
\begin{align}\label{genusg}
  \mc{N}_{g,n}(\mu_1,\ldots,\mu_n)
  & = \sum_{\lambda_1,\ldots,\lambda_g}  \mc{N}_{0,n}(\lambda_1,\lambda_1^*,\ldots,\lambda_g,\lambda_g^*,\mu_1,\ldots,\mu_n)\ ,\nonumber\\
  &= \sum_{\nu,\lambda_1,\ldots,\lambda_g} (\mc{S}_{\nu 0})^2 \frac{\mc{S}_{\lambda_1^*\nu}}{\mc{S}_{\nu 0}} \frac{\mc{S}_{\lambda_1\nu}}{\mc{S}_{\nu 0}} \cdots \frac{\mc{S}_{\lambda_g^*\nu}}{\mc{S}_{\nu 0}} \frac{\mc{S}_{\lambda_g\nu}}{\mc{S}_{\nu 0}} \frac{\mc{S}_{\mu_1\nu}}{\mc{S}_{\nu 0}} \frac{\mc{S}_{\mu_2\nu}}{\mc{S}_{\nu 0}} \frac{\mc{S}_{\mu_3\nu}}{\mc{S}_{\nu 0}}  \cdots \frac{\mc{S}_{\mu_{n-1}\nu}}{\mc{S}_{\nu 0}}  \frac{S_{\mu_n \nu}}{\mc{S}_{\nu 0}}\ ,\nonumber\\
  &=\sum_{\nu} (\mc{S}_{\nu 0})^{2-2g} \frac{\mc{S}_{\mu_1\nu}}{\mc{S}_{\nu 0}} \frac{\mc{S}_{\mu_2\nu}}{\mc{S}_{\nu 0}} \frac{\mc{S}_{\mu_3\nu}}{\mc{S}_{\nu 0}}  \cdots \frac{\mc{S}_{\mu_{n-1}\nu}}{\mc{S}_{\nu 0}}  \frac{S_{\mu_n \nu}}{\mc{S}_{\nu 0}}\ ,
\end{align}
where we have used that the Verlinde $\mc{S}$-matrix is unitary,
$\mc{S}^\dag \mc{S} = \mbb{1}$ to simplify the summations over
$\lambda_1,\ldots,\lambda_g$.

\subsection{$SU(2)$ characters on discretized
  eigenvalues}\label{sutch}

In this subsection we explicitly illustrate, for the special case of
$SU(2)_k$, the theorem \eqref{QCmaxconjec} that character of
integrable representations are maximized on the special eigenvalues
configuration $U^{(0)}$ \eqref{sutowmatmin}.

The character of $SU(2)$ spin $j$ representation evaluated on the SU(2) element 
$U(\theta)=\textrm{diag}(\te^{\i\theta},\te^{-\i\theta})$ is 
\begin{equation}\label{su2jchar}
  \chi_j(\theta) = \frac{\sin((2j+1)\theta)}{\sin\theta}\ .
\end{equation}
The discretized eigenvalues for $SU(2)_k$ corresponds to 
\begin{equation}
\theta_n = \frac{n\pi}{k+2} \quad \textrm{with} \quad n \in \{1,2, \ldots, k+1\}\ .
\end{equation}
Thus, according to our conjecture \eqref{QCmaxconjec} the eigenvalue
configuration with $n=1$ (and its $Z_2$ image i.e. $n=k+1$) maximizes
the character of all integrable representations i.e.
\begin{equation}\label{su2kevmax}
\chi_j(\theta_1) \geq \chi_j(\theta_n)  \quad \textrm{for all} \quad 1 < n \leq k+1\ ,
\quad \text{and} \quad j \in \{0,\tfrac{1}{2},1,\tfrac{3}{2},\ldots, \tfrac{k}{2})\ .
\end{equation}
Before proceeding further, let us note the following simple symmetry
properties of the discretized character $\chi_j(\theta_n)$
\begin{align}
& \chi_{\frac{k}{2}-j}(\theta_n) = \chi_j(\theta_n)\ ,\nonumber  \\
&  \chi_{j}(\theta_{k+2-n})  = \chi_j(\theta_n)\ .
\end{align}
Using these properties we can restrict the range of $n$ and $j$ to
$n \in \{1,2,\ldots, \left[\frac{k+1}{2}\right]\}$ and
$j \in \{0,\frac{1}{2},1,\ldots, \left[\frac{k}{4}\right]\}$. 

\begin{figure}[h!]
\begin{subfigure}{.49\textwidth}
\includegraphics[width=7cm]{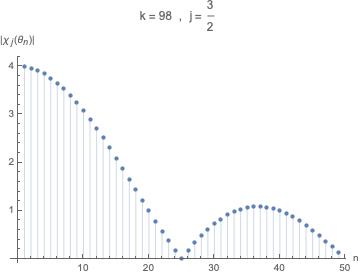}
\subcaption{}
\end{subfigure}
\begin{subfigure}{.49\textwidth}
\includegraphics[width=7cm]{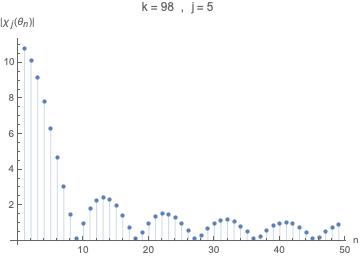}
\subcaption{}
\end{subfigure} \\
\vspace{1cm} \\
\begin{subfigure}{.49\textwidth}
\includegraphics[width=7cm]{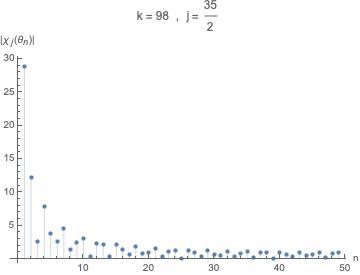}
\subcaption{}
\end{subfigure}
\begin{subfigure}{.49\textwidth}
\includegraphics[width=7cm]{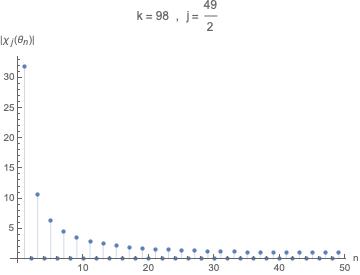}
\subcaption{}
\end{subfigure}
\caption{Figures (a),(b),(c) and (d) above show the variation with the
  discrete eigenvalue label $n$ of the absolute value of the character
  of four different integrable representations $|\chi_j(\theta_n)|$
  ($j=\frac{3}{2},5,\frac{35}{2},\frac{49}{2}$) for $k =98$.}
\label{fig:dcvsn}
\end{figure}

It would be interesting to find a direct analytic proof of \eqref{su2kevmax}. For the present 
we content ourselves with numerical verification. We have checked the
relation \eqref{su2kevmax} for many different values of $k$ and
integrable representations in Mathematica. As an illustration of the
general pattern, in Figure \ref{fig:dcvsn} we plot the absolute values
of the discretized characters $\chi_j(\theta_n)$ vs.~$n$ for various
values of $j$ corresponding to integrable representations while
keeping $k$ fixed. As can be seen in the plots, $|\chi_j(\theta_n)|$
is maximized at $n=1$ in all the cases.

\section{$q$-binomial identities} \label{qi}  

\subsection{A `fermionic' identity}\label{qident}
In this appendix, we prove the following $q$-binomial identity
\begin{equation}\label{qbi}
  A_{N_F,k_F}(x) \equiv \prod_{j=-(N_F-1)/2}^{(N_F-1)/2} (1+q^{j}x) = \sum_{n=0}^{N_F}  \binom{N_F}{n}_{q} x^{n}  \ . 
\end{equation}
This identity holds when $N_F$ is any integer, even or odd.  In each
case the product on the LHS is taken over the $N_F$ values for the
variable $j$ given by\footnote{Note that the above values are
  half-integral when $N_F$ is even and integral when $N_F$ is odd.}
\begin{equation}\label{jval}
\Bigg\{-\frac{N_F-1}{2}, \ -\frac{N_F-3}{2}, \ \cdots, \  \frac{N_F-3}{2}, \ \frac{N_F-1}{2} \Bigg\} \ . 
\end{equation} 
Let us define the expression on the LHS of \eqref{qbi} as $f(x)$,
i.e., we define
\begin{equation}\label{fx}
f(x) = \prod_{j=-(N_F-1)/2}^{(N_F-1)/2} (1+q^{j}x) \ . 
\end{equation}
It is then clear that 
\begin{equation}\label{fqx}
f(qx) = \prod_{j=-(N_F-1)/2}^{(N_F-1)/2} (1+q^{j+1}x)  \ .  
\end{equation}
We use the relabelling $j+1=j'$ to rewrite the RHS of \eqref{fqx} as 
\begin{equation}\label{fqxr}
f(qx) = \prod_{j'=-(N_F-3)/2}^{(N_F+1)/2} (1+q^{j'}x)  \ .  
\end{equation}
This can be written as 
\begin{equation}\label{fqxr2}
f(qx) = \frac{1+q^{(N_F+1)/2}x}{1+q^{-(N_F-1)/2}x}\prod_{j'=-(N_F-1)/2}^{(N_F-1)/2} (1+q^{j'}x)  \ . 
\end{equation}
We thus see that 
\begin{equation}\label{fqxefx}
f(qx) = \frac{1+q^{(N_F+1)/2}x}{1+q^{-(N_F-1)/2}x}  \  f(x) \ . 
\end{equation}
Alternatively, the above equation \eqref{fqxefx} can be rewritten as 
\begin{equation}\label{fqxefx2}
\Big(1+q^{-(N_F-1)/2}x\Big) f(qx) = \Big(1+q^{(N_F+1)/2}x\Big)   f(x) \ . 
\end{equation}
From the definition of the function $f(x)$ in \eqref{fx}, it is clear that it is a polynomial in $x$ of degree $N_F$. In other words, $f(x)$ can be written in the following general form 
\begin{equation}\label{fxp}
f(x) = \sum_{k=0}^{N_F} A_k x^k  \ , 
\end{equation}
where, $A_k$ is to be determined. It is also clear from the structure of the polynomial that $A_0=1$. Substituting \eqref{fxp} in \eqref{fqxefx2}, we see that 
\begin{equation}\label{fqxefx3}
\Big(1+q^{-(N_F-1)/2}x\Big) \sum_{k=0}^{N_F} A_k q^k x^k  = \Big(1+q^{(N_F+1)/2}x\Big)   \sum_{k=0}^{N_F} A_k x^k  \ . 
\end{equation}
Comparing the coefficients of the term $x^k$ from both sides, we get the following recursive relation for $k>0$, 
\begin{equation}\label{Akr}
A_k q^k+A_{k-1} q^{k-1} q^{-(N_F-1)/2} = A_k +A_{k-1} q^{(N_F+1)/2} \ . 
\end{equation}
Simplification of the above expression \eqref{Akr} gives 
\begin{equation}\label{Akr2}
A_k =A_{k-1}  \  \frac{q^{(N_F+1)/2}- q^{k} q^{-(N_F+1)/2} }{q^k-1} \ . 
\end{equation}
This can be written in a more symmetric form as below
\begin{equation}\label{Akr3}
A_k =A_{k-1}  \  \frac{q^{(N_F+1-k)/2}-  q^{-(N_F+1-k)/2} }{q^{k/2}-q^{-k/2}} \ . 
\end{equation}
In terms of $q$-numbers, \eqref{Akr3} can be written as
\begin{equation}\label{Akr4}
A_k =A_{k-1}  \  \frac{[N_F+1-k]_{q}}{[k]_{q}} \ , 
\end{equation}
where recall that
$[x]_q = \frac{q^{x/2} - q^{-x/2}}{q^{1/2} - q^{-1/2}}$. Applying the
successive recursion relations, we get
\begin{equation}\label{Akr5}
A_k =A_{0}  \ \frac{[N_F+1-k]_{q} [N_F+2-k]_{q}\cdots [N_F]_{q}}{[k]_{q}[k-1]_{q}\cdots [1]_{q}} \ . 
\end{equation}
The RHS of \eqref{Akr5} can be rewritten in a simple way
\begin{equation}\label{Akr6}
A_k =A_{0}   \frac{[N_F-k]_{q}! [N_F+1-k]_{q} [N_F+2-k]_{q}\cdots [N_F]_{q}}{[k]_{q}! } \cdot \frac{1}{[N_F-k]_{q}!} \ . 
\end{equation}
which is nothing but
\begin{equation}\label{Akf}
A_k = \frac{[N_F]_{q}!}{[k]_{q}! [N_F-k]_{q}!} = {N_F \choose k}_{q}  \ . 
\end{equation}
where we have used $A_0=1$. This proves the identity \eqref{qbi}.

\subsection{A `bosonic' identity} \label{tbi} 

In this subsection, we prove the identity
\begin{equation}\label{qbinb}
  B_{N_B, k_B}(x) \equiv
  \prod_{j=-(N_B-1)/2}^{(N_B-1)/2} \frac{1}{(1-q^{j}x)} ~\bigg|_{k_B} = \sum_{n=0}^{k_B} {N_B +n-1 \choose n}_q x^n\ ,
\end{equation} 
where $q = \te^{2\pi\i/\kappa_B}$ where we have assumed that $N_B$ and
$k_B$ are positive without loss of generality (for negative $k_B$, we
replace $k_B \to |k_B|$ in all expressions below). The symbol
$\big|_{k_B}$ tells us to expand the expression on the RHS in a power
series in $x$ and truncate this expansion at order $k_B$. In this
subsection, we will determine the expansion coefficients $B_n$ of the
function
\begin{equation}\label{qexpans}
  B_{N_B, k_B,  \alpha}(x) \equiv \sum_{n=0}^{k_B}
  B_n x^n \ . 
\end{equation} 
We will find it convenient to work with the function
\begin{equation}\label{qb}
{g}(x) = \prod_{j=-(N_B-1)/2}^{(N_B-1)/2} \frac{1}{1-q^{j}x} \equiv \sum_{n=0}^{\infty}
b_n x^n  \ . 
\end{equation}
Clearly $=B_n =b_n$ for $n \leq k$, but $B_n=0$ for $n>k$. Knowledge
of $b_n$ thus allows us to determine $B_n$.

We can find the quantities $b_n$ proceeding as in the previous
subsection. The analog of \eqref{fqxefx} is
\begin{equation}\label{fqxefxb}
g(qx) = \frac{1-q^{-(N_B-1)/2}x}{1-q^{(N_B+1)/2}x} g(x) \ ,
\end{equation}
or 
\begin{equation}\label{bfxb}
\left( 1-q^{(N_B+1)/2}x \right)  g(qx) = \left( 1-q^{-(N_B-1)/2}x \right) \ g(x) \ . 
\end{equation}
Plugging the series expansion \eqref{qb} into 
\eqref{bfxb} and equating powers of $x^n$ we find the recursion relation
\begin{equation}\label{recrre}
b_n q^n - b_{n-1} q^{\frac{N_B+2n -1}{2}}= 
b_n- b_{n-1} q^{- \frac{N_B-1}{2}}  \ . 
\end{equation} 
Equivalently, 
\begin{equation}\label{rearrange}
  b_n = \frac{ q^{ \frac{N_B+n -1}{2}}  - q^{- \frac{N_B+n -1}{2}}}{q^{\frac{n}{2}} - q^{-\frac{n}{2}}} \ b_{n-1} \ , 
\end{equation} 
that is,
\begin{equation}\label{rqno}
  b_n = \frac{[N_B+n -1]_q}{[n]_q} \ b_{n-1} \ . 
\end{equation} 
The above equation holds for $n \geq 1$. Note that that $b_0=1$. It
follows from \eqref{rqno} that
\begin{equation}\label{ansoln}
  b_n= \prod_{k=1}^n \frac{[N_B+k -1]_q}{[k]_q} = {N_B +n-1 \choose n}_q \ . 
\end{equation}
and hence
\begin{equation}\label{finbosans}
  B_{N_B, k_B}(x)=\sum_{n=0}^{k_B} {N_B +n-1 \choose n}_q x^n  \ . 
\end{equation} 

\subsubsection{$b_n$ for $n>k_B$}

Even though we do not need this for our paper, for completeness we
study the structure of the quantity $b_n$ for $n>k_B$. The reader who
is uninterested in this diversion should feel free to skip to the next
subsection. In order to proceed, let us recall that when
$q = \te^{\frac{2 \pi \i}{\kappa_B}}$, we have
\begin{equation}\label{propofqr} 
[m\,\kappa_B +a]_q= (-1)^m ~[a]_q \ , 
\end{equation} 
for any integer $m$. In particular
\begin{equation}\label{propofq}
  [m\, \kappa_B]_q = 0 \ .
\end{equation} 
Notice that the numerator in the $q$-binomial coefficient
\eqref{ansoln} has a factor of $[\kappa_B]_q$ whereas the denominator
has no such factor whenever $\kappa_B > n \geq k_B+1$. It follows that
\begin{equation}\label{anvan} 
  b_n=0 \quad{\rm for}\quad k_B+1 \leq n \leq \kappa_B-1 \ . 
\end{equation} 
(compare with e.g. the discussion around \eqref{condforvan}).

Moreover when $n= \kappa_B$, there is a factor of $[\kappa_B]$ in both
the numerator and the denominator and hence $b_{\kappa_B}$ has a
non-zero limiting value when we take $q \to \te^{2\pi\i/\kappa_B}$:
\begin{align}\label{anmanip}
  b_{\kappa_B} &= \frac{ \prod_{j=1}^{k_B}[ N_B+j-1]_q\ [\kappa_B]_q\ \prod_{r=1}^{N_B-1} [ \kappa_B + r ]_q}{\prod_{s=1}^{\kappa_B} [ s]_q}\ ,\nonumber\\
               &= \frac{ \prod_{j=1}^{k_B}[ N_B+j-1]_q\ [\kappa_B]_q\ \prod_{r=1}^{N_B-1} (- [r]_q)}{\prod_{s=1}^{\kappa_B} [s]_q}\ ,\nonumber\\
               &= (-1)^{N_B-1} \frac{ \prod_{j=1}^{k_B}[ N_B+j-1 ]_q\ [\kappa_B]_q\ \prod_{r=1}^{N_B-1} [r]_q }{\prod_{s=1}^{\kappa_B} [ s ]_q}\ ,\nonumber\\
               &= (-1)^{N_B-1} \frac{\prod_{j=1}^{\kappa_B} [ j]_q }{\prod_{s=1}^{\kappa_B} [ s]_q} =(-1)^{N_B-1} \ . 
\end{align}
More generally, the reader can easily convince herself that for
$j\leq 0 < |\kappa_B|$
\begin{equation}\label{genresofa}
  b_{m \kappa_B + j} = (-1)^{m(N_B-1)} \ b_j \ . 
\end{equation} 
It then follows from \eqref{anvan} that 
\begin{align}\label{finarresgx} 
  g(x)&=\Big( 1+ (-1)^{N_B-1} x^{\kappa_B} +(-1)^{2(N_B-1)} x^{2\kappa_B} + \ldots \Big)  \sum_{n=0}^{k_B}{N_B +n-1 \choose n}_q x^n\ ,\nonumber  \\
      &=\frac{1} {1-(-1)^{N_B-1} x^{\kappa_B}} \sum_{n=0}^{k_B}
      {N_B +n-1 \choose n}_q x^n  \ . 
\end{align}
In particular, when $N_B$ is odd we get
\begin{equation}\label{whennodd}
  g(x)=\frac{1}{1-x^{\kappa_B}} \sum_{n=0}^{k_B}{N_B +n-1 \choose n}_q x^n  \ , 
\end{equation} 
whereas, when $N_B$ is even we get 
\begin{equation}\label{whenneven}
g(x)=\frac{1}{1+x^{\kappa_B}} \sum_{n=0}^{k_B}
{N_B +n-1 \choose n}_q x^n \ . 
\end{equation}  

\subsection{A Bose-Fermi identity}
In this subsection, $k_B$ and $k_F$ can have either sign. We
demonstrate that the quantities $A_{N_F,|k_F|}$ \eqref{qbi} and
$B_{N_B,|k_B|}$ \eqref{qbinb} are equal to each other under the
Bose-Fermi duality map $N_B = |k_F|$, $N_F = |k_B|$. That is,
\begin{equation}\label{qequive}
  B_{N_B, |k_B|}(x)= A_{|k_B|, N_B}(x) \ ,
\end{equation} 
We reproduce the formulae for $A$ and $B$ below:
\begin{align}\label{ABdefr}
  B_{N_B, k_B}(x) &\equiv
                    \prod_{j=-(N_B-1)/2}^{(N_B-1)/2} \frac{1}{(1-q^{j}x)} ~\bigg|_{k_B} = \sum_{n=0}^{k_B} {N_B +n-1 \choose n}_q x^n\ ,\nonumber\\
    A_{N_F,k_F}(x) &\equiv \prod_{j=-(N_F-1)/2}^{(N_F-1)/2} (1+q^{j}x) = \sum_{n=0}^{N_F}  {N_F \choose n}_{q} x^{n}  \ . 
\end{align}
We show \eqref{qequive} in two ways. First, we show that the following
$q$-combinatorial identity holds:
\begin{equation}
  \binom{k}{n}_q = \binom{N + n-1}{n}_q\ ,
\end{equation}
for any two positive integers $k$ and $N$ which satisfy
$k + N = \kappa$ with $q = \te^{2\pi\i/\kappa}$. 

We have, by definition,
\begin{equation}\label{qbindef}
  {k \choose n}_{q}  = \frac{[k]_{q}!}{[n]_{q}![k-n]_{q}!} = \frac{[k]_{q}[k-1]_{q}\cdots [k-n+1]_{q}}{[n]_{q}!} \ . 
\end{equation}
Using $q = \te^{2\pi\i/\kappa}$ and $k + N = \kappa$, it is not hard
to see that
\begin{equation}\label{qnid}
  [k-i]_{q} = [N+i]_{q}\ ,\quad\text{for every}\quad i =0,\ldots,k \ ,
\end{equation}
which gives
\begin{equation}\label{qbindef1}
  {k \choose n}_{q} = \frac{[N]_{q}[N+1]_{q}\cdots [N+n-1]_{q}}{[n]_{q}!} = \binom{N+n-1}{n}_q\ .
\end{equation}
Since $\kappa = N_B + |k_B| = N_F + |k_F|$, the Bose-Fermi duality map
then gives
\begin{equation}
  \binom{N_F}{n}_q = \binom{|k_B|}{n}_q =   \binom{N_B+n-1}{n}_q\ ,
\end{equation}
from which \eqref{qequive} follows.

The second proof of \eqref{qequive} uses the product form of $A$ and
$B$. It is convenient to present the proof separately for four cases.

\subsubsection{$N_B$ odd}

In this case, the product in the expression for $B$ \eqref{ABdefr} is
taken over the $N_B$ values of $q^j$ that are $\kappa^{\rm th}$ roots
of unity. To process that formula we multiply both the numerator and
the denominator by $\prod_{q'} (1-q' x)$ where $q'$ run over the
remaining $|k_B|$ $\kappa^{\rm th}$ roots of unity:
\begin{equation}\label{qbinpro}
  B_{N_B, |k_B|}(x)= \left[\prod_{q'} (1-q' x) \prod_{j=0}^{\kappa-1}(1-q^{j}x)^{-1}\right]_{|k_B|} \ ,
\end{equation} 
Now, when $k_B$ is odd, then $\kappa = N_B + |k_B|$ is an even
integer. This implies that $-q'$ is also a $\kappa^{\rm th}$ root of
unity. Relabelling $q'$ as $-q'$ in the numerator gives a product over
$|k_B|$ $\kappa^{\rm th}$ roots of $-1$. When $k_B$ is an even
integer, $\kappa$ is an odd integer. Then $-q'$ is a $\kappa^{\rm th}$
root of $-1$ rather than of unity. Relabelling $q' \to -q'$ now gives
a product over $|k_B|$ $\kappa^{\rm th}$ roots of $-1$ in the
numerator. Thus, for either $|k_B|$ even or odd, we can write
\begin{align}\label{qbinpro1}
  B_{N_B, |k_B|}(x) &= \left[\prod_{q'} (1 + q' x) \prod_{j=0}^{\kappa-1}(1-q^{j}x)^{-1}\right]_{|k_B|}\ ,\nonumber\\
                    &= \left[\prod_{j=-(|k_B|-1)/2}^{(|k_B|-1)/2} (1+q^j x) \prod_{j=0}^{\kappa-1}(1-q^{j}x)^{-1}\right]_{|k_B|} \ ,
\end{align}
where one can indeed check that the product in the numerator is a
product over $|k_B|$ $\kappa^{\rm th}$ roots of unity when $k_B$ is
odd and $\kappa^{\rm th}$ roots of $-1$ when $k_B$ is even. Next,
using the identity
\begin{equation}\label{identprodroo}
  \prod_{j=0}^{\kappa-1}(1-q^{j}x)=1-x^\kappa \ , 
\end{equation} 
we find that \eqref{qbinpro} simplifies to
\begin{equation}\label{qbinpron}
B_{N_B, |k_B|}(x)= \left[ \prod_{j=-(|k_B|-1)/2}^{(|k_B|-1)/2} (1+q^j x) ( 1-x^\kappa)^{-1}\right]_{|k_B|} \ . 
\end{equation}
As in the discussion around \eqref{losnexen}, the denominator in
\eqref{qbinpron} admits a power series expansion in $x^\kappa$; none
of the non-trivial terms in this expansion contribute once we truncate
the expansion at order $|k_B|$ and so the denominator can simply be
replaced by unity. On the other hand the numerator is already a
polynomial of degree $|k_B|$. It follows that \eqref{qbinpron} further
simplifies to
\begin{equation}\label{qbinpronf}
B_{N_B, |k_B|}(x)= \prod_{j=-(|k_B|-1)/2}^{(|k_B|-1)/2} (1+q^j x)  \ ,
\end{equation}
which, upon using $|k_B| = N_F$ gives the product form for $A$ in
\eqref{ABdefr}.

\subsubsection{$N_B$ even}

In this case the product for $B$ in \eqref{ABdefr} is taken over $N_B$
values of $q^j$ which are all $\kappa^{\rm th}$ roots of $-1$. As in
the previous subsubsection we multiply both the numerator and the
denominator by $\prod_{q'} (1-q' x)$ where $q'$ run over the remaining
$|k_B|$ $\kappa^{\rm th}$ roots of $-1$ to get
\begin{equation}\label{qbinproee}
  B_{N_B, |k_B|}(x)= \left[\prod_{q'} (1-q' x) \prod_{j=0}^{\kappa-1}(1-q^{j+\frac{1}{2}}x)^{-1}\right]_{|k_B|} \ ,
\end{equation} 
When $k_B$ is an odd integer, $\kappa = N_B + |k_B|$ is an odd
integer. Thus, $-q'$ is a $\kappa^{\rm th}$ root of unity rather than
of $-1$. When $k_B$ is even, $\kappa$ is even and $-q'$ is also a
$\kappa^{\rm th}$ root of $-1$. Relabelling $q' \to -q'$ in the
numerator, we get a product over $|k_B|$ $\kappa^{\rm th}$ roots of unity when $k_B$ is odd and $\kappa^{\rm th}$ roots of $-1$ when $k_B$ is even. In either case, we can write
\begin{equation}\label{qbinproee1}
  B_{N_B, |k_B|}(x)= \left[\prod_{j=-(|k_B|-1)/2}^{(|k_B|-1)/2} (1+q^j x) \prod_{j=0}^{\kappa-1}(1-q^{j+\frac{1}{2}}x)^{-1}\right]_{|k_B|} \ ,
\end{equation} 
Next, using the identity
\begin{equation}\label{identprodroos}
  {\prod_{j=0}^{\kappa-1}(1-q^{j+\frac{1}{2}}x)} = 1 + x^\kappa \ , 
\end{equation} 
we find that \eqref{qbinproee1} simplifies to
\begin{equation}\label{qbinprons}
  B_{N_B, |k_B|}(x)= \left[ \prod_{j=-(|k_B|-1)/2}^{(|k_B|-1)/2} (1+q^j x)  (1+x^\kappa)^{-1}\right]_{|k_B|} \ . 
\end{equation}
As in the previous case the truncation to order $|k_B|$ allows us to
discard the contributions from $(1 + x^\kappa)^{-1}$ and simplify
\eqref{qbinproee1} to
\begin{equation}\label{qbinpronfs}
  B_{N_B, |k_B|}(x)= \ \prod_{j=-(|k_B|-1)/2}^{(|k_B|-1)/2} (1+q^jx')  \ . 
\end{equation}
Plugging in $|k_B| = N_F$ once again yields \eqref{qequive}.

\section{Entropy as a function of occupation numbers} \label{thermo} 

\subsection{Entropy as a function of occupation number for free
  fermions}

In the limit $k_F \to \infty$, the partition function of any number of
fundamental particles occupying the state with energy $E_a$ is given
by
\begin{equation}\label{pffree}
  \mf{z}_F(y_a) = \sum_{n=0}^{N_F} {N_F \choose n} y_a^n =(1+y_a)^{N_F}  \ ,
\end{equation} 
with $y_a = \te^{-\beta(E_a - \mu)}$. The average occupation number of
this state is given by
\begin{equation}\label{avoccno} 
  n_a = \frac{ \sum_{n=0}^{N_F}  n y_a^n ~{N_F \choose n}  }{\mf{z}_F} = y_a \partial_{y_a}  \log \mf{z}_F  =  \frac{N_F}{y_a^{-1} +1} \ . 
\end{equation} 
The entropy is then determined by \eqref{relgo}
i.e.
\begin{equation}\label{relgoapp}
  S_{F} = -n_a(y_a) \log y_a + \log \mf{z}_F(y_a)\ .
\end{equation}
Inverting \eqref{avoccno} to solve for $y_a$ in favour of $n_a$ we
find
\begin{equation} \label{solveforw} 
\frac{1}{y_a}= \frac{N_F}{n_a} -1 \ . 
\end{equation} 
Substituting \eqref{solveforw} into \eqref{pffree} and
\eqref{relgoapp}, we find
\begin{align}\label{newent}
  S_F(n_a) & = n_a \log \left( \frac{N_F-n_a}{n_a} \right)  + N_F \log \left( \frac{N_F}{N_F-n_a} \right)\ ,\nonumber \\
         &= -( N_F-n_a) \log ( N_F-n_a)  -n_a \log n_a  + N_F \log N_F   \ . 
\end{align}
In \eqref{newent} we have succeeded in rewriting the entropy of this
single state as a function of its occupation number. It is now easy to
take the derivative of the entropy with respect to $n_a$; we find
\begin{equation}\label{derent} 
  \partial_{n_a} S_F(n_a) = \log \left( \frac{N_F-n_a}{n_a} \right)  \ . 
\end{equation} 
As explained in Section \ref{entropyfun}, we can now extremize the
entropy of our system at fixed energy and charge. The quantity defined
in \eqref{whatext}, i.e.
\begin{equation}\label{whatextn}
  \sum_a (S_F(n_a) - \beta E_a n_a + \beta \mu n_a)\ ,
\end{equation} 
is now an explicit function of $n_a$. Extremizing with respect to
$n_a$ we find
\begin{equation}\label{howext}
\log \left( \frac{N_F-n_a}{n_a} \right) -\beta E_a +\beta \mu =0 \ . 
\end{equation} 
Solving \eqref{howext} for $n_a$ we obtain
\begin{equation}\label{solutionext}
  n_a=  \frac{N_F}{\te^{\beta (E_a - \mu)}+1} \ ,
\end{equation} 
in agreement with \eqref{avoccno}.

\subsection{Entropy as a function of occupation number for free bosons} 

In the case of free bosons, the single-state partition function of any
number of fundamental bosons occupying a single-particle state of
energy $E_{a}$ at finite chemical potential $\mu$, is given by
\begin{equation}\label{sspfb}
\mf{z}_B(y_a) = \sum_{n=0}^{\infty} \ {N_B+n-1 \choose n} \ y_a^{n}  = 
\frac{1}{(1-y_a)^{N_B}} \ , 
\end{equation}
where $y_a=\te^{-\beta(E_a-\mu)}$. The occupation number in this state
is given by
\begin{equation}\label{ocpnb}
  n_{a} = \frac{1}{\mf{z}_B(y_a)} \sum_{n=0}^{\infty} \ {N_B+n-1 \choose n} \ n y_a^{n}  = y_{a} \partial_{y_a} \log \mf{z}_B = \frac{N_B}{y_a^{-1}-1} \ . 
\end{equation}
It is easy to invert the final expression of \eqref{ocpnb} to write
$y_a$ as a function of $n_a$ as below
\begin{equation}\label{wasnab}
{y_a} =\frac{n_a}{N_B+n_a} \ . 
\end{equation}
We now set to compute the entropy of this single energy state by using
the expression \eqref{relgoapp}. Inserting $y_a$ as a function of
$n_a$ in \eqref{relgoapp}, we get
\begin{align}\label{entfrbos}
  S_B(n_a) & =- n_a \log \left( \frac{n_a}{N_B+n_a} \right) -N_B  \log  \left( \frac{N_B}{N_B+n_a}  \right) \\
           & = - n_a \log n_a + (N_B+n_a) \log (N_B+n_a) -N_B \log N_B  \ . 
\end{align}

\subsection{Explicit expressions in the large volume, large $N$ 't
  Hooft limit}
\subsubsection{For fermions}\label{fermentln}
In this section, we study the problem of determining the entropy as a
function of occupation numbers for fermions in the large $N_F$ 't
Hooft limit. Recall the formulae for the occupation numbers
\eqref{nexp} in the large volume limit,
\begin{align}\label{nexpapp}
n_F(y) &= \sum_{j=-(N_F-1)/2}^{(N_F-1)/2} \frac{1}{y^{-1}q^{-j}+1}\ ,\nonumber \\
n_B(y) &= \left( \sum_{j=-(N_B-1)/2}^{(N_B-1)/2} \frac{1}{ y^{-1} q^{-j}-1} \right)  -  \frac{|\kappa_B|}{1-y^{-|\kappa_B|}(-1)^{N_B-1} }\ ,
\end{align}
In the large $N$ limit, it is convenient to define the occupation
number per colour
\begin{equation}\label{avfermoc}
\bar{n}_F = \frac{n_a}{N_F} \ . 
\end{equation} 
At leading order in the large $N_F$ limit, the summation in
\eqref{nexpapp} turns into an integral, and we find (see Eq 5.6 of
\cite{Minwalla:2020ysu})
\begin{equation}\label{nfb}
\bar{n}_{F}(y) = \frac{1}{2} - \frac{1}{\pi|{\lambda}_F|} \tan^{-1} \Big(\frac{1-y}{1+y}\tan \frac{\pi|{\lambda}_F|}{2} \Big)\ . 
\end{equation}
This expression is easily inverted; making the $\tan^{-1}$ the
`subject of the formula' and taking the tan of both sides give
\begin{equation}\label{invone}
  \frac{1-y}{1+y} = \frac{\tan\big( \frac{\pi|\lambda_F|}{2}(1-2\bar{n}_{F} ) \big)}{\tan \big( \frac{\pi|{\lambda_F}|}{2} \big) }   \ . 
\end{equation}
Using \eqref{invone} it is easy to solve for $y$ in terms of
${\bar n}_F$. Simplifying the resultant expression we find
\begin{equation}\label{xnFi}
y = \frac{\sin \big(\pi|{\lambda}_F| \bar{n}_F\big)}{\sin \big(\pi|{\lambda}_F| (1-\bar{n}_F) \big)} = \frac{[n_F]_{q}}{[N_F-n_F]_{q}} \ . 
\end{equation}
Using the equation $n_a = y_a \partial_{y_a} \log \mf{z}_F(y_a)$, we find
\begin{equation}\label{intexpforpf}
  \log \mf{z}_F(y) = N_{F} \int \bar{n}_F \ \frac{dy}{y} \ .
\end{equation}
It follows from \eqref{xnFi} that 
\begin{equation}\label{differntialform} 
\frac{dy}{y}=  \pi |\lambda_F| \Big(\cot\big(\pi|\lambda_F| \bar{n}_F\big)+\cot \big(\pi|\lambda_F| (1-\bar{n}_F) \big) \Big) \ d \bar{n}_F \ . 
\end{equation} 
Substituting \eqref{differntialform} into \eqref{intexpforpf} and integrating 
with respect to ${\bar n}_F$ we find 
\begin{equation}\label{zint} 
\log \mf{z}_F =  N_{F} \pi |\lambda_F| \int \bar{n}_F \ \Big(\cot\big(\pi|\lambda_F| \bar{n}_F\big)+\cot \big(\pi|\lambda_F| (1-\bar{n}_F) \big) \Big) \ d \bar{n}_F \ . 
\end{equation}
Similarly, using \eqref{abder} for the entropy i.e.
\begin{equation}\label{abderapp}
\partial_{n}  S =- \log y\ ,
\end{equation}
and using \eqref{xnFi}, we get
\begin{equation}\label{enfaf1}
S_F  = - \int^{{n}_a}_{0}  \bigg(\log \big( [n'_a]_{q}\big) -\log \big( [N_F-n'_a]_{q}\big)  \bigg) \ d {n}'_a \ . 
\end{equation}
We could have gotten the above result \eqref{enfaf1} by using
\eqref{relgoapp} and the expression \eqref{zint} for $\log
\mf{z}_F$. Indeed, performing the integral in \eqref{zint} (and fixing
the integration constant using the condition that the entropy vanishes
when $n_a$ vanishes) we find
\begin{align}\label{enf}
S_F & = -N_F \Bigg[\bar{n}_F \log \frac{\sin \big(\pi|\lambda_F| \bar{n}_F\big)}{\sin \big(\pi|\lambda_F| (1-\bar{n}_F) \big)}\nonumber \\ 
    &\qquad - \pi |\lambda_F| \int^{\bar{n}_F}_{0}  \bar{n}'_F \ \Big(\cot\big(\pi|\lambda_F| \bar{n}'_F\big)+\cot \big(\pi|\lambda_F| (1-\bar{n}'_F) \big) \Big) \ d \bar{n}'_F\Bigg] \ . 
\end{align}
The expression on the RHS of \eqref{enf} can be manipulated as follows. Performing integration by parts in the second line of the above integral, \eqref{enf} can be rewritten as 
\begin{align}\label{enfa}
  S_F & = -N_F \Bigg[\bar{n}_F \log \frac{\sin \big(\pi|\lambda_F| \bar{n}_F\big)}{\sin \big(\pi|\lambda_F| (1-\bar{n}_F) \big)}\nonumber \\ 
      &\qquad -  \bar{n}_F \ \Big(\log \sin \big(\pi|\lambda_F| \bar{n}_F\big)-\log \sin  \big(\pi|\lambda_F| (1-\bar{n}_F) \big) \Big) \\
      &\qquad +  \int^{\bar{n}_F}_{0}  \Big(\log \sin \big(\pi|\lambda_F| \bar{n}'_F\big)-\log \sin  \big(\pi|\lambda_F| (1-\bar{n}'_F) \big) \Big) \ d \bar{n}'_F \Bigg] \ . 
\end{align}
The terms on the first and second line on the RHS of the above
expression \eqref{enfa} cancel each other so \eqref{enfa} takes the
simplified form
\begin{align}\label{enfaf}
S_F & = -N_F \int^{\bar{n}_F}_{0}  \Big(\log \sin \big(\pi|\lambda_F| \bar{n}'_F\big)-\log \sin  \big(\pi|\lambda_F| (1-\bar{n}'_F) \big) \Big) \ d \bar{n}'_F\ ,\nonumber \\
    &= - \int^{{n}_a}_{0}  \bigg(\log \sin \Big(\frac{\pi}{|\kappa_F|} {n}'_a\Big)-\log \sin  \Big(\frac{\pi}{|\kappa_F|} (N_F-{n}'_a) \Big) \bigg) \ d {n}'_a\ ,\nonumber \\
    & = - \int^{{n}_a}_{0}  \bigg(\log \big( [n'_a]_{q}\big) -\log \big( [N_F-n'_a]_{q}\big)  \bigg) \ d {n}'_a \ . 
\end{align}
As a check on the formula \eqref{enfaf1} we note that in the limit
$\lambda_F\to 0$, \eqref{enfaf1} gives
\begin{equation}
  S_F  = -N_F \bigg( \bar{n}_F \log \bar{n}_F+ (1-\bar{n}_F) \log
  (1-\bar{n}_F) \bigg) \ ,
\end{equation}
in perfect agreement with \eqref{newent}. 

\subsubsection{For bosons}\label{bosentln}
We repeat the analysis of the previous subsubsection \ref{fermentln}
here in the case of bosons. As in the case of fermions, it is useful
to define the average occupation number for bosons as
\begin{equation}\label{nbarbdef}
\bar{n}_B = \frac{n_a}{N_B} \ . 
\end{equation}
In the strict large $N_B$ limit, the average occupation number has
been worked out in detail in \cite{Minwalla:2020ysu} (see Eq $5.21$ of
\cite{Minwalla:2020ysu}). It is given by
\begin{equation}\label{nbarbln}
  \bar{n}_{B} = \frac{1-|\lambda_B|}{2|\lambda_B|} - \frac{1}{\pi|\lambda_B|} \tan^{-1} \bigg(\frac{1-y}{1+y}\cot \frac{\pi|\lambda_B|}{2} \bigg) \ . 
\end{equation}
It is useful to invert the above expression \eqref{nbarbln} to write
$y$ as a function of $\bar{n}_B$. Inverting \eqref{nbarbln}, we get
\begin{equation}\label{wfonbarb}
\frac{1-y}{1+y}  =  \frac{\cot\Big( \frac{\pi|\lambda_B|}{2}(1+2\bar{n}_{B} ) \Big)}{\cot \Big( \frac{\pi|\lambda_B|}{2} \Big) } \ . 
\end{equation}
Further simplification of \eqref{wfonbarb} gives 
\begin{equation}\label{wbos}
y = \frac{\sin \big(\pi|\lambda_B| \bar{n}_B \big)}{\sin \big(\pi|\lambda_B| (1+ \bar{n}_B)\big)} = \frac{[n_B]_q}{[n_B + N_B]_q} \ . 
\end{equation}
The entropy can again be obtained by \eqref{abderapp} and is given by
\begin{equation}\label{entbf}
S_B = - \int^{{n}_a}_{0}  \bigg(\log \big( [n'_a]_{q}\big) -\log \big( [N_B+n'_a]_{q}\big)  \bigg) \ d {n}'_a \ . 
\end{equation}
As a check on the formula \eqref{entbf} we note that in the limit
$\lambda_B\to 0$, \eqref{entbf} gives
\begin{align}\label{frlimentb}
  S_B & = N_B \Bigg( - \bar{n}_B \log \bar{n}_B+ (1+\bar{n}_B) \log (1+\bar{n}_B) \bigg) \ ,\nonumber \\
    & = - n_a \log n_a + (N_B+n_a) \log (N_B+n_a) -N_B\log N_B  \ , 
\end{align}
in perfect agreement with \eqref{entfrbos}.

\section{One-loop determinants in $U(N)_{k,k'}$ Chern-Simons theories}\label{oneloop}

Following Blau and Thompson \cite{Blau:1993tv} (see also Section 4.1
of \cite{Jain:2013py}) one may evaluate the path integral of pure
Chern-Simons theories on $\Sigma_g \times S^1$ as follows. Denote the
components of the gauge field along the two dimensional surface
$\Sigma$ as $A_\alpha$, $\alpha = 1,2$, and the time component as
$A_0$, and make the following choice of gauge:
\begin{itemize} 
\item[(1)] We demand $\partial_0 A_0 = 0$. This choice leaves time
  independent- so effectively two-dimensional - $U(N)$ gauge
  transformations unfixed.
\item[(2)] $A_0$ (which is now constant in time) is then chosen to be a
  diagonal matrix at each point on $\Sigma_g$. This choice is similar
  to one adopted by 't Hooft in his classic studies of confinement in
  gauge theories, but this time in two rather than four spacetime
  dimensions. 
\end{itemize}
The gauge invariance which is not fixed by (1) and (2) are diagonal
gauge transformations that are constant in time, corresponding to a
$U(1)^N$ gauge group in two dimensions. These affect only the constant
mode along $S^1$ of the gauge field $A_\alpha$ along $\Sigma_g$, which
we label $\mathring{A}_\alpha$. Further, only the diagonal components
of $\mathring{A}_\alpha$ transform as gauge fields of the $U(1)^N$
whereas the off-diagonal components transform in the `adjoint
representation' of $U(1)^N$ i.e.~the $ij^{\rm th}$ off-diagonal
element $(\mathring{A}_\alpha)_{ij}$ transforms with charge $-1$ under the
$i^{\rm th}$ $U(1)$ and with charge $+1$ under the $j^{\rm th}$
$U(1)$.
\begin{itemize}
\item[(3)] The $U(1)^N$ abelian gauge invariance can be fixed by
  imposing the Coulomb gauge on the diagonal components of
  $\mathring{A}_\alpha$: $\partial_\alpha \mathring{A}^\ut_\alpha = 0$. We have
  now completely fixed (small) gauge invariance.
\end{itemize}	
The gauge choices (1), (2) and (3) above leaves us with the path
integral over a gauge-fixed abelian two dimensional theory. The gauge
fixing procedure has singularities when two eigenvalues collide; as
explained in detail in section 4.1 of \cite{Jain:2013py}, these
singularities are codimension 1 in field space and divide field space
into sectors. From the point of view of the resultant 2 dimensional
abelian theory, the different sectors correspond to working with
different flux backgrounds. In other words the net effect of the 't
Hooftian singularities in field space is to force us to sum over
(correctly quantized) $U(1)^N$ fluxes. The summation over these flux
sectors played a central role in Sections \ref{suNpath} and
\ref{UNpath}. In particular the summation over $m_i$ \eqref{csmf} is
the summation over these fluxes, while the term
$w_i^{ -\kappa m_i - s \sum_{j} m_j}$ in the same equation is the
classical Chern-Simons action (in the dimensional regulation scheme)
of a configuration that contains both flux and nonzero holonomies. In
this Appendix we will explain that the remaining terms in
\eqref{csmf}, namely
\begin{equation}\label{whatwant} 
\prod_{i<j} |w_i-w_j|^{2-2g} \times (-1)^{(N-1)\sum_i m_i}\ ,
\end{equation}
make up the one-loop contribution to the Chern-Simons partition
function evaluated in a given flux sector\footnote{In \eqref{csmf}
  we specialized to the case $g=0$, but we will describe the
  generalization to a general genus $g$ surface in this Appendix.}.
	    
To proceed it is useful to define some notation. We use the
superscript ${\bf t}$ to denote a matrix that is purely diagonal
whereas a superscript ${\bf k}$ denotes a matrix that is purely off
diagonal. The $\delta$ function that implements the first two gauge
conditions (i.e. all but the Coulomb gauge fixing) is given by
\begin{equation}
  \int [d B] \exp\left(\i \int d^3x\, \tr(B A_0)\right)\ ,
\end{equation}
where $B = B^{\ut} + B^{\uk}$ is an adjoint-valued bosonic Lagrange
multiplier constrained so that the constant (on the $S^1$) part of
$B^{\bf t}$ vanishes. This constraint is needed to ensure that $B$
actually implements the first two gauge conditions above rather than
simply setting all of $A_0$ to zero. The Faddeev-Popov determinant
associated with the first two gauge conditions above is generated by
the path integral
\begin{equation}
  \int [dc d\bar{c}]\,\exp\left(\i\int d^3x\,\tr\left(\bar{c} D_0 c\right)\right)\ ,
\end{equation}
where $c$ and ${\bar c}$ are fermionic $U(N)$ matrices such that the
zero mode (on $S^1$) of $c^{\bf t}$ and $\bar{c}^{\ut}$ are
constrained to vanish\footnote{The constraint on $\bar{c}^\ut$ is
  required since it must obey the same constraints as the Lagrange
  multiplier $B$. The constraint on the zero mode (along $S^1$) of
  $c^{\ut}$ is needed since $c^{\ut}$ takes the place of an
  infinitesimal gauge parameter in BRST transformations, and the gauge
  conditions do not fix constant diagonal gauge transformations.}.
Completely explicitly our path integral is generated by the effective
action
\begin{multline}\label{gfact}
  \mc{S}_{\rm CS}[A,c,\bar{c},B] = \frac{\i k}{2\pi}\int_{\Sigma \times S^1} d t\, \tr\left(A_0^{\ut}d A^{\ut} + A^{\ut}\partial_0 A^{\ut} + A^{\uk} D_0 A^{\uk}\right) \\ + \frac{\i (s+1)}{2\pi}\int_{\Sigma \times S^1} d t\, \left(\tr(A_0^{\ut})d \tr(A^{\ut}) + \tr(A^{\ut})\partial_0 \tr(A^{\ut})\right) \\ + \i \int d^3x\, \tr(B A_0) + \i\int d^3x\,\tr\left(\bar{c}^{\uk} D_0 c^{\uk} + \bar{c}^{\ut}\partial_0 c^{\ut}\right)\ ,
\end{multline}
and also includes the delta function that imposes the $U(1)^N$ Coulomb
gauge condition\footnote{As the gauge invariance that is fixed by this
  condition is abelian, the corresponding Faddeev-Popov determinant is
  field independent and so can be ignored.}.

The zero mode (on $S^1$) of $A^{\bf t}$ contributes only to the first
terms on the first and second lines of \eqref{gfact}. As explained
under equation 4.4 in \cite{Jain:2013py}, the integral over the zero
modes of $A^{\bf t}$ yields a delta function that forces
$\nabla^2 A^{\ut}_0$ to vanish on the $S^2$ and consequently forces
$A^\ut_0$ to be a constant on the $S^2$. Invariance under large gauge
transformations i.e.~gauge transformations which are constant on $S^2$
but not connected to identity, forces $A^\ut_0$ to be periodic so that
the holonomy $U=\te^{\i \beta A^\ut_0}$ is the only physical
variable\footnote{This works as follows. Let us study consider
  following gauge transformations
\begin{equation}
g_\lambda = \exp\left(2\pi\i \beta^{-1} t \lambda\right)\ ,\quad\text{where}\quad \lambda = \diag\{\lambda_1,\ldots,\lambda_N\}\quad\text{with}\quad \lambda_i \in \mbb{Z}\ .\nonumber
\end{equation}
These are large gauge transformations since they are periodic under
$t \to t + \beta$ but are not continuously connected to the identity
transformation. The diagonal gauge field $A^\ut_0$ transforms under
$g_\lambda$ as
\begin{equation}
A^\ut_0 \to A^\ut_0 + \frac{2\pi}{\beta} \diag\{\lambda_1,\ldots,\lambda_N\}\ .\nonumber
\end{equation}
Due to this shift symmetry by the lattice $\mbb{Z}^N$, the components
of $A^\ut_0$ are compact scalars and the holonomy
$U = \te^{\i\beta A^\ut_0}$ is invariant under large gauge
transformations.}.

It now remains to perform the path integral over $c$ and ${\bar c}$,
as well as the integral over $A$. This path integral is quadratic and
so yields the ratio of determinants
\begin{equation}\label{1lpdet}
   \frac{\text{det}_{S} (D_0)}{\sqrt{\text{det}_{V}(D_0)}}\ ,
\end{equation}
where the numerator of this expression is the Faddeev-Popov
determinant and the denominator is the path integral over the gauge
fields $A$, omitting the zero modes (on $S^1$) of the diagonal parts
of $A$ since have already been integrated out (recall that they were
responsible for setting $A^\ut_0$ to be constant on
$S^2$).\footnote{Of course we are only interested in the field
  dependent contributions to \eqref{1lpdet}; only off diagonal modes
  give such contributions both to the numerator and for the
  denominator.}

As the operator $D_0$ has no spatial derivatives, the determinants in
both the numerator and denominator are highly divergent. At least
roughly, however, it is not difficult to convince oneself that these
divergences cancel between the numerator and denominator. A rough (and
as we will see below, importantly inaccurate) evaluation of the ratio
of determinants proceeds as follows (see Section 4.4 of
\cite{Jain:2013py}). Any vector field on a genus $g$ surface
$\Sigma_g$ is a linear combination over the $2g$ nontrivial harmonic
vector fields plus the gradient of a scalar $\phi$ plus the Hodge dual
of the gradient of a second scalar $\chi$. The zero modes (on
$\Sigma_g$) of the scalar fields $\chi$ and $\phi$ do not appear in
the formula for $A$ and so should be omitted in this construction. The
contribution to the determinant from the non-zero modes of $\phi$ and
$\chi$ cancel the part of the determinant in the numerator that comes
from the non-zero modes (on $\Sigma_g$) of the ghosts. Of course we
should really proceed by carefully regulating both determinants before
cancelling. Boldly (and unwisely) ignoring such subtleties, we find
that the numerator receives contributions only from zero modes on
$\Sigma_g$ (but from all modes along $S^1$) and so evaluates to
\begin{equation}\label{vandernum}
  \prod_{1\leq i<j \leq N} |w_i-w_j|^2\ .
\end{equation}
On the other hand the denominator receives contributions only from the
coefficients of the $2g$ nontrivial harmonic vector fields, one factor
of \eqref{vandernum} from each. Including the square root, the
contribution from the gauge field is
$$ \frac{1}{\prod_{1\leq i<j \leq N} |w_i-w_j|^{2g}}.$$
Putting the contributions of the numerator and denominator together,
we obtain
\begin{equation}\label{smg} 
\prod_{1 \leq i<j \leq N} |w_i-w_j|^{2-2g} \ ,
\end{equation} 
in partial agreement with \eqref{whatwant}.

The arguments of the previous paragraph were crude, and indeed turn
out not to be completely precise. Happily, Blau and Thompson have
evaluated the ratio of determinants \eqref{1lpdet} much more
carefully. The result they obtain using a heat-kernel regularization
scheme is \cite{Blau:1993tv}
\begin{equation}\label{heatreg}
\frac{\text{det}_{S} (D_0)}{\sqrt{\text{det}_{V}(D_0)}}\ =  \prod_{\alpha \in \Delta_+}\exp\left( \frac{1}{8\pi} \int_{\Sigma_g} R \log (M_\alpha M_{-\alpha}) - \frac{1}{2\pi} \int_\Sigma \alpha_i F_i \log \frac{M_\alpha}{M_{-\alpha}}\right)\ ,
\end{equation}
\footnote{There is an additional minus sign in the term with the field
  strengths compared to the expression eq.~(6.22) in
  \cite{Blau:1993tv} which can be traced to our convention that the
  integrand of the path integral is $\te^{-S_{\rm CS}}$ as opposed to
  $\te^{S_{\rm CS}}$ in \cite{Blau:1993tv}.} Here $\alpha$ runs over
the set of positive roots $\Delta_+$ of $SU(N)$, $R$ is the scalar
curvature of $\Sigma_g$, $F_i$ is the $i^{\rm th}$ component on the
diagonal of the gauge field strength along $\Sigma_g$ and
\begin{equation}
  M_{\alpha} = 1 - \te^{\i\alpha^i \phi_i}\ ,
\end{equation}
where $\phi = \{\phi_1,\ldots,\phi_N\}$ are the compact
diagonal elements of $A^\ut_0$ and $\alpha^i$ is the $i^{\rm th}$
component of the root $\alpha$. Let us write the above product
explicitly for $U(N)$. As discussed in Sections \ref{dteb},
\ref{unrep}, the positive roots for $SU(N)$ (and $U(N)$) are written
in terms of the orthonormal basis as
\begin{equation}
  \alpha_{ij} = \epsilon_i - \epsilon_j\ ,\quad 1 \leq i < j \leq N\ .
\end{equation}
This gives
\begin{equation}
  M_{\alpha_{(ij)}} = 1 - \te^{\i(\phi_i - \phi_j)}\ .
\end{equation}
Note that $\phi_i$ is a constant along $\Sigma_g$ due to integrating
out the fluctuating part of the abelian gauge field $A^\ut$ along
$\Sigma_g$. Using $\frac{1}{4\pi}\int_{\Sigma_g} d^2z R = 2 - 2g$, the
first term in the above exponential gives the usual Vandermonde factor
raised to $1-g$:
\begin{equation}
  \prod_{1 \leq i < j \leq N} \left[(1  - \te^{\i(\phi_i - \phi_j)})(1  - \te^{-\i(\phi_i - \phi_j)})\right]^{1-g} =   \prod_{1 \leq i < j \leq N} |\te^{\i\phi_i} - \te^{\i\phi_j}|^{2-2g}\ .
\end{equation}
Let us next focus on the second factor. We have
\begin{multline}
  \exp\left(-\sum_{\alpha \in \Delta_+} \frac{1}{2\pi}\int_{\Sigma_g} \alpha_i F_i
    \log (-\te^{\i\alpha_i \phi_i})\right) \\ =
  \exp\left(-\frac{1}{2\pi}\int_{\Sigma_g} 2\rho_i F_i \log (-1)\right)
  \exp\left(- \sum_{\alpha \in \Delta_+} \frac{\i}{2\pi}\int_{\Sigma_g} (\alpha_i
    F_i) (\alpha_j\phi_j)\right)\ ,
\end{multline}
where $\rho = \tfrac{1}{2}\sum_{\alpha \in \Delta_+} \alpha$ is the
Weyl vector. The integral of the field strength $F_i$ over $\Sigma$ is
$2\pi m_i$ where $m_i$ is an integer. Substituting this, we get
\begin{align} \label{phaexp}
  \exp\left(- 2\rho_i m_i \log (-1)\right)
  \exp\left(-\i \sum_{\alpha \in \Delta_+}(\alpha_i m_i)(\alpha_j\phi_j)\right)\ .
\end{align}
The expression in \eqref{phaexp} is a product of two terms. Let us
first study the second of these, namely
$$\exp\left(-\i \sum_{\alpha \in \Delta_+}(\alpha_i m_i)(\alpha_j\phi_j)\right).$$
Following \cite{Blau:1993tv} we now explain that this term effectively
implements the shift $k \rightarrow \kappa$.\footnote{This shift is
  expected because working with a heat kernel regulator gives the same
  result as regulating the Chern-Simons theory with an infinitesimal
  Yang-Mills term which provides the appropriate two-derivative
  operator required to regularize the determinants in the heat kernel
  regularization scheme.} Writing the sum over $\alpha \in \Delta_+$
explicitly, we have
\begin{align}
  &\exp\left(-\i \sum_{\alpha \in \Delta_+}(\alpha_i m_i)(\alpha_j\phi_j)\right)= 
    \exp \left(- \i \sum_{1 \leq i < j \leq N} (\alpha_{(ij)})_p m_p\, (\alpha_{(ij)})_u \phi_u\ \right)  ,\nonumber\\
  &\quad= \exp\left(- \i \sum_{1 \leq i < j \leq N}(m_i - m_j) (\phi_i - \phi_j) \right)  = \exp \left(- \frac{\i }{2}\sum_{i,j} (m_i - m_j) (\phi_i - \phi_j)\ \right)  ,\nonumber\\
  &\quad\ = \exp \left(- \i N \sum_j  \phi_j m_j + \i M \Phi\ \right) ,
\end{align}
where we have defined $M = \sum_i m_i$ and $\Phi =
\sum_i\phi_i$. Recall that the Chern-Simons action
$\tr(A^\ut_0 d A^\ut)$ gives $\exp(-k \phi_i m_i - (s+1) \Phi M)$ in
$\te^{-S_{\rm CS}}$ where $s$ is the integer that appeared in the
definition $k' = \kappa + s N = k + (s+1)N$. Notice that this contribution
is a pure phase. Adding the contribution of the classical action and
the one-loop determinant, we obtain the one-loop corrected
contribution to the path integral
\begin{equation}
  \exp \left(- \i (k+N) \phi_i m_i - \i s \Phi M \right) \ ,
\end{equation}
which anyway agrees with the contribution from the dimensionally
regularized classical action that we have used in Section
\ref{UNpath}.

Finally, we turn to the most tricky part of this ratio of determinants
i.e.~the term
$$\exp\left(- 2\rho_i m_i \log (-1)\right)\ .$$
This term is easy to simplify:
\begin{multline}\label{ms}
  \exp\left(-2\rho_i m_i \log(-1)\right) = \exp\left( -\sum_i(N-i+1)m_i \log (-1)\right) \\= (-1)^{(N-1)(m_1 + \cdots + m_N)} = (-1)^{(N-1)M}\ .
\end{multline}
Note that this term is also a phase; depending on the details of
fluxes and $N$, it is always $\pm 1$. In their original analysis
\cite{Blau:1993tv}, Blau and Thompson noted the possible presence of
such a term (see under equation 6.17 of that paper) but then also
correctly noted that this term reduces to unity when the gauge group
in question is semisimple (e.g. $SU(N)$). However, in the case of the
$U(N)$ theory (which Blau and Thompson did not consider) this term
makes an important contribution to the path integral -- one that is
crucial for obtaining agreement with the Verlinde formula or with the
predictions of level-rank duality.

Let us summarize. The modulus of the carefully evaluated determinant
\eqref{1lpdet} agrees with our simple minded guess
\eqref{smg}. However, the phase of this ratio is more subtle. One
piece of the phase implements the expected renormalization of the
Chern-Simons level and this is an effect that was easy to guess and
that we had already taken into account by using the level appropriate
for the dimensional regularization scheme. The second part of the
phase is the tricky minus sign \eqref{ms}. Accounting for this phase
gives the expression \eqref{whatwant}.

Thus, the integrand in the path integral is
\begin{equation}
  (-1)^{(N-1)M} \prod_{1\leq i\leq j < N} |w_i - w_j|^{2-2g}
  \prod_{i=1}^N w_i^{-\kappa m_i - s M}\ ,
\end{equation}
where we have defined $w_i = \te^{\i\phi_i}$, $i=1,\ldots,N$. Thus, we
have obtained the integrand that is written in \eqref{expse} and
\eqref{oneloopsign} in Section \ref{UNpath}.

\section{Transformation of the Verlinde formula measure under
   duality}\label{vander}
Let us start with the conjectured equality
\begin{equation}\label{conjdual}
  \frac{1}{\kappa^N}  \prod_{1 \leq i < j \leq N} |z_i - z_j|^2\ \stackrel{?}{=}\ \frac{1}{\kappa^k}  \prod_{1 \leq I < J \leq k} |z'_I - z'_J|^2\ ,
\end{equation}
where $z_i$, $i =1,\ldots,N$ are $N$ distinct $\kappa$-th roots of
unity and $z'_I$, $I=1,\ldots,k$ are the remaining $k$ $\kappa$-th
roots of unity. In Section \ref{typeIdual}, the quantities $z_i$ and
$z'_I$ are $\kappa^{\rm th}$ roots of $(-1)^{N-1}$ and $(-1)^{k-1}$
respectively, rather than of unity. However, since these are given by
$\kappa^{\rm th}$ roots of unity rotated by a common phase which is
unimportant when we take the absolute value as in \eqref{conjdual}, we
will prove \eqref{conjdual} for the case when $z_i$ and $z'_I$ are
$\kappa^{\rm th}$ roots of unity.

The expressions in \eqref{conjdual} without the prefactors
$\kappa^{-N}$ and $\kappa^{-k}$ can be written as the mod-squared of
the determinant of an $N \times N$ Vandermonde matrix and a
$k \times k$ Vandermonde matrix respectively:
\begin{equation}\label{NbyN}
  A = \begin{pmatrix} 1 & z_1 & z_1^2 & \cdots & z^{N-1}_{1} \\ 1 & z_2 & z_2^2 & \cdots & z^{N-1}_2 \\ \vdots & \vdots & \vdots & \cdots & \vdots \\ 1 & z_{N} & z_N^2 & \cdots & z_N^{N-1} \end{pmatrix}\ ,\quad   D' = \begin{pmatrix} 1 & z'_1 & z'^2_1 & \cdots & z'^{k-1}_{1} \\ 1 & z'_2 & z'^2_2 & \cdots & z'^{k-1}_2 \\ \vdots & \vdots & \vdots & \cdots & \vdots \\ 1 & z'_{k} & z'^2_k & \cdots & z'^{k-1}_k \end{pmatrix}\ ,
\end{equation}
\begin{equation}
  |\text{det}\, A|^2 = \prod_{1 \leq i < j \leq N} |z_i - z_j|^2\ ,\quad |\text{det}\, D'|^2 = \prod_{1 \leq I < J \leq N} |z'_I - z'_J|^2\ ,
\end{equation}
which allows us to rewrite the conjecture above as
\begin{equation}\label{conjequality}
  \frac{1}{\kappa^N} |\det A|^2\ \stackrel{?}{=}\ \frac{1}{\kappa^k} |\det D'|^2\ .
\end{equation}
We now look at a bigger Vandermonde matrix of which $A$ and $D'$
appear as $N \times N$ and $k \times k$ blocks. Consider the
$\kappa \times \kappa$ matrix
\begin{equation}\label{Vdef}
  V = \begin{pmatrix} 1 & z_1 & z_1^2 & \cdots & z_1^{\kappa-1} \\ 1 & z_2 & z_2^2 & \cdots & z_2^{\kappa-1} \\ \vdots & \vdots & \vdots & \cdots & \vdots \\ 1 & z_{\kappa} & z_\kappa^2 & \cdots & z_\kappa^{\kappa-1} \end{pmatrix}
\end{equation}
where $z_i$, $i = 1,\ldots,\kappa$, are the $\kappa$ distinct
$\kappa$-th roots of unity ordered such that the first $N$ roots
coincide with the $z_i$, $i=1,\ldots,N$, that appear in the matrix $A$
\eqref{NbyN}, and the remaining $k$ $\kappa$-th roots of unity are the
$z'_I$, $I=1,\ldots,k$. The mod-squared of the determinant of $V$ can
be calculated to be\footnote{To see this, observe that
  $V^\dag V = \kappa \mathbb{1}_{\kappa\times \kappa}$ which trivially
  gives $|\det V|^2 = \kappa^\kappa$. The determinant of $V$ itself
  can also be calculated quite easily though we do not need it
  here. We only have to determine the phase of $\det V$ since we have
  already calculated $|\det V|$. This is found to be
  $\exp\left(\i\pi\frac{(3\kappa - 2)(\kappa -1)}{4}\right)$.}
\begin{equation}
  |\det V|^2 = \kappa^\kappa\ .
\end{equation}
We then partition $V$ into blocks as follows:
\begin{equation}
  V = \begin{pmatrix} A_{N\times N} & B_{N \times k} \\ C_{k \times N} & D_{k \times k}\end{pmatrix}\ ,
\end{equation}
where we have indicated the sizes of the block matrices $A$, $B$, $C$
and $D$ in their subscripts; the entries of the various blocks can be
read off from definition of $V$ \eqref{Vdef}. Note that the block $A$
is the same matrix in \eqref{NbyN}; the block $D$ is related to the
matrix $D'$ in \eqref{NbyN} by an overall phase
\begin{equation}\label{DandDpr}
  D = (z_{N+1} z_{N+2}\cdots z_{\kappa})^N D'\ .
\end{equation}
Next, we write the formula for the determinant of a block matrix in
terms of its block entries:
\begin{equation}
  \det V = \det A \det (D - C A^{-1} B) \quad \Longrightarrow\quad \kappa^\kappa = |\det V|^2 = |\det A|^2\ |\det (D-C A^{-1} B)|^2\ .
\end{equation}
Let us simplify $|\det (D- C A^{-1} B)|^2$ further. We have
\begin{equation}\label{interdet}
  |\det (D- C A^{-1} B)|^2 = \det \big( (D^\dag - B^\dag  A^{-1}{}^\dag C^\dag)(D- C A^{-1} B)\big)\ .
\end{equation}
We need some identities on the block matrices that follow from
$V^\dag V = V V^\dag = \kappa \mathbb{1}_{\kappa \times \kappa}$:
\begin{align}\label{identities}
  &A^\dag A + C^\dag C = \kappa\mathbb{1}_{N \times N}\ ,\quad A^\dag B + C^\dag D = 0\ ,\quad B^\dag A + D^\dag C  = 0\ ,\quad B^\dag B + D^\dag D = \kappa \mathbb{1}_{k \times k}\ ,\nonumber\\
  &A A^\dag + B B^\dag = \kappa \mathbb{1}_{N \times N}\ ,\quad A C^\dag  + B D^\dag = 0\ ,\quad C A^\dag + D B^\dag = 0\ ,\quad C C^\dag + D D^\dag = \kappa \mathbb{1}_{k \times k}\ .
\end{align}
The $k \times k$ matrix on the right hand side of \eqref{interdet} can
be simplified as follows.
\begin{align}
  &(D^\dag - B^\dag  A^{-1}{}^\dag C^\dag)(D- C A^{-1} B)\ ,\nonumber\\
  &= D^\dag D - D^\dag C A^{-1} B - B^\dag A^{-1}{}^\dag C^\dag D + B^\dag A^{-1}{}^\dag C^\dag C A^{-1} B\ ,\nonumber\\
  &= D^\dag D + B^\dag B + B^\dag B + B^\dag A^{-1}{}^\dag (\kappa \mathbb{1}_{N \times N} - A^\dag A) A^{-1}B\ ,\nonumber\\
  &= D^\dag D + B^\dag B +\kappa B^\dag A^{-1}{}^\dag A^{-1}B\ ,\nonumber\\
  &= \kappa(\mathbb{1}_{k \times k} + B^\dag A^{-1}{}^\dag A^{-1}B)\ .
\end{align}
Next, using the identities \eqref{identities} we rewrite
$C C^\dag + D D^\dag = \kappa \mathbb{1}_{k \times k}$ as
\begin{equation}
  D (\mathbb{1}_{k \times k} + B^\dag A^{-1}{}^\dag A^{-1} B ) D^\dag = \kappa \mathbb{1}_{k \times k}\ ,
\end{equation}
which gives
\begin{equation}
  \det (\mathbb{1}_{k \times k} + B^\dag A^{-1}{}^\dag A^{-1} B) |\det D|^2 = \kappa^k\ .
\end{equation}
Using the various results derived above, we have
\begin{equation}
  \kappa^\kappa = |\det A|^2 \frac{\kappa^{2k}}{|\det D|^2}\ .
\end{equation}
Rearranging the terms above, using $\kappa = N + k$ and the relation between $D$ and $D'$ in \eqref{DandDpr}, we get
\begin{equation}
  \frac{1}{\kappa^N} |\det A|^2 = \frac{1}{\kappa^k} |\det D'|^2\ ,
\end{equation}
which is the conjectured identity \eqref{conjequality}.

\section{More about level-rank duality and the character formula}

\subsection{Map of characters under level rank duality for general representations in Type I theory at even $\kappa$} \label{levelranktoev}

In this subsection we will attempt to directly establish the identity
\eqref{charequivngen}.  In other words, given any choice of the
collection of eigenvalues $U = \diag\{w_i \}$, $i=1,\ldots,N$ (and so
a corresponding collection of dual eigenvalues
$\tl{U} = \diag\{z_I\}$, $I=1,\ldots,k$) we will attempt demonstrate
\begin{equation}\label{charident}
  \chi_{\vgap R}(U)=\chi_{\vgap R^T}(\tl{U})=\chi^*_{\vgap R^T}(\tl{U}^\dag)\ .
\end{equation} 
(the last equality is a triviality; the nontrivial part is to show the
first equality). Our attempt will be only partially successful.

\subsubsection{Representations with less than $\kappa$ boxes}

We will now demonstrate \eqref{charident} when $R$ is a representation
labelled by a Young tableau with the length of all rows lying between
\begin{equation}\label{lengthofrows} 
0 \leq \ell_i \leq k\ ,\quad\text{and}\quad \sum_i \ell_i < \kappa\ .
\end{equation} 
(i.e. the tableau has all row sizes positive and no larger than
$k$). $R^T$ is the transpose of the tableau $R$, and so has no more
than $k$ rows, each of which has a positive length that is no larger
than $N$.

The proof of \eqref{charident} follows immediately upon using the
$U(N)$ character formula for representations built out of tensor
products of fundamentals (and no antifundamentals)
\begin{equation}\label{chirchidag}
  \chi_{\vgap R}(U)= \frac{1}{n!} \sum_{g} \chi_{\vgap R}(g) \prod_m (\tr U^m)^{n^g_m}\ ,
\end{equation} 
where $R$ is any representation of $SU(N)$ that also carries with
$U(1)$ charge $n$, $n$ is the number of boxes in the Young tableau
corresponding to representation $R$, $g$ is a general element of
$\mc{S}_n$, the permutation group on $n$ objects, and $n_m^g$ is the
number of cycles of length $m$ in the group element $g$.

In the context at hand, we have
\begin{equation}\label{contath}
\tr U^m = \sum_{i=1}^N w_i^m\ ,
\end{equation} 
so that \eqref{chirchidag} becomes
\begin{equation}\label{chirchidagsp}
  \chi_{\vgap R}(U) = \frac{1}{n!} \sum_{g} \chi_{\vgap R}(g) \prod_m \left(\sum_i w_i^m \right)^{n^g_m}\ .
\end{equation} 
Now consider \eqref{chirchidag} applied to the level-rank dual theory
in the representation $R^T$, i.e. in the representation whose Young
tableau is the transpose of the original tableau. We have the identity
\begin{equation}\label{contath1}
  \tr\, U^m = \sum_i w_i^m = -\sum_I (w_I')^m = (-1)^{m+1}\left( \sum_i z_I^m \right) = (-1)^{m+1}\tr\, \tl{U}^m\ ,
\end{equation} 
where $w'_I$, $I=1,\ldots,k$, are the $k$ $\kappa^{\rm th}$ roots of
unity not in the set $\{w_i\}$, and $z_I = -w'_I$, and in the last
step we have used the identity which holds provided $m$ is not a
multiple of $\kappa$
\begin{equation}\label{identitynew}
  \sum_{\alpha=1}^\kappa q_\alpha^m = 0\ ,
\end{equation} 
where $q_\alpha$ runs over all $\kappa^{\rm th}$ roots of unity (when
$m$ is a multiple of $\kappa$ the RHS of \eqref{identitynew} is
$\kappa$ instead of zero).

\subsection{Total number of boxes in tableau less than $\kappa$}

Suppose $R$ is a representation with less than $\kappa$ boxes in its
Young tableau. In this case since the number of boxes $n < \kappa$ we
have cycle-lengths $m< \kappa$ and so \eqref{contath} holds.

It follows that in the level-rank dual theory
\begin{equation}\label{chirchidagspn}
  \chi_{\vgap R^T}(\tl{U}) = \frac{1}{n!} \sum_{g} \chi_{\vgap R^T}(g) \prod_m \left( (-1)^{m+1}\left(\sum_I z_I^m \right) \right) ^{n^g_m}\ .
\end{equation}
Next, using the identities
\begin{equation}\label{charide}
  \chi_{\vgap R^T}(g)= {\rm sgn}(g) \chi_{\vgap R}(g)\ ,
\end{equation} 
(where ${\rm sgn}(g)$ is the sign of the permutation $g$) and also 
\begin{equation}\label{charide1}
  {\rm sgn}(g) =\prod_m \left( (-1)^{m+1} \right)^{n_m^g}\ ,
\end{equation} 
it follows that the characters \eqref{chirchidagspn} equals that in
\eqref{chirchidagsp}, as we set out to show.

The result of this section applies, for example, to integrable $U(N)$
representations with boxes only in the first row of the
tableau. Recall that the number of boxes in a row is $\leq k$, and so
$< \kappa$ and the analysis of this subsubsection applies.  We know
that
\begin{align}\label{charactero}
  &\chi_{\vgap m-{\rm sym}}(U)=\chi_{\vgap { m_\kappa-{\rm asym}}}(\tl{U})\ ,\quad\text{for}\quad m\ \text{mod}\ \kappa \leq k\ ,\nonumber\\
  &\chi_{\vgap m-{\rm sym}}(U)=0\ ,\quad\text{for}\quad m\ \text{mod}\ \kappa > k\ .
\end{align}
Assuming for simplicity that $N$ and $k$ are both odd, it follows
immediately from this that for any choice of the collection $\{w_i\}$
\begin{equation}\label{symas}
  \left(1-x^\kappa\right) \prod_{i=1}^N \frac{1}{1-x w_i} 
  =    \prod_{I=1}^k (1+x z_I)\ .
\end{equation} 
This identity can also be understood as follows.  The LHS of
\eqref{symas} is
\begin{align}
  &(1-x^\kappa) \exp\left(-\sum_{i=1}^N\log (1-xw_i)\right)\ ,\nonumber\\
  &\quad=(1-x^\kappa)\exp\left(\sum_n \frac{x^n\sum_i w_i^n}{n}\right)\ ,\nonumber\\
  &\quad\quad = (1-x^\kappa)\exp\left(\sum_n (-1)^{n+1}\frac{x^n \sum_I z_I^n}{n} 
    + \sum_{p} \frac{x^{p\kappa}}{p}\right)\ ,\nonumber\\
  &\quad\quad\quad = (1-x^\kappa)\exp\left(\sum_n (-1)^{n+1}\frac{x^n \sum_I z_I^n}{n} 
    -\log(1-x^\kappa)\right)\ ,\nonumber\\
  &\quad\quad\quad\quad=\exp\left( \sum_n (-1)^{n+1}\frac{\sum_I z_I^n}{n} \right)\ ,\nonumber\\
  &\quad\quad\quad\quad\quad=\exp\left(\sum_{I=1}^k \log (1 + xz_I)\right)\ .
\end{align}
The last term in this manipulation is the RHS of \eqref{symas}.

\subsubsection{Comments on representations with more than $\kappa$ boxes} 

Once again in this section we assume, for simplicity, that $N$ and $k$
are both odd. When $R$ is a representation with $n>\kappa$ boxes in
its tableau, the sum over cycles of $\mc{S}_n$ in \eqref{chirchidagsp}
which have length $m = p \kappa$ for some nonzero integer values of
$p$. For such values of $m$ the analog of \eqref{contath} is
\begin{equation}\label{contathm}
\tr U^{p\kappa} = N= \kappa-k = \kappa - \tr  {\tilde U}^{p\kappa} \ .
\end{equation} 
For the rest of this discussion we will find it convenient to insert a
formal counting parameter $\epsilon$ into \eqref{contathm}
\begin{equation}\label{contathmep}
  \tr U^{p\kappa} = \epsilon \kappa - \tr  {\tilde U}^{p\kappa} \ .
\end{equation} 
$\epsilon$ is a formal counting parameter that will eventually be set
to unity. If we insert \eqref{contathmep} into \eqref{chirchidagsp}
and repeat the analysis that led to \eqref{chirchidagspn}, we find
\begin{equation}\label{transfchar}
\chi_{\vgap R}(z_i)= \sum_{p=0}^L \epsilon^p a_p(w_i)\ .
\end{equation} 
A little thought will convince the reader
that \begin{equation}\label{agon}
  a_0(w_i)= \chi_{\vgap R^T}(w_i)\ ,
\end{equation} 
as in the previous subsubsection. On the other hand $a_p(w)$ is a
linear sum of products of characters of representations such that the
total number of boxes, $n_1+n_2 + n_3$ in every term equals
$n-p\kappa$ boxes.

In order to establish \eqref{charident} for general representations we
must demonstrate that whenever the representation $R$ is integrable,
the sum over $a_i(w)$ for $i \geq 1$ must vanish. Given the
alternative argument for \eqref{charident} (see Section
\ref{typeonedualmap}), this result is presumably true, but we have not
been able to directly establish it. We leave the filling of his lacuna
to future work.


\section{Conventions on $S^2\times S^1$ and $\mc{N}=2$ supersymmetry
  transformations}\label{susyS2conv}
For convenience, we summarize the relevant conventions and
supersymmetry transformations in this appendix.  See
\cite{Benini:2013yva} for further details.
\begin{align}
\textrm{Coordinates}\ & : \quad x^\mu = (\theta, \phi, \tau)\ ,\nonumber \\
\textrm{Metric}\ & :\quad  ds^2  = g^{\mu\nu}dx^\mu dx^\nu = R^2 \left( d\theta^2 + \sin^2\theta d\phi^2 \right) + d\tau^2\ .
\end{align}
The radius of $S^2$ is $R$ while the perimeter of $S^1$ is
$\beta R$. With this choice of coordinates, we have
\begin{equation}
\dot x^\mu = \partial_\tau x^\mu = (0,0,1) \quad, \quad |\dot x| = 1\ .
\end{equation}
The $\gamma$-matrices, vielbeins and the spin connection 1-forms are
\begin{align}
& \gamma^a = \sigma^a\ ,\quad e^a = e^a_\mu dx^\mu = (R d\theta, R \sin\theta d\phi, d\tau)\ ,\quad \omega^{12}  = -\omega^{21} = -\cos\theta d\phi \ .
\end{align}
The supersymmetry transformation of the fields inside the ${\cal N}=2$
vector multiplet are
\begin{align}
\delta A_\mu & =  -\frac{\i}{2} \left( \bar\epsilon \gamma_\mu \lambda - \bar\lambda \gamma_\mu \epsilon \right)\ ,\nonumber  \\
\delta \lambda & =  \frac{1}{2} \gamma^{\mu\nu} \epsilon F_{\mu\nu} - D\epsilon 
+ \i\gamma^\mu \epsilon D_\mu\sigma  + \frac{2\i}{3} \sigma \gamma^\mu D_\mu \epsilon\ ,\nonumber \\
\delta \sigma & =  \frac{1}{2} (\bar\epsilon \lambda - \bar\lambda \epsilon)\ ,\nonumber \\
\delta D & = - \frac{\i}{2} \bar\epsilon\gamma^\mu D_\mu \lambda - \frac{\i}{2} D_\mu \bar\lambda \gamma^\mu \epsilon 
+ \frac{\i}{2}[\bar\epsilon \lambda, \sigma] + \frac{\i}{2}[\bar\lambda \epsilon, \sigma]  
- \frac{\i}{6} (D_\mu \bar\epsilon \gamma^\mu \lambda + \bar\lambda \gamma^\mu D_\mu \epsilon )\ .
\end{align}

\section{Some explicit results for the index in the presence of Wilson
  loops}\label{susywlresults}
In this appendix we collect results of some explicit Mathematica
computations of the superconformal index for $SU(2)_\kappa$ and
$U(2)_{\kappa,k}$ theory with integrable Wilson loop insertions. For
brevity and simplicity we will only quote results for Wilson loop
insertions at the north pole of $S^2$.

\subsection{$SU(2)_\kappa$ index}
The integrable representations are labelled by $su(2)$ spin $j$ in the
set $(0,\frac{1}{2},1,\frac{3}{2},\ldots, \frac{\kappa-2}{2})$,
assuming $\kappa$ positive.
\begin{align}
\langle W_{j} \rangle &= \delta_{j,0}\ ,\nonumber \\
\langle W_{j_1} W_{j_2} \rangle  &= \delta_{j_1,j_2}\ ,\nonumber \\ 
\langle W_{j_1} W_{j_2} W_{j_3} \rangle  & =
      1 \quad ~\textrm{when $j_1,j_2,j_3$ satisfy triangle rule and } j_1+j_2+j_3<\kappa-1\ ,\nonumber  \\
&= 1- \frac{1}{x^{2(\kappa-1)}}  \quad \textrm{when }  j_1+j_2+j_3 = \kappa-1\ ,\nonumber \\
                      &= 1 + \frac{1}{x^{2(\kappa +1)}} - \frac{2}{x^{2(\kappa-1)}}  \quad \textrm{when }  j_1+j_2+j_3 = \kappa\ ,\quad\text{and so on.}
\end{align}

\subsection{$U(2)_{\kappa,k}$ index}
We will consider here the integrable Wilson loops of type $W_{j,q}$,
labelled by $su(2)$ spin $j$ in the set
$(0,\frac{1}{2},1,\frac{3}{2},\ldots \frac{\kappa-2}{2})$ and the
$U(1)$ charge $q \in {\mathbb Z}$, with $q=\pm 2j$.
\begin{align}
  \langle W_{j,2j} \rangle &= \langle W_{j,-2j} \rangle = \delta_{j,0}\ ,\nonumber  \\ 
  \langle W_{j_1,2j_1} W_{j_2,-2j_2} \rangle  &= \delta_{j_1,j_2}\ ,\nonumber \\ 
  \langle W_{j_1,2j_1} W_{j_2,2j_2} \rangle  &= (-1)^\kappa \left( \frac{1}{x^{\kappa+1}} - \frac{1}{x^{\kappa-1}} \right)  \delta_{j_1+j_2,\frac{\kappa}{2}}\ .
\end{align}


\bibliography{ref}\bibliographystyle{JHEP}

\end{document}